\documentclass[12pt]{article}
\usepackage[a4paper, margin=0.75in]{geometry}
\usepackage[utf8]{inputenc}         
\usepackage[english]{babel}
\usepackage{lmodern}
\usepackage{graphicx}

\usepackage[colorinlistoftodos]{todonotes}

\usepackage[colorlinks=true]{hyperref}

\usepackage[dvipsnames]{xcolor}

\usepackage[]{color-edits}
\addauthor{UF}{blue}

\addauthor{VG}{red}
\newcommand{\vgc}[1]{{\VGcomment{#1}}}

\usepackage{amsthm}
\usepackage{amsmath}
\usepackage{amssymb}
\usepackage{amsfonts}
\usepackage{mathrsfs}
\usepackage{mathtools}
\usepackage{verbatim}
\usepackage{footnote}
\usepackage{lineno}
\usepackage{caption}
\usepackage{tikzsymbols}
\usepackage[capitalize, nameinlink]{cleveref}
\usepackage{tikz}
\usetikzlibrary{quantikz2}
\usepackage{nicematrix}
\usepackage{bm}

\usepackage{icomma}
\usepackage{enumitem}
\usepackage{array}
\usepackage{multirow}
\usepackage{setspace}
\usepackage{euscript}
\usepackage{indentfirst}
\usepackage{epigraph}
\usepackage{fancybox, fancyhdr}
\usepackage{titlesec}
\usepackage{dsfont}
\usepackage{csquotes}

\usepackage{algorithm}              
\usepackage{algorithmicx}           
\usepackage[noend]{algpseudocode}   
\usepackage{listings}

\floatname{algorithm}{Algorithm}

\crefname{starredthm}{\textbf{Theorem}}{}
\Crefname{starredthm}{\textbf{Theorem}}{}

\newtheorem{definition}{Definition}[section]

\newtheorem{assumption}{Assumption}
\newtheorem{theorem}{Theorem}[section]

\newtheorem{corollary}{Corollary}[section]
\newtheorem{lemma}{Lemma}[section]
\newtheorem{observation}{Observation}[section]

\newtheorem{remark}{Remark}[section]

\crefname{conjecture}{\textbf{Conjecture}}{}
\crefname{lemma}{\textbf{Lemma}}{}
\crefname{claim}{\textbf{Claim}}{}
\crefname{theorem}{\textbf{Theorem}}{}
\crefname{corollary}{\textbf{Corollary}}{}
\crefname{observation}{\textbf{Observation}}{}
\crefname{proposition}{\textbf{Proposition}}{}
\crefname{assumption}{\textbf{Assumption}}{}

\newcommand{\Z}{\mathbb{Z}}
\newcommand{\Qq}{\mathcal{Q}}

\newcommand{\R}{\mathbb{R}}
\newcommand{\B}{\mathbb{B}}

\usepackage{xparse}

\DeclarePairedDelimiterX\brackets[1]{(}{)}{
	
	#1
}
\DeclarePairedDelimiterX\squarebrackets[1]{[}{]}{
	
	#1
}
\DeclarePairedDelimiterX\figbrackets[1]{\{}{\}}{
	
	#1
}

\DeclarePairedDelimiterX\ntrs[1]{\langle}{\rangle}{
	
	#1
}
\DeclarePairedDelimiterX\kets[1]{\lvert}{\rangle}{
	
	#1
}
\DeclarePairedDelimiterX\bras[1]{\langle}{\rvert}{
	
	#1
}
\DeclarePairedDelimiterX\brakets[2]{\langle}{\rangle}{
	
	#1
    \delimsize\vert
    
    #2
}

\DeclarePairedDelimiterX\dprs[2]{\langle}{\rangle}{
	
	#1
    ,
    
    #2
}

\DeclarePairedDelimiterX\ketbras[2]{\lvert}{\rvert}{
	
	#1
    \delimsize\rangle\delimsize\langle
    
    #2
}
\DeclarePairedDelimiterX\qvals[3]{\langle}{\rangle}{
	
	#1
    \delimsize\vert
    #2
    \delimsize\vert
    
    #3
}
\DeclarePairedDelimiterX\qexps[2]{\langle}{\rangle}{
	
	#1
    \delimsize\vert
    #2
    \delimsize\vert
    
    #1
}

\let\P\undefined
\let\o\undefined
\let\O\undefined

\NewDocumentCommand{\P}{o g}{
	\IfNoValueTF{#1}{
		\IfNoValueTF{#2}{
			\mathop{\mathds{P}}
		}{
			\mathop{\mathds{P}}\squarebrackets*{#2}
		}
	}{
		\IfNoValueTF{#2}{
			\mathop{\mathds{P}}_{#1}
		}{
			\mathop{\mathds{P}}_{#1}\squarebrackets*{#2}
		}
	}
}

\NewDocumentCommand{\E}{o g}{
	\IfNoValueTF{#1}{
		\IfNoValueTF{#2}{
			\mathop{\mathds{E}}
		}{
			\mathop{\mathds{E}}\squarebrackets*{#2}
		}
	}{
		\IfNoValueTF{#2}{
			\mathop{\mathds{E}}_{#1}
		}{
			\mathop{\mathds{E}}_{#1}\squarebrackets*{#2}
		}
	}
}

\NewDocumentCommand{\D}{o g}{
	\IfNoValueTF{#1}{
		\IfNoValueTF{#2}{
			\mathop{\mathds{D}}
		}{
			\mathop{\mathds{D}}\squarebrackets*{#2}
		}
	}{
		\IfNoValueTF{#2}{
			\mathop{\mathds{D}}_{#1}
		}{
			\mathop{\mathds{D}}_{#1}\squarebrackets*{#2}
		}
	}
}

\NewDocumentCommand{\cov}{o g g}{
	\IfNoValueTF{#1}{
		\IfNoValueTF{#2}{
			\mathop{\mathrm{cov}}
		}{
			\IfNoValueTF{#3}{
				\mathop{\mathrm{cov}}
			}{
				\mathop{\mathrm{cov}}\brackets*{#2, #3}
			}
		}
	}{
		\IfNoValueTF{#2}{
			\mathop{\mathrm{cov}}_{#1}
		}{
			\IfNoValueTF{#3}{
				\mathop{\mathrm{cov}}_{#1}
			}{
				\mathop{\mathrm{cov}}_{#1}\brackets*{#2, #3}
			}
		}
	}
}

\NewDocumentCommand{\I}{g}{
	\IfNoValueTF{#1}{
		\mathop{\mathds{I}}
	}{
		\mathop{\mathds{I}}\figbrackets*{#1}
	}
}

\NewDocumentCommand{\V}{o g}{
	\IfNoValueTF{#1}{
		\IfNoValueTF{#2}{
			\mathop{\mathds{V}}
		}{
			\mathop{\mathds{V}}\squarebrackets*{#2}
		}
	}{
		\IfNoValueTF{#2}{
			\mathop{\mathds{V}}_{#1}
		}{
			\mathop{\mathds{V}}_{#1}\squarebrackets*{#2}
		}
	}
}

\NewDocumentCommand{\Tr}{o g}{
	\IfNoValueTF{#1}{
		\IfNoValueTF{#2}{
			\mathop{\mathrm{Tr}}
		}{
			\mathop{\mathrm{Tr}}\squarebrackets*{#2}
		}
	}{
		\IfNoValueTF{#2}{
			\mathop{\mathrm{Tr}}_{#1}
		}{
			\mathop{\mathrm{Tr}}_{#1}\squarebrackets*{#2}
		}
	}
}

\NewDocumentCommand{\o}{g}{\operatorname{\mathrm{o}}\IfNoValueTF{#1}{}{\brackets*{#1}}}
\NewDocumentCommand{\O}{g}{\operatorname{\mathrm{O}}\IfNoValueTF{#1}{}{\brackets*{#1}}}
\NewDocumentCommand{\Th}{g}{\operatorname{\Theta}\IfNoValueTF{#1}{}{\brackets*{#1}}}
\NewDocumentCommand{\om}{g}{\operatorname{\omega}\IfNoValueTF{#1}{}{\brackets*{#1}}}
\NewDocumentCommand{\Om}{g}{\operatorname{\Omega}\IfNoValueTF{#1}{}{\brackets*{#1}}}

\NewDocumentCommand{\ntr}{g}{\IfNoValueTF{#1}{}{\ntrs*{#1}}}
\let\ket\undefined
\NewDocumentCommand{\ket}{g}{\IfNoValueTF{#1}{}{\kets*{#1}}}
\let\bra\undefined
\NewDocumentCommand{\bra}{g}{\IfNoValueTF{#1}{}{\bras*{#1}}}
\NewDocumentCommand{\dpr}{g g}{\IfNoValueTF{#1}{}{\IfNoValueTF{#2}{}{\dprs*{#1}{#2}}}}
\let\braket\undefined
\NewDocumentCommand{\braket}{g g}{\IfNoValueTF{#1}{}{\IfNoValueTF{#2}{}{\brakets*{#1}{#2}}}}
\NewDocumentCommand{\ketbra}{g g}{\IfNoValueTF{#1}{}{\IfNoValueTF{#2}{}{\ketbras*{#1}{#2}}}}
\NewDocumentCommand{\qval}{g g g}{\IfNoValueTF{#1}{}{\IfNoValueTF{#2}{}{\IfNoValueTF{#3}{}{\qvals*{#1}{#2}{#3}}}}}
\NewDocumentCommand{\qexp}{g g}{\IfNoValueTF{#1}{}{\IfNoValueTF{#2}{}{\qexps*{#1}{#2}}}}

\newcommand{\minimize}{\mathop{\mathrm{minimize}}}
\newcommand{\subto}{\mathop{\mathrm{subject\ to}}}
\newcommand{\maximize}{\mathop{\mathrm{maximize}}}

\newcommand{\argmin}{\mathop{\mathrm{argmin}}}

\newcommand{\mms}{\mathrm{MMS}}

\newcommand{\wx}{\tilde{x}}

\newcommand{\wz}{\tilde{z}}

\newcommand{\fe}{\varphi}
\newcommand{\eps}{\varepsilon}

\newcommand{\One}{\mathds{1}}

\newcommand{\hB}{\hat{B}}
\newcommand{\hJ}{\hat{J}}
\newcommand{\hZ}{\hat{Z}}

\newcommand{\wM}{\widetilde{M}}

\newcommand{\Ll}{\mathcal{L}}

\newcommand{\Bb}{\mathcal{B}}

\newcommand{\Pp}{\mathcal{P}}

\newcommand{\Cc}{\mathcal{C}}
\newcommand{\Xx}{\mathcal{X}}
\newcommand{\Mm}{\mathcal{M}}

\newcommand{\Rr}{\mathcal{R}}

\newcommand{\Ee}{\mathcal{E}}
\newcommand{\Hh}{\mathcal{H}}
\newcommand{\Ii}{\mathcal{I}}

\newcommand{\Nn}{\mathcal{N}}

\newcommand{\hM}{\hat{M}}
\newcommand{\hbeta}{\hat{\beta}}

\newcommand{\wB}{\widetilde{B}}
\newcommand{\hx}{\hat{x}}
\newcommand{\hy}{\hat{y}}
\newcommand{\hz}{\hat{z}}

\newcommand{\ov}{\overline{v}}

\newcommand{\wX}{\widetilde{X}}

\newcommand{\hv}{\hat{v}}

\newcommand{\Jj}{\mathcal{J}}

\newcommand{\hX}{\hat{X}}

\newcommand{\hXx}{\hat{\mathcal{X}}}

\newcommand{\hpi}{\hat{\pi}}

\newcommand{\wv}{\tilde{v}}

\title{Approximate MMS allocations for large number of agents}
\date{}
\author{Uriel Feige\thanks{Weizmann Institute of Science, Israel. {\tt uriel.feige@weizmann.ac.il}} \and Vadim Grinberg\thanks{Weizmann Institute of Science, Israel. {\tt vadim.grinberg@weizmann.ac.il}}}
\begin{document}

\maketitle

\begin{abstract}
We consider allocation of indivisible items to agents with non-negative additive valuations, under the maximin share (MMS) fairness benchmark. We prove that for every $\rho < 1$, there exists some $n_{\rho}$ such that every allocation instance with
$n \ge n_{\rho}$ agents has a $\rho$-MMS allocation. This was explicitly stated as a conjecture in a recent paper of the authors, where it was shown to hold if the number of items is at most $(2 - \epsilon)n$ (and $n \ge n_{\rho, \epsilon}$). We build on some techniques from that earlier paper, including the construction of \textit{ordinal MMS} allocations that give $n - O(\log^2 n)$ agents their full MMS. Among the new techniques that we introduce is the definition of \textit{block-matchings} between the MMS partitions of two agents, and a proof that if the number of item types is small, then block-matchings of small width necessarily exist.
\end{abstract}

 


\section{Introduction}

\begin{remark}
    Parts of the introduction are copied verbatim from our own recent paper~\cite{FG26}, because in the long run we expect the current paper to replace~\cite{FG26}. 
\end{remark}

We consider fair allocations of $m$ indivisible goods to $n$ agents.
Let $M$ denote the set of $m$ items to allocate, and let the set of agents be $N := [n]$.
Every agent $i \in \Nn$ has a \textit{valuation function} $v_i$ defined on the subsets of $M$.
We assume that $v_i$ satisfies $v_i(\varnothing) = 0$, and for every $S_1 \subseteq S_2$ it holds $v_i(S_1) \leq v_i(S_2)$, i.e $v_i$ is monotone and non-decreasing. 
In addition, we assume that every $v_i$ is \textbf{additive}: for every $S\subseteq M$ it holds $v_i(S) = \sum_{e \in S}v_i(e)$.

An \textit{allocation} of items $M$ to agents $N$ is a partition $(P_1, \ldots, P_n)$ of $M$ into $n$ disjoint sets, where agent $i$ ``receives'' the set $P_i$.
We would like this allocation to be \textit{fair}, for a certain definition of fairness.
In this work, we will focus on a share-based notion of fairness called the \textit{maximin share}~\cite{Budish11}, or MMS for short.
\begin{definition}
    Let $\Pp_n$ be the set of all partitions $(P_1, \ldots, P_n)$ of items $M$ into $n$ disjoint sets.
    For an agent $i$ with valuation $v_i$, her maximin share is
    \[
        \mms(M, v_i, n) = \max_{(P_1, \ldots, P_n)\in \Pp_n}\min_{j \in [n]}v_i(P_j).
    \]
    We will use $\mms(n)$ to denote the MMS-value with respect to partitions of $M$ into $n$ sets.
\end{definition}
A partition $(A_1, \ldots, A_n)$ of $M$ is called an \textit{MMS-allocation} if for every agent $i \in N$ it holds $v_i(A_i)\geq \mms(M, v_i, n)$.
That is, each $i$ values the bundle $A_i$ given to her at least as high as her MMS.
We will also consider approximate allocations: for a value $\rho \in (0, 1]$, we say that allocation $(A_1, \ldots, A_n)$ is an $\rho$-MMS allocation, if for every agent $i \in N$ it holds $v_i(A_i) \geq \rho\cdot \mms(M, v_i, n)$.
In this work, we will focus on the existence of exact / approximate MMS allocations for additive agents, in two closely related settings.

\subsection{Our results, put in context}

\label{sec:results}

There are known examples showing that with additive valuations and $n \ge 3$ agents, MMS allocations need not exist. The first such examples were given in~\cite{KPW18-1}. Later examples gave explicit upper bounds on $\rho$ for $\rho$-MMS allocations, with $\rho \le \frac{39}{40}$ when $n=3$~\cite{FST21}, and $\rho \le \frac{20}{21}$ when $n = 4$~\cite{EG26}. For general $n$, \cite{EG26} showed an upper bound of $1-\Omega(\frac{1}{\log^2 n})$. On the positive side, following a long line of work, it is known that $\frac{7}{9}$-MMS allocations exist, and moreover, can be found in polynomial time~\cite{huang2025}. Better ratios are known to hold for $n=3$ (a ratio of $\frac{11}{12}$~\cite{FeNo22}), and for $n=4$ (a ratio of $\frac{4}{5}$~\cite{GhodsiHSSY22}).

A question that was left open in previous work is whether $\rho_n$ tends to~1 as $n$ grows. This was stated as a conjecture in a recent paper of ours~\cite{FG26}. In that paper we proved that this is true, but in a setting with agent redundancy.

\begin{theorem}[\cite{FG26}]
 \label{thm:largen}
    For every $\eps > 0$ and $\rho < 1$, there exists some $n_{\eps,\rho}$ such that every $n \ge n_{\eps,\rho}$, in every allocation instance with additive valuations and $d \ge (1 + \eps)n$ agents, there is a subset of $n$ agents that have a $\rho$-$\mms(n)$ allocation.    
\end{theorem}

A straightforward corollary of \cref{thm:largen} confirmed the conjecture when the number of items is small. 

\begin{corollary}[\cite{FG26}]
    \label{cor:largen}
For every $\epsilon > 0$ and $\rho < 1$, there exists some $n_{\epsilon, \rho}$ such that every allocation instance with additive valuations, $n \ge n_{\epsilon, \rho}$ agents $m \le (2 - \epsilon)n$ items has a $\rho$-MMS allocation. 
\end{corollary}

In the current paper, we build upon techniques that we developed in~\cite{FG26}, and some new techniques, and confirm the conjecture without any restriction on $m$.
Our main result is the following theorem. 

\begin{theorem}
\label{thm:main}
    For every $\rho < 1$, there exists some $n_\rho$ such that every allocation instance with additive valuations and $n \ge n_\rho$ agents has has a $\rho$-MMS allocation. 
\end{theorem}




Part of the proof of \cref{thm:main} builds on our earlier result in \cite{FG26} for a related topic of \textit{$1$-out-of-$d$ MMS} allocations for $d > n$, also known as \textit{ordinal MMS}, first introduced in \cite{babaioff2021competitive} (though also considered earlier in \cite{Budish11} with $d = n + 1$).
A partition $(A_1, \ldots, A_n)$ of $M$ is called a \textit{$1$-out-of-$d$ MMS allocation}, if for every agent $i \in N$ it holds $v_i(A_i) \geq \mms(M, v_i, d)$.
In this setting, the main question is how large should $d$ be (as a function of $n$) to guarantee that exact $\mms(d)$ allocations exist.
In our previous work \cite{FG26}, we showed that the number of additional bundles required for such allocations to exist is at most polylogarithmic in $n$.
\begin{theorem}[\cite{FG26}]\label{thm:dmms}
    Any instance with $n$ agents admits an $\mms(d)$ allocation with $d = n + O(\log^2n)$.
\end{theorem}
We repeat the proof of \cref{thm:dmms} in \cref{sec:dmms}, as it is relevant for our \cref{thm:main}.



\subsection{Overview of our algorithm}
\label{sec:newOverview}

Recall that in~\cite{FG26} we proved \cref{thm:largen}.
We shall briefly review some parts of the proof of that theorem, explaining how agent redundancy was used in that proof, and explain the main ideas that allow us in the current work to circumvent the need for agent redundancy.

We start with rather straightforward preliminaries.
Let $q$ be a large integer. Our goal is to show that there is some $n_q$ such that if $n > n_q$, there are $(1 - O(\frac{1}{q}))$-MMS allocations (with or without agent redundancy). By scaling, we may assume that the MMS of each agent is~1, and we may also assume that no items has value larger than~1. For an agent $i$, we refer to items $e$ of value $v_i(e) \ge \frac{1}{q}$ as {\em large} (or also as {\em head} items), and to other items as {\em small} (or also as {\em tail} items). We may assume that for every agent $i$, every bundle in her MMS partition has at most $q$ large items. For $Q \simeq q \log q$, consider the set of values $(x_1, x_2, \ldots, x_Q)$, where $x_1 = 1$, $x_{i+1} = (1 - \frac{1}{q})x_i$ rounded up to the nearest multiple of $\frac{1}{q^2}$ (for every $1 \le i \le Q-1$), and $Q$ is chosen largest possible so that $x_Q = \frac{1}{q}$. We modify valuation functions as follows. For every agent $i$ and large item $e$, we round $v_i(e)$ down to the nearest $x_i$, thus maintaining a value of at least $(1 - \frac{1}{q})v_i(e)$. Thus, the value of each bundle in the MMS partition of $i$ remains at least $(1 - \frac{1}{q})$. As we are satisfied with giving each agent $(1 - O(\frac{1}{q}))$-MMS, we can afford to round down item values as above. As a consequence, we have at most $Q$ different values such that for every agent and every item heavy for that agent, the value of the item is one of these $Q$ values.

The proof of \cref{thm:largen} considered several cases. The easy case is when the average (across agents) sum of values of small items is ``high", say  $\Omega(\log n)$. (This may not seem high compared to the total value of items which is at least $n$, but it suffices for the results.)  In that case, the proof of \cref{thm:largen} does not require agent redundancy. The need for agent redundancy arises only in the complementary case, when the average (across agents) sum of values of small items is ``low", $O(\log^2 n)$. In that case, the arguments in the proof can be broken into two components. One is how to handle the case when are no small items at all. That is, for every agent, all items are large. Already in this stage, agent redundancy is needed in the proof given in~\cite{FG26}. This is the main part for which our current work introduces new ideas.  The other component of the proof addresses the modifications that are needed to the proof when there are small items, but their total average value per agent is small. Though these modifications are non-trivial, the same type of modifications that work in the proof of \cref{thm:largen} (given in~\cite{FG26}) work also in our proof of Theorem~\ref{thm:main} (with some relatively minor modifications). For this reason, in this overview, we focus on the case in which all items are large for every agent.

Following our preliminaries and scaling item values by a factor of $q^2$, the case when all items are large can be represented as a setting in which there are $Q \simeq q \log q$ positive integer values such for every agent $i$ and item $e$, $v_i(e)$ is one of these $Q$ values, every MMS bundle has at most $q$ items, and the MMS of every agent is at most $q^2$. For this setting, we show that when $n$ is sufficiently large, there always is an MMS-allocation (full MMS, no need for approximation). Thus, the two reasons why in general we get only approximate MMS rather than full MMS are the rounding down of item values so that we only have $Q$ values for large items, and the existence of small items.

As is standard when valuations are additive~\cite{Bouveret2016}, we may assume an Identically Ordered instance (IDO), meaning the for every agent $i$ and two indices $1 \le j < k \le m$, $v_i(e_j) \ge v_i(e_k)$.

Partition the items into the smallest number of types, where two items $e_j$ and $e_k$ are of the same type if $v_i(e_j) = v_i(e_k)$ for every agent $i$ (though for a given type, different agents may disagree on the value of items of this type). Let $t$ denote the total number of types. We order the types from highest value to lowest value (thus, every agent values an earlier type at least as high as a later type, with strict inequality for at least one agent). For every $1 \le j \le t$, let $n_j$ denote the total number of items of all types from $1$ to $j$ (thus, $n_t = m$). 

To introduce the ideas in our proof with less clutter, we distinguish between the case that the number of types $t$ is a constant (this constant may depend on $q$), and the case that $t$ grows with $n$.

\subsubsection{Only large items, constant number of types}
\label{sec:constantTypesOverview}

Here and throughout, we consider IDO input instances. For fixed $q,t$, there are $Q = O(q \log q)$ permissible integer item values (in the range $[q, q^2]$) and $t$ items types. For every agent $i$, all items of the same type have the same value, which is one of the permissible values, and every bundle in her MMS partition contains at most $q$ items. The MMS value of each agent is $q^2$. We show that when $n$ is sufficiently large, and MMS allocation exists.

We partition agents into classes, such that all agents in the same class have the same valuation function. There are $t$ types of items, and each type can have one of $Q$ values. As the instance is ordered, there are at most $\binom{Q + t - 1}{Q} \le t^Q$ possible valuation functions. Thus, one the classes of agents, call it $U$, contains at least $\frac{n}{t^Q}$ agents. We would like $|U|$ to be fairly large, say, $|U| \ge n^{0.9}$, and this places a lower bound on the value of $n$ for which our proof applies.

Agents of the same class have the same MMS partition. A key contribution our current work (not present in~\cite{FG26}) is a lemma that connects between the MMS partitions of pairs of agents, $i$ and $j$, of different classes. This lemma involves the concept of a {\em block-matching}. An MMS partition of an agent contains $n$ bundles. We allow arranging this bundles in disjoint blocks, where each block has a positive number of bundles. The {\em width} of this block arrangement is the maximum number of bundles that are contained in a single block.

Given block arrangements for two agents $i$ and $j$, a {\em block-matching} is a perfect matching between the blocks, where two blocks can be matched if they each contain the same number of bundles, and furthermore, for every type of item, the number of items of this type is the same in both blocks. In other words, the items in the bundles in one block can be rearranged to give the bundles of the other block. The {\em width} of a block matching is the width of the respective block arrangements (if a block-matching exists, both block arrangements have the same width).

\begin{lemma}
    \label{lem:blockMatchingWidth}
    Consider an IDO instance with $t$ types of items in which for every agent, every bundle in her MMS partition has at most $q$ items. Then there is some constant $D_{t,q}$ (independent of $n$) such that every pair of agents has a block-matching of width at most $D_{t,q}$.
\end{lemma}

Lemma~\ref{lem:blockMatchingWidth} can be proved in a non-constructive way (without giving an explicit upper bound on $D_{t,q}$) using Dickson's lemma~\cite{Dickson13}. See statement of that lemma in~\cref{dicksontype}. (Dickson's lemma also played an important part in~\cite{FG26}, but for other purposes.) In \cref{appendix:constant} we prove that $D_{t,q} \le \binom{t+q}{q} (t+2) q^t$. 

We would like $D_{t,q}$ from Lemma~\ref{lem:blockMatchingWidth} to be small compared to $n$, say, $D_{t,q} \le n^{0.1}$. This places another lower bound on the values of $n$ for which our proof works.

Consider now the case that there are only two classes of agents, $U$ (the larger class) and $V$. In general, even with just two classes of agents, MMS allocations need not exist~\cite{KPW18-1, FST21}. However, armed with Lemma~\ref{lem:blockMatchingWidth}, it is relatively easy to prove the existence of MMS allocations if $n \ge 2q^3D_{t,q}$. Consider the MMS partition of an agent $u \in U$ and the MMS partition of an agent $v \in V$ (and recall that all agents in $U$ have the same MMS partition, and so do all agents in $V$). Consider a block-matching of width at most $D_{t,q}$ between them (as guaranteed to exist by Lemma~\ref{lem:blockMatchingWidth}). Starting with the MMS partition of $U$, we iteratively flip between matched blocks, replacing the MMS bundles of $u$ by MMS bundles of $v$, as long as the number of MMS bundles of $v$ that we create does not exceed $|V|$. The newly created MMS bundles of $v$ are each given to a distinct agent in $V$. After this, the number $n'$ of agents remaining in $V$ is less than $D_{t,q}$. Sort all remaining MMS bundles of $u$ in decreasing order of value, according to the valuation function of $v$. Give the top $n'$ bundles in this order to the remaining agents of $V$. Each such bundle is {\em acceptable} for $v$ (has value for $v$ at least as large as the MMS of $v$). This is because the average value for $v$ of a remaining bundle is at least her MMS $q^2$, no bundle has value above $q \cdot q^2 = q^3$, and every bundle not acceptable has value at most $q^2 - 1$ (as item values are integers). Thus, at least $\frac{n - |V|}{q^3} \ge n'$ bundles are acceptable  (the inequality holds because $|V| \le \frac{n}{2}$, $n \ge 2q^3D_{t,q}$ and $n' < D_{t,q}$). (The assumption that item values are integers is not an essential assumption in this proof, nor in other proofs. Even without this assumption it is not difficult to show that at least $\frac{n - |V|}{D_{t,q}} \ge n'$ bundles are acceptable, which makes our proofs work with somewhat larger $n$.)
Finally, the $n - |V| = |U|$ bundles that remain in the MMS partition of $u$ are given to distinct agents of $U$. 

The case that there are more than two classes is more complicated, because for each class $V$ of agents other than $U$, the block arrangement of the bundles of $u$ used in the respective block-matching might be different. To handle this, we set up a certain integer program that tells us which block in each the block arrangements of $u$ to flip. The integer program has constraints specifying that the flipped blocks are disjoint (this involves only $t$ constraints, stating that for each type of item, the total number of items of this type that all flipped blocks contain is not larger than the total number of items of this type), and specifying that after removing the flipped bundles, the bundles that remain in the MMS partition of $u$ contain sufficiently many bundles that are acceptable by those agents from other classes that did not get a bundle from a flipped block (this can be formulated using not more than $(t+1)^q$ constraints). We relax this integer program to a linear program (LP), and show that the LP has a feasible fractional solution. Then, because the LP has only a small number of constraints, any basic feasible solution for the LP is mostly integral, allowing us to give almost all those agents not in $U$ bundles from flipped blocks. The remaining agents are handled as in the case of two classes of agents, explained above.

For additional details, see \cref{intype}.


\subsubsection{Only large items, large number of types}

The proof in \cref{sec:constantTypesOverview} works when the number $t$ of types of items is a constant, or a slowly growing function of $n$ (say, $t = o(\log n)$). However, $t$ might be much larger, even linear in $n$. This requires major modifications to the proof. Many of the principles of how to do these modifications already appear in~\cite{FG26} (which allowed agent redundancy), and can be adapted, with some changes, to our setting (that does not allow agent redundancy).  

One principle is to dramatically reduce the number of types of items. This involves a change in the definition of the notion of type. To distinguish the new notion from the original notion, we refer in this overview to the new notion as a {\em pseudo-type}.  Recall that for each agent $i$, each item can take one of $Q$ values. Arranging the items from~1 to $m$ on a line, agent $i$ introduces at most $Q$ {\em breakpoints} along the line, where a breakpoint is an index of an item that is last of its value (the next item has strictly smaller value). As there are $n$ agents, we have $O(nQ)$ breakpoints, where the same item can serve as multiple breakpoints (for different agents). Observe that the number of items that serve as breakpoints (without multiplicities) is simply the number of types.

Based on the breakpoint locations (including multiplicities), we shall partition the line into $3t = O(q)$ disjoint intervals. This will be done by carefully selecting items $$(a_1 = 1, b_1, c_1, a_2, b_2, c_2, \ldots, a_t, b_t, c_t, a_{t+1} = m+1),$$ and the set of items between two consecutive selected items serves as an interval. More specifically, intervals are of the form $[a_i, b_i]$, $[b_i + 1, c_i]$ and $[c_i +1, a_{i+1} - 1]$. 

Our construction will be such that all intervals of the form $[b_i + 1, c_i]$ are of the same small length $\nu$, where $\nu = C \log n$ for some constant $C$ that depends only on $q$ and $Q$. The union of all these intervals will be denoted by $F$ (for ``forbidden"). Importantly, the set of agents that has breakpoints in these intervals (we call this set $W$) is small, of size $|W| \le \varepsilon n$ (for some small $\varepsilon > 0$). 

Intervals of type $[a_i, b_i]$ are somewhat larger, but still small, of size $O(\frac{C \cdot Q}{\varepsilon} \log n)$. The union of all these intervals will be denoted as $S$ {(for ``special").}

Intervals of type $[c_i +1, a_{i+1} - 1]$ may be of arbitrary length, but have the property that each such interval contains breakpoints of at most $\frac{n}{2}$ agents.

It will turn out for our proof that items within the same interval are essentially interchangeable with one another, even though they need not have the same value. Consequently, the new notion of type associates with each interval a distinct pseudo-type, and all items of an interval are of the corresponding pseudo-type. Hence, the number of pseudo-types of items is $3t = O(q)$.

A similar notion of pseudo-types was used also in~\cite{FG26}. The purpose there, and here, is to keep the intervals of $F$ completely free of breakpoints. If we manage to do so, then for every interval of $F$, all its items would be of the same type, not just the same pseudo-type. In~\cite{FG26} this property was achieved by simply discarding all agents of $W$. This was allowed there, because $|W|$ is small, and hence can be considered to be part of the agent redundancy. However, as we need to give MMS bundles to all agents, we cannot discard $W$. Instead, we keep all agents of $W$, but for each of them, discard those MMS bundles that contain items from $F$. As $|F|$ is small, each agent in $W$ keeps the vast majority of her MMS bundles. 

The vast majority of items lie in intervals of type $[c_i +1, a_{i+1} - 1]$. As each interval of this type has breakpoints for at most $\frac{n}{2}$ agents, there are at least $\frac{n}{2}$ ``good" agents for which all items in the interval are of the same type. In this sense, each such interval $I$ is a proxy to the original notion of type. Moreover, the interval from $F$ that precedes $I$ has no breakpoints at all, and thus types for good agents extend into $F$. Consequently, for the good agents, $I$ is not just an interval representing a type, but rather the suffix of such an interval. This fact generates slackness in some parameters in initial parts of the proof, which allows fitting in later parts of the proofs. 

The set $S$ is generated as an undesirable side effect in the construction of the desired set $F$. The set $S$ is problematic, because it can contain arbitrarily many breakpoints. Handling items within $S$ requires special treatment, which becomes possible because $|S|$ is small.

We refer to the set of items $M \setminus E$ as $E$. ($E$ may stand for {\em easy}, as these items are easier to handle, compared to items in $S$.)

As in~\cite{FG26}, our partition of the items into a small number of pseudo-types allows as to find a partial allocation in which a set $I$ of at least $n - \varepsilon |F| - 16q\log n$ agents get an acceptable bundle (of value at least their MMS). Moreover, this partial allocation has some build in slackness (in some technical sense that will not be defined here) that will be leveraged to fit in the remaining items.

Recall the notion of agent classes used in \cref{sec:constantTypesOverview}. It is defined with respect of types of items, and the value that an agent associates with the type. We  use a similar notion of agent classes, but it is defined with respect to pseudo-types of items, and the existence or in-existence of break points within the pseudo-type. For our classes, it is not true that all agents in a class have the same valuations function, and not even the same MMS partition. However, due to the slackness mentioned above, the MMS partitions of any two agents in a class will in a sense be equivalent, meaning that if (given the type of constraints involved in our proofs) we are able to allocate a set of items acceptable to one agent, we are also able to instead allocate a set (of possibly different) items to the other agent. 

Now we come to a major difference between our work and~\cite{FG26}. With agent redundancy, the set of agents that did not receive a bundle in the partial allocation is fairly large, of size $\Omega(n)$. Hence it contains a class $U$ of agents that is large, much larger than the number $\varepsilon |F| + 16q\log n$ of agents that we need in order to complete the partial allocation to a complete allocation. At a high level, the proof in~\cite{FG26} showed  that the partial allocation can be completed to an allocation by removing some agents from $I$ and releasing their items, and creating from these items (and possibly other items not used in the partial allocation) sufficiently many bundles acceptable by agents from $U$ so as to get a complete allocation. Technically, this is done in steps, where in each step the size of $I$ is increased by~1, and the proof that this step can be executed makes use of Dickson's lemma (that also helped in \cref{sec:constantTypesOverview}).

The approach described above makes essential use of agent redundancy, and hence does not apply in our case. Instead, our set $U$ is mostly or entirely a subset of $I$. As a technical point, it is important that $U$ is disjoint from the set $W$ that has breakpoints in $F$, because for agents in $U$, we need that for every interval of $F$, all items of that interval have the same value. Also, we need that members of $U$ agree on the locations of breakpoints within the set $S$. (This does not hurt the size of $U$ by too much, because every agent has only a constant $Q$ number of breakpoints, and $|S|$ is very small.) At a high level, we use $U$ and the notion of block-matching from \cref{sec:constantTypesOverview} so as to replace the bundles of some agents from $I$ by other acceptable bundles, in a way that makes it possible to in addition create acceptable bundles for the agents outside $I$. Also here, this is done in steps, where in each step the size of $I$ is increased by~1, and proofs involve the use of Dickson's lemma.

The full details of the proof appear in \cref{sec:integer} and \cref{sec:intsupplier}.

\subsubsection{Mixture of large and small items}

So far, we dealt with input instances that have only large items, referred also as head items. For this case, after an initial rounding of the item values, we could give each agent her full MMS with respect to the rounded valuation functions. We now address the case that some of the items are small for some of the agents. In this case, we no longer aim to give agents their full MMS, and are satisfied with an approximation of the MMS (with a ratio that tends to~1, as $n$ grows).

Small items, in combination with allowing each each agent to receive slightly less than her MMS, give us much flexibility in allocations. For example,  agents can afford transferring items that are small for them to other agent who need them more.   Nevertheless, handling the combination of large and small items involves many details, especially when the total value of small items is small, but not negligible. This makes our proofs considerably longer. We shall not review here how we handle such combinations, because the principles are very similar to those used in~\cite{FG26}. However, we will mention here one technical difference compared to~\cite{FG26}, that we found helpful.

In~\cite{FG26}, for every agent $i$, we kept a parameter $h_i$ that counts the number of items that are large for $i$ (head items). These parameters, one for each agent, were used in various parts of the proofs. In our current work, for the difficult case in which the total value of small items is small, we instead keep only one parameter $h$. {It is chosen such that for the vast majority of agents, the top $h$ items contain only head items, and moreover, contain the vast majority of the bundles of the MMS partition of the agent.} We prove the existence of a suitable value for $h$, and this makes our proofs more manageable (compared to the alternative of maintaining $n$ separate values of $h_i$).

\subsection{Related work}

The subject of fair division has been extensively studied in multiple different settings.
In our work, we focus specifically on indivisible goods and use MMS as our notion of fairness.
For work on divisible goods, see for example~\cite{Brams_Taylor_1996}.
For additional work on fair allocation of indivisible items, including envy-based notions like EF1 and EFX, see for example~\cite{AzLiMoWu22,  AmanatidisSurvey23}.
In addition, there has been a plenty of work done on fair allocations for valuations beyond additive, i.e submodular, XOS and subadditive classes from the complement free hierarchy \cite{LLN}, 
which is beyond the scope of this work.

For additive valuations and every $n \geq 3$, there are examples in which MMS allocations do not exist~\cite{KPW18-1}. {There has been a long line of work investigating the highest fraction of the MMS that can be guaranteed to each agent, and it is known to be between $\frac{7}{9}$~\cite{huang2025} and $\frac{20}{21}$~\cite{EG26}.}
For general $n$, \cite{FST21} showed an upper bound of $1 - \frac{1}{n^4}$, which was improved to $1-\Omega(\frac{1}{\log^2 n})$ by \cite{EG26}.


While we consider instances with large number of agents $n$, there has been additional results on instances with small $n$.
For additive valuations, for the case of $n = 3$ agents, it was shown in \cite{FeNo22} that $\frac{11}{12}$-MMS allocations exist, and for $n = 4$ agents, \cite{GhodsiHSSY22} proved an approximation ratio of $\frac{4}{5}$.

The concept of $1$-out-of-$d$ MMS allocations first appeared in \cite{Budish11}, and later was formally introduced in \cite{babaioff2021competitive}.
A series of works subsequently improved the upper bound on $d$, with notable milestones of $d = 4\lceil n/3\rceil$ by \cite{AGS24} and $d = n + O(\sqrt{n\log n})$ by \cite{EG26-2}.
As for the progress towards $d = n + 1$, \cite{schwer26} showed that $1$-out-of-$5$ allocations exist for $n = 4$ agents.

Regarding techniques, our algorithm utilizes the identical ordering reduction from \cite{Bouveret2016}.
When it comes to replacing MMS bundles with corresponding profiles of rounded item values, the idea is not new and has been utilized in various ways before.
In \cite{KulkarniMehtaTaki2021}, the authors compute MMS of a single agent by representing its MMS bundles via counts of its large items in rounded value classes, and completed selected bundle types using small items.
The construction somewhat resembling our head-tail profile framework was utilized in \cite{NR20} for welfare approximations.

{Our proofs make extensive use of proof techniques that we used in our recent paper \cite{FG26}. Some techniques in that paper were similar to techniques used earlier in~\cite{EG26-2}. A comparison between these techniques appears in the related work section of~\cite{FG26}.}

\subsection{Technical details}

In \cref{sec:intype} we prove \cref{thm:main} in the special case, when the there are at most $t$ types of items, every agent takes integral values in $[Q]$, and each bundle in MMS partitions is of size at most $q$.
Our sufficiently large $n$ depends on these parameters $t, Q, q$ and treats them as constants.
In \cref{sec:integer,sec:intsupplier}, we remove the assumption on the number of types $t$, and prove \cref{thm:main} in less restricted setting, obtaining sufficient $n$ depending on $q, Q$ only.
This section utilizes the core of our techniques, and demonstrates the key aspects of our approach for the general case. 
In \cref{sec:dmms}, we repeat the proof of \cref{thm:dmms} from our previous work \cite{FG26}, as we will need its methods/conclusions for the general case of \cref{thm:main}.
\cref{sec:dual,sec:partition,sec:supplier,sec:suggested,sec:tail} cover the proof of \cref{thm:main}.
In \cref{appendix:rounding}, we prove the relevant linear programming rounding results.
In \cref{appendix:supplier}, we present the proofs of arguments relying on Dickson's lemma.
In \cref{appendix:constant}, we give concrete estimates on all the relevant constants, as well as the final sufficient bound on $n$ in terms of approximation parameter $\rho$.

\subsection*{AI and other declarations}

This paper is a continuation of our earlier work~\cite{FG26}, posted on the arxiv on November 3, 2026 (due to the mandatory public posting of papers submitted to STOC 2027). Within a matter of days after posting that paper, we obtained the results reported in this paper, proving Conjecture~1 from~\cite{FG26}. This was done by a combination of new ideas of the authors (such as introducing the notion of block-matching and proving its existence using Dickson's lemma, or introducing the uniform common head $h$ and constructing it using LP-duality), and aided by the use of {OpenAI’s ChatGPT-6.1 Sol and ChatGPT-6.0 Astra}. (Without the new ideas of the authors, these AI assistants failed to prove the results of this paper, despite spending many hours on them.) The paper was mostly written by the human authors, and their limited availability for part of the time explains why it took a few weeks to post it. In the period since obtaining the results, Vadim Grinberg gave two public talks with the title ``On MMS allocations for large number of agents" (for the MPI reading group on fair division, via zoom, on November 11, and at the Weizmann Institute on November 18) in which the results were announced, though the technical content of these talks was based on~\cite{FG26}, and the new techniques of the current paper were not presented in these talks. Due to feedback from the talks, the authors rearranged and revised the current paper, starting with easier special cases and building up towards the full theorem.

On November 23, Uriel Feige received e-mail in which Biaoshuai Tao wrote him that together with Zehan Lin, Shengxin Liu, Xiaowei Wu, and Yuhao Zhang, they intend to post in the near future a paper that proves Conjecture~1 from~\cite{FG26}. Consequently, the authors of both papers agreed to submit their papers to the arxiv on the same day, November 24, even if the papers can still benefit from additional editing. None of the human authors of these papers has seen the paper of the other group, nor heard from the other group any technical details about the proofs. Cross references between the two papers and comparisons between them will be added in revised versions of these papers.

\section{Exact MMS for constant number of integer value types}

\label{sec:intype}

\begin{definition}
    An \textbf{item type} is a subset $M_s\subseteq M$ such that there exist values $(\alpha_{1,s}, \ldots, \alpha_{n,s})$ that for every $e \in M_s$  satisfy  $v_i(e) = \alpha_{i,s}$ for every agent $i \in N$.
    
    That is, items $e, e'\in M$ are of the same \textbf{type} if every agent $i \in N$ values $e$ the same as $e'$, i.e $v_i(e) = v_i(e')$.
\end{definition}
Note that for the same type $M_s\subseteq M$, values $\alpha_{i,s}$ may differ across the agents, and for some agents items of different types may have equal values.

First, we prove \cref{thm:main} in the special restricted case.
\begin{theorem}\label{intype}
    Let $t, Q, q\geq 1$ be integers.
    There exists $n(t, Q, q)$ such that for every $n \geq n(t, Q, q)$ the following holds.
    Consider a fair allocation instance with $m$ items $M$ and $n$ agents $N$, where for every $i \in N$ valuation $v_i$ is additive, and
    \begin{itemize}
        \item items $M$ have at most $t$ types;
        \item for every agent $i \in N$ and item $e \in M$ we have $v_i(e) \in \{1, \ldots, Q\}$;
        \item every agent $i \in N$ has an MMS partition $\{B_{i,1}, \ldots, B_{i,n}\}$ satisfying $|B_{i,j}| \leq q$ for all $j \in [n]$.
    \end{itemize}
    There exists an allocation $(B_1, \ldots, B_n)$ of $M$ to $N$ such that for every agent $i \in N$ we have \[
    v_i(B_i)\geq \mms(M, v_i, n).\]
\end{theorem}

For every $i \in N$, we denote $\mu_i := \mms(M, v_i, n)$.
We will assume that $m \geq n$, as otherwise all $\mu_i = 0$.
So, in the setting of \cref{intype}, for every $i \in N$ we have
\[
    1 \leq \mu_i \leq qQ,
\]
and there is a partition $\{B_{i,1}, \ldots, B_{i,n}\}$ satisfying $|B_{i,j}|\leq q$ and $v_i(B_{i,j}) \geq \mu_i$ for every $j \in [n]$.

The main idea of the proof of \cref{intype} is to choose a single agent $u_* \in N$ from a large class of agents that agree on all values of all item types.
We will use the MMS partition $\{B_{u_*,1}, \ldots, B_{u_*,n}\}$ of $u_*$ as a baseline, and then modify it in order to obtain MMS allocation for all agents.
The key ingredient is that for every other agent $i$ from a class different from $u_*$, we can take a small (i.e constant) number of MMS bundles of $u_*$ and repartition them into the same number of MMS bundles of $i$.
Moreover, we can do so in multiple different ways, and essentially transform the MMS partition of $u_*$ into an MMS partition of $i$ by repartitioning only a constant number of bundles at a time.
Then, the problem of finding an MMS allocation comes down to finding a way to coordinate these repartitions between agents of different classes, choosing which of them to use for which class of agents, so that in the end every agent will be able to get a distinct MMS bundle.

\begin{definition}
    Let $M_1, \ldots, M_t$ be nonempty item types.
    For each $s \in [t]$, choose an arbitrary \textbf{representative item} $e_s \in M_s$, and for $i \in N$ denote 
    \[
        \alpha_{i,s} := v_i(e_s).
    \]
    For every agent $i \in N$, define his \textbf{signature} as 
    \[
        \fe_i:= (\alpha_{i,1}, \ldots, \alpha_{i,t}).
    \]
    Two agents belong to the same class if they have the same signature.
    For signature $\fe$, denote
    \[
        N_\fe := \{i \in N: \fe_i = \fe\}.
    \]
\end{definition}
\begin{observation}\label{bigsignature}
    Let $\Phi$ denote the set of all occurring signatures.
    Then $|\Phi|\leq  \phi := Q^t$, (or at most $\binom{Q + t}{Q}$ in the ordered case) and there is a set $U = N_\fe$ for some $\fe\in \Phi$ satisfying $|U|\geq n/\phi$.
\end{observation}
\begin{proof}
    Since $v_i(e) \in \{1, \ldots, Q\}$ for every $i \in N, e \in M$, and $|\fe| = t$, there are at most $\phi:= Q^t$ possible signatures.
    Taking $\fe \in \Phi$ maximizing $|N_\fe|$, we get $|N_\fe|\geq n/\phi$.
\end{proof}
\begin{definition}
    Let $U = N_\fe$ for $\fe \in \Phi$ maximizing $|N_\fe|$.
    We choose an arbitrary agent $u_* \in U$ that we will refer to as \textbf{supplier}, and denote his signature by $\fe_*$.
\end{definition}
We show that every agent $i \in N$ ``likes'' a huge portion of supplier's MMS partition.
\begin{lemma}\label{acceptable}
    We call a bundle $B\subseteq M$ \textbf{acceptable} for agent $i \in N$, if $v_i(B) \geq \mu_i$.

    Every agent $i \in N$ finds at least $n/(qQ)$ bundles of the supplier's MMS partition acceptable.
\end{lemma}
\begin{proof}
    For a supplier $u_* \in U$, let $\{B_{u_*,1}, \ldots, B_{u_*,n}\}$ be his MMS partition.
    For agent $i \in N$, let $g_i := |\{j \in [n]:\; v_i(B_{u_*,j})\geq \mu_i\}|$ be the number of bundles acceptable for $i$ in $u_*$'s MMS partition.
    Since every MMS bundle has size at most $q$, every bundle $B_{u_*,j}$ has value at most $qQ$ for any agent $i \in N$.
    Since every unacceptable bundle for $i$ has value at most $\mu_i -1$, we have
    \[
        v_i(M) \leq g_i\cdot qQ + (n - g_i)\cdot (\mu_i - 1).
    \]
    On the other hand, for any $i \in N$ his MMS partition gives $v_i(M) \geq \mu_i n$.
    Consequently,
    \[
        n\mu_i \leq g_i\cdot qQ + (n - g_i)\cdot (\mu_i - 1) = n(\mu_i - 1) + g_i(qQ - \mu_i + 1),
    \]
    which implies
    \[
        g_i \geq \frac{n}{qQ - \mu_i + 1} \geq \frac{n}{qQ}.
    \]
\end{proof}

Note that any $u \in U$ in the supplier's signature class accepts any bundle of the supplier $u_*$'s MMS partition, as their valuations and MMS values are the same.
In addition, by \cref{acceptable} we can give supplier's bundles to at least $n/(qQ)$ agents of other classes.
We, however, would like \textit{every} agent $i \in N$ to receive some acceptable bundle.
Even though class $U$ is the largest, it can still be that most agents find the vast majority of $u_*$'s MMS bundles unacceptable, and have  very different MMS partitions, so we cannot just give supplier's bundles to everybody.

Our approach is to ask every agent $i \in N$ to modify the MMS partition $\{B_{u_*,1}, \ldots, B_{u_*,n}\}$ so as to obtain plenty acceptable bundles for $i$'s signature class.
Of course, $i$ can just trivially take all bundles $\bigcup_{j = 1}^nB_{u_*,j} = M$ and repartition them into $i$'s own MMS partition $\{B_{i,1}, \ldots, B_{i,n}\}$.
Then, we run into the exact same problem: most other agents do not accept most of $i$'s bundles.

Instead, we will ask $i$ to do something more peculiar.
We would like $i$ to break the supplier's MMS partition $\{B_{u_*,1}, \ldots, B_{u_*,n}\}$ into \textbf{blocks}, where each block contains only a constant number of bundles.
These blocks should have the following property: it is possible to repartition the corresponding bundles of $u_*$ into the exact same number of bundles from $i$'s MMS partition.

Given this ``bounded repartition'', we can now take a constant number of bundles of $u_*$, repartition them into bundles of $i$, and allocate them to the corresponding number of agents in the class of $i$.
Doing so repeatedly, we can obtain an MMS allocation for every agent of $i$'s class.
However, we cannot perform this procedure independently for every signature class, as different classes may use the same bundles in their repartitions.
In order to coordinate the replacements across all agents, we will use a Linear Programming relaxation that maximizes the total number of non-$u_*$-class agents that get allocated using these repartitions, and chooses which blocks to repartition for each signature class.
We will show that the rounded solution leaves only a small number of agents unallocated, and then use \cref{acceptable} to show that for sufficiently large $n$, the coordinated repartitioning leaves sufficiently many acceptable supplier bundles for every remaining agent.
Then, every remaining agent gets an acceptable supplier bundle, and we obtain a full MMS allocation.

\subsection{Bounded repartition and Dickson's lemma}

It will be convenient for us to describe bundles without referring to their particular items, so instead we will consider patterns of their item-type counts.
Note that the total number of such patterns is only a constant.

\begin{definition}
    For bundle $B\subseteq M$, its \textbf{pattern} counts the number of items of each type:
    \[
        p(B) := (|B \cap M_1|, \ldots, |B\cap M_t|).
    \]
    For a pattern $p$, write 
    \[|p|:= \sum_{s = 1}^tp_s,
    \]
    and $P := \{p \in \Z^t_{\geq 0} : 1 \leq |p|\leq q\}$ as the set of all possible patterns of size at most $q$.
    Note that
    \[
        |P|= \binom{t + q}{q} - 1. 
    \]
    If bundle $B$ has pattern $p$, then we can express $v_i(B)$ as
    \[
        v_i(B) = v_i(p) := \sum_{s = 1}^t \alpha_{i,s}\cdot p_s.
    \]
\end{definition}
\begin{definition}
    For agent $i \in N$ with MMS partition $\{B_{i,1}, \ldots, B_{i,n}\}$, for every $p \in P$, let
    \[
        \lambda_i(p) := |\{j \in [n]:\; p(B_{i,j}) = p\}|
    \]
    denote the number of bundles in $i$'s MMS partition with pattern $p$.
    Note that $\sum_{p \in P}\lambda_i(p) = n$.
\end{definition}

While MMS partitions of the supplier $u_*$ and agent $i \in N$ may contain very different patterns, they must contain the same total number of items of each type.
So, replacing the entire partition of $u_*$ by a partition with patterns of agent $i$ is always possible.
The following theorem shows that full repartition can be done in blocks of constant size, while preserving the number of bundles and items of each type in every block.
\begin{definition}
    A \textbf{repartition block} is a pair $\beta = (x, y)$, where $x, y\in \Z_{\geq 0}^P$, satisfying
    \[
        \sum_{p \in P}x(p) = \sum_{p \in P}y(p) \geq 1\qquad\text{and}\qquad \sum_{p \in P} x(p)\cdot p = \sum_{p \in P}y(p) \cdot p.
    \]
    Here $x(p)$ counts how many bundles of pattern $p$ are used, and $y(p)$ counts how many bundles of pattern $p$ are produced, during the repartition process.
    The first condition preserves the number of the bundles during repartition, and the second preserves the number of items of each type.

    Define the \textbf{size} of the block by 
    \[
        |\beta| = \sum_{p \in P}x(p) = \sum_{p \in P}y(p).
    \]
\end{definition}

It is important to note that a repartition will produce bundles with the same pattern as in agent $i$'s MMS partition, but the actual used items may be different (though type-equivalent). 

\begin{theorem}[Block repartition]\label{blockpartition}
    For every pair of integers $t, q \geq 1$, there exists a global constant $D_{t,q} \geq 1$, depending only on $t$ and $q$, such that the following holds.

    For a fixed supplier $u_*$ and every agent $i\in N$, there exists a sequence of $K_i \leq n$ blocks
    \[
        \beta_{i,1} = (x_{i,1}, y_{i,1}), \ldots, \beta_{i, K_i} = (x_{i, K_i}, y_{i, K_i}),
    \]
    such that for every $k \in [K_i]$ we have $|\beta_{i,k}|\leq D_{t,q}$, and for every pattern $p \in P$ blocks satisfy
    \[
        \sum_{k = 1}^{K_i}x_{i,k}(p) = \lambda_{u_*}(p),\qquad \text{and}\qquad \sum_{k = 1}^{K_i}y_{i,k}(p) = \lambda_i(p).
    \]
    Thus, we can group MMS bundles of $u_*$ into disjoint blocks of size at most $D_{t,q}$, so that every block can be repartitioned into the same number of bundles with patterns from $i$'s MMS partition.
\end{theorem}
\begin{proof}
    Consider the set of all possible repartition blocks with patterns from $P$:
    \[
        \Bb(t, q) := \left\{ (x, y) \in \Z_{\geq 0}^{P}
         \times \Z_{\geq 0}^P:\; \begin{array}{l}
        \sum_{p \in P}x(p) = \sum_{p \in P}y(p) \geq 1;\\
        \sum_{p \in P} x(p)\cdot p = \sum_{p \in P}y(p) \cdot p;
        \end{array}
        \right\}.
    \]
    Note that, since $P = \{p \in \Z^t_{\geq 0} : 1 \leq |p|\leq q\}$ where $|p| = \sum_{s = 1}^tp_s$, this set depends only on $t, q$.

    We would like to repartition some bundles of $u_*$'s MMS partition in order to obtain a collection of bundles with patterns of agent $i$'s MMS partition.
    A trivial repartition always exists, and corresponds to $\beta^* = (x^*, y^*)$ such that for every $p \in P$, 
    \[
        x^*(p) = \lambda_{u_*}(p)\qquad \text{and} \qquad y^*(p) = \lambda_i(p),
    \] i.e taking all bundles of $u_*$ at once and replacing them with MMS partition of $i$.
    Observe that $\beta^* \in \Bb(t, q)$: since both partitions contain $n$ bundles, we have $\sum_{p \in P}x^*(p) = n = \sum_{p \in P}y^*(p)$, and since both partitions cover all items of $M$,  we have
    \[
        \sum_{p \in P}x^*(p)\cdot p = (|M_1|, \ldots, |M_t|) = \sum_{p \in P}y^*(p)\cdot p.
    \]
    The problem is that this repartition consists of one block of size $n$.
    We would like to find a block $\beta = (x, y) \in \Bb(t, q)$, such that 1) both $x$ and $y$ match corresponding MMS partitions of agents $u_*$ and $i$ respectively, and 2) the size $|\beta|$ of this block is only a constant.

    The existence of such element of $\Bb(t, q)$ is guaranteed due to the following lemma by Dickson, a short proof of which is available in \cref{appendix:supplier} (\cref{dicksonproof}).
    \begin{lemma}[\cite{Dickson13, Cox15}]\label{dicksontype}
            For $K \geq 1$, every subset of $\Z^K_{\geq0}$ has finitely many component-wise minimal elements, and every element of the subset dominates some minimal element.
    \end{lemma}
    Since $\Bb(t, q) \subseteq \Z_{\geq 0}^{2|P|}$, by \cref{dicksontype} set $\Bb(t, q)$ has finitely many component-wise minimal elements, let them be $\Ee$.
    Since $\Bb(t, q)$ depends only on $t, q$, the number of minimal elements $|\Ee|$ as well as the coordinates of every minimal element $(\hx, \hy) \in \Ee$ and the sum $\sum_{p \in P}\hx(p)$, are also bounded by some function of $t, q$ (otherwise we would obtain an infinite sequence of minimal elements).
    We take
    \[
        D_{t,q} := \max_{(\hx, \hy) \in \Ee}\sum_{p \in P}\hx(p),
    \]
    which is a constant depending only on $t, q$.
    Now, $\sum_{p \in P}\hx(p) \leq D_{t,q}$ for every minimal $(\hx, \hy) \in \Ee$.

    Now, take the element $\beta^* = (x^*, y^*) \in \Bb(t, q)$ we considered earlier, corresponding to the full replacement of $u_*$'s MMS partition with $i$'s MMS partition.
    By \cref{dicksontype}, there exists a minimal element $(\hx, \hy) \in \Ee \subseteq \Bb(t, q)$ such that $(\hx, \hy) \leq (x^*, y^*)$ component-wise, and $\sum_{p \in P}\hx(p) \leq D_{t,q}$.
    Since $\hx \leq x^*$, for every pattern $p \in P$ we have $\hx(p) \leq x^*(p)$.
    This means that, if $\hx$ takes a bundle with pattern $p$, this pattern $p$ must exist in $u_*$'s MMS partition (as zeros stay zeros), and it cannot repartition more bundles with pattern $p$ than there are such bundles in the MMS partition of $u_*$.
    The inequality $\hy \leq y^*$ implies the exact same property for patterns $p$ of agent $i$'s MMS partition.

    At the same time, $(\hx, \hy) \in \Bb(t, q)$ ensures that the number of repartitioned bundles of the supplier $u_*$ is the exact same as the number of resulting bundles with patterns of $i$, and each group of bundles uses the exact same number of items of each type.
    Consequently, if we remove the corresponding bundles and items of $(\hx, \hy)$ from all the patterns of both $u_*$ and $i$, and consider an element $(x^*, y^*) - (\hx, \hy) = (x^* - \hx, y^* - \hy)$, this element will still be a valid replacement block, and thus $(x^* - \hx, y^* - \hy) \in \Bb(t, q)$ (as long as it is not $(0, 0)$).

    We will use the last observation and apply the argument above repeatedly, and instead of just one block of constant size, obtain a sequence of blocks that cover the whole MMS partitions of $u_*$ and $i$.
    We do this in iterations.
    Initially, let 
    \[
        x^1 := \lambda_{u_*}, \qquad \text{and}\qquad y^1 := \lambda_i.
    \]
    At iteration $k \geq 1$, suppose that both $x^k, y^k > 0$, and $(x^k, y^k) \in \Bb(t, q)$.
    Let $\beta_{i,k} := (x_{i,k}, y_{i,k}) \in \Ee$ be the minimal element of $\Bb(t, q)$ such that $(x_{i,k}, y_{i,k}) \leq (x^k, y^k)$, by \cref{dicksontype} such element exists.
    Define the residual replacement as 
    \[
        x^{k + 1} := x^k - x_{i,k}, \qquad\text{and}\qquad y^{k + 1} := y^k - y_{i,k},
    \]
    these vectors are non-negative.
    It is easy to see that (using induction on $k$),
    \[
         \sum_{p \in P}x^{k + 1}(p) =  \sum_{p \in P}x^k(p) - \sum_{p \in P}x_{i,k}(p) = \sum_{p \in P}y^k(p) - \sum_{p \in P}y_{i,k}(p) = \sum_{p \in P}y^{k + 1}(p),
    \]
    and completely analogously,
    \[
        \sum_{p \in P}x^{k + 1}(p) \cdot p = \sum_{p \in P}y^{k + 1}(p)\cdot p.
    \]
    Therefore, either $(x^{k + 1}, y^{k + 1}) = (0, 0)$, or $(x^{k + 1}, y^{k  +1}) \in \Bb(t, q)$ once again, and we proceed to iteration $k + 1$.
    At every iteration $k$, we remove $1 \leq |\beta_{i,k}| \leq D_{t,q}$ bundles from both $u_*$ and $i$.
    As a result, after at most $n$ iterations we will repartition all bundles.
    The theorem follows.
\end{proof}
We will now translate the result of \cref{blockpartition} from patterns to actual items.
\begin{corollary}\label{bundleblock}
    Let $\beta_{i,k} = (x_{i,k}, y_{i,k})$ be a block obtained from \cref{blockpartition} applied to a supplier $u_*$ and agent $i \in N$.
    We can choose $|\beta_{i,k}|\leq D_{t,q}$ MMS bundles of $u_*$ and repartition their union into $|\beta_{i,k}|$ acceptable bundles for $i$ of size at most $q$, with exactly $y_{i,k}(p)$ bundles for each $p \in P$.

    Furthermore, for each fixed $i$, these repartitions can be performed simultaneously for all blocks, using every bundle of supplier $u_*$'s MMS partition exactly once.
\end{corollary}
\begin{proof}
    For every pattern $p \in P$, by \cref{blockpartition} we have 
    \[
        \sum_{k = 1}^{K_i}x_{i,k}(p) = \lambda_{u_*}(p).
    \]
    So, we can choose $x_{i,k}(p) \leq \lambda_{u_*}(p)$ bundles of each pattern $p \in P$ from the supplier's MMS partition.
    For every item type $s \in [t]$, their union contains exactly
    \[
        \sum_{p \in P}x_{i,k}(p)\cdot p_s = \sum_{p \in P}y_{i,k}(p)\cdot p_s
    \]
    items of type $s$.
    We create $y_{i,k}(p)$ bundles with pattern $p$ for agent $i$, as follows.
    For each item type $s \in [t]$, distribute the selected type-$s$ items among these bundles, giving exactly $p_s$ items to each bundle.
    The equality above guarantees that we will have exactly enough items, and every selected item is used once.
    Since $\sum_{p \in P}y_{i,k}(p) = \sum_{p \in P}x_{i,k}(p) = |\beta_{i,k}|$, we will get the same number of bundles as chosen from the supplier.
    By construction, all patterns have size at most $q$, thus every produced bundle will have size at most $q$.
    In addition, if $y_{i,k}(p) > 0$, then by \cref{blockpartition} $\lambda_{i}(p) > 0$, meaning that pattern $p$ occurs in agent $i$'s MMS partition, and thus $v_i(p) \geq \mu_i$, i.e corresponding bundle is acceptable.

    Finally, since $\sum_{k = 1}^{K_i}x_{i,k}(p) = \lambda_{u_*}(p)$ for every $p \in P$, we can distribute all supplier bundles of pattern $p$ among the blocks $\beta_{i,1}, \ldots, \beta_{i, K_i}$, assigning exactly $x_{i,k}(p)$ of them to block $k$.
    Performing the construction above separately within these disjoint collections finishes the proof.
\end{proof}

Note that if some agent $i'\neq i$ belongs to the same signature class as $i$, then every block of the partition from \cref{blockpartition} contains acceptable bundles also for agent $i'$.
Hence, performing one repartition $\beta_{i,k}$ already provides $|\beta_{i,k}|$ acceptable bundles for the corresponding number of agents of $i$'s class.
In the following subsection we will show how to coordinate these repartitions between different classes, so that all agents can simultaneously receive an acceptable bundle.

\subsection{Coordinating repartitions via LP}

It is important that \cref{bundleblock} applies to \textit{any} collection of supplier bundles, as long as they match the prescribed patterns of repartition block, as swapping items of the same type does not change the value of the bundles.
Because of this, we can treat each repartition block as a proposed \textit{repartition rule} for corresponding bundle patterns, and choose how many copies of each rule to perform.
After selecting a set of repartition rules, each individual rule then will be realized via a disjoint collection of supplier bundles.

We will do this using a Linear Program (LP) that maximizes the number of agents outside $U$, i.e with a class different from the supplier, that will get an acceptable bundle after realizing the chosen collection of repartition rules.
By rounding its optimal solution and realizing the selected rules, we will allocate almost all non-supplier-class agents, except for at most a constant number.
As for the remaining unallocated agents outside $U$, our plan is to give each of these agents a distinct bundle from the supplier's MMS partition that they find acceptable.
It order to make sure that enough acceptable bundles remain, we will preserve a certain fraction of bundles for every pattern $p$ of the supplier's MMS partition.

We assume that $N\setminus U \neq \varnothing$, as otherwise supplier's partition already gives every agent an acceptable bundle.
For supplier $u_* \in U$, let $\fe_*$ denote his signature (so $U = N_{\fe_*}$), and for every $\fe \in \Phi \setminus \fe_*$ write $n_\fe := |N_\fe|$.
\begin{definition}
    For every $\fe \in \Phi\setminus \fe_*$, choose a representative $i_\fe \in N_\fe$ and apply \cref{blockpartition} to $u_*$ and $i_\fe$.
    Let $\Bb(\fe) = \{\beta_{i_\fe,1}, \ldots, \beta_{i_\fe, K_{i_\fe}}\}$ be the resulting collection of blocks, and denote
    \[
        \Bb = \bigsqcup_{\fe \in \Phi\setminus\fe_*}\Bb(\fe).
    \]
    Each block $\beta \in \Bb$ is associated with a corresponding signature class, as well as its corresponding index within that class, i.e we treat $\beta = \beta_{i_\fe,k} = (x, y)$ and $\beta'= \beta_{i_{\fe'}, k'} = (x', y')$ as different blocks when $\fe \neq \fe'$ or $k \neq k'$, even if $x = x'$ and $y = y'$.
\end{definition}
\begin{definition}
    For each $\beta \in \Bb$ where $\beta = \beta_{i_\fe,k}$, define for every $p \in P$:
    \[
        \lambda_\beta(p) := x_{i_\fe,k}(p).
    \]
    So, $\lambda_\beta(p)$ is the number of supplier bundles of pattern $p$ that is required by one copy of block $\beta$.
\end{definition}
\begin{observation}\label{lambdab}
    By definition and by \cref{blockpartition}, it holds for every $\beta \in \Bb$ that
    \[
        \sum_{p \in P}\lambda_\beta(p) = |\beta|, \qquad\text{and}\qquad 1 \leq |\beta| \leq D_{t,q}.
    \]
    In addition, for every $\fe \in \Phi\setminus\fe_*$,
    \[
        \forall p \in P:\; \sum_{\beta \in \Bb(\fe)}\lambda_\beta(p) = \lambda_{u_*}(p)\qquad \implies \qquad \sum_{\beta \in \Bb(\fe)}|\beta| = n.
    \]
\end{observation}
By \cref{bundleblock}, every block (or in this case, repartition rule) $\beta \in \Bb(\fe)$ produces $|\beta|$ acceptable bundles for agent $i_\fe$.
These bundles are also acceptable for every agent in class $N_\fe$, as agents with the same signature have identical valuations and MMS values.

For every pattern $p$ of the supplier's MMS partition, we will reserve
\[
    \eta := \frac{|U|}{n}.
\]
fraction of all supplier's bundles with pattern $p$.
Note that ``in expectation'' this reserves exactly $|U|$ supplier bundles, i.e one per every agent of supplier-class.

For every block $\beta \in \Bb$, we introduce a variable $z_\beta \geq 0$, which corresponds to the number of copies of repartition rule $\beta$ that we perform.
Consider the following linear program (\ref{patlp}).
\[
\begin{aligned}
    &\maximize&& \sum_{\beta \in \Bb}|\beta|\cdot z_\beta\\
    &\subto && \sum_{\beta \in \Bb}\lambda_\beta(p)\cdot z_\beta  \leq (1 - \eta)\lambda_{u_*}(p),&\forall p \in P;\\
    &&& \sum_{\beta \in \Bb(\fe)} |\beta|\cdot z_\beta \leq n_\fe,&\forall \fe \in \Phi \setminus \fe_*;\\
    &&& z_\beta \geq 0,&\forall \beta \in \Bb.
\end{aligned}\label{patlp}\tag{LP}
\]
The objective maximizes the total number of \textit{non-supplier-class} agents that are allocated by performing the selected rules $\beta \in \Bb$.
The first constraint ensures that for every pattern $p \in P$, selected blocks use no more than $(1 - \eta)$-fraction of the total number of such patterns in the supplier $u_*$'s MMS partition, i.e $\lambda_{u_*}(p)$.
The second constrain says that the number of produced bundles for class $\fe \in \Phi\setminus\phi_*$ should not exceed the total number of agents $n_\fe$ of this class.

Observe that (\ref{patlp}) has only $|P| + |\Phi| - 1$ constraints (i.e constant number), even though the number of variables $|\Bb|$ can depend on $n$.
We will use this in estimating the loss incurred from rounding its optimal solution.
\begin{lemma}[Block rounding]\label{introunding}
    The optimum of (\ref{patlp}) is $n - |U|$.
    Furthermore, there exists an optimal solution $z^*$ with at most $|P| + |\Phi| - 1$ positive coordinates, and the integral integral solution $\hz$ such that $\hz_\beta := \lfloor z^*_\beta\rfloor$ for all $\beta \in \Bb$ is feasible and satisfies
    \[
        \hZ := n - |U| - \sum_{\beta \in \Bb}|\beta|\cdot \hz_\beta \leq D_{t,q}(|P| + |\Phi| - 1).
    \]
    That is, rounded solution $\hz$ allocates all but at most $D_{t,q}(|P| + |\Phi| - 1)$ non-supplier-class agents.
\end{lemma}
\begin{proof}
    For every $\fe \in \Phi\setminus\fe_*$ and every $\beta \in \Bb(\fe)$, set
    \[
        z^0_\beta := \frac{n_\fe}{n}.
    \]
    We will show that this is a feasible solution.
    For every pattern $p \in P$, by \cref{lambdab}
    \[
        \sum_{\beta \in \Bb}\lambda_\beta(p)\cdot z_\beta^0 = \sum_{\fe\in \Phi\setminus \fe_*}\frac{n_\fe}{n}\sum_{\beta \in \Bb(\fe)}\lambda_\beta(p) = \frac{n - |U|}{n}\cdot \lambda_{u_*}(p) = (1 - \eta)\lambda_{u_*}(p).
    \]
    Similarly, for every class $\fe \in \Phi\setminus \fe_*$, \cref{lambdab} gives
    \[
        \sum_{\beta\in \Bb(\fe)}|\beta|\cdot z_\beta^0 = \frac{n_\fe}{n}\sum_{\beta \in \Bb(\fe)}|\beta| = n_\fe \qquad \text{and} \qquad \sum_{\beta \in \Bb}|\beta|\cdot z_\beta^0 = n - |U|.
    \]
    Thus, $z^0$ is feasible with objective value $n- |U|$.
    It is easy to see by summing up the constraints for $\fe \in \Phi\setminus \fe_*$ that no feasible solution can have larger value.
    Thus, $z^0$ is also an optimal solution, furthermore, every constraint $\sum_{\beta \in \Bb(\fe)} |\beta|\cdot z_\beta \leq n_\fe$ \textit{must} be tight at every optimal solution.

    Choose an optimal solution $z^*$ that is also a basic feasible solution of (\ref{patlp}).
    Since the LP has $|P| + |\Phi| - 1$ constraints, $z^*$ has at most $|P| + |\Phi| - 1$ positive coordinates.

    Finally, the rounded solution $\hz$, where $\hz_\beta := \lfloor z^*_\beta\rfloor$ for all $\beta \in \Bb$, is feasible.
    Since every constraint for $\fe \in \Phi\setminus\fe_*$ is tight for $z^*$, the difference between the values of $z^*$ and $\hz$ is
    \[
        \hZ := (n - |U|) - \sum_{\beta\in \Bb}|\beta|\cdot \hz_\beta = \sum_{\beta \in \Bb}|\beta|(z^*_\beta - \hz_\beta) \leq  \sum_{\substack{\beta \in \Bb\\z^*_\beta > 0}}|\beta| \leq D_{t,q}(|P| + |\Phi| - 1).
    \]
    The last inequality follows from $|\beta|\leq D_{t,q}$ for all $\beta \in \Bb$ by construction.
\end{proof}

It is important that the blocks selected by \cref{introunding} can be simultaneously realized using distinct bundles of the supplier, while leaving intact at least $\eta$-fraction of every supplier pattern.
\begin{corollary}\label{realization}
    There exists a partial allocation $\{B_i\}_{i \in I}$ of $M$ to agents $I\subseteq N\setminus U$, such that 
    \begin{itemize}
        \item for every $i \in I$, bundle $B_i$ is acceptable, i.e $v_i(B_i) \geq\mu_i$;
        \item $|I| = n - |U| - \hZ$, and the unallocated items $M\setminus \bigcup_{i \in I}B_i$ form a union of exactly $|U| + \hZ$ intact bundles of the supplier's MMS partition $\{B_{u_*,1}, \ldots, B_{u_*,n}\}$;
        \item for every pattern $p \in P$, at least $\eta \lambda_{u_*}(p)$ of these intact supplier bundles have pattern $p$.
    \end{itemize}
    Here $\hZ$ is some integer satisfying $0 \leq \hZ \leq D_{t,q}(|P| + |\Phi|- 1).$
\end{corollary}
\begin{proof}
    For the integral solution $\hz$ from \cref{introunding}, put 
    \[
        \hZ = n - |U| - \sum_{\beta \in \Bb}|\beta|\cdot \hz_\beta.
    \]
    For every pattern $p \in P$, it holds that
    \[
        \sum_{\beta \in \Bb}\lambda_\beta(p)\cdot \hz_\beta \leq (1 - \eta)\lambda_{u_*}(p),
    \]
    so for each repartition rule $\beta$ we can select $\hz_\beta$ disjoint sub-collections of supplier bundles, each corresponding to $\beta$.
    Furthermore, these sub-collections will be disjoint also across different $\beta$.

    For each such sub-collection, we apply \cref{bundleblock}.
    Each copy of $\beta = \beta_{i_\fe,k}$ repartitions $|\beta|$ supplier bundles into $|\beta|$ bundles acceptable for the agents of the corresponding class $\fe$.
    Since all sub-collections are disjoint, all resulting bundles will also be pairwise disjoint.
    Since for every  $\fe \in \Phi\setminus\fe_*$ we have $\sum_{\beta \in \Bb(\fe)}|\beta|\cdot \hz_\beta \leq n_\fe$, we can assign the resulting bundles to  distinct agents of the corresponding class.

    Denote this partial allocation $\{B_i\}_{i \in I}$, with allocated agents $I \subseteq N\setminus U$.
    It is clear that 
    \[
        |I| = \sum_{\beta \in \Bb}|\beta|\cdot \hz_\beta = n - |U| - \hZ.
    \]
    The allocation uses exactly $|I|$ bundles of the supplier's MMS partition $\{B_{u_*,1}, \ldots, B_{u_*,n}\}$ and allocates all their items.
    Hence, the unallocated items form exactly $n - |I| = |U| + \hZ$ intact supplier bundles.
    Finally, by construction, for every $p \in P$ the number intact supplier bundles with pattern $p$ is precisely
    \[
        \lambda_{u_*}(p) - \sum_{\beta \in \Bb}\lambda_\beta(p)\cdot \hz_\beta \geq \eta \lambda_{u_*}(p).
    \]
\end{proof}

We are now ready to prove the original \cref{intype}.
\begin{proof}[Proof of \cref{intype}]
    Construct the partial allocation $\{B_i\}_{i \in I}$ from \cref{realization}, and consider an arbitrary unallocated agent $i \notin U$.
    Recall that by \cref{acceptable}, the supplier's MMS partition $\{B_{u_*,1},\ldots, B_{u_*,n}\}$ contains at least $n/(qQ)$ bundles acceptable for $i$.
    Each of this acceptable bundles corresponds to some pattern $p$ with $v_i(p) \geq \mu_i$.
    Therefore by \cref{realization}, the number of supplier bundles, acceptable to $i$, that were not allocated to $I$, is at least
    \[
        \eta\cdot \sum_{\substack{p \in P\\v_i(p)\geq \mu_i}}\lambda_{u_*}(p) \geq \frac{\eta n}{qQ} = \frac{|U|}{qQ}.
    \]
    The last equality follows by definition of $\eta = |U|/n$.
    By \cref{bigsignature}, $|U|\geq n/\phi$, hence the number of remaining bundles acceptable for $i \notin U$ is at least
    \[
        \frac{|U|}{qQ}\geq \frac{n}{\phi qQ}.
    \]
    Now, suppose that $n$ satisfies
    \[
        n \geq \phi qQD_{t,q}(|P| + \phi - 1).
    \]
    Since $|\Phi| \leq \phi$, by \cref{introunding} the number $\hZ$ of unallocated agents outside $U$ satisfies
    \[
        \hZ \leq D_{t,q}(|P| + |\Phi| - 1) \leq \frac{n}{\phi q Q}.
    \]
    Hence, every unallocated agent $i\notin U$ accepts at least $\hZ$ of the remaining supplier bundles.

    Allocate these $\hZ$ agents in $i \in N\setminus (U\cup I)$ greedily, giving each $i$ an intact supplier bundle that $i$ finds acceptable.
    Since every such agent $i$ finds at least $\hZ$ remaining bundles acceptable, we will always be able to allocate a bundle.

    Afterwards, by \cref{realization} exactly $|U|$ intact supplier bundles remain.
    Since every agent $u \in U$ has the same signature as the supplier $u_*$, every remaining bundle is acceptable to every agent in $U$.
    Allocate them to $U$ arbitrarily.
    Together with $\{B_i\}_{i \in I}$ and earlier allocation of $N\setminus (U\cup I)$, we obtain  a full allocation, where every agent gets an acceptable bundle.

    Finally, since $\phi, |P|, D_{t,q}$ all depend only on $t, Q, q$, taking $n(t, Q, q) = \phi qQD_{t,q}(|P| + \phi - 1)$ finishes the proof.
    To be more explicit, we can take $D_{t,q} = |P|(t + 2)q^t$, which can be estimated similarly to \cref{blockpartitionconst} in \cref{appendix:constant}.
\end{proof}

\section{Exact MMS for integer values and unbounded types}

\label{sec:integer}

In this section, we prove the extension of \cref{intype}, where we remove the assumption that the number of types of items in $M$ is at most some constant $t$.
We still keep the assumptions that each item has a value in $\{1, \ldots, Q\}$, and that each bundle in every agent's MMS partition has size at most $q$.
This way, for every individual agent items split into at most $Q$ types, but the total number of types across different agents may depend on $n$ and $m$.

We will assume that $m\geq n$, as otherwise the statement is trivial.
\begin{theorem}\label{integer}
    Let $Q, q\geq 1$ be integers.
    There exists $n(Q, q)$ such that for every $n \geq n(Q, q)$ the following holds.
    Consider a fair allocation instance with $m$ items $M$ and $n$ agents $N$, where for every $i \in N$ valuation $v_i$ is additive, and
    \begin{itemize}
        \item for every agent $i \in N$ and item $e \in M$ we have $v_i(e) \in \{1, \ldots, Q\}$;
        \item every agent $i \in N$ has an MMS partition $\{B_{i,1}, \ldots, B_{i,n}\}$ satisfying $|B_{i,j}| \leq q$ for all $j \in [n]$.
    \end{itemize}
    There exists an allocation $(B_1, \ldots, B_n)$ of $M$ to $N$ such that for every agent $i \in N$ we have \[
    v_i(B_i)\geq \mms(M, v_i, n).\]
\end{theorem}
We will additionally assume in the setting of \cref{integer} that the instance is \textit{ordered}.
In particular, we associate $M$ with the set $[m]$, and for every agent $i \in N$ we simultaneously have
\[
    v_i(1) \geq v_i(2) \geq \ldots \geq v_i(m).
\]
By \cite{Bouveret2016}, when all valuations are additive, one can reduce the fair division instance to an ordered one, while preserving every agent's MMS value and sizes of MMS bundles.
This is a standard assumption when designing MMS-related algorithms for the additive case.

The core of our approach to proving \cref{integer} lies in the following observation.
Let $\{B_1, \ldots, B_n\}$ be some MMS allocation of $M$ to $N$.
Then, it must hold for every prefix $0 \leq \ell \leq m$:
\[
\sum_{i \in N}|B_i\cap [\ell]| \leq \ell.
\]
In other words, within every prefix $\{1, \ldots, \ell\}$ the allocation cannot use more than $\ell$ available items.
At the same time, agent $i$ need not receive some specific bundle $B_i$, he actually would be satisfied by any bundle with items of the same \textit{value}.

Our approach is, instead of allocating agent $i \in N$ some specific bundle $B_i$, first assign $i$ a \textit{profile of values} of the desired $B_i$, i.e a multiset $\{v_i(e) :\; e \in B_i\}$.
For each such multiset, we can measure how many items within the first $\ell$ can meet the prescribed values, and capture the condition above in an analogous manner.
Note that since for every $i \in N$, $v_i$ takes values in $\{1, \ldots Q\}$, and we consider only bundles of size at most $q$, the number of possible profiles across all agents will be only a constant, even if the number of types of items is unbounded.
This gives us additional flexibility when designing an allocation, as now we can work with just multisets of values, not associating them with any specific item.
After obtaining a feasible collection of profiles, assigned to corresponding agents, we will then translate it into a valid allocation.

\subsection{Demand profiles}

We will now give formal definitions of profiles and related terms, needed in our approach.

\begin{definition}
    A \textbf{demand} is a value $\alpha > 0$.
    We say that item $e \in M$ \textbf{satisfies }the demand $\alpha$ for agent $i \in N$ if $v_i(e) \geq \alpha$.
\end{definition}
\begin{definition}
    A \textbf{demand profile} is a finite multiset $\pi = \{\alpha_1, \ldots, \alpha_{|\pi|}\}$ of demands.
    A subset $B\subseteq M$ \textbf{satisfies} the demand profile $\pi$ for agent $i \in N$ if each demand $\alpha \in \pi$ can be assigned a distinct item $e\in B$ that satisfies it.
    That is, there exists $\{e_1, \ldots, e_{|\pi|}\}\subseteq B$ such that \[
        v_i(e_1) \geq \alpha_1,\ldots,  v_i(e_{|\pi|})\geq \alpha_{|\pi|}.
    \]
    A (guaranteed) \textbf{value of a profile} $\pi$ is 
    \[
        v(\pi) := \sum_{\alpha \in \pi}\alpha.
    \]
    If agent $i$ is assigned profile $\pi$, satisfying $\pi$ with some bundle $B\subseteq M$ guarantees $v_i(B) \geq v(\pi)$.
\end{definition}

Instead of measuring how many items a particular bundle $B\subseteq M$ uses in a given prefix $\{1, \ldots, \ell\}$, i.e  the value $|B\cap [\ell]|$, we will measure an analogous quntity with respect to demand profiles.
Given some profile $\pi$, we want to know for each agent $i$ how many items from each prefix $\{1, \ldots, \ell\}$ agent $i$ needs to satisfy $\pi$.
We will do so using a notion of \textit{breakpoints}.
\begin{definition}\label{breakpoint}
    For $i \in N$ and demand $\alpha$, an $\alpha$-\textbf{breakpoint} for agent $i$ is the number $m_i(\alpha)$ of items in $M=[m]$ with value at least $\alpha$.
    That is, for every $e \leq m_i(\alpha)$ it holds $v_i(e) \geq \alpha$.
\end{definition}

By definition, agent $i$ can satisfy a demand $\alpha$ only by the items $\{1, \ldots, m_i(\alpha)\}$.
Hence, if $m_i(\alpha) \leq \ell$, then agent $i$ must use an item in prefix $\{1, \ldots, \ell\}$ to satisfy $\alpha$.
Therefore, it is natural to define the \textit{$\ell$-prefix load of a profile} $\pi$ as a total number of breakpoints of $\pi$ before $\ell$.
\begin{definition}\label{prefixload}
    Let $\pi = \{\alpha_1, \ldots, \alpha_{|\pi|}\}$ be a demand profile.
    For an integer $0 \leq \ell \leq m$ and agent $i \in N$, denote by $\Pp_i(\pi, \ell)$ the total number of demands in $\pi$ with breakpoints at most $\ell$:
    \[
        \Pp_i(\pi, \ell) = |\{\alpha \in \pi : m_i(\alpha) \leq \ell\}|.
    \]
    That is, $\Pp_i(\pi, \ell)$ counts how many demands in $\pi$ can be satisfied only by the first $\ell$ items.
    We refer to $\Pp_i(\pi, \ell)$ as an \textbf{$\ell$-prefix load} of $\pi$ for $i$.
\end{definition}
We extend the definition of prefix load to a collection of profiles.
\begin{definition}
    Let $I\subseteq N$, let $\Pi_I := \{\pi_i\}_{i \in I}$ be a collection of demand profiles, where $\pi_i$ is assigned to agent $i \in I$.
    For $0 \leq \ell \leq m$, the \textbf{total $\ell$-prefix load} of $\Pi_I$ is
    \[
        \Pp(\Pi_I, \ell) := \sum_{i \in I}\Pp_i(\pi_i, \ell),
    \]
    the total number of demands across all agents in $I$ that can be satisfied only by the first $\ell$ items.
\end{definition}

The following lemma shows that finding an allocation of $M$ to some agents $I\subseteq N$ is equivalent to finding a feasible collection of appropriate demand profiles $\Pi_I$.
\begin{lemma}\label{loadequiv}
    Let $I\subseteq N$, let $\pi_i = \{\alpha_{i,1}, \ldots, \alpha_{i,|\pi_i|}\}$ be some demand profile of agent $i \in I$, and let $\Pi_I := \{\pi_i\}_{i \in I}$.
    For any set $E\subseteq M$ of items, we can simultaneously satisfy all profiles of $\Pi_I$ using distinct items of $E$ if and only if 
    \[
        \Pp(\Pi_I, \ell) \leq |E\cap [\ell]|,\qquad\qquad\forall 0 \leq \ell \leq m.
    \]
    Furthermore, if $K := \sum_{i \in I}|\pi_i|$, it suffices to use only the first $K$ items of $E$.
    In particular, when $E = M$, we can satisfy all profiles of $\Pi_I$ using only items $\{1, \ldots, K\}$ as long as
    \[
        \Pp(\Pi_I, \ell) \leq \ell,\qquad\qquad\forall 0 \leq \ell \leq m.
    \]
\end{lemma}
\begin{proof}
    Suppose that we can satisfy all profiles of $\Pi_I$ using distinct items of $E$.
    By definition,
    \[
        \Pp(\Pi_I, \ell) = \sum_{i \in I}\Pp_i(\pi_i, \ell) =\sum_{i \in I}|\{\alpha \in \pi_i : m_i(\alpha) \leq \ell\}|.
    \]
    Consider all $\alpha \in \Pi_I$ with breakpoints at most $\ell$ (for corresponding agents), each of these values must be satisfied by an item in $\{1,\ldots, \ell\}$.
    Every such demand gets a distinct item in $E$ satisfying it, so there are at most $|E\cap [\ell]|$ demands with breakpoints at most $\ell$.
    Thus, $\Pp(\Pi_I, \ell) \leq |E\cap [\ell]|$.

    Now, suppose that $\Pp(\Pi_I, \ell) \leq |E\cap [\ell]|$ for every $0 \leq \ell \leq m$.
    Taking $\ell = 0$ we obtain $\Pp(\Pi_I, 0) \leq |E\cap[0]| = 0$, so no demand in the assigned profiles has breakpoint $0$.
    Consequently, every demand has at least one eligible item.
    
    List all individual demands $\alpha \in \Pi_I$ as $(i, \alpha)$ where $\alpha\in \pi_i$, including repetitions, and sort them according to breakpoint values $m_i(\alpha)$.
    We get a list of $K = \sum_{i \in I}|\pi_i| =\Pp(\Pi_I, m)$ pairs
    \[
        (i_1, \alpha_1), \ldots, (i_K, \alpha_K), \qquad \qquad 1\leq m_{i_1}(\alpha_1) \leq \ldots \leq m_{i_K}(\alpha_K)\leq m.
    \]
   Note that $K = \Pp(\Pi_I, m) \leq |E\cap [m]|$ implies $K \leq |E|$.

    Denote $m_k := m_{i_k}(\alpha_k)$, $m_1 \leq \ldots \leq m_K$.
    By definition, for each $k$ there are at least $k$ pairs $(i, \alpha)$ with breakpoint $m_i(\alpha) \leq m_k$.
    This means that there are at least $k$ different demands that want items among the first $m_k$, so $\Pp(\Pi_I, m_k) \geq k$.
    
    By assumption, $\Pp(\Pi_I, m_k) \leq |E\cap [m_k]|$.
    It follows that, if $E= \{e_1 < \ldots < e_{|E|}\}$, then $e_k \leq m_k$.
    Take this $m_k = m_{i_k}(\alpha_k)$ and allocate item $e_k$ to agent $i_k$ to satisfy $\alpha_k$.
    As $m_k \geq e_k$, it must be that $v_{i_k}(e_k) \geq \alpha_k$.
    We do so for every $1 \leq k \leq K$.
    Since we will always use distinct items, and $K \leq |E|$, we will use only the first $K$ items of $E$.
    The lemma follows.
\end{proof}

So, to find an MMS allocation of items $M$ to $N$, it suffices to find a collection $\{\pi_i\}_{i \in N}$ of demand profiles, where agent $i \in N$ gets assigned a profile $\pi_i$, so that for every prefix $0 \leq \ell \leq m$,
\[
    \sum_{i \in N}\Pp_i(\pi_i, \ell) \leq \ell,
\]
and for every $i \in N$ we have $v(\pi_i) \geq \mu_i$.
This will be our goal for the rest of the section.

Our approach is to first find a sufficiently large set $I\subseteq N$ of agents and assign to them  some profiles $\Pi_I = \{\pi_i\}_{i \in I}$, such that $v(\pi_i) \geq \mu_i$ for every $i \in I$.
We will use this assignment as a baseline, and then complete it to a full one.
We will need two additional conventions.
\begin{definition}
    An \textbf{empty profile} is $\varnothing$.
    If agent $i \in N$ is assigned an empty  $\pi = \varnothing$, it means that he may receive no items.
    It always holds $\Pp_i(\varnothing, \ell) = 0$ for all $0 \leq \ell \leq m$.
\end{definition}
For now, this definition is mostly a formality, however it will become useful in the future.
\begin{definition}
    A \textbf{no-profile} is $\bot$.
    If agent $i \in N$ is assigned $\bot$, it means that we do not select this agent, and he receives no items at all.
    It always holds $\Pp_i(\bot, \ell) = 0$ for all $0 \leq \ell \leq m$.
\end{definition}
A ``no-profile'' $\bot$ corresponds to an option of not choosing $i \in N$ into the initial selection $I$.

We will also define a collection of MMS profiles for every agent $i \in N$.
Our allocation procedure will first try to find a large subset $I\subseteq N$, such that every agent $i \in I$ receives a profile $\pi_i$ corresponding to some bundle of $i$'s MMS partition.
We will not be able to guarantee such an assignment for every agent $i \in N$, however, the set $I$ will contain ``almost all'' agents.
\begin{definition}
    For every agent $i \in N$, let $\{B_{i,1},\ldots, B_{i,n}\}$ denote the partition of $M$ witnessing $\mms(M, v_i, n)$, for every $B_{i,j}$, it holds $v_i(B_{i,j}) \geq \mu_i$ and $|B_{i,j}|\leq q$.
    For every $j \in [n]$, define 
    \[
        \pi_{i,j} := \left\{v_i(e) :\; e \in B_{i,j}\right\}.
    \]
    Let $\Pi := \{\varnothing, \bot\} + \bigcup_{i \in N}\{\pi_{i,j} : j \in [n]\}$.
    Note that for all $\pi \in \Pi$ we have $|\pi|\leq q$ by assumption.
\end{definition}
Note that in the final assignment of profiles, there may be some agents $i \in N$ that receive a profile $\pi$ such that $\pi \neq \pi_{i,j}$ for any $j \in [n]$.

\subsection{$E/S$-partition and the buffer set}

\label{sec:ES}

In our previous work \cite{FG26}, we showed that it is possible to obtain an MMS allocation for at least $n - O(\log^2 n)$ agents, and we repeat the proof of this result later in \cref{sec:dmms} (\cref{dmms}).
The main question is, whether it possible to complete this allocation to a full one, and give acceptable bundles to the remaining $O(\log^2 n)$ agents, while keeping the already allocated agents satisfied?

We will translate this question into the language of demand profiles.
Suppose for now that we already obtained a large subset of agents $I\subseteq N$ (say, $|I| \geq n - O_{Q,q}(\log^2 n)$) with assigned profiles $\Pi_I = \{\pi_i\}_{i \in I}$, where each $\pi_i$ satisfies $\pi_i = \pi_{i,j}$ for some $j \in [n]$, and for all $0\leq \ell \leq m$,
\[
    \sum_{i \in I}\Pp_i(\pi_i, \ell)\leq \ell.
\]
Let $U = N \setminus I$ be the remaining unassigned agents.
We would like to give every $u \in U$ some new profile $\pi_u$ (not necessarily of the form $\pi_{u,j}$ for some $j \in [n]$), so that for every $0 \leq \ell \leq m$,
\[
    \sum_{i \in I}\Pp_i(\pi_i, \ell) + \sum_{u \in U}\Pp_u(\pi_u, \ell) \leq \ell.
\]
Satisfying this condition for every $\ell$ while keeping all profiles $\pi_i$ of $i \in I$ as they are may be impossible.
Instead, we will allow some small number of already assigned agents $i \in I$ to ``give up'' their current profiles $\pi_i$, and instead receive a new profile $\hpi_i$.
That is, we want to find a large subset $A\subseteq I$ of agents that will keep their assigned profiles, while agents $I \setminus A$ and $U$ will receive new profiles, so that for every $0 \leq \ell \leq m$ it holds
\[
    \sum_{i \in A}\Pp_i(\pi_i, \ell) + \sum_{i \in I \setminus A}\Pp_i(\hpi_i, \ell) + \sum_{u \in U}\Pp_u(\pi_u, \ell) \leq \ell.
\]

In order to tackle these constraints, we will impose some additional structure that 1) will guide us how to select this large set $A\subseteq I$ of agents who keep their assigned profiles, and 2) will ensure that it is possible to create replacement profiles $\hpi_i$ and new profiles $\pi_u$ so that the total prefix load is still bounded.
We will now give a high-level overview of our approach.

We will partition the items $M$ into $M = E \sqcup S$, where set $E$ contains most of items, and set $S$ is very small.
Observe that, for demands of agents $i \in N$ to fit together into a prefix $\ell$, it suffices that the $E$-parts of their profiles fit in $E\cap [\ell]$, and the $S$-parts of their profiles fit in $S\cap [\ell]$.
To be more concrete, we will consider decompositions into $E$- and $S$-sub-profiles.
\begin{definition}
    Consider an arbitrary partition $M = E \sqcup S$.
    An $E/S$-decomposition of a demand profile $\pi$ is a pair $\{\pi^E, \pi^S\}$ of sub-profiles such that \[
        \pi = \pi^E \uplus \pi^S,
    \] where $\uplus$ is union with multiplicities.
    Demands in $\pi^E$ are designated to be satisfied exclusively by items of $E$, and demands in $\pi^S$ are designated to be satisfied exclusively by items of $S$.
\end{definition}
\begin{remark}
    Note that, since we work with demand profiles and not actual bundles, in reality each demand of sub-profiles $\pi^E, \pi^S$ may be also satisfied by many items outside of $E, S$ respectively.
    However, for the purpose of the approach, our intension is to try and restrict $\pi^E, \pi^S$ only to items from their corresponding sets $E, S$, and find the collection of appropriate sub-profile pairs that indeed can be satisfied exclusively by respectively $E, S$ only.
\end{remark}
Observe that, for profiles $\{\pi_i\}_{i \in A}$, $\{\hpi_i\}_{i \in I \setminus A}$, $\{\pi_u\}_{u \in U}$ of agents $A$, $I \setminus A$, $U$ respectively to satisfy all prefix-load constraints, it suffices that each of these profiles has some $E/S$-decomposition, such that for every $0 \leq \ell \leq m$ it holds
\begin{gather}
    \sum_{i \in A}\Pp_i(\pi_i^E, \ell) + \sum_{i \in I \setminus A}\Pp_i(\hpi_i^E, \ell) + \sum_{u \in U}\Pp_u(\pi_u^E, \ell) \leq |E\cap [\ell]|;\tag{E}\label{E}\\
    \sum_{i \in A}\Pp_i(\pi_i^S, \ell) + \sum_{i \in I \setminus A}\Pp_i(\hpi_i^S, \ell) + \sum_{u \in U}\Pp_u(\pi_u^S, \ell) \leq |S\cap [\ell]|.\tag{S}\label{S}
\end{gather}
Indeed, $\Pp_i(\pi, \ell) = \Pp_i(\pi^E, \ell) + \Pp_i(\pi^S, \ell)$, and the same holds for $S$, so together (\ref{E}) and (\ref{S}) imply
\[
     \sum_{i \in A}\Pp_i(\pi_i, \ell) + \sum_{i \in I \setminus A}\Pp_i(\hpi_i, \ell) + \sum_{u \in U}\Pp_u(\pi_u, \ell)\leq |E\cap [\ell]| + |S\cap [\ell]| = \ell.
\]

We will first focus on how to ensure condition (\ref{E}).
Our choice of the small set $S$ will be dictated by the sufficient conditions that guarantee that we can add new profiles of agents $I \setminus A$ and $U$ without violating any prefix-load constraints with respect to $E$.

For convenience, we will use $\pi_u$ to denote profiles of both initially unassigned agents $U$ or the agents $I \setminus A$ who give up their profile from rounding.
So, 
\[
    N \setminus A = (I \setminus A)\cup U.
\]
We want to ensure that (\ref{E}) holds for an appropriate choice of agents $A\subseteq I$ who keep their profiles from rounding, and choice of new profiles $\pi_u$ for $u \in N \setminus A$.
That is, for every $0 \leq \ell \leq m$,
\begin{equation}
    \sum_{i \in A}\Pp_i(\pi_i^E, \ell) +  \sum_{u \in N \setminus A}\Pp_u(\pi_u^E, \ell) \leq |E\cap [\ell]|.\tag{E}\label{E'}
\end{equation}
Observe that the total $E$-load introduced by new profiles $\pi_u$ for $u \in N\setminus A$ can be upper bounded by the total number of demands in these profiles.
That is,
\[
    \sum_{u \in N \setminus A}\Pp_u(\pi_u^E, \ell) \leq \sum_{u \in N \setminus A}|\pi_u|.
\]
In addition, since $A\subseteq I$, the total load of assigned profiles $\pi_i$ for $i \in A$ is at most the total load of $\pi_i$ for all initially assigned $i \in I$, i.e including the old profiles of $I \setminus A$.
Consequently,
\[
    \sum_{i \in A}\Pp_i(\pi_i^E, \ell) +  \sum_{u \in N \setminus A}\Pp_u(\pi_u^E, \ell) \leq \sum_{i \in I}\Pp_i(\pi_i^E, \ell) +  \sum_{u \in N \setminus A}|\pi_u|.
\]
In particular, for (\ref{E'}) to hold, it suffices that for some integer $\nu \geq 1$, the old profiles $\pi_i$ for $i \in I$ and the new profiles  $\pi_u$ for $u \in N\setminus A$ satisfy
\begin{align}
    &&&\sum_{i \in I}\Pp_i(\pi_i^E, \ell) \leq |E\cap [\ell]| - \frac{\nu}{2},&\forall 0\leq \ell\leq m\tag{E-$\nu$}\label{Ebad}\\
    &&&\sum_{u \in N \setminus A}|\pi_u|\leq \frac{\nu}{2} \tag{A-$\nu$}\label{Ubad}.
\end{align}
We cannot expect uniform slackness (\ref{Ebad}) to hold at every prefix $0 \leq \ell \leq m$.
For example, if we keep increasing $\ell$, the available capacity $|E\cap [\ell]|$ stays the same even if $\ell \in S$.
Meanwhile, the $E$-load $\Pp_i(\pi_i^E, \ell)$ of some agent $i \in I$ may increase.
Indeed, even though a demand $\alpha \in \pi_i^E$ is designated to $E$, its breakpoint $m_i(\alpha)$ may be in $S$, and if $\ell$ passes $S$, the prefix load increases.

To deal with this issue, in addition to the set $S$, we will construct a small \textit{buffer set} $F\subseteq E$.
We will establish condition (\ref{Ebad}) at every prefix $\ell \in E\setminus F$.
For prefixes $\ell \in S\cup F$, we will instead enforce a \textit{stronger} prefix load constraint for a \textit{constant} number of carefully selected prefixes (which we will refer to as \textit{protected prefixes}).
These stronger load conditions will imply prefix feasibility over all $S\cup F$, and combined with (\ref{Ebad}) for $\ell \in E\setminus F$ will guarantee that the original condition (\ref{E'}) holds for every prefix $0 \leq \ell \leq m$.

\subsubsection{Buffer set construction}

Our set $S\subseteq M$ will be of the form
\[
    S = \{a_1, a_1 + 1,\ldots, b_1\} \cup \ldots \cup \{a_t, a_t + 1, \ldots, b_t\},
\]
where each $[a_s, b_s]$ for $s \in [t]$ is a maximal consecutive interval of items from $S$.
We will have $t \leq 2Q + 1$,  and the length of each consecutive interval and the total size of $S$ are both at most $O_{Q, q}(\log n)$.
In addition, it will hold that $a_1 = 1$ and $b_t = m$, i.e $S$ includes some prefix $\{1, \ldots, b_1\}$ and some suffix $\{a_t, \ldots, m\}$ of items $M$.

Take $E = M\setminus S$.
What is arguably more important for slackness bound is not necessarily the $S$ itself, but a \textbf{buffer set} $F\subseteq E$ defined by $S$.
Specifically, after every interval $[a_s, b_s]\subset S$, $s \in [t]$, we will have another interval $\{b_s + 1, \ldots, b_s + \nu\}$ of length $\nu = O_{Q, q}(\log n)$, and
\[
    F = E\cap (\{b_1 + 1, \ldots, b_1 + \nu\} \cup \ldots \cup \{b_t + 1, \ldots, b_t + \nu\}),
\]
or equivalently, taking $c_s := \min\{b_s + \nu, a_{s + 1} - 1\}$ for $s < t$, and using $b_t = m$,
\[
    F = \{b_1 + 1, \ldots, c_1\} \cup \ldots \cup \{b_{t - 1} + 1, \ldots, c_{t -1}\}.
\]
The set $S$ and parameter $\nu \geq 1$ are picked in such a way that
\begin{itemize}
    \item for every $\ell \in E \setminus F$, the whole interval $(\ell - \nu, \ell] = \{\ell - \nu + 1, \ldots, \ell\}\subseteq E$;
    \item at least $n/2$ agents have \textit{no}  breakpoints in $(\ell - \nu, \ell]$.
\end{itemize}
For the agents with no breakpoints in $(\ell - \nu, \ell]$, the prefix $E$-load stays constant over the whole interval $(\ell - \nu, \ell]$ of length $\nu$.
At the same time, the available $E$-capacity at $\ell$ increases by $\nu$.
Averaging over all agents $N$ gives slackness at least $\nu/2$ at this prefix $\ell$.

We will now describe the construction of $S$.

\begin{lemma}[Buffer set]\label{buffer}
    Let $\Qq: = [Q]$.
    Let $\nu \geq 1$ be an integer and $\eps \in (0, 1)$.
    There exist sets $S\subseteq M$ and $F\subseteq E := M\setminus S$ of items, and set $W\subseteq N$ of agents, with the following properties.
    \begin{itemize}
        \item $1, m \in S$, and $S$ is a union of at most $t_0=2Q + 1$ maximal consecutive intervals;
        \item sets $S$ and $F$ will be of the form
        \[
    S = \bigcup_{s = 1}^t\{a_s, \ldots, b_s\},\quad\quad F = \bigcup_{s = 1}^{t - 1}\{b_s + 1, \ldots, c_s\},\quad \text{where}\quad c_s := \min\{b_s + \nu, a_{s + 1} - 1\};
\]
        \item the sizes of the sets satisfy
        \[
            |S|\leq t_0\left(1 + \frac{Q\nu}{\eps}\right),\qquad |F|\leq t_0\nu,\qquad |W|\leq \eps n;
        \]
        \item set $W$ contains every agent $i \in N$ with a breakpoint in $F$, i.e
        \[
            W = \{i \in N: \; \exists \alpha \in \Qq\ \text{  s.t  }\ m_i(\alpha) \in F\};
        \]
        \item for every $\ell \in E\setminus F$, $(\ell - \nu, \ell]\subseteq E$, and the total number of breakpoints in $(\ell - \nu, \ell]$ is
        \[
            \big|\big\{(i, \alpha) \in N\times \Qq :\; \ell - \nu < m_i(\alpha) \leq \ell\big\}\big| \leq \frac{n}{2}.
        \]
    \end{itemize}
\end{lemma}
\begin{proof}
    For every agent $i \in N$, we list all possible positive breakpoints of $i$, i.e all $ m_i(\alpha) > 0$ for $\alpha \in \Qq$, and combine them into a multiset $\Mm(\Qq)$:
\[
    \Mm(\Qq) := \biguplus_{i \in N}\biguplus_{\substack{\alpha \in Q\\m_i(\alpha) > 0}}\{(i, m_i(\alpha))\}.
\]
    Note that $|\Mm(\Qq)| \leq Qn$.
    We sort $\Mm(\Qq)$ in non-decreasing order w.r.t $m_i(\alpha)$, and mark every $\kappa$-th breakpoint, where
    \[
        \kappa := \left\lfloor \frac{n}{2}\right\rfloor + 1.
    \]
    Note that any sub-interval of $\Mm(\Qq)$ with no marked breakpoints has length at most $\kappa - 1$.
    Let set $S_0$ contain $\{1, m\}$ and every unique position in $M$ corresponding to a marked breakpoint, i.e
\[
    S_0 = \{1, m\}\cup \big\{s \in \{1, \ldots, m\} : \exists i \in N \text{ s.t }(i, s) = (i_{j\kappa},    m_{j\kappa}) \in \Mm(\Qq) \text{ for some }j\geq 1\big\}.
\]
    Since $\kappa > n / 2$, the number of marked breakpoints is less than $2Q$, therefore, after adding $\{1, m\}$,
\[
    |S_0| \leq 2Q + 1 =:t_0.
\]
Any subinterval of $M$ disjoint from $S_0$ contains at most $\kappa - 1$ breakpoints across all agents.
Indeed, (possibly repeating) breakpoints that fall into a subinterval of $M$  form a consecutive interval in $\Mm(\Qq)$, and such interval without any marked breakpoint has length less than $\kappa$.

We construct a sequence of \textbf{protected regions}.
For integer radius $r \geq 1$, protected region $S(r)$ will contain all items within distance $r$ from $S_0$ (to the right):
\[
        S(r) := M\cap \bigcup_{s \in S_0}\{s, s + 1, \ldots, s + r -1\}.
\]
We grow the radius in increments of $\nu$.
For $k = 0, 1, \ldots, K$, where $K := \lfloor Q/\eps\rfloor + 1$, put
\[
    r_k := 1 + k\nu.
\]
By construction, sets $S(r_{k + 1}) \setminus S(r_{k})$ are pairwise disjoint for all $k \geq 0$.
Observe that there must exist some $0 \leq k^* \leq  K - 1$ such that at most $\eps n$ breakpoints of $\Mm(\Qq)$ fall into $S(r_{k^* + 1})\setminus S(r_{k^*})$.
Indeed, otherwise the total number of breakpoints falling into any $S(r_{k + 1})\setminus S(r_k)$ is more than
\[
    K\cdot \eps n = \left(\left\lfloor \frac{Q}{\eps}\right\rfloor + 1\right) \cdot \eps n > Qn.
\]
Take this $k^*$ and define sets
\[
    S:= S(r_{k^*}),\qquad F := S(r_{k^* + 1})\setminus S(r_{k^*}),\qquad E := M\setminus S.
\]
In addition, let $W$ be the agents who have a breakpoint in $F$, then $|W|\leq \eps n$.

Clearly, every $s \in S_0$ contributes to $S$ an interval of length at most $1 + (K - 1)\nu$, and growing this interval once more contributes at most $\nu$ items to $F$.
Hence,
\[
    |S|\leq |S_0|(1 + (K - 1)\nu) \leq t_0\left(1 + \frac{Q\nu}{\eps}\right),\qquad\qquad |F|\leq |S_0|\nu \leq t_0\nu.
\]
It follows that $S$ is a union of $t\leq t_0$ maximal consecutive intervals, i.e
\[
    S = \bigcup_{s = 1}^t\{a_s, \ldots, b_s\},\quad \text{where}\quad a_1= 1 \text{ and }b_t = m,
\]
and the buffer set $F$ is precisely
\[
    F = E\cap \bigcup_{s = 1}^{t}\{b_s + 1, \ldots, b_s + \nu\} = \bigcup_{s = 1}^{t - 1}\{b_s + 1, \ldots, c_s\}\quad\text{where}\quad c_s := \min\{b_s + \nu, a_{s + 1} - 1\}.
\]

Finally, for $\ell \in E\setminus F$, let $b_s \leq \ell$ be the endpoint of an interval of $S$ immediately preceding $\ell$, such endpoint exists as $1 \in S$.
Since $\ell \notin F$, we must have $\ell > b_s + \nu$.
Hence,  $(\ell - \nu, \ell] \subseteq (b_s, \ell]\subseteq E$.
This interval does not intersect $S_0\subseteq S$, so it must contain at most $\kappa - 1$ breakpoints of $\Mm(\Qq)$.
The lemma follows.
\end{proof}

We will apply \cref{buffer} with some integer $\nu \geq 1$ and $\eps \in (0, 1)$, and obtain a partition $M = E\sqcup S$, as well as buffer set $F\subseteq E$ and a small set $W\subseteq N$ of agents whose breakpoints fall into $F$.
These agents $W$ will require extra attention during the initial profile assignment.
Specifically, we want to ``forbid'' the selected profiles of $i \in W$ from using any items of $F$, in order to guarantee enough slackness at every prefix.

Each MMS profile $\pi_{i,j}$ now has a natural $E/S$-decomposition.
\begin{definition}
    Consider the partition $M = E \sqcup S$ from \cref{buffer}.
    For an agent $i \in N$ with partition $\{B_{i,1}, \ldots, B_{i,n}\}$ of $M$, for every MMS profile $\pi_{i,j} = \{v_i(e) : \; e \in B_{i,j}\}$ define
    \[
        \pi_{i,j}^E := \{v_i(e) :\; e \in B_{i,j} \cap E\}, \qquad\qquad
        \pi_{i,j}^S := \{v_i(e) :\; e \in B_{i,j}\cap S\}.
    \]
    Clearly, for all $\pi_{i,j}$ it holds $\pi_{i,j} = \pi_{i,j}^E \uplus \pi_{i,j}^S$.
\end{definition}
\begin{remark}
    The partition $M = E\sqcup S$ imposes additional restriction on the selected profiles.
    From now on, for a profile $\pi$ to be equal to $\pi_{i,j}$, it must also coincide on both $E$- and $S$-parts:
    \[
        \pi = \pi_{i,j} \quad \iff \quad (\pi^E, \pi^S) = (\pi_{i,j}^E, \pi_{i,j}^S).\]
    So, if $\pi$ and $\pi_{i,j}$ are the same as multisets, but have a different $E/S$-decomposition, we treat them as two different profiles.
    With this stronger equality notion, we define the common collection
    \[\Pi := \{\varnothing, \bot\} \cup \bigcup_{i \in N}\{\pi_{i,j} : j \in [n]\}.\]
\end{remark}

\begin{observation}\label{mmspart}
    For every $i \in N$ and every prefix $0 \leq \ell \leq m$ it holds
    \[
        \sum_{j = 1}^n\Pp_i(\pi_{i,j}^E, \ell) \leq |E\cap [\ell]|,\qquad\qquad \sum_{j = 1}^n\Pp_i(\pi_{i,j}^S, \ell) \leq |S\cap [\ell]|.
    \]
\end{observation}
\begin{proof}
    Observe that for every agent $i \in N$ and every $0 \leq \ell \leq m$ it holds that
    \[
        \sum_{j = 1}^n\Pp_i(\pi_{i,j}, \ell) \leq \ell.
    \]
    By definition $\Pp_i(\pi_{i,j}, \ell)$ counts how many demands in $\pi_{i,j}$ can be satisfied only by the first $\ell$ items, and the sum above counts the total number of such demands for agent $i$.
    Note that every demand in $\pi_{i,j}$ comes from a unique item of $B_{i,j}$, and MMS partition $\{B_{i,1},\ldots, B_{i,n}\}$ cannot use more than $\ell$ items in every $\ell$-prefix.
    Hence, are at most $\ell$ such demands across all profiles $\{\pi_{i,1},\ldots, \pi_{i,n}\}$, and $\sum_{j = 1}^n\Pp_i(\pi_{i,j}, \ell) \leq\ell$.
    By the same reasoning, there are at most $|E\cap [\ell]|$ $E$-demands and at most $|S\cap[\ell]|$ $S$-demands across all profiles $\{\pi_{i,j} : j \in [n]\}$ in every $\ell$-prefix, proving the two inequalities above.
\end{proof}

\subsubsection{Protected prefixes}

As mentioned earlier, the buffer $F$ will allow us to obtain enough slackness at every prefix $\ell \in E\setminus F$ in order to absorb the load of new profiles $\pi_u$ for $u \in N\setminus A$.
At the remaining prefixes, i.e $\ell \in S\cup F$, we will preserve a constant number of stronger prefix-load constraints.

We will now define the set of \textit{protected prefixes}.
Recall that
\[
    S = \bigcup_{s = 1}^t[a_s, b_s],\quad \text{where}\quad a_1= 1,\; b_t = m \quad\text{and}\quad t \leq 2Q+ 1.
\]
In addition, 
\[
    F = \bigcup_{s = 1}^{t - 1}(b_s, c_s]\quad\text{where}\quad c_s := \min\{b_s + \nu, a_{s + 1} - 1\}.
\]

Consider an interval $[a_s, b_s]\subset S$ for $s < t$.
Throughout this interval, the $E$-capacity $|E\cap [\ell]|$ stays constant, but the $E$-load of old/new profiles may increase.
However, we can control the load at the whole interval $[a_s, b_s]$ by controlling the load at a later \textit{protected prefix}.
Concretely, we can include the following buffer $(b_s, c_s]\subset F$, and preserve the $E$-load on the entire interval $[a_s, c_s]$ using a stronger prefix-load condition at the endpoint of $F$:
\[
    \sum_{i \in A}\Pp_i(\pi_i^E, c_s) + \sum_{u \in N\setminus A}\Pp_u(\pi_u^E, c_s) \leq |E\cap [b_s]|.
\]
Indeed, for every $\ell \in [a_s, c_s]$, the prefix-load at $\ell$ is at most the prefix-load at $c_s$.
Meanwhile
\begin{itemize}
    \item if $\ell \leq b_s$, we have $[\ell, b_s]\subset S$ and $E\cap [\ell] = E\cap [b_s]$,
    \item if $\ell \geq b_s$, then $[b_s]\subseteq [\ell]$, and $|E\cap [\ell]| \geq |E\cap [b_s]|$,
\end{itemize}
so the single constraint above enforces $E$-load feasibility at every prefix $\ell \in [a_s, c_s]$.

When $s = t$, i.e $[a_t, b_t] = [a_t, m]\subset S$, we need a slightly different endpoint to preserve the $E$-load on $[a_t, m]$.
Since this interval ends at $m$, every $e \in E$ satisfies $e \leq a_t$.
Hence, we can put
\[
    \sum_{i \in A}\Pp_i(\pi_i^E, m) + \sum_{u \in N\setminus A}\Pp_u(\pi_u^E, m) \leq |E|,
\]
i.e with endpoint $c_t = m$, to ensure that the total number of breakpoints of $E$-parts of the profiles does not exceed $|E|$.
This constraint controls the load at every $\ell \in [a_t, b_t]$.

\begin{definition}
    For the given sets $S, F \subseteq M$ of the form above, obtained from \cref{buffer} applied to some integer $\nu \geq 1$ and $\eps \in (0, 1)$, define the set $\Ll_S$ of $t$ \textbf{protected prefixes} as
    \[
        \Ll_S := \{(b_1, c_1), \ldots, (b_{t - 1}, c_{t - 1}), (b_t, c_t)\},\qquad \text{where}\qquad (b_t, c_t) = (m, m).
    \]
    Note that $|\Ll_S| \leq 2Q + 1$.
\end{definition}
For agents $A\subseteq I$, retaining their old profiles, and agents $N\setminus A$ receiving new profiles, the \textit{protected conditions} for every protected prefix $(b_s, c_s) \in \Ll_S$ are
\begin{equation}
    \sum_{i \in A}\Pp_i(\pi_i^E, c_s) + \sum_{u \in N\setminus A}\Pp_u(\pi_u^E, c_s) \leq |E\cap [b_s]|.\tag{$\mathrm{E}_{\Ll_S}$}\label{ES'}
\end{equation}

\begin{corollary}\label{protpref}
    Consider a partition $M = E\sqcup S$ obtained from \cref{buffer} applied to some integer $\nu \geq 1$ and $\eps \in (0, 1)$.
    Suppose that agents $I\subseteq N$ are assigned profiles $\{\pi_i\}_{i \in I}$ satisfying 
    \[
        \sum_{i \in I}\Pp_i(\pi_i^E, \ell) \leq \frac{1}{n}\sum_{i \in N}\sum_{j = 1}^n\Pp_i(\pi_{i,j}^E, \ell),\qquad\qquad \forall 0 \leq \ell \leq m.
    \]
    Let $A\subseteq I$, and to every agent $u \in N\setminus A$ give a new profile $\pi_u$.
    If $\{\pi_i\}_{i \in A}$ and $\{\pi_u\}_{u \in N\setminus A}$ satisfy
    \begin{align}
    &&&\sum_{i \in A}\Pp_i(\pi_i^E, c_s) + \sum_{u \in N\setminus A}\Pp_u(\pi_u^E, c_s) \leq |E\cap [b_s]|,&\forall (b_s, c_s) \in \Ll_S;\tag{$\mathrm{E}_{\Ll_S}$}\label{ES}\\
    &&&\sum_{u \in N \setminus A}|\pi_u|\leq \frac{\nu}{2} \tag{A-$\nu$}\label{Anu};
\end{align}
then for every prefix $0 \leq \ell \leq m$ it holds
\begin{equation}
    \sum_{i \in A}\Pp_i(\pi_i^E, \ell) +  \sum_{u \in N \setminus A}\Pp_u(\pi_u^E, \ell) \leq |E\cap [\ell]|.\tag{E}\label{Eprot}
\end{equation}
In particular, if $|\pi_u|\leq q$ for every $u \in N\setminus A$, it suffices that (\ref{ES}) holds with $|N\setminus A|\leq \nu/(2q)$.
\end{corollary}
\begin{proof}
    First, take $\ell \in E\setminus F$, and consider the interval $(\ell - \nu, \ell]$.
    By \cref{buffer}, $(\ell - \nu, \ell]\subseteq E$, and at least $n/2$ agents have no breakpoint in $(\ell - \nu, \ell]$.
    Denote these agents by $N_\ell$, then for every $i \in N_\ell$ \cref{mmspart} gives
    \[
        \sum_{j = 1}^n\Pp_i(\pi_{i,j}^E, \ell) = \sum_{j = 1}^n\Pp_i(\pi_{i,j}^E, \ell - \nu) \leq |E\cap [\ell - \nu]| = |E\cap[\ell]| - \nu.
    \]
    For every other $i\in N\setminus N_\ell$, the total $\ell$-prefix load of $E$-subprofiles $\pi_{i,j}^E$ is at most $|E\cap [\ell]|$.
    Hence,
    \begin{multline*}
        \frac{1}{n}\sum_{i \in N}\sum_{j = 1}^n\Pp_i(\pi_{i,j}^E, \ell) = \frac{1}{n}\sum_{i \in N_\ell}\sum_{j = 1}^n\Pp_i(\pi_{i,j}^E, \ell) + \frac{1}{n}\sum_{i \in N\setminus N_\ell}\sum_{j = 1}^n\Pp_i(\pi_{i,j}^E, \ell)\\
        \leq \frac{|N_\ell|}{n}\cdot (|E\cap[\ell]| - \nu) + \frac{|N\setminus N_\ell|}{n}\cdot |E\cap [\ell]| = \frac{|N|}{n}\cdot |E\cap[\ell]| - \frac{|N_\ell|}{n}\cdot \nu \\
        \leq |E\cap [\ell]| - \frac{\nu}{2}.
    \end{multline*}
    It follows from assumptions that, for any $A\subseteq I$, under condition (\ref{Anu}) it holds for this prefix $\ell$:
    \begin{multline*}
        \sum_{i \in A}\Pp_i(\pi_i^E, \ell) +  \sum_{u \in N \setminus A}\Pp_u(\pi_u^E, \ell) \leq \sum_{i \in I}\Pp_i(\pi_i^E, \ell) + \sum_{u \in N \setminus A}|\pi_u|\\
        \leq \frac{1}{n}\sum_{i \in N}\sum_{j = 1}^n\Pp_i(\pi_{i,j}^E, \ell) + \frac{\nu}{2} \leq |E\cap [\ell]| - \frac{\nu}{2} + \frac{\nu}{2} = |E\cap [\ell]|.
    \end{multline*}
    It remains to consider prefixes $\ell \in S\cup F$.
    Denote $\Pp^E(\ell) := \sum_{i \in A}\Pp_i(\pi_i^E, \ell) +  \sum_{u \in N \setminus A}\Pp_u(\pi_u^E, \ell)$ for short.
    Take the interval representation of $S$ and $F$.
    If $\ell \in [a_s, c_s]$ for  $s < t$, (\ref{ES}) ensures
    \[
        \Pp^E(\ell) \leq \Pp^E(c_s) \leq |E\cap [b_s]|\leq |E\cap [\ell]|,
    \]
    as $[\ell, b_s]\subset S$ for $\ell \leq b_s$, and $[b_s]\subseteq [\ell]$ for $\ell \geq b_s$.
    If $a_t \leq \ell \leq m$, the last protected prefix gives
    \[
        \Pp^E(\ell) \leq  \Pp^E(m) \leq |E| = |E\cap [\ell]|.
    \]
    Finally, since $1 \in S$ and $[b_1] \subseteq S$, we have $E\cap [b_1] = \varnothing$ and the first protected prefix gives $\Pp^E(c_1) \leq |E\cap [b_1]| = 0$.
    So, for $\ell = 0$ we have $\Pp^E(0)\leq \Pp^E(c_1) = 0$.
    The corollary follows.
\end{proof}

It follows that in order to guarantee (\ref{Eprot}), we need to find
\begin{itemize}
    \item a set of agents $I\subseteq N$ with profiles $\{\pi_i\}_{i \in I}$ such that their total $E$-load is at most the average total $E$-load of MMS profiles $\{\pi_{i,j} : j \in [n]\}$ for $i \in N$.
    We will find such set $I$ by rounding a specific collection of \textit{fractional profiles} using Linear Programming.
    \item a set $A\subseteq I$ of agents who keep old profiles, and new profiles $\pi_u$ for every $u \in N\setminus A$, satisfying (\ref{ES}) and (\ref{Anu}).
    This will be done later.
\end{itemize}

\subsection{Fractional profiles}

Instead of considering the profiles $\{\pi_{i,j} :\; j \in [n]\}$ directly, for every agent $i \in N$ we will define a \textit{distribution} over profiles $\Pi$.
We will refer to these distributions as \textit{fractional profiles}.
Our goal then will be first to define a collection of fractional profiles that, for every $0 \leq \ell \leq m$, fractionally satisfy the prefix load conditions
\[
    \sum_{i \in I}\Pp_i(\pi_i^E, \ell) \leq |E\cap [\ell]|,\qquad \text{and}\qquad \sum_{i \in I}\Pp_i(\pi_i^S, \ell) \leq |S\cap [\ell]|
\]
and then round it to an integral collection while preserving every prefix constraint.

\begin{definition}
    Let $\Pi$ be a collection of profiles, containing $\varnothing$ and $\bot$.
    A \textbf{fractional profile} of agent $i \in N$ is a pair $(x_i, \Pi)$, where $x_i(\pi) \geq 0$ for every $\pi \in \Pi$ and $\sum_{\pi \in \Pi}x_i(\pi) = 1$.
\end{definition}

We naturally extend the definition of $E/S$-prefix loads to fractional profiles.

\begin{definition}
    For agent $i \in N$ with fractional profile $(x_i, \Pi)$, his \textbf{fractional $\ell$-prefix $E/S$-load} of $\Pi$ for $0 \leq \ell \leq m$ is respectively
    \[
        \Hh_i^E(x_i, \Pi, \ell) := \sum_{\pi\in \Pi}x_i(\pi)\cdot \Pp_i(\pi^E, \ell),\qquad\text{and}\qquad \Hh_i^S(x_i, \Pi, \ell) := \sum_{\pi\in \Pi}x_i(\pi)\cdot \Pp_i(\pi^S, \ell).
    \]
    Assume every agent $i \in N$ has fractional profile $(x_i, \Pi)$ for the shared set of profiles $\Pi$, and let $x = \{x_i\}_{i \in N}$.
    For $0 \leq \ell \leq m$, the \textbf{total fractional $\ell$-prefix $E/S$-load} of $\Pi$ for $x$ is respectively
    \begin{gather*}
        \Hh^E(x, \Pi, \ell) := \sum_{i \in N}\Hh_i^E(x_i, \Pi, \ell) = \sum_{i \in N}\sum_{\pi\in \Pi}x_i(\pi)\cdot \Pp_i(\pi^E, \ell);\\
        \Hh^S(x, \Pi, \ell) := \sum_{i \in N}\Hh_i^S(x_i, \Pi, \ell) = \sum_{i \in N}\sum_{\pi\in \Pi}x_i(\pi)\cdot \Pp_i(\pi^S, \ell).
    \end{gather*}
\end{definition}
 We will be interested in the expected number of agents $i \in N$ that are assigned a non-$\bot$ profile according with respect to fractional profiles $(x_i, \Pi)$.
\begin{definition}
    Assume that every agent $i \in N$ has fractional  profile $(x_i, \Pi)$ for the shared set of profiles $\Pi$, and let $x = \{x_i\}_{i \in N}$.
    The \textbf{fractional number of assigned profiles} with respect to $(x, \Pi)$ is
    \[
        \Ii(x, \Pi) := \sum_{i \in N}\sum_{\pi\neq\bot}x_i(\pi).
    \]
\end{definition}

For every agent $i \in N$ we will define a fractional profile $(\hx_i, \Pi)$, that we will later round to an integral profile.
A natural way to do it would be to simply take a uniform distribution over all profiles, and set for every $\pi \in \Pi$
\[
    \hx_i(\pi):= \frac{|\{j \in [n]:\; \pi_{i,j} = \pi\}|}{n}
\]
It is easy to see by \cref{mmspart} that these profiles actually satisfy for all $0 \leq \ell \leq m$
\[
    \Hh^E(\hx, \Pi, \ell) \leq |E\cap [\ell]|,\qquad \text{and}\qquad \Hh^S(\hx, \Pi, \ell) \leq |S\cap [\ell]|.
\]

However, this will not be good enough for us.
In addition to these conditions, we would like guarantee that the fractional $(\hx, \Pi)$ also provides slackness, satisfying a stronger prefix load condition for every protected prefix $(b_s, c_s) \in \Ll_s$:
\[
    \Hh^E(\hx, \Pi, c_s) \leq |E\cap [b_s]|.
\]
This will allow us to satisfy the condition (\ref{ES}) fractionally, so that rounding gives us $I\subseteq N$ also satisfying
\[
    \sum_{i \in I}\Pp_i(\pi_i^E, c_s) \leq |E\cap [b_s]|,
\]
for corresponding collection $\{\pi_i\}_{i \in I}$.
It will be important for us later that agents $I$ and their profiles satisfy the condition above.

In order to obtain these stronger fractional $E$-load bounds at protected prefixes, we will omit some profiles from the fractional selection.
Concretely, if agent $i \in W$ and his MMS bundle $B_{i,j}$ contains items from the buffer $F$, we will not use the corresponding profile $\pi_{i,j}$, and also move its uniform weight $1/n$ to $\hx_i(\bot)$.
This will allow us to satisfy the protected prefix conditions (\ref{ES}) fractionally for $(\hx, \Pi)$.
By \cref{buffer}, both sets $W$ and $F$ are small, so omitting the corresponding profiles still leaves a lot of total weight $\Ii(\hx, \Pi)$.

\begin{definition}
    For every agent $i \in N$, define the following fractional profile $(\hx_i, \Pi)$.
    \begin{itemize}
        \item If $i \in N \setminus W$, 
        set $\hx_i(\varnothing) = 0$, $\hx_i(\bot) = 0$, and for every $\pi \in \Pi\setminus\{\varnothing, \bot\}$ let
        \[
            \hx_i(\pi) = \frac{|\{j \in [n]:\; \pi_{i,j} = \pi\}|}{n}.
        \]
        \item if $i \in W$, set $\hx_i(\varnothing) = 0$, for every $\pi \in \Pi\setminus \{\varnothing, \bot\}$, let
        \[
            \hx_i(\pi) = \frac{|\{j \in [n]:\; \pi_{i,j} = \pi \text{ and }B_{i,j}\cap F = \varnothing\}|}{n},
        \]
        as well as
        \[
            \hx_i(\bot) = \frac{|\{j \in [n] :\; B_{i,j} \cap F\neq \varnothing\}|}{n}.
        \]
    \end{itemize}
\end{definition}
Thus, in the fractional collection $(\hx, \Pi)$ every $\pi_{i,j}$ that corresponds to a bundle $B_{i,j}$ of some agent $i \in W$ that contains items from the buffer $F$, contributes its weight $1/n$ to $\bot$ instead.

We will next show that the fractional profiles $(\hx, \Pi)$ satisfy the requirements of \cref{protpref}, as well as condition (\ref{S}), fractionally.

\begin{lemma}\label{fracload}
    The number $\hX$ of strictly fractional coordinates in $\hx$ satisfies $\hX \leq n^2$. 
    Moreover,
    \[
        \Ii(\hx, \Pi) \geq  n  - \eps |F|,
    \]
    In addition, for every prefix $0 \leq \ell \leq m$ it holds
    \begin{gather*}
        \Hh^E(\hx, \Pi, \ell) \leq \frac{1}{n}\sum_{i \in N}\sum_{j = 1}^n\Pp_i(\pi_{i,j}^E, \ell) \leq |E\cap [\ell]|;\\
        \Hh^S(\hx, \Pi, \ell) \leq \frac{1}{n}\sum_{i \in N}\sum_{j = 1}^n\Pp_i(\pi_{i,j}^S, \ell) \leq|S\cap [\ell]|;
    \end{gather*}
    and for every protected prefix $(b_s, c_s)  \in \Ll_S$ it holds
    \[
        \Hh^E(\hx, \Pi, c_s) \leq |E\cap [b_s]|.
    \]
\end{lemma}
\begin{proof}
    For every agent $i \in N$, there are at most $n$ positive weights $\hx_i(\pi)$ for $\pi = \pi_{i,j}$.
    If $\hx_i(\bot) > 0$, at least one profile $\pi_{i,j}$ contributes its individual weight to $\bot$ (note that some $\hx_i(\pi_{i,j})$ may still be positive, as multiple $B_{i,j}$ can induce the same profile).
    There are at most $n - 1$ other $\pi_{i,j}$ contributing to non-$\bot$ coordinates, with a total of at most $n$ positive $\hx_i(\pi)$.
    Thus, $\hX \leq  n^2$.
    
    Next, by construction, we have
    \[
        \Ii(\hx, \Pi) = \sum_{i \in N}\sum_{\pi\neq \bot}\hx_i(\pi) = |N| - \sum_{i \in N}\hx_i(\bot) 
         = n   - \frac{1}{n}\sum_{i \in W}|\{j \in [n] :\; B_{i,j} \cap F \neq \varnothing\}|.
    \]
    For every agent $i \in W$, there are at most $|F|$ MMS bundles that contain any item from $F$.
    Hence, using the bound on $|W|$ from \cref{buffer},
    \[
        \frac{1}{n}\sum_{i \in W}|\{j \in [n] : \; B_{i,j} \cap F \neq \varnothing\}| \leq \frac{|W|}{n}\cdot |F| \leq \frac{\eps  n}{n}\cdot |F| = \eps |F|. 
    \]
    Thus,
    \[
        \Ii(\hx, \Pi) \geq n - \eps|F|.
    \]

    Moving to the prefix load, each $\pi_{i,j}$ contributes at most $1/n$ to fractional load, so
    \[
         \Hh^E(\hx, \Pi, \ell) = \sum_{i \in N}\sum_{\pi \neq \bot}\hx_i(\pi)\cdot \Pp_i(\pi^E, \ell)\leq \frac{1}{n}\sum_{i \in N}\sum_{j = 1}^n\Pp_i(\pi_{i,j}^E, \ell) \leq |E\cap [\ell]|,
    \]
    where the last inequality follows from \cref{mmspart}.
    The bounds for $S$ are analogous.

    Finally, fix a protected prefix $(b_s, c_s)  \in \Ll_s$ for $s < t$.
    For agent $i \in N$, consider his MMS profile $\pi_{i,j} \neq \varnothing$, and let $\alpha \in \pi_{i,j}^E$ be its demand counted at prefix $c_s$, i.e $m_i(\alpha) \leq c_s$.
    If $e \in B_{i,j}\cap E$ is the originating item of $\alpha$, then $\alpha = v_i(e)$ and $e \leq m_i(\alpha) \leq c_s$.
    Hence,
    \begin{itemize}
        \item if $i \notin W$, breakpoint $m_i(\alpha)$ cannot belong to $(b_s, c_s]\subset F$, so $m_i(\alpha)\leq b_s$ and thus $e \leq b_s$;
        \item if $i \in W$ and $B_{i,j}\cap F = \varnothing$, then $e\notin F$, and as $e \leq c_s$ and $(b_s, c_s]\subset F$, necessarily  $e \leq b_s$.
    \end{itemize}
    Thus, whenever whenever $i \notin W$ or $i \in W$ and $B_{i,j}\cap F = \varnothing$, we must have $e\leq b_s$.
    Therefore, by construction of $(\hx_i, \Pi)$, every $E$-demand counted at prefix $c_s$, for a profile $\pi_{i,j}$ that does not contribute to $\bot$, must originate from an item in $E\cap [b_s]$.
    And, all these items must be distinct, since $\pi_{i,j}$ are induced by different MMS bundles $B_{i,j}$, $j \in [n]$.
    It follows that
    \[
        \sum_{\pi \neq \bot}\hx_i(\pi)\cdot \Pp_i(\pi^E, c_s) = \frac{1}{n}\sum_{j= 1}^n\Pp_i(\pi_{i,j}^E, c_s)\One\{i \notin W \text{ or }B_{i,j}\cap F = \varnothing\} \leq \frac{1}{n}\cdot |E\cap [b_s]|,
    \]
    and, summing over all $i \in N$,
    \[
        \Hh^E(\hx, \Pi, c_s) \leq |E\cap [b_s]|.
    \]
    For the last protected prefix $(b_t, c_t) = (m, m)$, the earlier bound already gives
    \[
        \Hh^E(\hx, \Pi, c_t) = \Hh^E(\hx, \Pi, m) \leq |E| = |E\cap[b_t]|.
    \]
    This finishes the proof.
\end{proof}

We will now round the fractional profiles $(\hx, \Pi)$ to integral $(y, \Pi)$, that will define a large subset $I\subseteq N$ of selected agents with assigned profiles $\{\pi_i\}_{i \in I}$, satisfying the requirements of \cref{protpref}.
We will do it using the following theorem.

\begin{theorem}[Joint rounding]\label{rounding}
    For a partition $M = E\sqcup S$, let $\Pi$ be a shared set of profiles with well-defined $E/S$-decomposition, such that for every $\pi \in \Pi\setminus \bot$ it holds $|\pi|\leq q$ for some integer $q \geq 1$.
    For every $i \in N$, let $(x_i, \Pi)$ be a fractional profile, denote $x = \{x_i\}_{i \in N}$. 

    There exists a collection $y = \{y_i\}_{i \in N}$ of integeral profiles $(y_i, \Pi)$, such that for every $i, \pi$ we have $y_i(\pi) \in \{0, 1\}$ and $y_i(\pi) = 1$ for $\pi \neq \bot$ only if $x_i(\pi) > 0$, and
    \[\begin{aligned}
        & && \Hh^E(y, \Pi, \ell) \leq \Hh^E(x, \Pi, \ell),  & \forall 0 \leq \ell \leq m,\\
        & && \Hh^S(y, \Pi, \ell) \leq \Hh^S(x, \Pi, \ell),  & \forall 0 \leq \ell \leq m,\\
        &  && \Ii(y, \Pi)\geq \Ii(x, \Pi) - \rho_{ES}(q, X), 
    \end{aligned}\]
    where, for $X= |\{(i, \pi):0 < x_i(\pi) <1 \}|$,
    \[
        \rho_{ES}(q, X) =  8q\log \max\{1, X\}.
    \]
\end{theorem}
Its proof is  presented in \cref{appendix:rounding} (\cref{roundingproof}).

\begin{corollary}\label{ESrounding}
    Let $M =E\sqcup S$, $F\subseteq E$ and $W\subseteq N$ be obtained from \cref{buffer} applied with some integer $\nu \geq 1$ and $\eps \in (0, 1)$, and let $(\hx, \Pi)$ be the fractional profiles defined earlier.

    There exist a set $I\subseteq N$ and profiles $\{\pi_{i}\}_{i \in I}$, where every $\pi_i$ has an $E/S$ decomposition $\pi_i = \pi_i^E \uplus \pi_i^S$, and satisfies $|\pi_i|\leq q$, $\alpha \in \Qq$ for all $\alpha \in \pi_i$, and $v(\pi_i) \geq \mu_i$.
    In addition,
    \[
        |I|\geq n - \eps|F| - 16q\log n
    \]
    and for every prefix $0 \leq \ell \leq m$ it holds
    \[
        \sum_{i \in I}\Pp_i(\pi_i^E, \ell) \leq \frac{1}{n}\sum_{i \in N}\sum_{j = 1}^n\Pp_i(\pi_{i,j}^E, \ell),\qquad \qquad \sum_{i \in I}\Pp_i(\pi_i^S, \ell) \leq |S\cap [\ell]|,
    \]
    as well as for every protected prefix $(b_s, c_s) \in \Ll_S$,
    \[
        \sum_{i \in I}\Pp_i(\pi_i^E, c_s) \leq |E\cap [b_s]|.
    \]
\end{corollary}
\begin{proof}
    By construction, profile $\pi \in \Pi\setminus \{\varnothing, \bot\}$ has a well-defined $E/S$ decomposition, and since every MMS bundle has size at most $q$, it holds $|\pi|\leq q$.
    We apply \cref{rounding} to $(\hx, \Pi)$ and obtain integral profiles $(y, \Pi)$.
    Let $I\subseteq N$ contain all $i \in N$ with $y_i(\bot)=  0$, and take $\pi_i \in \Pi\setminus \bot$ satisfying $y_i(\pi_i) = 1$ for this $i$.
    Since $y_i(\pi_i) > 0$ only if $\hx_i(\pi_i) > 0$, it must hold that $\pi_i = \pi_{i,j}$ for some $j \in [n]$, therefore $v(\pi_i) \geq \mu_i$, for all  $i \in I$.

    Next, using the bounds of \cref{fracload} on $\hX\leq n^2$ and $\Ii(\hx, \Pi)$, \cref{rounding} gives us
    \[
        |I| = \Ii(y, \Pi) \geq \Ii(\hx, \Pi) - \rho_{ES}(q, \hX) \geq n - \eps|F| - \rho_{ES}(q, n^2) = n - \eps|F| - 16q\log n.
    \]

    Next, for  every prefix $0 \leq \ell \leq m$, we have by \cref{rounding}:
    \[
        \sum_{i \in I}\Pp_i(\pi_i^E, \ell) = \Hh^E(y, \Pi, \ell) \leq \Hh^E(\hx, \Pi, \ell); \qquad
        \sum_{i \in I}\Pp_i(\pi_i^S, \ell) = \Hh^S(y, \Pi, \ell) \leq \Hh^S(\hx, \Pi, \ell).
    \]
    Combining these inequalities with \cref{fracload}  proves
    \[
        \sum_{i \in I}\Pp_i(\pi_i^E, \ell) \leq \frac{1}{n}\sum_{i \in N}\sum_{j = 1}^n\Pp_i(\pi_{i,j}^E, \ell),\qquad \qquad \sum_{i \in I}\Pp_i(\pi_i^S, \ell) \leq |S\cap [\ell]|,
    \]
    and for every protected prefix $(b_s, c_s)  \in \Ll_S$,
    \[
        \sum_{i \in I}\Pp_i(\pi_i^E, c_s) \leq  \Hh^E(\hx, \Pi, c_s) \leq |E\cap [b_s]|.
    \]
\end{proof}

\section{Single supplier}

\label{sec:intsupplier}

Recall our original goals, described in \cref{sec:ES}.
Given a large subset of agents $I\subseteq N$ with assigned integral profiles $\Pi_I = \{\pi_i\}_{i \in I}$, we want to find a subset $A\subseteq I$ of agents that will keep their assigned profiles, and give a new profile $\pi_u$ to every agent $u \in N \setminus A$, so that for every $0 \leq \ell \leq m$ it holds
\[
    \sum_{i \in A}\Pp_i(\pi_i, \ell) + \sum_{u \in N\setminus A}\Pp_u(\pi_u, \ell) \leq \ell.
\]
As discussed earlier, to get this condition for every prefix it suffices that, for an arbitrary partition $M= E\sqcup S$, for every $0 \leq \ell \leq m$ we have
\begin{align}
    &&&\sum_{i \in A}\Pp_i(\pi_i^E, \ell) + \sum_{u \in N\setminus A}\Pp_u(\pi_u^E, \ell) \leq |E\cap [\ell]|,&\forall 0 \leq \ell \leq m. \tag{E}\label{E2}\\
    &&&\sum_{i \in A}\Pp_i(\pi_i^S, \ell) + \sum_{u \in N\setminus A}\Pp_u(\pi_u^S, \ell) \leq |S\cap [\ell]|,&\forall 0 \leq \ell \leq m. \tag{S}\label{S2bad}
\end{align}

Suppose that we applied \cref{buffer} with some integer $\nu \geq 1$ and $\eps \in (0, 1)$, and obtained a partition $M = E\sqcup S$, as well as protected prefixes $\Ll_S$.
Assume that the originally selected integral profiles $\Pi_I = \{\pi_i\}_{i \in I}$ are the ones from \cref{ESrounding}.
As shown in \cref{sec:ES}, \cref{buffer} and \cref{protpref}, to satisfy the conditions (\ref{E2}) and (\ref{S2bad}) above for every prefix $0 \leq \ell \leq m$, it suffices to find $A\subseteq I$ and profiles $\pi_u$ for $u \in N\setminus A$ such that
\begin{align}
    &&&\sum_{i \in A}\Pp_i(\pi_i^E, c_s) + \sum_{u \in N\setminus A}\Pp_u(\pi_u^E, c_s) \leq |E\cap [b_s]|,&\forall (b_s, c_s) \in \Ll_S;\tag{$\mathrm{E}_{\Ll_S}$}\label{ES2}\\
    &&&\sum_{u \in N \setminus A}|\pi_u|\leq \frac{\nu}{2} \tag{A-$\nu$}\label{Anu2};\\
    &&&\sum_{i \in A}\Pp_i(\pi_i^S, \ell) + \sum_{u \in N\setminus A}\Pp_u(\pi_u^S, \ell) \leq |S\cap [\ell]|,&\forall 0 \leq \ell \leq m. \tag{S}\label{S2}
\end{align}
By \cref{protpref}, conditions (\ref{ES2}) and (\ref{Anu2}) applied to $I\subseteq N$ from \cref{ESrounding} give (\ref{E2}).

By \cref{ESrounding}, the number of already assigned agents is $|I| \geq n -  \eps|F| - 16q\log n$, which, assuming that $|F| = O_{Q ,q}(\log n)$, is at least $n - O_{Q ,q}(\log n)$.
We would like to retain most of these profiles $\{\pi_i\}_{i \in I}$, and replace only a small number of them, say $O_{Q ,q}(\log n)$, and assign new profiles to the remaining $O_{Q ,q}(\log n)$ ones, to satisfy (\ref{ES2}), (\ref{Anu2}) and (\ref{S}).
This way, each agent $i \in N$ will receive a profile, so that the whole collection of profiles is simultaneously satisfiable by distinct items of $M$.

We will inevitably encounter several issues that we briefly mentioned in \cref{sec:integer}, but set aside earlier.
First, we would like to assign to every agent $i \in N$ a profile that is acceptable, i.e of value at least $\mu_i$.
The already assigned profiles $\pi_i$ for $i \in I$ satisfy $v(\pi_i) \geq \mu_i$, and we need to guarantee the same for every new profile $\pi_u$, satisfying just (\ref{ES2}), (\ref{Anu2}) and (\ref{S}) is not enough.

This leads us to the second major issue: coordination.
Choosing a new profile for every agent independently provides little control over the combined load, as well as which items are required to simultaneously satisfy these profiles.
For example, if we create a new profile for $i \in N$ using agent $i$'s personal MMS partition, different agents may require the same items in order to satisfy their profiles, which may violate (\ref{ES2}) and (\ref{S}).

\subsection{Suggested profiles}

We will resolve both of these issues using the approach we used for the proof in \cref{sec:intype}.
We will pick a special agent $u_* \in N$, which we will refer to as \textbf{supplier}.
This agent will use his personal MMS partition $\{B_{u_*,1}, \ldots, B_{u_*, n}\}$ in order to create a \textbf{suggested profile} $\hpi_{u_*,i}$ for every agent $i \in N$.
The supplier's MMS partition will provide a point of reference, which will allow us to simultaneously control the load associated with different bundles.
The suggested profiles $\hpi_{u_*,i}$ may reorganize the items in the partition $\{B_{u_*,1}, \ldots, B_{u_*, n}\}$ and adapt to every individual agent $i \in N$, while preserving the prefix feasibility constraints.

\begin{definition}
    Fix a supplier $u_* \in N$.
    For each agent $i \in N$, his \textbf{suggested profile} is
    \[
        \hpi_{u_*,i} := \hpi_{u_*,i}^E \uplus \hpi_{u_*,i}^S,
    \]
    where each part is intended to be satisfied using exclusively distinct items of $E, S$ respectively.
\end{definition}

We will now partially explain how supplier $u_*$ constructs suggested profiles for agents $i \in N$.

\begin{definition}\label{excep}
    For a supplier $u_* \in N$ with MMS partition $\{B_{u_*,1}, \ldots, B_{u_*, n}\}$, we call a bundle $B_{u_*,j}$, $j \in [n]$, \textbf{exceptional}, if
    \[
        B_{u_*,j}\cap S \neq \varnothing.
    \]
    Note that the total number of exceptional bundles is at most $|S|$.
\end{definition}
When constructing suggested profiles for agents $i \in N$, actions of the supplier $u_*$ differ based on which bundles of $\{B_{u_*,1}, \ldots, B_{u_*, n}\}$ are exceptional and which are not.
Concretely:
\begin{itemize}
    \item For every exceptional bundle $B_{u_*,j}$, we will choose a distinct recipient $i \in N$ that agrees with $u_*$ on values of every item in $S$, i.e $v_{u_*}(e) = v_i(e)$ for all $e \in S$.
    We will give $i$ a suggested profile $\hpi_{u_*,i} := \pi_{u_*,j} = \{v_{u_*}(e) :\; e \in B_{u_*,j}\}$, and make sure that it is acceptable for agent $i$, in particular, that $v(\hpi_{u_*,i}^E) + v(\hpi_{u_*,i}^S) \geq \mu_i$.
    \item All non-exceptional bundles $B_{u_*,j}$ satisfy $B_{u_*,j}\subseteq E$.
    For the remaining agents, we will reorganize the $E$-parts $B_{u_*,j}\cap E$ of non-exceptional bundles, obtaining a collection of suggested profiles $\hpi_{u_*,i}$ such that $\hpi_{u_*,i}^S = \varnothing$, i.e $\hpi_{u_*,i} = \hpi_{u_*,i}^E$, and for every recipient $i$ we have $v(\hpi_{u_*,i}^E)\geq \mu_i$.
    This will be done via an analogue of \cref{sec:intype}.
\end{itemize}

It will be convenient for us to deal with $E$- and $S$-parts of suggested profiles separately.
We will first describe the $S$-part construction formally, assuming that we already know which exceptional bundle goes to which agent.
\begin{definition}\label{suggestpart}
    Fix a supplier $u_* \in N$, and designate exceptional bundles of supplier's MMS partition $\{B_{u_*, 1}, \ldots, B_{u_*,n}\}$ to distinct recipients $i \in N$, call these agents \textbf{exceptional}.

    For every exceptional agent $i$ with designated bundle $B_{u_*,j}$, define
    \[
        \hpi_{u_*,i}^S:=  \{v_{u_*}(e):\; e \in B_{u_*,j}\cap S\}.
    \]
    For every non-exceptional agent $i$, set $\hpi_{u_*,i}^S = \varnothing $.
\end{definition}
Note that the values in suggested profile $\hpi_{u_*,i}$ are according to $u_*$ and not to $i$.
A natural question is, whether assigning such a profile to agents $i \in N$ is still feasible, i.e they will be able to simultaneously satisfy all profiles using distinct items.
We show that, for every suggested profile $\hpi_{u_*,i}$, as long as the recipient $i$ is ``compatible'' with its designated exceptional bundle $B_{u_*,j}$, the $S$-parts the suggested profiles together are feasible in $S$ for any subset of agents $N$.

\begin{observation}\label{compatible}
    For the suggested parts $\hpi_{u_*,i}^S$, $i \in N$, defined above, for every $0 \leq \ell \leq m$:
    \[
        \sum_{i \in N}\Pp_{u_*}(\hpi_{u_*,i}^S, \ell) \leq |S\cap [\ell]|.
    \]
    Furthermore, if additionally every agent $i \in N$ satisfies \textbf{$S$-compatibility} condition
    \[
        \hpi_{u_*,i}^S \neq \varnothing \quad \implies \quad \forall e \in S:\; v_i(e) = v_{u_*}(e),
    \]
    then for every prefix $0 \leq \ell \leq m$ we have
    \[
        \sum_{i \in N}\Pp_i(\hpi_{u_*,i}^S, \ell) \leq |S \cap [\ell]|.
    \]
\end{observation}
\begin{proof}
    Since $\hpi_{u_*,i}^S$ are constructed from supplier's MMS partition, every demand $\alpha$ counted by supplier's prefix load $\Pp_{u_*}(\hpi_{u_*,i}^S, \ell)$ has an originating item $e \in S$ satisfying $e \leq m_{u_*}(\alpha) \leq \ell$.
    And, all these items are distinct across all suggested $S$-profiles.
    This proves the the first inequality.

    If agent $i \in N$ has $v_i(e) = v_{u_*}(e)$ for every $e \in S$, then any demand of the form $\alpha = v_{u_*}(e)$ satisfies $S\cap [m_{u_*}(\alpha)] = S\cap [m_i(\alpha)]$.
    Consequently, any item of $S$ satisfying $\alpha$ for $u_*$ also satisfies it for $i$.
    So, if we match every demand in $\hpi_{u_*,i}^S$ to its originating item in $B_{u_*,j}\cap S$ for corresponding $j \in [n]$, the matched distinct items will also simultaneously satisfy their demands for $i$.
    
    Now, if demand $\alpha$ is counted in $\Pp_i(\hpi_{u_*,i}^S, \ell)$, its matched item $e$ satisfies $e \leq m_i(\alpha) \leq \ell$.
    Thus, every counted demand in prefix $\ell$ is matched to a distinct item of $S\cap [\ell]$.
    The proof follows.
\end{proof}
We will pick a supplier $u_*$ and designated recipients of his exceptional bundles so that the compatibility condition of \cref{compatible} holds.
The construction of $E$-parts of suggested profiles is done as in \cref{intype}, and will be covered later.

\subsubsection{Bounded replacement}

Suppose for now that we chose some supplier $u_* \in N$ and constructed some collection $\{\hpi_{u_*,i}\}_{i \in N}$ of suggested profiles (with $S$-parts according to \cref{compatible}), where agent $i \in N$ is suggested $\hpi_{u_*,i}$.
Now, as discussed earlier in the section, our task is to choose a set $A\subseteq I$ of agents that retain their old profiles $\pi_i$, and give a suggested profile $\hpi_{u_*,i}$ to every other agent $i \in N\setminus A$, so that conditions (\ref{ES2}), (\ref{Anu2}) and (\ref{S2}) hold.
More precisely, we want that
\begin{align}
    &&&\sum_{i \in A}\Pp_i(\pi_i^E, c_s) + \sum_{i \in N\setminus A}\Pp_i(\hpi_{u_*,i}^E, c_s) \leq |E\cap [b_s]|,&\forall (b_s, c_s) \in \Ll_S;\tag{$\mathrm{E}_{\Ll_S}$}\label{ES3}\\
    &&&\sum_{i \in N \setminus A}|\hpi_{u_*,i}|\leq \frac{\nu}{2} \tag{A-$\nu$}\label{Anu3};\\
    &&&\sum_{i \in A}\Pp_i(\pi_i^S, \ell) + \sum_{i \in N\setminus A}\Pp_i(\hpi_{u_*,i}^S, \ell) \leq |S\cap [\ell]|,&\forall 0 \leq \ell \leq m. \tag{S}\label{S3}
\end{align}

The constraints of (\ref{S}) are more problematic, as we need to simultaneously control load at $m + 1$ different prefixes. 
Fortunately, we can obtain a similar reduction as we did earlier with conditions (\ref{E2}) and (\ref{ES2}), and preserve (\ref{S}) at only a constant number of different prefixes, corresponding to the breakpoints of the supplier $u_*$.
\begin{definition}
    Let $I\subseteq N$ and $\{\pi_i\}_{i \in I}$ be obtained from \cref{ESrounding}.
    An \textbf{$S$-certificate} for profiles $\{\pi_i^S\}_{i \in I}$ is a family of pairwise-disjoint sets $S_i\subseteq S$ for $i \in I$, such that $|S_i| = |\pi_i^S|$ and $S_i$ satisfies profile $\pi_i^S$ for agent $i$.
    By \cref{ESrounding} and \cref{loadequiv}, such certificate exists.
\end{definition}

The sets $S_i$ for $i \in I$ record which items of set $S$ satisfy the $S$-part of the old profile $\pi_i^S$.
If the agent $i \in I$ keeps his old profile, i.e $i \in A$, we ``reserve'' its certificate $S_i$, otherwise the items of $S_i$ become available for supplier $u_*$'s suggested profiles.

Note that for every $i \in I$ it holds $\Pp_i(\pi_i^S, \ell) \leq |S_i\cap [\ell]|$ for every prefix $\ell$.
We will use this bound in order to reduce condition (\ref{S}) on both profiles of agents $A$ and suggested supplier profiles of $N\setminus A$, to a condition on supplier profiles $\hpi_{u_*,i}$ only.
This way, instead of checking (\ref{S}) at every relevant breakpoint of profiles $\pi_{i}^S$ for $i \in A$ and $\hpi_{u_*,i}^S$ for $i \in N\setminus A$, the number of which can depend on the number of agents/items, it will suffice to check a supplier-only version of (\ref{S}) at every possible supplier breakpoint, which is a constant amount since $|\Qq| = Q$.
This will add an additional set of \textit{protected prefix} constraints w.r.t items $S$.

\begin{lemma}\label{sprot}
    Fix a supplier $u_*\in N$ and construct suggested profiles $\hpi_{u_*,i}^S$, $i \in N$, and suppose they satisfy the $S$-compatibility condition of \cref{compatible}.
    If choice $A\subseteq I$ satisfies
    \begin{equation}
        \sum_{i \in A}|S_i\cap [m_{u_*}(\alpha)]| + \sum_{i\in N\setminus A}\Pp_{u_*}(\hpi_{u_*,i}^S, m_{u_*}(\alpha)) \leq |S\cap [m_{u_*}(\alpha)]|\tag{S$*$}\label{S*}
    \end{equation}
    for every possible supplier breakpoint $m_{u_*}(\alpha)$ for $\alpha \in \Qq$, then for every prefix $0 \leq \ell \leq m$,
    \begin{equation}
        \sum_{i \in A}\Pp_i(\pi_i^S, \ell) + \sum_{i \in N\setminus A}\Pp_i(\hpi_{u_*,i}^S, \ell) \leq |S\cap [\ell]|.\tag{S}\label{Sstar}
    \end{equation}
\end{lemma}
\begin{proof}
    Reserve bundle $S_i$ for every $i \in A$, these agents will retail their old $S$-profiles.
    We will show that the suggested $S$-profiles $\hpi_{u_*,i}^S$ of all other agents $i\in N\setminus A$ can be simultaneously satisfied using distinct remaining items.

    Consider the condition (\ref{S*}) for an arbitrary prefix $\ell$.
    Observe that we can rearrange it as 
    \begin{equation}
        \sum_{i\in N\setminus A}\Pp_{u_*}(\hpi_{u_*,i}^S, \ell) \leq |S\cap [\ell]|- \sum_{i \in A}|S_i\cap [\ell]|.\tag{$*$}\label{*}
    \end{equation}
    The right-hand side of (\ref{*}) counts the items of $S$ in prefix $[\ell]$ that remain available after reserving the certificates for $A$.
    Indeed, since all $S_i$ are pairwise-disjoint subsets of $S$, we have
    \[
        |S\cap [\ell]|- \sum_{i \in A}|S_i\cap [\ell]| = \left|\left(S\setminus\bigcup_{i \in A}S_i\right)\cap [\ell]\right|.
    \]
    In particular, this quantity is non-negative and non-decreasing in $\ell$, and when we increase $\ell$ by $1$, the remaining capacity in (\ref{*}) increases by $1$ only if the item $\ell + 1$ is an unreserved item of $S$.

    On the other hand,  the left-hand side of (\ref{*}) increases only at supplier's breakpoints $m_{u_*}(\alpha)$, $\alpha \in \Qq$.
    Therefore, in order to satisfy (\ref{*}) for every prefix $0 \leq \ell \leq m$, it suffices to presetve the inequality only at prefixes $\ell = m_{u_*}(\alpha)$ for $\alpha \in \Qq$.
    Indeed, if load at $\ell$ is positive, and $m_{u_*}(\alpha) \leq \ell$ is the last $S$-breakpoint of $\alpha \in \hpi_{u_*,i}^S$ preceding $\ell$, then load at  $m_{u_*}(\alpha)$ and $\ell$ is the same, while the remaining capacity at $\ell$ is at least the one at $m_{u_*}(\alpha)$.
    Hence, enforcing (\ref{*}) at $m_{u_*}(\alpha)$, which is exactly (\ref{S*}), implies it at $\ell$ as well.
    If the load is $0$, then inequality holds regardless.

    By assumption, (\ref{S*}) holds for every supplier breakpoint $m_{u_*}(\alpha)$, therefore (\ref{*}) holds at every prefix $0 \leq \ell \leq m$.
    Then, by \cref{loadequiv} we can simultaneously satisfy all suggested profiles $\{\hpi_{u_*,i}^S\}_{i \in N\setminus A}$ w.r.t valuation $v_{u_*}$ using distinct items of 
    \[
        S\setminus\bigcup_{i \in A}S_i.
    \]
    
    We can view this allocation as a matching between demands $\alpha \in \hpi_{u_*,i}^S$ and corresponding items $e$ satisfying $\alpha$, i.e with $v_{u_*}(e)\geq \alpha$.
    Under $S$-compatibility of \cref{compatible}, for every agent $i \in N\setminus A$ and every demand $\alpha \in \hpi_{u_*,i}^S$ matched to item $e$, it also holds $v_i(e) \geq \alpha$.
    
    It follows that, together with certificates $S_i$ for $i \in A$, we obtain a matching between items of $S$ and demands of all profiles $\{\pi_i^S\}_{i \in A}$ and $\{\hpi_{u_*,i}^S\}_{i\in N\setminus A}$, such that for every item $e$ matched to a demand $\alpha$ in the profile of agent $i$ it holds $v_i(e)\geq \alpha$.
    But then, every demand counted in the total load  at prefix $\ell$ of agents $A$ and $N\setminus A$ together must be matched to a distinct item of $S\cap [\ell]$.
    As a result, for every prefix $0\leq \ell \leq m$ it must hold
    \[
        \sum_{i \in A}\Pp_i(\pi_i^S, \ell) + \sum_{i \in N\setminus A}\Pp_i(\hpi_{u_*,i}^S, \ell) \leq |S\cap [\ell]|.
    \]
\end{proof}

It follows from \cref{compatible} and \cref{sprot} that, for the chosen subset $A\subseteq I$ of agent who keep their assigned profiles $\{\pi_{i}\}_{i \in A}$, and for the remaining agents $i \in N\setminus A$ who receive a supplier suggestion $\hpi_{u_*,i}$, to satisfy for every prefix $0 \leq \ell \leq m$
\[
    \sum_{i \in A}\Pp_i(\pi_i, \ell) + \sum_{i \in N\setminus A}\Pp_i(\hpi_{u_*,i}, \ell) \leq \ell,
\]
it suffices that the following conditions hold on $A\subseteq I$ and suggested collection $\{\hpi_{u_*,i}\}_{i \in N}$.
\begin{align}
    &&&\sum_{i \in A}\Pp_i(\pi_i^E, c_s) + \sum_{i \in N\setminus A}\Pp_i(\hpi_{u_*,i}^E, c_s) \leq |E\cap [b_s]|,&\forall (b_s, c_s) \in \Ll_S;\tag{$\mathrm{E}_{\Ll_S}$}\label{ES4}\\
    &&&\sum_{i \in N \setminus A}|\hpi_{u_*,i}|\leq \frac{\nu}{2} \tag{A-$\nu$}\label{Anu4};\\
    &&& \sum_{i \in A}|S_i\cap [m_{u_*}(\alpha)]| + \sum_{i\in N\setminus A}\Pp_{u_*}(\hpi_{u_*,i}^S, m_{u_*}(\alpha)) \leq |S\cap [m_{u_*}(\alpha)]|, & \forall \alpha \in \Qq.\tag{S$*$}\label{S4}
\end{align}

Note that now the total number of constraints that we need to take care of is $|\Ll_S| + |\Qq| + 1$, which is only a \textbf{constant} depending on $Q$, as $|\Ll_S| \leq 2Q + 1$.

Let us first focus on trying to find a set $A\subseteq I$ that satisfies just (\ref{ES4}) and (\ref{S4}).
Suppose that the suggested supplier collection $\{\hpi_{u_*,i}\}_{i \in N}$ satisfies (\ref{ES4}) and (\ref{S4}) with  $A = \varnothing$, that is,
\begin{align}
    &&&\sum_{i \in N}\Pp_i(\hpi_{u_*,i}^E, c_s) \leq |E\cap [b_s]|,&\forall (b_s, c_s) \in \Ll_S;\tag{$\mathrm{E}_{\Ll_S}$-$u$}\label{ESu}\\
    &&& \sum_{i\in N}\Pp_{u_*}(\hpi_{u_*,i}^S, m_{u_*}(\alpha)) \leq |S\cap [m_{u_*}(\alpha)]|, & \forall \alpha \in \Qq.\tag{S$*$-$u$}\label{Su}
\end{align}
This way, if we replace every old profile $\{\pi_i\}_{i \in I}$ with suggested profile, we obtain a feasible collection with respect to protected prefix constraints (\ref{ES4}) and (\ref{S4}).
The problem with such replacement is that the total number of added demands $\sum_{i \in N}|\hpi_{u_*,i}|$ may be unbounded, while we seek a replacement of size at most $\nu/2$, which will be roughly $O_{Q,q}(\log n)$.
Our aim is to show that only a bounded number of old profiles need to be replaced with suggested profiles, in order to get a feasible $A\subseteq I$ satisfying (\ref{ES4}) and (\ref{S4}).

The reader may ask, why should such a bounded exchange exist?
Consider inserting a suggested profile for just one unassigned agent from $N\setminus I$.
We know that some feasible exchange exists: replace every old profile with supplier's suggestion.
Now, each replacement $\pi_i \to \hpi_{u_*,i}$ affects only a constant number of constraints $|\Ll_S| + |\Qq|$, and its contribution to each constraint is an integer bounded by $\max\{|\pi_i|, |\hpi_{u_*,i}|\}$.
Therefore, if all suggested profiles have bounded size, say $|\hpi_{u_*,i}|\leq q$, then in total there is only a bounded number of possible load configurations in (\ref{ES4}) and (\ref{S4}), for old and suggested profiles together.
Then, Dickson's lemma \cite{Dickson13} will allow us to extract a replacement of only constant size from a potentially unbounded one.
This way, adding a new unassigned agent forces us to swap only a constant number of old profiles to the corresponding suggested ones.

\begin{theorem}[Bounded replacement]\label{supplierchange}
    For every triple of integers $t, Q, q \geq 1$, there exists a global constant $C_{t,Q, q} \geq 1$, depending only on $t, Q, q$, such that the following holds.

    Let $I\subseteq N$ and $\{\pi_i\}_{i \in I}$ be obtained from \cref{ESrounding}.
    For every $i \in I$, fix an $S$-certificate $\{S_i\}_{i \in I}$, and let the number of protected $E$-prefixes be $|\Ll_S| = t$.
    Fix a supplier $u_* \in N$, and for every agent $i \in N$ a corresponding suggested profile $\hpi_{u_*,i} = \hpi_{u_*,i}^E \uplus \hpi_{u_*,i}^S$.
    
    Suppose that for every $i \in N$, $|\hpi_{u_*,i}|\leq q$, every $S$-demand $\alpha \in \hpi_{u_*,i}^S$ belongs to $\Qq$, as well as 
    \begin{align}
    &&&\sum_{i \in N}\Pp_i(\hpi_{u_*,i}^E, c_s) \leq |E\cap [b_s]|,&\forall (b_s, c_s) \in \Ll_S;\tag{$\mathrm{E}_{\Ll_S}$-$u$}\label{ESu2}\\
    &&& \sum_{i\in N}\Pp_{u_*}(\hpi_{u_*,i}^S, m_{u_*}(\alpha)) \leq |S\cap [m_{u_*}(\alpha)]|, & \forall \alpha \in \Qq.\tag{S$*$-$u$}\label{Su2}
    \end{align}

    Then there exists $A\subseteq I$ such that $|N\setminus A|\leq C_{t,Q,q}(|N| - |I|)$ and
    \begin{align}
    &&&\sum_{i \in A}\Pp_i(\pi_i^E, c_s) + \sum_{i \in N\setminus A}\Pp_i(\hpi_{u_*,i}^E, c_s) \leq |E\cap [b_s]|,&\forall (b_s, c_s) \in \Ll_S;\tag{$\mathrm{E}_{\Ll_S}$}\label{ES5}\\
    &&& \sum_{i \in A}|S_i\cap [m_{u_*}(\alpha)]| + \sum_{i\in N\setminus A}\Pp_{u_*}(\hpi_{u_*,i}^S, m_{u_*}(\alpha)) \leq |S\cap [m_{u_*}(\alpha)]|, & \forall \alpha \in \Qq.\tag{S$*$}\label{S5}
\end{align}
    In particular,
    \[
        \sum_{i \in N\setminus A}|\hpi_{u_*,i}|\leq qC_{t,Q,q}(|N| - |I|),
    \]
    so if integer $\nu$ satisfies $\nu \geq 2qC_{t,Q,q}(|N| - |I|)$, condition (\ref{Anu4}) holds for the same choice of $A$.
\end{theorem}
We prove \cref{supplierchange} in \cref{appendix:supplier} (\cref{supplierchangeproof}).

\cref{supplierchange} allows to complete the initial collection $\{\pi_i\}_{i \in I}$ of profiles for agents $I\subseteq N$ to a full collection of profiles for all agents $N$, while replacing only a constant number of old profiles per every new agent added.
We will now show how we can combine this theorem with previously established reductions in \cref{sec:integer} in order obtain feasibility at every prefix $0 \leq \ell \leq m$.

Recall that \cref{ESrounding} leaves at most 
\[
    |N| - |I| \leq \eps |F| + 16q\log n
\]
agents unassigned.
Therefore, \cref{supplierchange} introduces at most
\[
    \sum_{i \in N\setminus A}|\hpi_{u_*,i}|\leq qC_{t,Q,q}( \eps |F| + 16q\log n)
\]
additional demands.
We want this quantity to be at most $\nu/2$, so that \cref{protpref} applies.
Since $|F|\leq (2Q + 1)\cdot \nu$, increasing $\nu$ also increases our bound on the number of unassigned agents caused by the buffer $F$.
So, in order to control this contribution, we will first choose sufficiently small $\eps \in (0, 1)$, and then choose integer $\nu \geq 1$ to accommodate the remaining loss.

\begin{definition}
    For integer $q \geq 1$, define the global constant $C_{Q,q}$ as
    \[
        C_{Q,q} := \max\left\{C_{t,Q,q} : 1 \leq t \leq 2Q + 1\right\},
    \]
    where $C_{t,Q,q}$ are obtained from \cref{supplierchange}.
\end{definition}
\begin{corollary}\label{epsnuconst}
    Choose $\eps \in (0, 1)$ and integer $\nu \geq 1$ satisfying
    \[
        \eps \leq \frac{1}{4qC_{Q,q}\left(2Q + 1\right)}, \qquad\qquad \nu \geq 64q^2C_{Q,q}\log n.
    \]
    Apply \cref{buffer} with these values of $\eps, \nu$ to obtain a partition $M = E\sqcup S$, sets $F \subseteq E$, $W\subseteq N$ and protected prefixes $\Ll_S$, and then use \cref{ESrounding} to get the set $I\subseteq N$ of agents with assigned profiles $\{\pi_i\}_{i \in I}$, with the following properties:
    \begin{itemize}
        \item for every $i \in I$, $\pi_i$ has a decomposition $\pi_i = \pi_i^E \uplus \pi_i^S$, with $v(\pi_i) \geq \mu_i$, $|\pi_i|\leq q$ and 
        \[
            |N| - |I|\leq \eps |F| + 16q\log n;
        \]
        \item their $E$-prefix loads satisfy
        \begin{align}
        &&&\sum_{i \in I}\Pp_i(\pi_i^E, \ell) \leq |E\cap [\ell]| - \frac{\nu}{2},&\forall \ell \in E\setminus F;\tag{$\mathrm{E}$-$\nu$}\label{Enu6}\\
        &&& \sum_{i \in I}\Pp_i(\pi_i^E, c_s) \leq |E\cap [b_s]|, & \forall (b_s, c_s)\in \Ll_S.\tag{$\mathrm{E}_{\Ll_S}$}\label{ES6}
        \end{align}
        \item there are disjoint sets $S_i\subseteq S$, $i \in I$, such that $S_i$ satisfies $\pi_i^S$ for agent $i$ and $|S_i| = |\pi_i^S|$.        
    \end{itemize}    
    Suppose there is a supplier $u_*\in N$ and profiles $\hpi_{u_*,i} = \hpi_{u_*,i}^E\uplus \hpi_{u_*,i}^S$, $i \in N$, such that
    \begin{itemize}
        \item for every $i \in N$, $v(\hpi_{u_*,i}) \geq \mu_i$, $|\hpi_{u_*,i}|\leq q$, and $\hpi_{u_*,i}^S$ are as in \cref{suggestpart};
        \item the $E$-parts of $\{\hpi_{u_*,i}\}_{i \in N}$ satisfy protected prefix inequalities:
        \begin{align}
        &&&\sum_{i \in N}\Pp_i(\hpi_{u_*,i}^E, c_s) \leq |E\cap [b_s]|,&\forall (b_s, c_s) \in \Ll_S;\tag{$\mathrm{E}_{\Ll_S}$-$u$}\label{ESu6}
        \end{align}
        \item every $i \in N$ is \textbf{$S$-compatible}: if $\hpi_{u_*,i}^S \neq \varnothing$ then for every $e \in S$  it holds $v_i(e) = v_{u_*}(e)$.
    \end{itemize}
    Then there exists set $A\subseteq I$ such that $|N\setminus A| \leq \nu/(2q)$ and for every prefix $0 \leq \ell \leq m$,
    \[
        \sum_{i \in A}\Pp_i(\pi_i, \ell) + \sum_{i \in N\setminus A}\Pp_i(\hpi_{u_*,i}, \ell) \leq \ell.
    \]
    As a corollary, these profiles $\{\pi_i\}_{i \in A}$ and $\{\hpi_{u_*,i}\}_{i \in N\setminus A}$ can be simultaneously satisfied by distinct items of $M$, with their $E$-, $S$-parts satisfied within $E, S$ respectively, and every agent $i \in N$ receives a bundle $B_i$ such that $v_i(B_i) \geq \mu_i$.
\end{corollary}
\begin{proof}
    Since $|F|\leq (2Q + 1)\cdot \nu$, by the choice of parameters $\eps, \nu$ we have
    \[
        |N| - |I|\leq 16q\log n + \eps\left(2Q + 1\right)\nu \leq \frac{\nu}{4qC_{Q,q}} + \frac{\nu}{4qC_{Q,q}} = \frac{\nu}{2qC_{Q,q}}.
    \]
    By definition, all demands in $\hpi_{u_*,i}^S$ are in $\Qq$, and by \cref{compatible},
    \begin{align}
    &&& \sum_{i\in N}\Pp_{u_*}(\hpi_{u_*,i}^S, m_{u_*}(\alpha)) \leq |S\cap [m_{u_*}(\alpha)]|, & \forall \alpha \in \Qq.\tag{S$*$-$u$}\label{Su5}
    \end{align}
    Hence, we can apply \cref{supplierchange}, getting $A\subseteq I$ with $|N\setminus A| \leq C_{Q,q}(|N|- |I|) \leq\nu/(2q)$ and
    \begin{align}
    &&&\sum_{i \in A}\Pp_i(\pi_i^E, c_s) + \sum_{i \in N\setminus A}\Pp_i(\hpi_{u_*,i}^E, c_s) \leq |E\cap [b_s]|,&\forall (b_s, c_s) \in \Ll_S;\tag{$\mathrm{E}_{\Ll_S}$}\label{ES7}\\
    &&& \sum_{i \in A}|S_i\cap [m_{u_*}(\alpha)]| + \sum_{i\in N\setminus A}\Pp_{u_*}(\hpi_{u_*,i}^S, m_{u_*}(\alpha)) \leq |S\cap [m_{u_*}(\alpha)]|, & \forall \alpha \in \Qq.\tag{$\mathrm{S}*$}\label{Sstar7}
\end{align}
In particular,
\[
    \sum_{i \in N\setminus A}|\hpi_{u_*,i}|\leq q|N\setminus A| \leq \frac{\nu}{2}.
\]
Applying \cref{protpref} to $A$ and $\{\hpi_{u_*,i}\}_{i \in N\setminus A}$ with (\ref{ES7}), we get that for every $0 \leq \ell \leq m$,
\begin{equation}
    \sum_{i \in A}\Pp_i(\pi_i^E, \ell) +  \sum_{i \in N \setminus A}\Pp_i(\hpi_{u_*,i}^E, \ell) \leq |E\cap [\ell]|.\tag{E}\label{E7}
\end{equation}
Next, (\ref{Sstar7}) together with $S$-compatibility allows us to apply \cref{sprot}, giving for $0 \leq \ell \leq m$:
\begin{equation}
        \sum_{i \in A}\Pp_i(\pi_i^S, \ell) + \sum_{i \in N\setminus A}\Pp_i(\hpi_{u_*,i}^S, \ell) \leq |S\cap [\ell]|.\tag{S}\label{S7}
\end{equation}
Combining (\ref{E7}) and (\ref{S7})  gives for every $0 \leq \ell \leq m$:
\[
        \sum_{i \in A}\Pp_i(\pi_i, \ell) + \sum_{i \in N\setminus A}\Pp_i(\hpi_{u_*,i}, \ell) \leq |E\cap[\ell]| + |S\cap[\ell]|= \ell.
\]
\cref{loadequiv} realizes each sub-profile within its respective set, and distributing unused items arbitrarily gives us a full allocation of $M$.
By assumption, every $i \in I$ had $v(\pi_i) \geq \mu_i$, and for every $i \in N$ it holds $v(\hpi_{u_*,i}) \geq \mu_i$.
Hence, satisfying all the selected profiles of $A$ and $N\setminus A$ simultaneously guarantees that every agent receives a bundle of value at least their MMS.
\end{proof}

\subsection{Selecting the supplier and recipients}

\label{sec:intselecting}

\cref{epsnuconst} identifies the conditions that supplier $u_*$ and its suggested profiles $\{\hpi_{u_*,i}\}_{i \in N}$ must satisfy in order to complete initial assignment $\{\pi_i\}_{i \in I}$ for agents $I\subseteq N$ to a full feasible assignment of all agents $N$ (potentially replacing some old profiles with suggested ones).
We will describe how to choose a supplier and recipients of his suggested profiles, that are \textit{compatible}, i.e giving these agents corresponding profiles satisfies condition (\ref{ESu6}), as well as $S$-compatibility conditions of \cref{compatible}.

Recall that for supplier $u_*$ with partition $\{B_{u_*,1}, \ldots, B_{u_*,n}\}$, a bundle $B_{u_*,j}$ is \textit{exceptional} if
\[
    B_{u_*,j}\cap S \neq \varnothing.
\]
Each of these exceptional bundles will be designated to a distinct \textit{exceptional recipient} $i \in N$, and corresponding suggested $S$-parts of $\hpi_{u_*,i}$ are constructed simply as
\[
    \hpi_{u_*,i}^S = \{v_{u_*}(e) :\; e \in B_{u_*,j}\cap S\}.
\]
According to \cref{compatible}, for a supplier $u_*$ and recipient $i$ of $\hpi_{u_*,i}$, we require:
\[
    \hpi_{u_*,i}^S \neq \varnothing \quad \implies \quad \forall e \in S:\; v_i(e) = v_{u_*}(e).
\]
This suggests partitioning all agents $i \in N$ into \textit{equivalence classes} based on their agreement on the values of items in $S$.
Since every item can take only values in $\Qq = \{1, \ldots, Q\}$, this will induce at most $\binom{|S| + Q - 1}{Q - 1}$ equivalence classes, which is small when $|S| = O_{Q,q}(\log n)$.

For every exceptional bundle of the supplier $u_*$, we will pick a distinct recipient that belongs to the same $S$-equivalence class as $u_*$.
Note that this is only possible when this equivalence class is large enough, i.e contains at least as many agents as there are exceptional bundles.
In particular, it suffices that the $S$-equivalence class of $u_*$ has size at least $|S|$.

Observe that this way, we can choose the supplier and the recipients of its exceptional bundles within any sufficiently large set of supplier candidates.
This motivates us to employ the following three-stage process of selecting the supplier and recipients and constructing suggested profiles.
\begin{enumerate}
    \item Choose a sufficiently large set $U$ of supplier candidates, such that for any candidate $u \in U$ we can always construct a collection $\{\hpi_{u,i}^E\}_{i \in N}$ of suggested $E$-profiles satisfying
    \begin{align}
        &&&\sum_{i \in N}\Pp_i(\hpi_{u,i}^E, c_s) \leq |E\cap [b_s]|,&\forall (b_s, c_s) \in \Ll_S.\tag{$\mathrm{E}_{\Ll_S}$-$u$}\label{ESu7}
\end{align}
\item Partition all agents of $U$ into $S$-equivalence classes, and choose a supplier $u_* \in U$ from a class of size at least $|S|$.
\item Having chosen the supplier $u_*$, designate every exceptional bundle of this supplier to a distinct eligible recipient, and construct $S$-profiles according to \cref{suggestpart}.
Then, construct the corresponding suggested $E$-profiles $\{\hpi_{u_*,i}^E\}_{i \in N}$ satisfying (\ref{ESu7}).
\end{enumerate}

The construction of suggested $E$-profiles $\{\hpi_{u,i}^E\}_{i \in N}$ for a given supplier candidate $u$ will be explained in the next subsection.
The rest of this subsection will be dedicated to choosing the large set of supplier candidates $U$ and subsequent selection of the supplier $u_* \in U$.

\subsubsection{$E$-classes and supplier candidates}

In order to find a large set $U$ of supplier candidates, such that for any $u \in U$ there are suggested $E$-profiles  $\{\hpi_{u,i}^E\}_{i \in N}$ satisfying protected prefix inequalities (\ref{ESu7}), that also are acceptable, we will partition all agents $N$ into classes, so that within each class, all agents are ``equivalent'' with respect to protected prefixes $(b_s, c_s) \in \Ll_S$.
We do it as follows.
Recall that set $S$ from \cref{buffer}, as well as internal protected prefix $c_s$, $s \in [t - 1]$, have the following form:
\[
    S = \bigcup_{s = 1}^t\{a_s, \ldots, b_s\},\qquad\text{and}\qquad c_s = \min\{b_s + \nu, a_{s + 1} - 1\},\; \forall s \in [t - 1].
\]
The last protected prefix is $(b_t, c_t) = (m, m)$ and will not affect our classification.
The set $S$ partitions items $E = M\setminus S$ into $t - 1$ intervals between consecutive components of $S$.
\begin{definition}
    For every $s \in [t - 1]$, define the $s$-th interval of $E$ as
    \[
    E_s := \{b_s + 1,\ldots, a_{s + 1} - 1\}.
    \]
    We will associate each $e \in E$ with a \textbf{representative item} $c_s + 1$ of interval $E_s$ containing $e$.
    Then, we will classify agents $N$ based off their rounded values of these $t - 1$ representative items.
    
    For every agent $i \in N$ and every $s \in [t - 1]$, define the \textbf{$s$-th representative demand} as
    \[
        \alpha_{i,s} := v_{i}(c_s + 1).
    \]
    Note that $m_{i}(\alpha_{i,s}) \geq c_s + 1$, so demand $\alpha_{i,s}$ contributes no load at protected prefix $c_s$.
\end{definition}
\begin{definition}
    For every agent $i \in N$, define his \textbf{$E$-signature} as 
    \[
        \fe_i :=  (\alpha_{i,1}, \ldots, \alpha_{i,t - 1}, \mu_i).
    \]
    Two agents belong to the same class if they have the same $E$-signature.
    For signature $\fe$, denote
    \[
        N_\fe := \{i \in N: \; \fe_i = \fe\}.
    \]
\end{definition}

\begin{lemma}\label{signature}
    Let $\Phi := \{\fe_i : i \in N\}$ denote the set of all occurring $E$-signatures.
    Then
    \[
        |\Phi| \leq \phi := qQ\binom{3Q}{Q}.
    \]
    Moreover, there exists a signature $\fe$, such that the set $U= (N\setminus W)\cap N_\fe$ satisfies 
    \[
        |U|\geq \frac{n}{2\phi},
    \]
    and for every $u \in U, s \in [t -1]$ and $e \in E_s$,
    \[
        v_u(e) \leq \alpha_{u,s}.
    \]
    That is, value of every item $e \in E_s$ does not exceed the value of $c_s + 1$ for every agent $u \in U$.
\end{lemma}
\begin{proof}
    Since every MMS bundle has size at most $q$ and values in $\{1, \ldots, Q\}$, for every $i \in N$ we have $1 \leq \mu_i \leq qQ$.
    Next, since positions of representative items $c_s + 1$ increase with with $s$, for every agent $i \in N$ we have $\alpha_{i,1} \geq \alpha_{i,2} \geq \ldots \geq \alpha_{i, t - 1}$.
    Hence, first $t - 1$ coordinates of every signature $\fe_i$ is a non-increasing vector of length $t - 1$ with entries in $\Qq\cup \{0\}$.
    Every such vector is determined by the number of entries of each type, so by stars and bars the total number of possible signatures is (using the upper bound $t \leq 2Q + 1$ from \cref{buffer}):
    \[
        qQ\binom{t - 1 + Q}{Q} \leq qQ\binom{2Q + Q}{Q} = \phi.
    \]

    Next, we find a signature class containing many supplier candidates.
    By \cref{buffer} we have $|W|\leq\eps n$.
    By the choice of parameters in \cref{epsnuconst} we have $\eps \leq 1/2$, so
    \[
        |N\setminus W| \geq |N| - |W| \geq n - \eps n \geq \frac{n}{2}.
    \]
    Choose $\fe$ maximizing $|(N\setminus W)\cap N_\fe|$, and take $U := (N\setminus W)\cap N_\fe$.
    Since there are at most $\phi$ classes, it must hold that $|U|\geq n/(2\phi)$.

    Fix $u \in U, s \in [t - 1]$ and item $e \in E_s$, put $\alpha := v_u(e)$.
    If $e \geq c_s + 1$, then we immediately have $v_u(e) \leq v_u(c_s + 1) = \alpha_{u,s}$.
    Otherwise, $b_s < e \leq c_s$ and $m_u(\alpha) \geq e > b_s$.
    Since $u \notin W$, none of $u$'s breakpoints belongs to $F$, in particular, $m_u(\alpha)\notin (b_s, c_s]$.
    Hence, $m_u(\alpha) \geq c_s + 1$ and therefore $\alpha_{u,s} = v_u(c_s + 1) \geq \alpha = v_u(e)$.
\end{proof}

We will show in the next subsection  that for every agent $u \in U$, where $U$ is chosen in \cref{signature}, it is possible to construct a collection $\{\hpi_{u,i}^E\}_{i \in N}$ of suggested $E$-profiles satisfying protected prefix inequalities (\ref{ESu7}).
This will be done in a similar way to \cref{sec:intype}.

\subsubsection{$S$-classes and choosing the supplier}

Fix signature $\fe$ and set $U = (N\setminus W)\cap N_\fe$, obtained in \cref{signature}.
All agents $u \in U$ agree on representative demands corresponding to the intervals $E_s$, $s \in [t - 1]$.
However, their values on $S$ may differ.
We will partition $U$ into equivalence classes, so that within each class all agents agree on their values of items from $S$.
Then, we will pick a supplier $u_*$ from the largest $S$-equivalence class.
Provided that the set $U$ is sufficiently large, the size of $u_*$'s class will be at least $|S|$, hence he will be able to designate all his exceptional to distinct recipients, while preserving compatibility conditions of \cref{compatible}.

The number of these equivalence classes may depend on $|S|$.
However, every item in $S$ has value in $\Qq = \{1, \ldots, Q\}$ for every agent $i \in N$, so the total number of classes is not too large.
\begin{definition}
    Enumerate $S := \{e_1 <  \ldots < e_{|S|}\}$.
    For agent $i \in U$, define his \textbf{$S$-signature}
    \[
        \sigma_i = (v_i(e_1), \ldots, v_i(e_{|S|})).
    \]
    Two agents belong to the same $S$-class, if they have the same $S$-signature.
    For signature $\sigma$, let
    \[
        U_\sigma := \{i \in U:\; \sigma_i = \sigma\}.
    \]
\end{definition}
\begin{lemma}[Supplier choice]\label{choice}
    Suppose that
    \[
        |U| \geq |S| \binom{|S| + Q - 1}{Q - 1}.
    \]
    Then there exists a supplier $u_* \in U$ and a designation of his exceptional bundles $B_{u_*,j}$ to distinct recipients $i(j) \in U$ such that $v_{i(j)}(e) = v_{u_*}(e)$ for every $e \in S$, and the full suggested profile 
    \[
        \hpi_{u_*,i} := \{v_{u_*}(e) : \; e \in B_{u_*,j}\}
    \]
    satisfies $|\hpi_{u_*,i(j)}|\leq q$ and $v(\hpi_{u_*,{i(j)}})\geq \mu_{i(j)}$.
    In particular, the size condition above holds whenever
    \[
        n \geq 2\phi|S| \binom{|S| + Q - 1}{Q - 1}.
    \]
\end{lemma}
\begin{proof}
    Every $S$-signature is a non-increasing vector of length $|S|$ with entries in $\{1, \ldots, Q\}$, so the number of possible signatures is at most
    \[
        \binom{|S| + Q - 1}{Q - 1}.
    \]
    Choose the largest class $U_\sigma$, by assumption it holds
    \[
        |U_\sigma| \geq \frac{|U|}{\binom{|S| + Q - 1}{Q - 1}} \geq |S|.
    \]
    Take any supplier $u_* \in U_\sigma$, as mentioned in \cref{excep}, it has at most $|S|$ exceptional bundles.
    Hence, we can designate all exceptional bundles of $u_*$ to distinct agents from $U_\sigma$.
    By definition, all recipients agree with $u_*$ on every item of $S$.

    In addition, supplier and all designated recipients belong to $N_\fe$, therefore $\mu_{i(j)} = \mu_{u_*}$.
    Consequently, choosing $\hpi_{u_*,i(j)} := \{v_{u_*}(e) : \; e \in B_{u_*,j}\}$ gives $|\hpi_{u_*,i(j)}| = |B_{u_*,j}| \leq q$ and $v(\hpi_{u_*,i(j)}) = v_{u_*}(B_{u_*,j}) \geq \mu_{u_*} = \mu_{i(j)}$.

    Finally, by \cref{signature} we have $|U|\geq n/(2\phi)$, which finishes the proof.
\end{proof}

It follows that for sufficiently large $n$, we can choose a supplier $u_* \in U$ and construct $S$-parts of suggested profile collection $\{\hpi_{u_*,i}\}_{i \in N}$, satisfying conditions of \cref{epsnuconst}.
\begin{corollary}\label{suplargen}
    Fix integers $Q, q\geq 1$, and choose parameters $\eps, \nu$ in \cref{epsnuconst} as:
    \[
        \eps := \frac{1}{4qC_{Q,q}\left(2Q + 1\right)}, \qquad\qquad \nu := \lceil 64q^2C_{Q,q}\log n\rceil.
    \]
    For sufficiently large $n$, there exists a supplier $u_* \in U$ and a designation of its exceptional bundles to distinct recipients in $U$, such that the suggested profiles
    \[
        \hpi_{u_*,i}^E:= \{v_{u_*}(e):\; e \in B_{u_*,j}\cap E\},\qquad\hpi_{u_*,i}^S:= \{v_{u_*}(e):\; e \in B_{u_*,j}\cap S\}
    \]
    for every exceptional agent $i$ with designated bundle $B_{u_*,j}$, and $\hpi_{u_*,i}^S=\varnothing$ for every non-exceptional agent $i$, satisfy 
    \[
        \hpi_{u_*,i}^S \neq \varnothing \quad \implies \quad \forall e \in S:\; v_i(e) = v_{u_*}(e),
    \]
    and every exceptional agent $i$ satisfies $|\hpi_{u_*,i}|\leq q$ and $v(\hpi_{u_*,i}) \geq \mu_i$.
\end{corollary}
\begin{proof}
    Since $C_{Q,q}$ depend only on $Q,q$, parameter $\eps > 0$ is a constant depending only on $q, Q$.
    Hence our choice gives $\nu = O_{Q,q}(\log n)$.
    Then, by \cref{buffer}, we have
    \[
    |S| \leq \left(2Q + 1\right)\left(1 + \frac{Q\nu}{\eps}\right) =  O_{Q,q}(\log n).
    \]
    By \cref{signature} $|U|\geq n/(2\phi)$, 
    and since $\phi$ depends only on $Q, q$, for all sufficiently large $n$ 
    \[
        2\phi|S|\binom{|S| + Q - 1}{Q - 1} = O_{Q,q}(\log^Q n) < n.
    \]
    It follows by \cref{choice} that there exists a supplier $u_* \in U$ and distinct eligible recipients $i(j) \in U$ of all exceptional bundles $B_{u_*,j}$, satisfying $S$-compatibility conditions.
\end{proof}

\subsection{Suggested $E$-profiles for protected prefixes}

It remains to show that for sufficiently large $n$, for the chosen supplier $u_* \in U$ from \cref{choice} we can always construct a collection $\{\hpi_{u,i}^E\}_{i \in N}$ of suggested $E$-profiles satisfying
\begin{align}
        &&&\sum_{i \in N}\Pp_i(\hpi_{u,i}^E, c_s) \leq |E\cap [b_s]|,&\forall (b_s, c_s) \in \Ll_S.\tag{$\mathrm{E}_{\Ll_S}$-$u$}\label{ESu8}
\end{align}

We will do essentially the same way as in \cref{sec:intype}.
\begin{definition}
    For each agent $i \in N$, define \textbf{representative valuation} on $E$ by
    \[
        \hv_i(e) := \alpha_{i,s}\quad \text{where}\quad e \in E_s,\; s \in [t - 1],
    \]
    and extending it additively to subsets of $E$.
    Observe that for $e \in E_s\setminus F$ we have
    \[
        e\geq c_s + 1 \quad \implies \quad \hv_i(e) = v_i(c_s+1) \geq v_i(e),
    \]
    while for supplier $u_* \in U$ \cref{signature} gives $\hv_{u_*}(e) \geq v_{u_*}(e)$ for all $e \in E$.
\end{definition}

\begin{definition}
    We partition items of $M$ into the following $t$ types $M_1, \ldots, M_{t - 1}, M_{t}$:
    \[
        M_t := S\cup F,\qquad \qquad \forall s \in [t - 1]:\; M_s := E_s \setminus F.
    \]
\end{definition}
As in \cref{sec:intype}, we will consider patterns of bundles $B\subseteq M$:
\[
    p(B) := (|B\cap M_1|,\ldots, |B\cap M_t|),\quad \text{with}\quad |p|= \sum_{s = 1}^{t}p_s,
\]
and consider all possible patterns of size at most $q$:
\[
    P := \{p  \in \Z_{\geq 0}^t :\; 1 \leq |p|\leq q\}. 
\]

Recall \cref{blockpartition} from \cref{sec:intype}, its proof uses only bundle patterns and their counts in corresponding MMS partitions.
Hence, it also applies to our types $M_1,\ldots, M_t$ and patterns $p$, even though though these need not be types for the original valuations.
\begin{theorem}[\cref{blockpartition} restated]\label{intblock}
    For every pair of integers $t, q \geq 1$, there exists a global constant $D_{t,q} \geq 1$, depending only on $t$ and $q$, such that the following holds.

    For a fixed supplier $u_*$ and every agent $i\in N$, there exists a sequence of $K_i \leq n$ blocks
    \[
        \beta_{i,1} = (x_{i,1}, y_{i,1}), \ldots, \beta_{i, K_i} = (x_{i, K_i}, y_{i, K_i}),
    \]
    such that for every $k \in [K_i]$ we have $|\beta_{i,k}|\leq D_{t,q}$, and for every pattern $p \in P$ blocks satisfy
    \[
        \sum_{k = 1}^{K_i}x_{i,k}(p) = \lambda_{u_*}(p),\qquad \text{and}\qquad \sum_{k = 1}^{K_i}y_{i,k}(p) = \lambda_i(p).
    \]
    Thus, we can group MMS bundles of $u_*$ into disjoint blocks of size at most $D_{t,q}$, so that every block can be repartitioned into the same number of bundles with patterns from $i$'s MMS partition.
\end{theorem}

We will denote by $\fe_*$ the $E$-signature of supplier $u_*$, and denote the full $E$-class of $u_*$ by
\[
    V:= N_{\fe_*},
\] note that $U = (N\setminus W)\cap V$.
For every $\fe \in \Phi \setminus \fe_*$ write $n_\fe := |N_\fe|$.
\begin{definition}
    For every $\fe \in \Phi\setminus \fe_*$, choose a representative $i_\fe \in N_\fe$ and apply \cref{intblock} to $u_*$ and $i_\fe$.
    Let $\Bb(\fe) = \{\beta_{i_\fe,1}, \ldots, \beta_{i_\fe, K_{i_\fe}}\}$ be the resulting collection of blocks, and denote
    \[
        \Bb = \bigsqcup_{\fe \in \Phi\setminus\fe_*}\Bb(\fe).
    \]
    Each block $\beta \in \Bb$ is associated with a corresponding signature class, as well as its corresponding index within that class, i.e we treat $\beta = \beta_{i_\fe,k} = (x, y)$ and $\beta'= \beta_{i_{\fe'}, k'} = (x', y')$ as different blocks when $\fe \neq \fe'$ or $k \neq k'$, even if $x = x'$ and $y = y'$.
\end{definition}

For every pattern $p$ of supplier $u_*$'s MMS partition, we will reserve
\[
    \eta := \frac{|V|}{n}
\]
fraction of all supplier's bundles with pattern $p$.
As in \cref{sec:intype}, we consider the following linear program (\ref{intlp}).
\[
\begin{aligned}
    &\maximize&& \sum_{\beta \in \Bb}|\beta|\cdot z_\beta\\
    &\subto && \sum_{\beta \in \Bb}\lambda_\beta(p)\cdot z_\beta  \leq (1 - \eta)\lambda_{u_*}(p),&\forall p \in P;\\
    &&& \sum_{\beta \in \Bb(\fe)} |\beta|\cdot z_\beta \leq n_\fe,&\forall \fe \in \Phi \setminus \fe_*;\\
    &&& z_\beta \geq 0,&\forall \beta \in \Bb.
\end{aligned}\label{intlp}\tag{LP}
\]
It follows from the proof of \cref{introunding} that there exists an integral solution $\hz$ to (\ref{intlp}) satisfying
\[
    n - |V| - \sum_{\beta \in\Bb }|\beta|\cdot \hz_\beta \leq D_{t,q}(|P| + |\Phi| - 1),
\]
so only a few non-supplier-class agents are not covered by the selected repartition blocks.

As in \cref{realization}, we first realize the blocks selected by $\hz$ using a disjoint collection of supplier MMS bundles.
However, we will omit every block copy that uses bundles containing items of $M_t$, they will be instead used for exceptional agent allocation.
Every remaining block copy will provide acceptable bundles w.r.t representative valuations.
\begin{corollary}\label{partcon}
    There exists a set $J\subseteq N\setminus V$ and disjoint bundles $B_i\subseteq M\setminus M_t$ for $i \in J$, that we will call \textbf{source bundles}, such that for every $i \in J$, $|B_i|\leq q$ and $\hv_i(B_i) \geq \mu_i$, as well as
    \begin{itemize}
        \item the number of agents receiving these \textbf{source bundles} $B_i$ satisfies $|J| = n - |V| - \hZ$, where
        \[
            0 \leq \hZ \leq D_{t,q}(|P| + |\Phi| - 1 + |M_t|);
        \]
        \item the remaining items $M\setminus_{i \in J}B_i$ form a union of exactly $|V| + \hZ$ intact supplier bundles, including every supplier bundle intersecting $M_t$;
        \item for every pattern $p \in P$, at least $\eta\lambda_{u_*}(p)$ of the remaining supplier bundles have pattern $p$.
    \end{itemize}
\end{corollary}
\begin{proof}
    Take integral solution $\hz$, obtained above.
    As in \cref{realization}, we can choose disjoint collections of supplier bundles realizing the corresponding patterns of all selected block copies.

    Now, discard a selected copy of block $\beta$ if its corresponding bundles contain an item from $M_t$.
    Since different block copies use disjoint supplier bundles, there are at most $|M_t|$ discarded copies.
    Each copy is a repartition block of size at most $D_{t,q}$, therefore by discarding these copies we remove at most $D_{t,q}|M_t|$ proposed bundles for non-supplier-class agents.
    Hence, there will be at most
    \[
        D_{t,q}(|P| + |\Phi| - 1 + |M_t|)
    \]
    agents outside $V$ without an allocated bundle.

    Now, consider a block copy $\beta \in \Bb(\fe)$.
    Since the supplier MMS bundles corresponding to $\beta$ contain no items of $M_t$, every output pattern $p$ used by $\beta$ satisfies $p_t = 0$.
    Furthermore, since \cref{intblock} preserves the number of items of each type during repartition, such pattern $p$ must occur in the MMS partition of agent $i_\fe$ representing $N_\fe$, i.e for some bundle $B_{i_\fe,j}$ disjoint from $M_t$.

    We use the item-count construction from the proof of \cref{bundleblock} to this $\beta$ and obtain a collection of source bundles $B_i$ for the agents of type $N_\fe$.
    By the reasoning above, if a resulting bundle $B_i$ of pattern $p$, produced by block $\beta$, is allocated to agent $i \in N_\fe$, then
    \[
        \hv_i(B_i) = \sum_{s = 1}^{t - 1}p_s\cdot \alpha_{i,s} = \hv_{i_\fe}(B_{i_\fe,j}) \geq v_{i_\fe}(B_{i_\fe,j}) \geq \mu_{i_\fe} = \mu_{i}.
    \]
    Repeating this process for every block $\beta$ selected by $\hz$, we get a collection of agents $J\subseteq N\setminus V$ with allocated source bundles.
    By the same reasoning as in \cref{realization}, this construction uses exactly $|J|$ original supplier bundles, leaving exactly $n - |J| = |V| + \hZ$ bundles intact.

    Finally, before discarding block copies, the selection $\hz$ left at least $\eta\cdot \lambda_{u_*}(p)$ supplier bundles of pattern $p$ intact, for every $p \in P$.
    Discarding block copies only leaves more bundles intact, so the same lower bound holds.
\end{proof}

These source bundles obtained in \cref{partcon} will define the suggested supplier profiles.
The partial construction of \cref{partcon} leaves only $\hZ$ agents outside $V$ without an allocated source bundle.
As in \cref{sec:intype}, we will complete these agents greedily from the intact bundles of supplier's MMS partition.
First, we will deal with bundles disjoint from $M_t$.
\begin{lemma}\label{clean}
    Call a supplier MMS bundle $B_{u_*,j}$ \textbf{clean} if $B_{u_*,j}\cap M_t = \varnothing$.
    For every agent $i \in N$, at least
    \[
        \frac{n - qQ|M_t|}{qQ}
    \]
    clean bundles of the supplier's MMS partition satisfy $\hv_i(B_{u_*,j})\geq \mu_i$.
\end{lemma}
\begin{proof}
    At most $|M_t|$ supplier bundles intersect $M_t$, and their union contains at most $q|M_t|$ items, hence has value at most $qQ|M_t|$ for every agent $i \in N$.

    Note that for every item $e$ of the clean bundle, by definition $\hv_i(e) \geq v_i(e)$.
    Therefore,
    \[
        \sum_{\substack{j\in[n]\\B_{u_*,j}\text{ clean}}}\hv_i(B_{u_*,j})\geq v_i(M) - qQ|M_t| \geq n\mu_i - qQ|M_t|.
    \]
    Let $g_i$ be the number of clean bundles that have representative value $\hv_i$ at least $\mu_i$.
    Since the representative value of each bundle cannot exceed $qQ$, and every bundle not counted by $g_i$ has value at most $\mu_i - 1$, 
    \[
       \sum_{\substack{j\in[n]\\B_{u_*,j}\text{ clean}}}\hv_i(B_{u_*,j})\leq g_i\cdot qQ + n(\mu_i - 1).
    \]
    The lemma follows.
\end{proof}

We can now complete the source bundles for every agent $i \in N$ and use them to define all suggested profiles.
The resulting partition of $M$ will also automatically give the protected prefix conditions, as every demand has a distinct originating item.

\begin{lemma}\label{complete}
    Fix a supplier $u_*$, and suppose that
    \[
        n \geq 4\phi q Q D_{t,q}(|P| + |\Phi| - 1 + |M_t|).
    \]
    Then the source bundles from \cref{partcon} extend to a partition $\{B_1, \ldots, B_n\}$ of $M$ such that
    \begin{itemize}
        \item every exceptional agent $i$ is assigned a corresponding exceptional bundle $B_i = B_{u_*,j}$ of $u_*$;
        \item every other agent $i$ has $B_i\subseteq E$, $|B_i|\leq q$ and $\hv_i(B_i) \geq \mu_i$.
    \end{itemize}
    In addition, with suggested profiles $\hpi_{u_*,i}$ for exceptional $i$ as in \cref{suplargen}, and for every other agent $i$ defined as
    \[
        \hpi_{u_*,i}^E := \{\hv_i(e) :\; e \in B_i\},\qquad\qquad \hpi_{u_*,i}^S = \varnothing,
    \]
    suggested profiles satisfy $v(\hpi_{u_*,i}) \geq \mu_i$ and $|\hpi_{u_*,i}|\leq q$ for all $i \in N$, and  
    \begin{align}
        &&&\sum_{i \in N}\Pp_i(\hpi_{u,i}^E, c_s) \leq |E\cap [b_s]|,&\forall (b_s, c_s) \in \Ll_S.\tag{$\mathrm{E}_{\Ll_S}$-$u$}\label{ESu9}
\end{align}
\end{lemma}
\begin{proof}
    The assumption on $n$ implies $n\geq 2qQ|M_t|$.
    By \cref{clean}, for every $i \in N$ there are at least $n/(2qQ)$ clean bundles of the supplier's MMS partition that have $\hv_i$-value at least $\mu_i$.

    Note that for a clean bundle $B_{u_*,j}$, its value $\hv_i(B_{u_*,j})$ depends only on its pattern.
    By \cref{partcon}, we keep at least $\eta$-fraction of every supplier's pattern intact, therefore every unassigned agent $i\notin V$ has at least
    \[
        \eta\cdot \frac{n}{2qQ}\geq \frac{n}{4\phi qQ} \geq D_{t,q}(|P| + |\Phi| - 1 + |M_t|) \geq \hZ
    \]
    clean intact supplier bundles left of value at least $\mu_i$.
    Therefore, we can give to each of the $\hZ$ remaining non-supplier-class agents a distinct clean supplier bundle greedily, and they will each receive an $\hv_i$-value at least $\mu_i$.

    Now, exactly $|V|$ intact supplier bundles remain, including those intersecting $M_t$, as we only used clean bundles so far.
    Note that every designated exceptional bundle is still available, as each of them intersects $M_t$.

    Every exceptional agent $i$ comes from a supplier's equivalence class $U_\sigma\subseteq V$.
    Give every exceptional bundle of $u_*$ to the corresponding agent in $V$, according to \cref{choice}.
    For every remaining agent, assign the remaining supplier bundles arbitrarily.
    Note that every such bundle is disjoint from $S$, but may intersect $F$.
    However, since every $i \in V$ has the same $E$-signature as the supplier $u_*$, for every non-exceptional agent $i \in V$ receiving supplier bundle $B_{u_*,j}$ it holds
    \[
        \hv_i(B_{u_*,j}) = \hv_{u_*}(B_{u_*,j}) \geq v_{u_*}(B_{u_*,j}) \geq \mu_{u_*} = \mu_i.
    \]
    Thus, for the obtained partition $\{B_1, \ldots, B_n\}$ of $M$,  for every non-exceptional $i$, $\hv_i(B_i) \geq \mu_i$.

    By \cref{choice}, all exceptional suggested profiles already have size at most $q$ and value at least the MMS of corresponding recipient.
    For a non-exceptional agent $i$, profile $ \hpi_{u_*,i}^E := \{\hv_i(e) :\; e \in B_i\}$ satisfies $v(\hpi_{u_*,i}^E) = \hv_i(B_i) \geq \mu_i$ by construction.

    It remains to verify that these suggested profiles satisfy the protected prefix conditions.
    Consider a demand $\alpha \in \hpi_{u_*,i}^E$, and let $e \in B_i\cap E$ be its originating item.
    Suppose that $e \in E_s$.
    If $i$ is not exceptional, then $\alpha = \alpha_{i,s}$, and $m_i(\alpha) \geq c_s + 1$.
    If $i$ is exceptional, then $\alpha = v_{u_*}(e)$ and $i \in N_{\fe_*}$, and by \cref{signature} it holds $v_{u_*}(e) \leq \alpha_{u_*,s} = \alpha_{i,s}$.
    Thus, we also have $m_i(\alpha) \geq c_s + 1$.

    Consider a protected prefix $(b_r, c_r)$ for $r < t$.
    A demand counted at prefix $c_r$ must originate from an item in some $E_s$ with $s < r$.
    Since source bundles $B_i$ partition $M$, all such originating items are distinct.
    As a result,
    \[
        \sum_{i \in N}\Pp_i(\hpi_{u_*,i}^E, c_r) \leq \sum_{s < r}|E_s| = |E\cap [b_r]|.
    \]
    As for the last protected pair $(b_t, c_t) = (m, m)$, there are exactly $|E|$ suggested $E$-demands, one for each item of $E$.
    The lemma follows.
\end{proof}

Combining all of the results, we can finally prove \cref{integer}.
\begin{proof}[Proof of \cref{integer}]
    Choose parameters $\eps, \nu$ as in \cref{suplargen}.
    Since $t \leq 2Q + 1$, all of $D_{t,q}, |P|, |\Phi|$ are bounded by constants depending only on $Q, q$.
    Thus, for sufficiently large $n$ the supplier-selection condition of \cref{choice} holds.
    In addition,
    \[
        |M_t| = |S| + |F| = O_{Q,q}(\log n),
    \]
    hence for $n$ large enough (depending only on $Q, q$) we have as well
    \[
        n \geq 4\phi q Q D_{t,q}(|P| + |\Phi| - 1 + |M_t|).
    \]

    Use \cref{ESrounding} in order to obtain the initial collection $\{\pi_i\}_{i \in I}$ for agents $I\subseteq N$.
    Then,  apply \cref{choice} and \cref{complete} in order to obtain a collection of suggested profiles $\{\hpi_{u_*,i}\}_{i \in N}$ satisfying the conditions of \cref{epsnuconst}.
    Hence, using the bounded replacement of \cref{epsnuconst} yields an exact MMS allocation via \cref{loadequiv}.
\end{proof}
\section{$1$-out-of-$d$ MMS for $d = n + O(\log^2 n)$}
\label{sec:dmms}
In this section, we  repeat our proof of \cref{thm:dmms} from \cite{FG26}, as this it will play crucial role in our proof of the main result \cref{thm:main}.
\begin{theorem}[\cref{thm:dmms}, \cite{FG26}]\label{dmms}
    Consider a fair allocation instance with items $M$ and $n$ additive agents $N$ with valuations $\{v_i\}_{i \in N}$.
    There always exists an $\mms(d)$ allocation when
    \[
        d \geq n + 46(1 + \log n)^2,
    \]
\end{theorem}
We make the following assumptions, standard in designing approximate MMS allocations.
\begin{assumption}\label{ass:wlog}
    Without loss of generality, in the setting of \cref{thm:largen} we can assume that
    \begin{itemize}
        \item for every agent $i \in N$ and item $e \in M$ it holds $v_i(e) \in [0, 1]$;
        \item for every $i \in N$, $d$-partition $(B_{i,1}, \ldots, B_{i,d})$ of $M$ witnessing $\mms(M, v_i, d)$ satisfies
        \[
            \forall j \in [d]:\; v_i(B_{i,j}) = 1;
        \]
        We refer to the bundles in this $d$-partition as the {\em MMS bundles} of agent $i$.
        \item (\cite{Bouveret2016}) items $M = \{e_1, \ldots, e_m\}$ are ordered so that
        for every agent $i \in N$ it holds that
        \[
            v_i(e_1) \geq v_i(e_2) \geq \ldots \geq v_i(e_m).
        \]
    \end{itemize}
\end{assumption}
Notably, $v_i(M) = d$ for every $i \in N$.

We will start with several straightforward observations.
\begin{observation}\label{leftovers}
    Let $U\subseteq N$, $S\subseteq M$ be such that $v_u(S) \geq 2|U| - 1$ for every $u \in U$.
    There exists an allocation of $S$ to $U$ such that every agent $u \in U$ receives a bundle of value at least $1$.
\end{observation}
\begin{proof}
    If $|U| = 1$, give all of $S$ to a single agent in $U$.
    For $|U| > 1$, add items in arbitrary order to an initially empty bundle $B$, until some agent $u \in U$ values $v_u(B) \geq 1$.
    Note that just before the last item was added to $B$, every agent valued $B$ less than $1$, and for every $u'\neq u$, $v_{u'}(B) < 2$.
    Hence, $v_{u'}(S - B) \geq 2|U| - 3 = 2(|U|-1) - 1$.
    The lemma follows by induction on $|U|$.
\end{proof}

If $d \geq 2n - 1$, we can take $U = N$ and $S = M$ and apply \cref{leftovers}.
For every agent $i \in N$, $v_i(S) = v_i(M) = d \geq 2n - 1 = 2|U| - 1$.
From now on, assume $n + 1 \leq d \leq 2n - 1$.

\cref{leftovers} suggests the following.
Suppose that we allocated almost all agents, using some set of items $E\subseteq M$ such that every agent $i \in N$ values $E$ at most $\eta d$ for some parameter $\eta \in (0, 1]$.
Then, as long as the number of the remaining agents is sufficiently small, we can get their MMS from only the remaining items.

\begin{corollary}\label{leftoveralloc}
    Let $\eta \in (0, 1]$, and suppose that $E\subseteq M$ satisfies $v_i(E) \leq \eta d$ for every $i \in N$.
    For any set $U\subseteq N$ of $|U|\leq \frac{(1 - \eta)d + 1}{2}$ agents, there exists an allocation of $M\setminus E$ to $U$ such that every agent $u \in U$ receives a bundle of value at least $1$.

    In particular, when $\eta =n/d$ or $\eta = (n + 1)/d$, such set $E$ should satisfy respectively $v_i(E) \leq n$ or $v_i(E) \leq n + 1$ for every $i \in N$, and we should allocate all but $(d - n)/2$ agents.
\end{corollary}
\begin{proof}
    If $U = \varnothing$, there is nothing to do.
    Otherwise, as $v_u(M) = d$ for every $u \in U$,
    \[
    v_u(M \setminus E)  = v_u(M) - v_u(E) \geq (1-\eta)d \geq 2|U| - 1.
    \]
    This is exactly the condition of \cref{leftovers} with sets $S = M \setminus E$ and $U$.
\end{proof}

We will now focus on finding such set $E\subseteq M$ of total value $v_i(E)\leq \eta d$ for all $i \in N$ (with $\eta$ to be chosen as $(n + 1)/d$), that we can use to allocate all but $\frac{(1-\eta)d + 1}{2} = (d - n)/2$ agents.

By assumption, we associate $M$ with $[m]$, and for every agent $i \in N$ we simultaneously have
\[
    v_i(1) \geq v_i(2) \geq \ldots \geq v_i(m).
\]
We observe that, if one wants to bound the value of items $E\subseteq M$, it suffices to bound the total number of items from $E$ in every \textit{prefix} of $M = \{1, \ldots, m\}$.
\begin{lemma}\label{prefixbound}
    Let $E\subseteq M$, and suppose that for every $0 \leq \ell \leq m$ it holds $|E\cap [\ell]| \leq \eta \ell$ for some value $\eta \in (0, 1]$.
    Then for every agent $i \in N$ it holds $v_i(E)\leq \eta d$.
\end{lemma}
\begin{proof}
    Observe that for every $0 \leq \ell \leq m$, $\ell\in E$ if and only if $|E\cap [\ell]| - |E\cap [\ell - 1]| = 1$.
    Hence,
    \begin{multline*}
        v_i(E) = \sum_{\ell = 1}^mv_i(\ell)\cdot \big(|E\cap [\ell]| - |E\cap [\ell - 1]|\big) = \sum_{\ell = 1}^m\big(v_i(\ell) - v_i(\ell + 1)\big)\cdot |E\cap[\ell]|
        \\
         \leq \eta\sum_{\ell = 1}^m\big(v_i(\ell) - v_i(\ell + 1)\big)\cdot \ell = \eta\sum_{\ell = 1}^mv_i(\ell) =  \eta d.
    \end{multline*}
    We used a conventions $[0] = \varnothing$ and $v_i(m + 1) = 0$.
\end{proof}

We will refer to $|E\cap [\ell]|$ as a \textit{$\ell$-prefix load} of the set $E$.
Our approach in \cite{FG26} to finding a set $E$ with bounded prefix loads is to instead consider demand profiles, similar to  \cref{sec:integer}.
In that section, we proved that finding an allocation of $M$ to some agents $I\subseteq N$ is equivalent to finding a feasible collection of appropriate demand profiles $\Pi_I$.
\begin{lemma}[\cref{loadequiv} restated]\label{loadequiv2}
    Let $I\subseteq N$, let $\pi_i = \{\alpha_{i,1}, \ldots, \alpha_{i,|\pi_i|}\}$ be some demand profile of agent $i \in I$, and let $\Pi_I := \{\pi_i\}_{i \in I}$.
    For any set $E\subseteq M$ of items, we can simultaneously satisfy all profiles of $\Pi_I$ using distinct items of $E$ if and only if 
    \[
        \Pp(\Pi_I, \ell) \leq |E\cap [\ell]|,\qquad\qquad\forall 0 \leq \ell \leq m.
    \]
    Furthermore, if $K := \sum_{i \in I}|\pi_i|$, it suffices to use only the first $K$ items of $E$.
    In particular, when $E = M$, we can satisfy all profiles of $\Pi_I$ using only items $\{1, \ldots, K\}$ as long as
    \[
        \Pp(\Pi_I, \ell) \leq \ell,\qquad\qquad\forall 0 \leq \ell \leq m.
    \]
\end{lemma}

In \cref{prefixbound}, however, we seek for a set $E\subseteq M$ such that prefix load of $E$ satisfies $|E\cap [\ell]|\leq \eta \ell$, while \cref{loadequiv2} guarantees only a trivial bound of $\ell$.
A slight modification of the proof of \cref{loadequiv2} allows us to get the desired set $E\subseteq M$, provided that the given collection of profiles $\Pi_I$ satisfies a stronger prefix load condition.

\begin{corollary}[Refinement of \cref{loadequiv2}]\label{etaload}
    Let $I\subseteq N$, let $\pi_i$ be some demand profile of agent $i \in I$, and let $\Pi_I := \{\pi_i\}_{i \in I}$.
    Let $\eta \in (0, 1]$, and suppose that for every $0 \leq \ell \leq m$ it holds
    \[
        \Pp(\Pi_I, \ell) \leq \eta\ell.
    \]
    Then, all profiles of $\Pi_I$ can be simultaneously satisfied using distinct items of $E\subseteq M$ such that
    \[
        |E\cap [\ell]|\leq \eta \ell,\qquad \forall 0 \leq \ell \leq m.
    \]
\end{corollary}
\begin{proof}
    List all demands $\alpha \in \Pi_I$ as pairs $(i, \alpha)$ where $\alpha\in \pi_i$, including repetitions, and sort them according to breakpoint values $m_i(\alpha)$.
    We obtain a sorted list of $K = \sum_{i \in I}|\pi_i| =\Pp(\Pi_I, m)$ pairs
    \[
        (i_1, \alpha_1), \ldots, (i_K, \alpha_K), \qquad \qquad 1\leq m_{i_1}(\alpha_1) \leq \ldots \leq m_{i_K}(\alpha_K)\leq m.
    \]
    Denote $m_k := m_{i_k}(\alpha_k)$, $m_1 \leq \ldots \leq m_K$.
    Following the proof of \cref{loadequiv2}, we have for every $k \in [K]$ that $k \leq \Pp(\Pi_I, m_k) \leq\eta m_k$.
    Take this $m_k = m_{i_k}(\alpha_k)$, and allocate item $e_k := \lceil k/\eta\rceil$ to agent $i_k$.
    Note that $1 \leq e_k \leq m_k \leq m$, so $e_k \in M$ and satisfies demand $\alpha_k$.
    Since $\eta \leq 1$, it holds $e_{k + 1} \geq e_k + 1$, hence such allocation will use distinct items for every demand.

    Let $E = \{e_k\}_{k \in [K]}$ be the set of allocated items.
    We immediately obtain for every $0\leq \ell \leq m$:
    \[
        |E\cap [\ell]| = |\{ k \in [K] : \lceil k/\eta\rceil \leq \ell\}| = |\{k \in [K] : k \leq \eta\ell\}| =  \min\{K, \lfloor \eta \ell\rfloor\} \leq \eta \ell.
    \]
\end{proof}

So, to find a set $E\subseteq M$ with bounded prefix loads, it suffices to find a collection $\Pi_I$ of demand profiles of some agents $I \subseteq N$ with bounded prefix loads.
A natural starting point would be the profiles of $i \in N$ induced by its MMS bundles $\{B_{i,1}, \ldots, B_{i, d}\}$.
That is, some $\pi_{i,j} = \{v_i(e) : e \in B_{i, j}\text{ and }v_i(e) > 0\}$.
The issue here is that the total number of demands in $\pi_{i,j}$ can be as large as $m$.
For this reason, we will split each profile $\pi$ into two: a \textit{head} profile with bounded number of demands, and a \textit{tail} profile.

\subsection{Head and tail profiles}

As per \cref{etaload}, we would like to find a set $I\subseteq N$ of agents and assign to them some value-$1$ profiles with bounded total prefix load.
If $|I|$ is sufficiently large, \cref{leftoveralloc} guarantees that unassigned agents will be able to receive a bundle of value at least $1$.

Instead of finding a collection of full profiles of value $1$ for some subset $I$ of agents, at once, we will split the task and separately find a collection $\Pi_I$ of \textit{head profiles} $\pi_i \in \Pi_I$ and a collection $\Psi_I$ of \textit{tail profiles} $\psi_i \in \Psi_I$.
Agent $i \in I$ in this case receives a full profile $\{\pi_i, \psi_i\}$.

\begin{definition}
    For $h \in [m]$, we call $\{1, \ldots, h\}$ \textbf{head items}, and $\{h + 1, \ldots, m\}$ \textbf{tail items}.
    For agent $i \in N$, the \textbf{tail value} of $i$ is
    \[
        T_i = \sum_{e > h}v_i(e).
    \]
\end{definition}
\begin{definition}
    For an agent $i \in N$, we call a demand profile $\pi$ a \textbf{head profile} for $i$, if every $\alpha \in \pi$ satisfies $\alpha \leq v_i(e)$ for some head item $e \leq h$.
    
    Similarly, we call a demand profile $\psi$ a \textbf{tail profile} for $i$, if every $\sigma \in \psi$ satisfies $\sigma \leq v_i(e)$ for some tail item $e > h$.
\end{definition}

We will have the same conventions as before.
\begin{definition}
    An \textbf{empty profile} is $\varnothing$.
    If agent $i \in N$ is assigned an empty head profile $\pi = \varnothing$ or an empty tail profile $\psi = \varnothing$, it means that he may receive no items from head or tail.
    It always holds $\Pp_i(\varnothing, \ell) = 0$ for all $0 \leq \ell \leq m$.
\end{definition}
We allow an agent $i \in N$ to be one of $I$ assigned ones, but get an empty profile $\pi_i = \varnothing$ or an empty tail profile $\psi_i = \varnothing$.

We will also define a ``no-profile'' $\bot$, which corresponds to an option of not choosing $i \in N$ into the final set of agents $I$.
\begin{definition}
    A \textbf{no-profile} is $\bot$.
    If agent $i \in N$ is assigned $\bot$, it means that we do not select this agent, and he receives no items at all.
    It always holds $\Pp_i(\bot, \ell) = 0$ for all $0 \leq \ell \leq m$.
\end{definition}

Our approach is to first find a sufficiently large set $I\subseteq N$ of agents and assign to them \textit{only head profiles} $\Pi_I = \{\pi_i\}_{i \in I}$, that use only items of $[h]$.
All the other $N - I$ agents will get $\bot$, and will get a bundle from \cref{leftoveralloc} later.
After each $i \in I$ gets a head profile $\pi_i$, we will find a corresponding tail profile $\psi_i$ so that $v(\pi_i) + v(\psi_i) \geq 1$.
\cref{etaload} then guarantees that $i$ will get a bundle of value at least $1$.

Since profiles $\Pi_I$ are not using all items of $M$, many profiles $\pi_i \in \Pi_I$ may have $v(\pi_i) < 1$.
We introduce a notion of profile deficit, to measure much value $i \in I$ misses from $\pi_i$.

\begin{definition}
    For a head profile $\pi \neq \bot$, its \textbf{deficit} is
    \[
        \Delta(\pi) := (1 - v(\pi))_+.
    \]
    Here $(\cdot)_+ := \max\{0, \cdot\}$.
\end{definition}

Suppose that agent $i \in N$ gets assigned a head profile $\pi_i\neq \bot$.
If $\Delta(\pi_i) > 0$, then satisfying $\pi_i$ with a subset of head items $[h]$ does not guarantee that $i$ gets a bundle of value $1$.
So, we will need to use some items from the tail $\{h + 1, \ldots, m\}$ to compensate for the missing value.

We do so using an approach inspired by faithful rounding method, which also appeared in \cite{BEF22BoBW,EG26-2}.
Suppose that, in addition to head items satisfying $\pi_i$, agent $i$ could take the same $\frac{\Delta(\pi_i)}{T_i}$-fraction from every tail item $e > h$.
Then, $i$ is guaranteed to receive a value at least
\[
    v(\pi_i) + \frac{\Delta(\pi_i)}{T_i}\sum_{e > h}v_i(e) = v(\pi_i) + \Delta(\pi_i) \geq 1.
\]

In essence, value $\frac{\Delta(\pi_i)}{T_i}$ defines a ``fractional allocation'' of the tail items to agent $i$.
While it is known  how to round a fractional allocation by sacrificing at most one item, we would like to give each agent total value $1$ exactly, so giving up items is not an option here.
We will provide sufficient conditions on the prefix load of head profiles $\pi_i$ that allows ``rounding'' this fractional allocation without any loss.

\begin{definition}
    For agent $i \in N$ and head profile $\pi$, let
    \[
        \tau_i(\pi) := \begin{cases}
            \Delta(\pi)/T_i,& \pi\neq \bot\text{ and }T_i > 0;\\
            0,& \text{otherwise}.
        \end{cases}
    \]
\end{definition}
Consider some prefix $\ell > h$, where we potentially may use tail items.
Taking the same $\tau_i(\pi_i)$-fraction from every tail item $e > h$ can be viewed as generating only a $\tau_i(\pi_i)$-fraction of a demand per item taken.
So, for a prefix $\ell > h$, the total additional load  agent $i$ generated via this process, i.e the amount of added demands, is ``essentially'' $\tau_i(\pi_i)(\ell - h)$.
This gives an idea on how much slackness the head profiles will need to allow completing the tail profiles.
\begin{lemma}[Tail completion]\label{etatailcomplete}
    Let $I\subseteq N$, let $\pi_i$ be a head profile assigned to agent $i \in I$, and let $\Pi_I = \{\pi_i\}_{i \in I}$.
    Assume that $T_i = 0$ implies $\Delta(\pi_i) = 0$ for every $i \in I$.
    For $\eta \in (0, 1]$, suppose that for every $0 \leq \ell \leq h$ we have $\Pp(\Pi_I, \ell) \leq \eta \ell$, and for every $h + 1 \leq \ell \leq m$ we have
    \[
         \Pp(\Pi_I, \ell) \leq \eta\ell -(\ell - h)\sum_{i \in I}\tau_i(\pi_i) - |I|.
    \]
    There exists a collection $\Psi_I := \{\psi_i\}_{i \in I}$ of tail profiles, where agent $i$ is assigned $\psi_i$, such that
    \[
        \forall 0 \leq \ell \leq m:\; \Pp(\Pi_I, \ell) + \Pp(\Psi_I, \ell) \leq \eta \ell,
    \]
    and for every $i \in I$ it holds $v(\pi_i) + v(\psi_i) \geq 1$.
\end{lemma}
\begin{proof}
    Consider $i \in I$, and denote $\tau_i := \tau_i(\pi_i)$.
    Construct the tail profile $\psi_i$ as follows.
    We break the tail $\{h + 1, \ldots, m\}$ into segments of length roughly $ 1/\tau_i$, and in each segment add the value of the first item to $\psi_i$.
    That is, for every $t = 1, \ldots, \lceil \tau_i\cdot  (m - h)\rceil$ add $v_i(e_{i,t})$ to $\psi_i$, as long as it is positive, where
    \[
        e_{i,t} = h + 1 + \left\lfloor \frac{t - 1}{\tau_i}\right\rfloor.
    \]
    The intuition behind this construction is straightforward: after taking the first item of every segment of length $1/\tau_i$, the total value of selected items is at least as large as if we took a $\tau_i$-fraction of every tail item.
    If $\tau_i = 0$, then we simply put $\psi_i = \varnothing$.
    
    We can describe this construction equivalently as follows.
    For $1 \leq k \leq m - h$, let $\sigma_{i,k} := v_i(h + k)$, i.e the value of $k$-th tail item for agent $i$.
    Then, for every $k \in [m - h]$, we add exactly
    \[
        \lceil \tau_i \cdot k \rceil - \lceil \tau_i\cdot (k - 1)\rceil 
    \]
    copies of $\sigma_{i,k}$ to the profile $\psi_i$ (as long as it is positive).
    To see this, observe that
    \[
        e_{i,t} \leq h + k \iff   \left\lfloor \frac{t - 1}{\tau_i}\right\rfloor \leq k - 1 \iff t - 1 < \tau_i\cdot k \iff t \leq \lceil \tau_i\cdot k\rceil.
    \]
    So, the first $k$ tail items contain at most $\lceil \tau_i\cdot k\rceil$ selected items from $\psi_i$, and the number of copies of tail item $k$ is exactly $\lceil \tau_i \cdot k \rceil - \lceil \tau_i\cdot (k - 1)\rceil $.
    Note that both constructions allow $\tau_i > 1$, in which case agent $i$ adds to $\psi_i$ multiple copies of the same demand.
    In our actual applications of this lemma, however, we will have $\tau_i \leq 1$, so each item will add at most one demand copy to $\psi_i$.

    Let $\sigma_{i,m - h + 1} = 0$.
    It is easy to see that such choice of $\psi_i$ covers the entire deficit of agent $i$:
    \begin{multline*}
        v(\psi_i) = \sum_{k = 1}^{m - h}\sigma_{i,k}\cdot \big(\lceil \tau_i \cdot k \rceil - \lceil \tau_i\cdot (k - 1)\rceil \big)
        =\sum_{k = 1}^{m - h}\lceil \tau_i\cdot k\rceil\cdot (\sigma_{i,k} - \sigma_{i, k + 1})\\ \geq \tau_i\sum_{k = 1}^{m - h}k\cdot (\sigma_{i,k} - \sigma_{i, k + 1}) = \tau_i\sum_{k = 1}^{m - h}\sigma_{i,k} = \tau_iT_i = \Delta(\pi_i).
    \end{multline*}
    Since $\Delta(\pi_i)= (1 - v(\pi_i))_+$, it immediately follows that $v(\pi_i) + v(\psi_i) \geq 1$.

    It remains to bound the additional prefix load generated by $\Psi_I = \{\psi_i\}_{i \in I}$.
    Note that for every $i \in I$, the breakpoint of $\sigma_{i,k} \in \psi_i$ is at least $h + k$.
    Hence, for every $0 \leq \ell \leq h$ we have 
    \[
        \Pp_i(\psi_i, \ell) = \sum_{\sigma \in \psi_i}\One\{m_i(\sigma) \leq \ell\} = 0,
    \]
    for every $i \in I$, and thus $\Pp(\Psi_I, \ell) = 0$.
    For $\ell > h$, as mentioned earlier the number of selected items from $\psi_i$ within the first $k = \ell - h$ tail items is at most $\lceil \tau_i\cdot (\ell - h)\rceil\leq \tau_i\cdot (\ell - h) + 1$.
    So, the additional load generated by tail profiles $\Psi_I$ at prefix $\ell > h$ is at most
    \[
        \Pp(\Psi_I, \ell) \leq \sum_{i \in I}\big(\tau_i\cdot (\ell - h) + 1) = (\ell - h)\sum_{i \in I}\tau_i + |I|.
    \]
    The lemma follows.
\end{proof}

Thus, in order to get a set $I\subseteq N$ of agents with a collection of \textit{full profiles} satisfying \cref{etaload}, it suffices to find $I\subseteq N$ and a collection of \textit{only head profiles} $\Pi_I = \{\pi_i\}_{i \in I}$ satisfying \cref{etatailcomplete}.
We explain how to choose a sufficiently large set $I\subseteq N$ (of size roughly $n - O(\log^2 n)$) and corresponding profiles $\Pi_I$ with these properties in the following subsection.

\subsection{Fractional head profiles}
\label{sec:fracpro}

We will first define a collection of head profiles that agents $i \in N$ will choose from.
\begin{definition}
    For every agent $i \in N$, let $\{B_{i,1},\ldots, B_{i,d}\}$ denote the $d$-partition of $M$ witnessing $\mms(M, v_i, d)$, for every $B_{i,j}$, $j \in [d]$, it holds $v_i(B_{i,j}) = 1$.
    For every $j \in [d]$, define 
    \[
        \pi_{i,j} := \left\{v_i(e) :\; e \in B_{i,j}\cap [h], v_i(e) > 0\right\}.
    \]
    Let $\Pi := \{\varnothing, \bot\} + \bigcup_{i \in N}\{\pi_{i,j} : j \in [d]\}$.
\end{definition}

Instead of considering these profiles directly, for every agent $i \in N$ we will define a \textit{distribution} over profiles $\Pi$.
We will refer to these distributions as \textit{fractional profiles}.
Our goal then will be first to define a collection of fractional profiles that satisfy the prefix load condition of \cref{etatailcomplete}, and then round it to an integral collection while preserving every prefix constraint.

\begin{definition}
    Let $\Pi$ be a collection of head profiles, containing $\varnothing$ and $\bot$.
    A \textbf{fractional head profile} of agent $i \in N$ is a pair $(x_i, \Pi)$, where $x_i(\pi) \geq 0$ for every $\pi \in \Pi$ and $\sum_{\pi \in \Pi}x_i(\pi) = 1$.
\end{definition}

We naturally extend the definition of (head) prefix load to fractional profiles.

\begin{definition}
    For agent $i \in N$ with fractional head profile $(x_i, \Pi)$, his \textbf{fractional $\ell$-prefix head load} of $\Pi$ for $0 \leq \ell \leq m$ is 
    \[
        \Hh_i(x_i, \Pi, \ell) := \sum_{\pi\in \Pi}x_i(\pi)\cdot \Pp_i(\pi, \ell).
    \]
    Assume that every agent $i \in N$ has fractional head profile $(x_i, \Pi)$ for the shared set of profiles $\Pi$, and let $x = \{x_i\}_{i \in N}$.
    For $0 \leq \ell \leq m$, the \textbf{total fractional $\ell$-prefix head load} of $\Pi$ for $x$ is 
    \[
        \Hh(x, \Pi, \ell) := \sum_{i \in N}\Hh_i(x_i, \Pi, \ell) = \sum_{i \in N}\sum_{\pi\in \Pi}x_i(\pi)\cdot \Pp_i(\pi, \ell).
    \]
\end{definition}

In addition to the fractional head load, we will also consider the fractional \textit{tail load}.
As all agents share a common tail $\{h + 1, \ldots, m\}$, instead of defining the fractional tail load for every prefix $\ell$, we omit the multiplier $(\ell - h)$ from the definition and consider one common term.
\begin{definition}
    For agent $i \in N$ with fractional head profile $(x_i, \Pi)$, his \textbf{fractional tail load} of $\Pi$ is
    \[
        \tau_i(x_i, \Pi) := \sum_{\pi \in \Pi}x_i(\pi)\cdot \tau_i(\pi).
    \]
    Assume that every agent $i \in N$ has fractional head profile $(x_i, \Pi)$ for the shared set of profiles $\Pi$, and let $x = \{x_i\}_{i \in N}$.
    The \textbf{total fractional tail load} of $\Pi$ for $x$ is 
    \[
        \tau(x, \Pi) := \sum_{i \in N}\tau_i(x_i, \Pi) = \sum_{i \in N}\sum_{\pi\in \Pi}x_i(\pi)\cdot \tau_i(\pi).
    \]
\end{definition}
Finally, we will be interested in the expected number of agents $i \in N$ that are assigned a non-$\bot$ profile according with respect to fractional profiles $(x_i, \Pi)$.
\begin{definition}
    Assume that every agent $i \in N$ has fractional head profile $(x_i, \Pi)$ for the shared set of profiles $\Pi$, and let $x = \{x_i\}_{i \in N}$.
    The \textbf{fractional number of assigned profiles} with respect to $(x, \Pi)$ is
    \[
        \Ii(x, \Pi) := \sum_{i \in N}\sum_{\pi\neq\bot}x_i(\pi).
    \]
\end{definition}

For every agent $i \in N$, our starting fractional head profile $(x_i, \Pi)$ will simply be a uniform distribution over agent $i$'s MMS profiles.
\begin{definition}
    Let $\Pi := \{\varnothing, \bot\} + \bigcup_{i \in N}\{\pi_{i,j} : j \in [d]\}$.
    For every agent $i \in N$, consider the following fractional head profile $(\hx_i, \Pi)$: for every $\pi \in \Pi$ let
    \[
        \hx_i(\pi) = \frac{|\{j \in [d]: \pi_{i,j} = \pi\}|}{d}.
    \]
    We will refer to $(\hx_i, \Pi)$ as a \textbf{fractional MMS head profile} of agent $i$.
\end{definition}

We will now show that the MMS profiles $\pi_{i,j}$ of agents $i \in N$ satisfy the conditions of \cref{etatailcomplete} fractionally with respect to profiles $(\hx_i, \Pi)$.

\begin{lemma}\label{etafracload}
    Consider the fractional MMS head profiles $(\hx, \Pi)$ of agents $i \in N$.
    It holds that
    \[
        \forall 0 \leq\ell\leq m:\; \Hh(\hx, \Pi, \ell) \leq \frac{n}{d}\min\{\ell, h\};\quad\quad \tau(\hx, \Pi) \leq \frac{n}{d},\quad\quad \Ii(\hx, \Pi) = n.
    \]
    As a corollary, for $\eta = (n + 1)/d$ and $h = \min\{nd, m\}$ we have for every $0 \leq \ell \leq h$ that $\Hh(\hx, \Pi, \ell) \leq \eta \ell$, and for every $h + 1 \leq \ell \leq m$ that
    \[
        \Hh(\hx, \Pi, \ell) \leq \eta \ell - (\ell - h)\cdot \tau(\hx, \Pi) - \Ii(\hx, \Pi).
    \]
\end{lemma}
\begin{proof}
    We first observe that for every agent $i \in N$ and every $0 \leq \ell \leq m$ it holds that
    \[
        \sum_{j = 1}^d\Pp_i(\pi_{i,j}, \ell) \leq \min\{\ell, h\}.
    \]
    By definition $\Pp_i(\pi_{i,j}, \ell)$ counts how many demands in $\pi_{i,j}$ can be satisfied only by the first $\ell$ items, and the sum above counts the total number of such demands for agent $i$.
    Note that every demand in $\pi_{i,j}$ comes from a unique item of $B_{i,j}$, and MMS partition $\{B_{i,1},\ldots, B_{i,d}\}$ cannot use more than $\ell$ items in every $\ell$-prefix.
    Hence, are at most $\ell$ such demands across all profiles $\{\pi_{i,1},\ldots, \pi_{i,d}\}$.
    Since demands in every $\pi_{i,j}$ correspond only to head items in $[h]$, there cannot be more than $h$ total demands for $i$.
    Combining two bounds gives $\sum_{j = 1}^d\Pp_i(\pi_{i,j}, \ell) \leq \min(\ell, h)$.

    It immediately follows that
    \[
        \Hh(\hx, \Pi, \ell) = \frac{1}{d}\sum_{i \in N}\sum_{j = 1}^d\Pp_i(\pi_{i,j}, \ell) \leq \frac{n}{d}\min\{\ell, h\}.
    \]
    Next, for every $i \in N$ and $j \in [d]$ we have \[
        \Delta(\pi_{i,j}) = 1 - v(\pi_{i,j}) = v(B_{i,j} \cap \{h + 1, \ldots, m\}),
    \]
    so by additivity it holds $\sum_{j = 1}^d\Delta(\pi_{i,j}) = T_i$ and
    \[
         \tau(\hx, \Pi) = \frac{1}{d}\sum_{\substack{i \in N\\T_i > 0}}\sum_{j = 1}^d\frac{\Delta(\pi_{i,j})}{T_i} \leq \frac{n}{d}.
    \]
    The equality $\Ii(\hx, \Pi) = n$ follows from $\sum_{\pi\neq \bot}\hx_i(\pi) = 1$ for every $i \in N$.

    Now, put $\eta = (n + 1)/d$ and $h = \min\{nd, m\}$.
    If $h = m$, there is no tail, and $\Hh(\hx, \Pi, \ell)\leq \eta \ell$ for all $0 \leq \ell \leq m$ is automatic.
    If $h = nd < m$, then for every $\ell > h$ it holds $n < \ell/d$.
    For this $\ell$,
    \begin{multline*}
        \Hh(\hx, \Pi, \ell) + (\ell - h)\cdot \tau(\hx, \Pi) + \Ii(\hx, \Pi) \leq \frac{n}{d}\cdot h + (\ell - h)\cdot \frac{n}{d} + n \\
        < \frac{n}{d}\cdot \ell + \frac{\ell}{d} = \frac{n + 1}{d}\cdot \ell = \eta \ell.
    \end{multline*}
    The lemma follows.
\end{proof}

The path forward is now clear: set $h = \min\{nd, m\}$ and round the fractional MMS profiles $(\hx, \Pi)$ to some integral collection of profiles $\Pi_I = \{\pi_i\}_{i \in I}$ for sufficiently large set of agents $I\subseteq N$.
Specifically, to prove \cref{dmms} via \cref{leftoveralloc}, we will need $|I| \geq n - O(\log^2 n)$.
So, starting from $\Ii(\hx, \Pi) = n$, we are allowed to lose at most $O(\log^2 n)$ agents in the rounding.

Specifically, we establish the following theorem.
\begin{theorem}[Head rounding]\label{etarounding}
    Let $\Pi$ be a shared set of head profiles, such that $|\pi|\leq h$ for some integer $h \geq 1$.
    For every $i \in N$, let $(x_i, \Pi)$ be a fractional profile, denote $x = \{x_i\}_{i \in N}$.

    There exists a collection $y = \{y_i\}_{i \in N}$ of integer head profiles $(y_i, \Pi)$, such that for every $i, \pi$ we have $y_i(\pi) \in \{0, 1\}$ and $y_i(\pi) = 1$ for $\pi \neq \bot$ only if $x_i(\pi) > 0$, and
    \[\begin{aligned}
        & && \Hh(y, \Pi, \ell) \leq \Hh(x, \Pi, \ell),  & \forall 0 \leq \ell \leq m,\\
        & && \tau(y, \Pi) \leq\tau(x, \Pi),\\
        & && \Ii(y, \Pi) \leq \Ii(x, \Pi),\\
        &  && \Ii(y, \Pi)\geq \Ii(x, \Pi) - \rho(h, X), 
    \end{aligned}\]
    where $X = |\{(i, \pi) :\; 0 < x_i(\pi) < 1\}|$ and
    \[
        \rho(h, X) := 8 + 4\log \max\{1, X\}\cdot (1 + \ln h + \ln \max\{1, X\}).
    \]
\end{theorem}
We prove \cref{etarounding} in \cref{appendix:rounding} (\cref{etaroundingproof}), and use it to prove \cref{dmms}.
\begin{proof}[Proof of \cref{dmms}]
    Take $h = \min\{nd, m\}$ and construct MMS head profiles $\pi_{i,j}$ for every $i \in N$, $j \in [d]$, and let $\Pi := \{\varnothing, \bot\} + \bigcup_{i \in N}\{\pi_{i,j} : j \in [d]\}$.
    Consider the fractional MMS head profiles $(\hx, \Pi)$.
    By definition of $\pi_{i,j}$, we automatically have for every $\pi \in \Pi$ that $|\pi|\leq h$.
    
    Note that the total number of strictly fractional entries in $\hx$ is at most $\hX\leq nd$.
    Apply \cref{etarounding} to $(\hx, \Pi)$.
    By \cref{etafracload}, obtain integral head profiles $(y, \Pi)$ such that 
    \[\begin{aligned}
        & && \Hh(y, \Pi, \ell) \leq \Hh(\hx, \Pi, \ell) \leq \frac{n}{d}\min\{\ell, h\},  & \forall 0 \leq \ell \leq m,\\
        & && \tau(y, \Pi) \leq\tau(\hx, \Pi) \leq \frac{n}{d}\\
         & && \Ii(y, \Pi) \leq \Ii(\hx, \Pi) = n,\\
        &&&\Ii(y, \Pi)\geq \Ii(\hx, \Pi) - \rho(h, \hX) \geq n - \rho(h, nd).
    \end{aligned}\]
    Let $I\subseteq N$ be a set of size $\Ii(y, \Pi)$, where $i \in I$ if and only if $y_i(\bot)  = 0$.
    Let $\pi_i \in \Pi - \bot$ be the profile such that $y_i(\pi_i) = 1$, and $\Pi_I := \{\pi_i\}_{i \in I}$.

    If $h = m$, taking $\eta = n/d$ automatically gives us for every $0 \leq \ell \leq m$ that $\Pp(\Pi_I, \ell)= \Hh(y, \Pi, \ell) \leq \Hh(\hx, \Pi, \ell) \leq \eta \ell$.
    If $h < m$, then with $\eta = (n + 1)/d$, on top of $\Pp(\Pi_I, \ell)\leq \eta \ell$ for all $0\leq \ell \leq h$, we have for every $h + 1 \leq \ell \leq m$ that
    \begin{multline*}
        \Pp(\Pi_I, \ell) + (\ell - h)\sum_{i \in I}\tau_i(\pi_i) + |I| = \Hh(y, \Pi, \ell) + (\ell - h)\tau(y, \Pi) + \Ii(y, \Pi) \\
        \leq \Hh(\hx, \Pi, \ell) + (\ell - h)\cdot \tau(\hx, \Pi) + \Ii(\hx, \Pi) \leq\frac{n}{d}\cdot h + (\ell - h)\cdot \frac{n}{d} + n \\
        < \frac{n}{d}\cdot \ell + \frac{\ell}{d} = \frac{n + 1}{d}\cdot \ell = \eta \ell.
    \end{multline*}
    By construction, for every $i \in I$ we have that $T_i = 0$ implies $\Delta(\pi_i) = 0$, as $y_i(\pi_i) = 1$ implies $x_i(\pi_i) > 0$, hence $\pi_i$ must be some MMS profile of agent $i$.
    Then, we can apply \cref{etatailcomplete} and obtain a collection $\Psi_I = \{\psi_i\}_{i \in I}$ of tail profiles such that  for every $0 \leq \ell \leq m$  we have  \[
        \Pp(\Pi_I, \ell) + \Pp(\Psi_I, \ell) \leq \eta \ell,
    \]
    and for every $i \in I$ it holds $v(\pi_i) + v(\psi_i) \geq 1$.
    Hence, \cref{etaload} applied to $(I, \Pi_I, \Psi_I)$ produces some set $E\subseteq M$, such that all profiles $\{\pi_i, \psi_i\}_{i \in I}$ are simultaneously satisfied using distinct items of $E$, and for every $0 \leq \ell \leq m$ it holds $|E\cap [\ell]|\leq \eta \ell$.

    Every agent $i \in I$ receives a disjoint bundle $E_i\subseteq E$ such that $v_i(E_i) \geq v(\pi_i) + v(\psi_i) \geq 1$.
    It remains to allocate the agents in $U := N - I$.
    Since $|I| = \Ii(y, \Pi) \geq n - \rho(h, nd)$, we have $|U| \leq \rho(h, nd)$.
    At the same time, \cref{prefixbound} guarantees us that for every $u \in U$ it holds $v_u(E) \leq \eta d$.
    Hence, as long as $|U|\leq (d - n)/2$, \cref{leftoveralloc} allocates agents $U$ with items $M \setminus E$ so that every $u \in U$ receives a bundle of value at least $1$.
    Then, to guarantee $|U|\leq (d - n)/2$ it suffices to take $d$ such that 
    \[
        \rho(h, nd) \leq (d - n)/2 \quad \iff \quad d \geq n + 2\rho(h, nd).
    \]
    We now estimate $\rho(h, nd)$.
    Since $h , \hX \leq nd$ and $d \leq 2n - 1$ (\cref{leftovers} solves $d > 2n-1$),
    \begin{multline*}
        \rho(h, nd) \leq 8 + 4\log (nd)\cdot (1 + \ln (nd) + \ln (nd))\\
        \leq 8 + 4\log (nd) + 8\ln2\log^2(nd) \leq 8 + 4\log(2n^2) + 8\ln2 \log^2(2n^2).
    \end{multline*}
    Setting $z = 1 + \log n$, the bound above is
    \begin{multline*}
        \rho(h, nd) \leq 8 + 4(2z - 1) + 8\ln2 (2z - 1)^2\\
        = 32\ln 2z^2  + (8 - 32\ln 2)z + 4 + 8\ln 2 < 32\ln2 z^2 < 23(1 + \log n)^2.
    \end{multline*}
    As a result, taking $d \geq n + 46(1 + \log n)^2$ suffices.
\end{proof}
\section{$(1 - \delta)$-MMS allocation for sufficiently large $n$}

\label{sec:dual}

In this section, we prove \cref{thm:main}.
We assume $m \geq n$, otherwise pad items with zeroes.
We will consider $\rho$-MMS allocations with $\rho > 3/4$ only, as \cite{huang2025} guarantees $\frac{3}{4}$-MMS allocations for additive valuations with no restrictions on $n$.

\begin{theorem}[\cref{thm:main} restated]\label{deltamms}
    For every fixed $0 < \delta < 1/4$, there exists $n_{\delta}$ such that for every $n \geq n_{\delta}$, the following holds.
    Consider a fair allocation instance with $m$ items $M$ and $n$ agents $N$, where for every $i \in N$ valuation $v_i$ is additive. 
    There exists an allocation $(B_1, \ldots, B_n)$ of $M$ to $N$ such that for every agent $i \in N$ we have $v_i(B_i) \geq (1 - \delta)\cdot \mms(M, v_i, n)$.
    Furthermore, it suffices to take $n_\delta$ as
    \[
        n_\delta =  2^{16\delta^{-2}\log^2(8/\delta)}.
    \]
\end{theorem}

Instead of proving this theorem directly, we will prove it under additional assumption on the total number of items.
Specifically, we will assume that $m \leq \Gamma n$ for some integer $\Gamma \geq 1$.
\begin{theorem}\label{deltammsproof}
    Let $\Gamma \geq 1$ and $q \geq 9$ be integers.
    There exists $n(\Gamma , q)$ such that for every $n \geq n(\Gamma , q)$ the following holds.
    Consider a fair allocation instance with $m$ items $M$ and $n$ agents $N$, where for every $i \in N$ valuation $v_i$ is additive.
    If $m \leq \Gamma n$, there exists an allocation $(B_1, \ldots, B_n)$ of $M$ to $N$ such that for every agent $i \in N$ we have $v_i(B_i) \geq (1 - 2/q)\cdot \mms(M, v_i, n)$.
    Furthermore, it suffices to take $n(\Gamma , q)$ as
    \[
        n(\Gamma , q) = 2^{4q^2\log^2(2q)}\cdot (\log(2\Gamma ))^{2q}.
    \]
\end{theorem}

Assuming \cref{ass:wlog} (with $d =n$), we show how \cref{deltammsproof} implies \cref{deltamms}.

\begin{proof}[Proof of \cref{deltamms} using \cref{deltammsproof}]
    For a fixed $0 < \delta < 1/4$, take
    \[
        q = \left\lceil \frac{2}{\delta} + \frac{1}{4}\right\rceil,\qquad\text{and}\qquad \Gamma  = 4q^2.
    \]

    We will transform our initial instance with $m$ items and $n$ additive valuations $\{v_i\}_{i \in N}$ into an instance with $m \leq \Gamma n$ \textit{meta-items} and $n$ additive valuations $\{\wv_i\}_{i \in N}$, defined on meta-items.

    We do it as follows.
    Let $K = 2q^2n$.
    We keep the items $\{1, \ldots, K - 1\}$ individually (or all items if $m < K$), and then group the remaining items into $K$ buckets according to round-robin partition.
    That is, for a bucket $E_k$ for $k \in K$, we take
    \[
        E_k := M\cap \bigcup_{s \geq 1}\{s\cdot K + k - 1\} = M\cap\{K + k - 1, 2K + k - 1, \ldots\}.
    \]
    We will treat each bucket $E_k$ as a single \textit{meta-item}.
    Let $\wM = \{1, \ldots, K- 1, E_1, E_2, \ldots, E_K\}$ be the set of meta-item, and for every agent $i \in N$ define valuations $\wv_i$ on $\wM$ by
    \[
        \forall e \in [K - 1]:\;\wv_i(e) = v_i(e),\qquad\qquad \forall k \in [K]:\; \wv_i(E_k) = v_i(E_k) =\sum_{e \in E_k}v_i(e),
    \]
    extending additively to subsets of $\wM$.
    It is easy to see that $|\wM|\leq 2K - 1 < \Gamma n$.
    \begin{lemma}\label{metamms}
        For every agent $i \in N$ it holds $\mms(\wM, \wv_i, n) \geq 1 - 1/(2q^2)$.
    \end{lemma}
    \begin{proof}
    It suffices to consider only the case $m \geq K + 1$, otherwise $\wM = M$.
    To see the MMS bound, fix some $i \in N$ and consider agent $i$-th MMS partition $\{B_{i,1}, \ldots, B_{i,n}\}$.
    We create a new partition $\{\wB_{i,1},\ldots, \wB_{i,n}\}$ as follows.
    First, for each MMS-bundle we keep the items within the first $[K-1]$ untouched, call these subsets
    \[
        \forall j \in [n]:\; B_{i,j}^K := B_{i,j}\cap [K - 1].
    \]
    
    We will now greedily distribute the whole meta-items among $B_{i,j}^K$, starting with $j = 1$ and moving up to $n$.
    Suppose that currently we consider $B_{i,j}^K$, $j \in[n]$.
    Let $b_j := 1 - v_i(B_{i,j}^K)$ denote the missing value.
    If $b_j \leq 1/(2q^2)$, there is noting to add, as $\wv_i(B_{i,j}^K) \geq 1 - 1/(2q^2)$.
    Otherwise, we keep adding available meta-items $E_k$ in arbitrary order, until the total value of $B_{i,j}^K$ reaches $1 - 1/(2q^2)$ (or equivalently, until we add total value at least $b_j - 1/(2q^2)$).
    Let $K_j$ be the meta-items added to $B_{i,j}^K$, and set
    \[
        \wB_{i,j} = B_{i,j}^K\cup \{E_k : k \in K_j\}.
    \]
    The meta-items $K_j$ are unavailable, and we proceed to part $j + 1$.
    Once $j = n$ is completed, assign the remaining meta-items arbitrarily.
    Without loss of generality and for convenience of the proof, assume that by $j = n$ all meta-items were allocated.

    To guarantee that the process above always succeeds, it suffices to show that at the beginning of every iteration $j$, the total value of available meta-items is at least $\sum_{k = j}^nb_k$.
    We show this by induction on $j$, the base case $j = 1$ follows by 
    \[
        \sum_{j = 1}^nb_{j} = n - v_i([K - 1])  = v_i(M - [K -1]) = \sum_{k = 1}^Kv_i(E_k).
    \]
    Suppose that the value lower bound holds up to iteration $j$.
    If $b_j > 1/(2q^2)$, the available meta-items have value at least $b_j$, so we can guarantee adding at least $b_j - 1/(2q^2)$.
    Observe that for every prefix $\ell \geq 0$, for every $k \in [K]$ we have $|E_k\cap [\ell]|\leq \ell/K$.
    Then, by \cref{prefixbound}, we get $\wv_i(E_k)  \leq n/K = 1/(2q^2)$ for every agent $i \in N$.
    Since just before adding the last meta-item into $K_j$, the total added value was below $b_j - 1/(2q^2)$, it must hold that
    \[
        \sum_{k \in K_j}v_i(E_k) < b_j,
    \]
    implying that at the beginning of iteration $j+1$, the total available value will be at least
    \[
        \sum_{k = j}^n b_k - \sum_{k \in K_j}v_i(E_k) \geq \sum_{k = j+1}^nb_k.
    \]
    Thus, the meta-item allocation algorithm successfully finishes.

    By construction, all items are allocated, and for every $j \in[n]$ either $b_{j} \leq 1/(2q^2)$ and $ \wv_i(\wB_{i,j}) \geq \wv_i(B_{i,j}^K) \geq 1 - 1/(2q^2)$, or $b_j > 1/(2q^2)$ and we have
    \[
        \wv_i(\wB_{i,j}) \geq v_i(B_{i,j}^K) + b_j - 1/(2q^2) = 1 - 1/(2q^2).
    \]
    It follows that $\mms(\wM, \wv_i, n) \geq 1 - 1/(2q^2)$.
\end{proof}
If $n \geq n(\Gamma , q)$, applying \cref{deltammsproof} to meta-items $\Mm$ and valuations $\{\wv_i\}_{i \in N}$ gives an allocation $(\wB_1, \ldots, \wB_n)$ such that, by \cref{metamms}, every agent $i \in N$ receives value 
\[
    \wv_i(\wB_i) \geq \left(1 - \frac{2}{q}\right)\mms(\wM, \wv_i, n) \geq \left(1 - \frac{2}{q}\right)\left(1 - \frac{1}{2q^2}\right) \geq 1 - \frac{2}{q} - \frac{1}{2q^2} \geq 1 - \delta,
\]
where we used $\delta \geq 2/q + 1/(2q^2)$ for $q = \lceil 2/\delta + 1/4\rceil \geq 9$.
Expanding the allocated meta-items into regular items gives a $(1-\delta)$-MMS allocation of $M$.

It remains to show that $n(\Gamma , q) \leq 2^{16\delta^{-2}\log^2(8/\delta)}$. 
Since for $q\geq 9$ we have $\log(2q)\leq q/2$, 
\[
    \log(2\Gamma) = \log(8q^2) = 2\log(2q) + 1 \leq  2q,
\]
therefore
\[
    \log n(\Gamma , q) = \log\left(2^{4q^2\log^2(2q)}\cdot (\log(2\Gamma))^{2q}\right) \leq 4q^2\log^2(2q) + 2q\log(2q) < 4\left(q\log(2q) + \frac{1}{4}\right)^2.
\]
Take the convex function $f(x) := x\log(4x)$.
By the choice of $q$, $2/\delta > q - 5/4$, then by convexity
\begin{multline*}
    \frac{2}{\delta}\log\frac{8}{\delta} = f(2/\delta) \geq f(q) -\frac{5}{4}f'(q)\\
    \geq q\log(2q) + q -\frac{5}{4}(\log(2q) + 1 + \log e) \geq q\log(2q) + \frac{3q - 25}{8} \geq q\log(2q) + \frac{1}{4}.
\end{multline*}
It follows that $n(\Gamma , q) \leq 2^{16\delta^{-2}\log^2(8/\delta)} =: n_\delta$.
\end{proof}

\subsection{Common head/tail from LP duality}

The rest of the paper is dedicated to proving \cref{deltammsproof}.
By assumption, $m \leq \Gamma n$, so our first naive attempt to get a $(1-2/q)$-MMS allocation is simply trying to use already existing approach from \cref{sec:dmms} with $h = m$ and $d = n$.
More specifically, for every agent $i \in N$ and its MMS bundle $j \in [n]$, consider the full MMS profiles
\[
    \pi_{i,j} := \{v_i(e) :\; e \in B_{i,j}, v_i(e) > 0\},
\]
set $\Pi = \{\varnothing, \bot\} + \bigcup_{i \in N}\{\pi_{i,j} : j \in [n]\}$, and create fractional MMS head profiles
\[
    \forall \pi \in \Pi: \; \hx_i(\pi) = \frac{|\{j \in [n]: \pi_{i,j} = \pi\}|}{n}.
\]
By \cref{etafracload}, for every prefix $0\leq \ell \leq m$ it holds $\Hh(\hx, \Pi, \ell) \leq \ell$, and $\Ii(\hx, \Pi) = n$.
Then, directly applying the rounding of \cref{etarounding} produces a set of at least $n - O(\log^2 n)$ agents with assigned integral profiles of value $1$, which can be simultaneously satisfied by distinct items of $M$.
Unfortunately, this allocation would provide no guarantees for the remaining $O(\log^2 n)$ agents with respect to the set of remaining unallocated items: as $d = n$ and $\eta = 1$ in \cref{prefixbound}, \cref{leftoveralloc} gives us nothing as it applies only to agent subsets of size zero.

However, the fact that the rounding of \cref{etarounding} fails to assign profiles to all agents gives us some important information about the prefix load of the MMS profiles.

To get some intuition, let us first consider a ``standard'' approach of finding a $(1-2/q)$-MMS allocation using configuration LP.
We introduce a natural notion of \textit{acceptable} bundles.
\begin{definition}
    Let $i \in N$.
    A bundle $B\subseteq M$ is called \textbf{acceptable} for agent $i$, if $v_i(B) \geq 1 - 2/q$.
    Let $\Bb_i = \{B \subseteq M: v_i(B) \geq 1 - 2/q\}$ denote the set of all acceptable bundles for agent $i$.
\end{definition}
For $i \in N, B\subseteq M$ let $x_i(B)$ be the indicator variable corresponding to allocating bundle $B$ to agent $i$.
Consider the following variation of the configuration LP:
\[
\begin{aligned}
    &\minimize &&\eta\\
    &\subto && \sum_{B \in \Bb_i}x_i(B) = 1,&\forall i \in N;\\
    &&& \sum_{i \in N}\sum_{\substack{B\in \Bb_i\\B\ni e}}x_i(B) \leq \eta,&\forall e \in M;\\
    &&& x_i(B) \geq 0,&\forall i \in N, B\subseteq M.
\end{aligned}\label{clp}\tag{C-LP}
\]
Here variable $\eta$  corresponds to the maximum fractional usage of an item.
It plays essentially the same role as in \cref{sec:dmms}, where it depicted the maximum fractional load imposed on every item of an $\ell$-prefix.
Clearly, giving weight $x_i(B_{i,j}) = 1/n$ to every bundle of agent $i$'s MMS partition makes (\ref{clp}) feasible with $\eta = 1$.

If the optimum $\eta^*$ of (\ref{clp}) is sufficiently below $1$, the fractional solution provides a lot of slackness that can compensate for the agents lost during rounding.
In \cref{sec:dmms} when we had $d = n + O(\log^2 n)$ and MMS partitions into $d$ bundles, and a feasible solution giving $x_i(B_{i,j}) = 1/d$ to every MMS bundle would guarantee optimum $\eta^* \leq \frac{n}{d} = 1 - \frac{d - n}{d}$.
Then, if some rounding omits at most $\rho(n)$ agents, and produces an allocation using a set of items of total worth at most $\eta^* d$ to every agent, every omitted agent values the remaining items at least $(1 - \eta^*)d$.
Our sufficient condition to be able to allocate the omitted agents (i.e \cref{leftoveralloc}) was
\[
    \rho(n)\leq \frac{(1 - \eta^*)d}{2} \quad \iff \quad d - n \geq 2\rho(n).
\]
So, when $\rho(n) = O(\log^2 n)$ and $d = n + O(\log^2n)$ as in \cref{dmms}, we can allocate all agents.

However, in the setting of \cref{deltammsproof} we have $d = n$, and the starting uniform solution with $x_i(B_{i,j}) = 1/n$ gives $\eta = 1$, i.e no guaranteed slackness.
Let us look at the optimal fractional load $\eta^*$ instead.
By the same reasoning as above, if some rounding of (\ref{clp}) allocates all but $\rho(n)$ agents using a set $E$ satisfying $v_i(E) \leq \eta^*n$ for every agent $i$, we can always complete the allocation using the remaining items as long as
\[
    \rho(n) \leq \frac{(1 - \eta^*)n}{2}.
\]
If $\eta^*$ is sufficiently separated from $1$, we have a lot of slackness and can complete the allocation.

The main issue is what to do when $\eta^*$ is very close to $1$.
It turns out that the \textit{dual} of (\ref{clp}) provides useful structural information in this regime.
After rescaling, the dual  can be written as
\[
\begin{aligned}
    &\maximize && \frac{1}{n}\sum_{i \in N}\lambda_i\\
    &\subto && \sum_{e \in B}w(e) \geq \lambda_i,&\forall i \in N, B \in \Bb_i;\\
    & && \sum_{e \in M}w(e) = n;\\
    &&& w(e) \geq 0,&\forall e \in M.
\end{aligned}\tag{D-LP}\label{dlp}
\]
Here, $w$ is an additive function on $M$, defined by the \textit{common prices} $w(e)$ for every item $e \in M$, and $\lambda_i$ for $i \in N$ is a lower bound on the price of every acceptable bundle for agent $i$.

Let $(\lambda^*, w^*)$ be the optimal dual solution.
Note that optimality forces $\lambda_i^* = \min_{B \in \Bb_i}w^*(B)$, the minimum price of an acceptable bundle for agent $i \in N$, otherwise we could increase $\lambda_i^*$.
Since every MMS bundle $B_{i,j}$ is acceptable for $i$, it follows that for every $i \in N$ that
\[
    0 \leq \lambda_i^* \leq \min_{j \in [n]}w^*(B_{i,j})\leq \frac{1}{n}\sum_{j = 1}^nw^*(B_{i,j}) = \frac{w^*(M)}{n} = 1.
\]
Define $\Lambda^* := \sum_{i \in N}(1 - \lambda_i^*)$.
By strong duality we have
\[
       \sum_{i \in N}\lambda_i^* = \eta^*n \quad \iff \quad  \Lambda^*=\sum_{i \in N}(1 - \lambda_i^*) = (1 - \eta^*)n .
\]
In particular, since every $(1 - \lambda_i^*)$ is non-negative, for any $\eps > 0$ at least $(1 - \eps)n$ agents $i \in N$ have the common price of their acceptable bundles at least 
\[
    \lambda_i^* \geq 1 -  \frac{\Lambda^*}{\eps n} = 1 - \frac{1 - \eta^*}{\eps}.
\]

Why is all this particularly useful?
By construction, (\ref{dlp}) guarantees that for every agent $i \in N$ and bundle $B\subseteq M$, if $B$ is acceptable for $i$ then necessarily $w^*(B) \geq \lambda_i^*$.
Then, when $\eta^*$ is very close to $1$, most of the agents have $\lambda_i^*$ very close to $1$, which implies that acceptable bundles of all these agents have a high common price.

This observation allows us to produce a \textit{common prefix} that consists of items that \textit{most} of the agents value highly, as follows.
For agent $i \in N$, feasible $w^*$ must satisfy $\sum_{j = 1}^nw^*(B_{i,j}) = w^*(M) = n$.
If some subset $D_{i,j}\subseteq B_{i,j}$ has value at most $2/q$, then $v_i(B_{i,j}\setminus D_{i,j}) \geq 1 - 2/q$, so $B_{i,j}\setminus D_{i,j}$ is acceptable.
Then, its common price $w^*(B_{i,j}\setminus D_{i,j}) \geq \lambda_i^*$, and
\[
    w^*(D_{i,j}) \leq w^*(B_{i,j}) - \lambda_i^*.
\]
Hence, if agent $i$ has $\lambda_i^* \geq 1 - \Lambda^*/(\eps n)$, the total common price of these small-valued subsets of agent $i$'s MMS bundles is at most
\[
    \sum_{j = 1}^nw^*(D_{i,j}) \leq \sum_{j = 1}^n\big(w^*(B_{i,j}) - \lambda_i^*\big) = n(1 - \lambda_i^*) \leq \frac{\Lambda^*}{\eps}.
\]
Thus, for every $\eps > 0$, at least $(1 - \eps)$-fraction of agents have only $\Lambda^*/\eps$ of total common price in the small-valued subsets of their MMS bundles.
Therefore, if the condition on $\eta^*$ fails, we get
\[
    \rho(n) > \frac{(1 - \eta^*)n}{2} \quad \implies \quad \Lambda^* = (1 - \eta^*)n < 2\rho(n).
\]
In this case, at least $(1 - \eps)n$ agents have the common price of their small-valued subsets at most $2\rho(n)/\eps$, while the total price of items is $n$.
Then, for a fixed $\eps> 0$, if $\rho(n) \ll n$, these subsets account only for a small fraction of the total common price.

By itself, this observation does not give much, as different agents can have different low-valued parts, and these low-valued parts could collectively cover all items $M$.
However, we will show that as long as $w^*$-prices of items are \textit{ordered}, i.e $w^*(1) \geq w^*(2) \geq \ldots \geq w^*(m)$, then there must be a common prefix of high-valued items for most agents, provided that $n$ is large enough.
\begin{lemma}\label{commonprice}
    For a feasible solution $(\lambda, w)$ to (\ref{dlp}), denote $\Lambda := \sum_{i \in N}(1 - \lambda_i)$.
    Suppose that common prices $w$ satisfy $w(1) \geq w(2) \geq \ldots \geq w(m)$.
    Then, for any $\eps \in (0, 1]$ there exists an integer $h \in \{0, 1, \ldots, m\}$ such that $qn \geq h \geq n - 2q\Lambda/\eps - q$ and
    \begin{itemize}
        \item every item $e \in \{1, \ldots, h\}$ has price $w(e) \geq 1/q$;
        \item the subset of agents
        \[
            G_\eps = \left\{i \in N: \; n(1 - \lambda_i) \leq \frac{\Lambda}{\eps}\right\}  
        \] has 
        $|G_\eps|\geq (1 - \eps)n$, and for every $i \in G_\eps$ and item $e \in \{1, \ldots, h\}$ it holds $v_i(e) \geq 1/q$;
        \item for every agent $i \in N$ with $\lambda_i \geq 1/2$ it holds 
        \[v_i(\{h + 1, \ldots, m\}) \leq \frac{4q\Lambda}{\eps} + q.\]
    \end{itemize}
\end{lemma}
\begin{proof}
    As discussed earlier, since for every agent $i \in N$ every MMS bundle $B_{i,j}$, $j \in [n]$ is acceptable, for a feasible solution $(\lambda, w)$ it holds
    \[
        n = w(M) = \sum_{j = 1}^nw(B_{i,j}) \geq n\lambda_i \quad \implies \quad \lambda_i \leq 1.
    \]
    In particular, $1 - \lambda_i \geq 0$ for all $i \in N$.
    Consequently, as $\sum_{i \in N}(1 - \lambda_i) = \Lambda$, the set
    \[
            G_\eps = \left\{i \in N: \; n(1 - \lambda_i) \leq \frac{\Lambda}{\eps}\right\}  
        \]
    has size at least $(1 - \eps)n$.
    Note that this set is non-empty even if $\eps = 1$, as there always exists an agent $i$ with $(1 - \lambda_i) \leq \Lambda/n$, and if $\Lambda = 0$ then $G_\eps = N$.

    First suppose that $n \leq \frac{2q\Lambda}{\eps} + q-3$.
    In this case, take $h = 0$, and for every $i \in N$ it holds
    \[
        v_i(\{1, \ldots, m\}) = n \leq \frac{2q\Lambda}{\eps} + q-3 \leq \frac{4q\Lambda}{\eps} + q.
    \]
    From now on assume $n > \frac{2q\Lambda}{\eps} + q-3$, since $q \geq 9$ it holds $n>\frac{2q\Lambda}{\eps}$.
    
    For each agent $i \in N$, let $L_i := \{e \in M:\; v_i(e) < 1/q\}$.
    We will show that these low-valued items must have small price, precisely $w(L_i) \leq qn(1 - \lambda_i)$.
    To see this, take $B_{i,j}$ for some $j \in [n]$ and partition $B_{i,j}\cap L_i$ greedily into sub-bundles of $v_i$-value at most $2/q$, adding items into a bundle as long as adding another item does not raise the total value beyond $2/q$.
    Since every item in $B_{i,j}\cap L_i$ has value less than $1/q$, every sub-bundle except for the last one will have value greater than $1/q$.
    Then, since $v(B_{i,j}) = 1$, there will be at most $q$ such sub-bundles.
    Denote them by $D_{i,j,k}$ for $k \leq q$.

    For every $k$, removing $D_{i,j,k}$ from $B_{i,j}$ keeps the bundle acceptable, as $v_i(B_{i,j}\setminus D_{i,j,k}) \geq 1 - 2/q$.
    Hence, we must have $w(D_{i,j,k}) \leq w(B_{i,j}) - \lambda_i$.
    Summing over all $k$ and over all parts of MMS partition, we get
    \[
        w(L_i) = \sum_{j = 1}^n\sum_{k}w(D_{i,j,k}) \leq q\sum_{j = 1}^n(w(B_{i,j}) - \lambda_i) = qn(1 - \lambda_i).   
    \]

    Next, for every $i \in G_\eps$,
    \[
        \lambda_i \geq 1 - \frac{\Lambda}{\eps n}> 1 - \frac{1}{2q},
    \]
    in particular $\lambda_i \geq 1/2$ and $w(L_i) \leq q\Lambda/\eps$.
    Observe that every MMS bundle of $i \in G_{\eps}$, contained entirely in $L_i$, still has price at least $1/2$.
    Then, the total number of such bundles is at most
    \[
        |\{j \in [n]: \; B_{i,j}\subseteq L_i\}| \leq 2w(L_i) \leq \frac{2q\Lambda}{\eps}.
    \]
    Every other MMS bundle of $i$ contains at least one item of value at least $1/q$.
    Define
    \[
        h_i := |\{e \in M :\; v_i(e) \geq 1/q\}|\qquad \text{and}\qquad h_\eps := \min_{i \in G_\eps}h_i.
    \]
    It follows that $h_\eps \geq n - \frac{2q\Lambda}{\eps}$, and every agent of $G_\eps$ values every item of $[h_\eps]$ at least $1/q$. 
   
    Take $h := h_\eps - q + 3$, clearly $qn \geq h \geq n - \frac{2q\Lambda}{\eps} - q + 3$.
    We will now establish the lower bound on common price.
    Consider the $q - 2$ items $\{h, h + 1, \ldots, h_\eps\}$, note that for any $i \in G_\eps$ we have $v_i(\{h, \ldots, h_\eps\}) \geq (q-2)/q = 1 - 2/q$, so $\{h, \ldots, h_\eps\}$ is acceptable.
    Then, as $w(h) \geq w(h +1) \geq \ldots \geq w(h_\eps)$, and using $\lambda_i \geq 1 - 2/q$ for $i \in G_\eps$, we get
    \[
        (q - 2)\cdot w(h) \geq w(\{h, \ldots, h_\eps\}) \geq \lambda_i \geq (q - 2)/q.
    \]
    Thus, $w(h) \geq 1 /q$, implying that for every $e \leq h$ we also have $w(e) \geq 1/q$.

    Finally, we bound the tail value.
    Let $i_\eps \in G_\eps$ be some agent attaining $h_{i_\eps} = h_\eps$.
    Then, $\{h_\eps + 1,\ldots, m\} = L_{i_\eps}$, and $w(\{h_\eps + 1, \ldots, m\}) \leq qn(1 - \lambda_{i_\eps}) \leq q\Lambda/\eps$.
    
    Take any agent $i \in N$ with $\lambda_i \geq 1/2$, and partition $\{h_\eps + 1, \ldots, m\}$ greedily into sub-bundles of $v_i$-value at least $1 - 2/q$.
    That is, we keep adding items into a bundle until it reaches at least $1-2/q$, and then move to the next, while the value of the remaining items is at least $1-2/q$.
    Clearly, every sub-bundle except for possibly the last one is acceptable and has value at most $2$.

    Let $k$ be the number of acceptable sub-bundles.
    Their total common price is at least $k\cdot \lambda_i$, and at most $w(\{h_\eps + 1, \ldots, m\})$, therefore
    \[
        k \leq \frac{w(\{h_\eps + 1, \ldots, m\})}{\lambda_i} \leq 2w(\{h_\eps + 1, \ldots, m\}).
    \]
    Since the remaining items have total value less than $1$, it follows that
    \[
        v_i(\{h_\eps + 1, \ldots, m\}) \leq 2k + 1\leq 4w(\{h_\eps + 1, \ldots, m\}) + 1\leq \frac{4q\Lambda}{\eps} + 1.
    \]
    At last, $\{h + 1, \ldots, h_\eps\}$ contains exactly $q - 3$ items, each worth at most $1$, so
    \[
        v_i(\{h +  1, \ldots, m\})\leq v_i(\{h_\eps + 1, \ldots, m\}) + q -3 \leq \frac{4q\Lambda}{\eps} + q - 2.
    \]
    The lemma follows.
\end{proof}

Recall that in the optimal solution $(\lambda^*, w^*)$ it holds $\Lambda^* = (1 - \eta^*)n$.
As mentioned earlier,
\[
    \rho(n) > \frac{(1 - \eta^*)n}{2} \quad \implies \quad \Lambda^* = \sum_{i \in N}(1 - \lambda_i^*) = (1 - \eta^*)n < 2\rho(n),
\]
where we assume $\rho(n) \ll n$.
Then, if $w^*$ is \textit{ordered}, by \cref{commonprice} for any $\eps > 0$ by there exists a large common head $\{1, \ldots, h\}$ such that 1) the common prices $w^*$ of every item in $[h]$ are at least $1/q$, 2) at least $(1 - \eps)n$ agents (i.e $G_\eps$) value every item of $[h]$ at least $1/q$, and 3) for every agent $i \in N$ with $\lambda_i^*\geq 1/2$ the value of the common tail $\{h + 1,\ldots, m\}$ is very small.
This will allow us to successfully utilize our head/tail profile framework from \cref{sec:dmms} w.r.t $h$.

\subsection{The dual of the profile configuration LP}

Earlier we described the dichotomy between the optimal primal and dual solutions for the configuration LP.
We will now formulate the same idea using demand profiles and prefix load constraints, so that to apply our rounding algorithm \cref{etarounding} and get the value $\rho(n) = O(\log^2 n)$, as in \cref{sec:dmms}.
Furthermore, this will allow us to obtain the common prices that are \textit{ordered}, so that \cref{commonprice} applies.
These prices will arise directly from the dual weights on the prefix-load constraints, where each item will receive the total weight of the prefixes containing it.
We will then use complementary slackness to show that the common prices actually encode which prefix load constraints in the optimal profile LP solution are tight, and which have a lot of spare capacity to accumulate extra demands.

First and foremost, we will consider the demand profiles of all acceptable bundles, and not just MMS bundles (recall that rounding of \cref{etarounding} does not restrict us on which specific common profiles are in use).
For now, there is no head nor tail, as those will be determined later from the common prices and \cref{commonprice}.
So, in the setting of \cref{etarounding} we have $h = m$.
\begin{definition}
    For $i \in N$ and $B\subseteq M$, define induced demand profile $\pi_i(B)$ as
    \[
        \pi_i(B) := \{v_i(e):\; e \in B, v_i(e) > 0\}.
    \]
    Profile $\pi$ is \textbf{acceptable} for agent $i$ if $\pi = \pi_i(B)$ for some $B \in \Bb_i$.
    Let $\Pi_i := \{\pi_i(B) : B\in \Bb_i\}$.
\end{definition}
\begin{definition}
    Let $\Pi := \{\varnothing, \bot\} + \bigcup_{i \in N}\Pi_i$.
    The \textbf{profile configuration LP} is
    \[
    \begin{aligned}
        &\minimize && \eta\\
        &\subto && \sum_{\pi \in \Pi_i}x_i(\pi) = 1,&\forall i \in N;\\
        &&&\Hh(x, \Pi, \ell) \leq \eta \ell, &\forall 1 \leq \ell \leq m;\\
        &&& x_i(\pi) \geq 0,&\forall i \in N, \pi \in \Pi_i.
    \end{aligned}\tag{P-LP}\label{plp}
    \]
    Here $\Hh(x, \Pi, \ell)$ is the fractional $\ell$-prefix load from \cref{sec:dmms}.
    We extend each $x_i$ to the shared $\Pi$ by setting $x_i(\pi) = 0$ for $\pi \notin \Pi_i$, so that every agent fractionally gets an acceptable profile in the feasible solution.
    Rounded solutions may violate these conditions and assign $\bot$ to some agents.
\end{definition}
Note that in any feasible solution to (\ref{plp}) the load at prefix $0$ is always $0$, as every agent has non-zero weight only at acceptable profiles.
Because of this  $\Hh(x, \Pi, 0)\leq 0$ is automatic.

\begin{theorem}[Profile LP duality]\label{dualprofile}
    Let $\eta^*$ be the optimum of (\ref{plp}).
    Then $\eta^* \in (0, 1]$, and
    \begin{itemize}
        \item if  $n(1 - \eta^*) \geq 2\rho(m, n + m)$, then there exists a $(1 - 2/q)$-MMS allocation of $M$ to $N$;
        \item otherwise, there exists a feasible solution $(\lambda, w)$ to (\ref{dlp}) such that $w(1) \geq \ldots \geq w(m)$, $\lambda_i \in [0, 1]$ for every $i \in N$, and
        \[
            \Lambda := \sum_{i \in N}(1 - \lambda_i) < 2\rho(m, n + m),
        \]
        where $\rho(h, X) := 8 + 4\log\max\{1, X\} \cdot (1 + \ln h + \ln \max\{1, X\})$
        as in \cref{etarounding}.
    \end{itemize}
\end{theorem}
\begin{proof}
    Recall notation $\pi_{i,j} = \pi_i(B_{i,j})$ for MMS bundle $B_{i,j}$ of agent $i \in N$.
    The fractional profiles 
    \[
        x_i(\pi) = \frac{|\{j \in [n]: \pi_{i,j} = \pi\}|}{n},
    \]
    produce a feasible solution to (\ref{plp}) with $\eta = 1$.
    This follows by applying \cref{etafracload} with $d = n$ to get $\Hh(x, \Pi, \ell) \leq \ell$ for all $0\leq \ell \leq m$.
    Then, if $x^*$ is the optimal solution to (\ref{plp}) achieving value $\eta^*$, it holds $\eta^* \leq 1$.
    Conversely, every acceptable profile is nonempty, so any feasible solution satisfies 
    \[
        \Hh(x, \Pi, m) = \sum_{i \in N}\sum_{\pi \in \Pi_i}x_i(\pi)\cdot |\pi| \geq n,
    \]
    thus $\eta^* \geq n/m > 0$.

    Observe that we can choose $x^*$ with at most $n + m$ positive coordinates, by taking $x^*$ which is basic feasible solution of (\ref{plp}), as there are $n + m$ constraints in the LP.

    Suppose first that $n(1 - \eta^*) \geq 2\rho(m, n + m)$.
    This case is essentially the argument of \cref{sec:dmms} and the proof of \cref{dmms}: each prefix $0 \leq \ell \leq m$ has at most $\eta^*\ell$ total fractional load, which leaves enough value to allocate the agents omitted by the rounding of \cref{etarounding}.

    Concretely, apply the rounding of \cref{etarounding} to solution $x^*$, treating it as a fractional profile on the shared profiles $\Pi$ (by extending it with $x_i^*(\pi) = 0$ for $\pi \notin \Pi_i$) with $h = m$.
    Every profile has size at most $m$, and the number $X^*$ of strictly fractional coordinates in $x^*$ is at most $n + m$.
    We obtain integral profiles $y$ such that $\Ii(y, \Pi) \geq n - \rho(m, n + m)$ and 
    \[
        \Hh(y, \Pi, \ell) \leq \Hh(x^*, \Pi, \ell) \leq \eta^*\ell,\qquad \forall 0 \leq \ell \leq m.
    \]
    Rounding may assign $\bot$ to some of the agents, however, every $\pi \in \Pi\setminus\bot$ such that $y_i(\pi) = 1$ for agent $i \in N$ must be acceptable, i.e $\pi \in \Pi_i$.
    Note that these shall not necessarily be the original MMS profiles, as we round an optimal solution $x^*$, not just arbitrary feasible one.

    Let $I\subseteq N$ be the agents selected by the rounding, i.e $I = \{i \in N: y_i(\bot) =0\}$.
    Applying \cref{etaload}, we allocate all agents $I$ using distinct items from a set $E\subseteq M$ satisfying 
    \[
        |E\cap [\ell]|\leq \eta^*\ell,\qquad \forall 0 \leq \ell \leq m.
    \]
    By \cref{prefixbound}, for every $i \in N$ we have $v_i(E) \leq \eta^* n$.
    The number of omitted agents is at most
    \[
        \rho(m, n + m) \leq \frac{(1 - \eta^*)n}{2},
    \]
    thus applying \cref{leftoveralloc}, we get an allocation of $M - E$ to $N - I$ such that every agent $u \in N- I$ gets a bundle of value at least $1$.
    The agents selected by the rounding get a bundle of value at least $1 - 2/q$, which gives us a $(1 - 2/q)$-MMS allocation of $M$ to $N$.

    Now suppose that $n(1 - \eta^*)< 2\rho(m, n + m)$.
    In order to construct the feasible solution to (\ref{dlp}), we first consider the dual of (\ref{plp}).
    In this dual, we associate a multiplier $\omega_\ell /n$ with every prefix load constraint $1 \leq \ell \leq m$, and a multiplier $\lambda_i/n$ with every agent weight constraint $i \in N$.
    After rescaling, we obtain the following linear program.
    \[
    \begin{aligned}
        &\maximize&& \frac{1}{n}\sum_{i \in N}\lambda_i\\
        &\subto && \sum_{\ell = 1}^m\omega_\ell \cdot \Pp_i(\pi, \ell) \geq \lambda_i,& \forall i \in N, \pi \in \Pi_i;\\
        &&& \sum_{\ell = 1}^m\omega_\ell \cdot \ell = n;\\
        &&& \omega_\ell \geq 0,& \forall 1 \leq \ell \leq m.
    \end{aligned}\tag{DP-LP}\label{dplp}
    \]
    Variables $\omega_\ell$ correspond to the price of one load unit of prefix $\{1, \ldots, \ell\}$.
    If, to satisfy profile $\pi$, agent $i$ needs to use $\Pp_i(\pi, \ell)$ items from $\{1, \ldots, \ell\}$, the agent must pay the total price of $\omega_\ell \cdot \Pp_i(\pi, \ell)$ for using prefix $\ell$.
    Consequently, the total price of satisfying profile $\pi$ for agent $i$ is $\sum_{\ell = 1}^m\omega_\ell \cdot \Pp_i(\pi, \ell)$.
    As we will see from complementary slackness, if at optimal solution we have $\omega_\ell^* > 0$, then $\Hh(x^*, \Pi, \ell) = \eta^*\ell$, i.e the fractional load at prefix $\ell$ is tight and leaves no spare capacity.
    Similarly, if $\Hh(x^*, \Pi, \ell) < \eta^*\ell$, then $\omega^*_\ell = 0$ and using prefix $\ell$ comes ``for free''.

    The dual load prices naturally define the common item prices.
    Every item $e \in  M$ belongs to the prefixes $\{1, \ldots, \ell\}$ precisely when $\ell \geq e$.
    Then, if we want to use item $e$, we use one load unit in every prefix $\ell \geq e$, and have to pay $\omega_\ell$ for each of these units.
    Thus, for the optimal dual solution $(\lambda^*, \omega^*)$ to (\ref{dplp}), we define
    \[
        w^*(e) := \sum_{\ell = e}^m\omega_\ell^*,
    \]
    Clearly, $w^*$ is additive, and it holds automatically that $w^*(1)\geq \ldots \geq w^*(m)\geq 0$, and
    \[
        w^*(M) = \sum_{e = 1}^m\sum_{\ell = e}^m\omega_\ell^* = \sum_{\ell = 1}^m\omega_\ell^*\cdot \ell = n.
    \]
    This construction reflects how each demand profile contributes to each prefix load.
    For profile $\pi \in \Pi_i$ of agent $i \in N$, define its total dual load price by
    \[
        \Lambda_i^*(\pi) := \sum_{\ell = 1}^m\omega_\ell^* \cdot \Pp_i(\pi, \ell).
    \]
    A demand $\alpha$ with breakpoint $m_i(\alpha)$ contributes one load unit to every prefix $\ell \geq m_i(\alpha)$, so the dual price of this breakpoint is $w^*(m_i(\alpha))$.
    Therefore,
    \[
        \Lambda_i^*(\pi) = \sum_{\ell = 1}^m\omega_\ell^* \cdot |\{\alpha \in \pi : m_i(\alpha)\leq \ell\}| = \sum_{\alpha \in \pi}w^*(m_i(\alpha)).
    \]
    In particular, for the optimal solution $(\eta^*, x^*)$ to (\ref{plp}), the total fractional dual load price is
    \[
        \sum_{i \in N}\sum_{\pi \in \Pi_i}x_i^*(\pi)\cdot \Lambda_i^*(\pi) = \sum_{\ell = 1}^m\omega_\ell^* \cdot \Hh(x^*, \Pi, \ell).
    \]
    By strong duality,
    \[
        \sum_{i \in N}\lambda_i^* = n\eta^*,
    \]
    and at the optimal solution $(\lambda^*, \omega^*)$, we must have 
    \[
        \lambda_i^* = \min_{\pi \in \Pi_i}\Lambda_i^*(\pi) \geq 0.
    \]
    Therefore,
    \[
        n\eta^* = \sum_{i \in N}\lambda_i^* \leq \sum_{i \in N}\sum_{\pi \in \Pi_i}x_i^*(\pi)\cdot \Lambda_i^*(\pi)
        = \sum_{\ell = 1}^m\omega_\ell^* \cdot \Hh(x^*, \Pi, \ell)\leq \eta^*\sum_{\ell = 1}^m\omega_\ell^* \cdot \ell = \eta^*n,
    \]
    forcing
    \[
        \sum_{\ell = 1}^m\omega_\ell^* \cdot (\eta^*\ell - \Hh(x^*, \Pi, \ell)) = 0.
    \]
    This is exactly the complementary slackness: since all the summands are non-negative, $\omega_\ell^* > 0$ forces $\Hh(x^*, \Pi, \ell) = \eta^*\ell$, so we pay positive price only for prefixes where the fractional load is tight.
    In addition, by definition of $w^*$, for every $1 \leq \ell \leq m$ (with $w^*(m + 1) = 0$) we have
    \[
        w^*(\ell) - w^*(\ell + 1) = \omega^*_\ell.
    \]
    So, the common price of an item strictly decreases only if the item passes a tight prefix constraint.
    In turn, if the prefix constraint is not tight, the item price stays the same.

    Having discussed the intuition behind $w^*$, we can finally show that $(\lambda^*, w^*)$ is a feasible solution to (\ref{dlp}).
    Take any agent $i \in N$ and his acceptable bundle $B \in \Bb_i$, by definition $\pi_i(B) \in \Pi_i$.
    Furthermore, for every $e \in B$ with $v_i(e) > 0$ we have $m_i(v_i(e))\geq e$.
    Therefore,
    \[
        \lambda_i^* \leq \Lambda_i^*(\pi_i(B)) = \sum_{\substack{e \in B\\v_i(e) > 0}}w^*(m_i(v_i(e))) \leq \sum_{\substack{e \in B\\v_i(e) > 0}}w^*(e) \leq w^*(B).
    \]
    Thus, every acceptable bundle for $i$ has price at least $\lambda_i^*$, so $(\lambda^*, w^*)$ is feasible.
    In addition,
    \[
        n\lambda_i^* \leq \sum_{j = 1}^nw^*(B_{i,j}) = w^*(M) = n,
    \]
    so $\lambda_i^* \in [0, 1]$ for every agent $i \in N$.
    Finally, optimality forces
    \[
        \Lambda^* := \sum_{i \in N}(1 - \lambda_i^*) = n - \sum_{i \in N}\lambda_i^* = n(1- \eta^*) < 2\rho(m, n + m).
    \]
\end{proof}

Applying \cref{dualprofile} and \cref{commonprice} with $\eps = 1/4$, we get the corollary below.
\begin{corollary}\label{commonhead}
    Let $\eta^*$ be the optimum of (\ref{plp}).
    If $n(1 - \eta^*) < 2\rho(m, n + m)$, there exist common additive prices $w$ satisfying $w(1) \geq \ldots \geq w(m) \geq 0$ and numbers $\lambda_i \in [0, 1]$ for $i\in N$, as well as an integer $h \in \{0, 1, \ldots, m\}$ with the following properties.
    Let $\Lambda := \sum_{i \in N}(1 - \lambda_i)$ and 
    \[
        G := \{i \in N : \; n(1 - \lambda_i) \leq 4\Lambda\}.
    \]
    Then $\Lambda < 2\rho(m, n + m)$, integer $h$ satisfies $qn \geq h \geq n - q(8\Lambda + 1)$, and
    \begin{itemize}
        \item $w(M) = n$, and every bundle $B\subseteq M$ acceptable to agent $i \in N$ satisfies $w(B) \geq \lambda_i$;
        \item for every item $e \in \{1, \ldots, h\}$ it holds $w(e) \geq 1/q$;
        \item $|G|\geq 3n/4$, and for every agent $i \in G$ and item $e \in \{1, \ldots, h\}$ it holds $v_i(e) \geq 1/q$;
        \item for every agent $i \in N$ with $\lambda_i \geq 1/2$ it holds
        \[
            T_i := v_i(\{h + 1, \ldots, m\}) \leq q(16\Lambda + 1),
        \]
    \end{itemize}
    If also $n \geq q^2(16\Lambda + 1)$, then $G\subseteq \{i \in N:\lambda_i \geq 1/2\}$, and for every $i \in N$ with $\lambda_i \geq 1/2$,
    \[
        v_i(\{1, \ldots, h\}) \geq n - q(16\Lambda + 1) .
    \]
\end{corollary}

From now on, assume that $n(1-\eta^*)< 2\rho(m, n + m)$, i.e \cref{dualprofile} did not get sufficient condition to produce an allocation giving every agent value at least $1 - 2/q$.
So, instead of performing rounding, we obtained the prices $w$ and numbers $\lambda_i$, as well as the common head $h$.

\subsection{Preallocation of head-poor agents}

The common head $[h]$ from \cref{commonhead} contains almost all the value for every agent $i \in N$ with $\lambda_i \geq 1/2$.
In addition, for every agent $i \in G$, which we will refer to as \textit{good agents}, every item $e \leq h$ individually is worth at least $1/q$.

We would like to continue working only with agents that value the common head $[h]$ highly.
We will achieve this by \textit{preallocating} a small subset of tail items $\{h + 1,\ldots, m\}$ to all the agents that value $[h]$ poorly, giving every such agent an acceptable bundle.
The only question is, how exactly to define ``highly'' and ``poorly'' in order to make the approach valid.

By \cref{commonhead}, every $i \in N$ with $\lambda_i \geq 1/2$ has $v_i([h]) \geq n - q(16\Lambda + 1)$ (for $n$ large enough), and it is easy to see that the number of agents failing this bound is at most $2\Lambda$, i.e small.
However, this lower bound is too strong, as we may not guarantee enough tail value per poor agent while keeping the used subset of items sufficiently small.
Instead, we will say that $i \in N$ values $[h]$ highly if $v_i([h]) \geq (1 - 1/q)n$, a slightly weaker bound.
Note that it holds automatically for all agents $i$ with $\lambda_i \geq 1/2$ if $n \geq q^2(16\Lambda + 1)$.
\begin{definition}
    We call agent $u\in N$ \textbf{head-poor} if $v_u([h]) < (1 - 1/q)n$.
    Denote the set of head-poor agents by $U= \{u \in N: v_u([h]) < (1 - 1/q)n\}$.
\end{definition}
\begin{observation}\label{poorsize}
    If $n \geq q^2(16\Lambda + 1)$, then $\lambda_u < 1/2$ for every head-poor agent $u \in U$.
    In particular, we have $G\cap U = \varnothing$ and $|U|\leq 2\Lambda$.
\end{observation}
\begin{proof}
    When $n \geq q^2(16\Lambda + 1)$, every $u \in U$ satisfies
    \[
        T_u = n - v_u([h]) > n/q \geq q(16\Lambda + 1),
    \]
    which implies $\lambda_u < 1/2$ by \cref{commonhead}.
    Therefore, $G\cap U = \varnothing$ and
    \[
    \Lambda = \sum_{i \in N}(1-\lambda_i) \geq \sum_{u \in U}(1-\lambda_u) \geq |U|/2.\]
\end{proof}
We will \textit{preallocate} all head-poor agents $U$ using a subset of common tail items $\{h + 1, \ldots, m\}$.
This subset will be chosen in such a way that every remaining agent with $\lambda_i \geq 1/2$ loses very little value, i.e at most $1/(2q^2)$.
After this is done, we will remove all preallocated agents and their items, and complete the allocation of the remaining agents using only the remaining items.
\begin{lemma}\label{poor}
    Suppose that $n \geq \max\{q^2(16\Lambda + 1), 8q^4\Lambda(16\Lambda + 1) + 2q\}$.
    There are disjoint sets $D_u \subseteq \{h + 1, \ldots, m\}$ for $u\in U$, such that $v_u(D_u) \geq 1$ for every head-poor agent $u \in U$, and
    \[
        D := \bigcup_{u \in U}D_u
    \] satisfies $v_i(D) \leq 1/(2q^2)$ for every agent $i \in N$ with $\lambda_i \geq 1/2$.
\end{lemma}
\begin{proof}
    The key intuition here is that if agent $u \in U$ values the head $[h]$ low, then a substantial portion of agent $u$'s value must from the tail items $\{h + 1, \ldots, m\}$.
    We will use this and construct an allocation for $U$ using a small sparse subset of tail items.
    We will do it similarly to the tail profile construction of \cref{etatailcomplete}, but without introducing additional tail-profile terminology.

    If $|U| = 0$, take $D_u = \varnothing$ for all $u \in U$ and we are done.
    Otherwise, put
    \[
        L := \left\lfloor \frac{n}{2q}\right\rfloor,
    \]
    by assumption $L\geq 2q^3|U|(16\Lambda + 1)$.
    We break the tail $\{h +1 ,\ldots, m\}$ into consecutive blocks $B_1, \ldots, B_K$ of size $|B_k| = L$ each, with an additional block $B_{K + 1}$ of size less than $L$, for an appropriate number of blocks $K$.
    There exist at least two full blocks, as $m - h \geq T_u$ and
    \[
        T_u = n - v_u([h]) > n/q \geq 2L.
    \]
    For every $k \in [K]$, we take the \textit{last} $|U|$ items of block $B_k$ and allocate them to agents $U$, so that each $u \in U$ receives exactly one of these items from $B_k$.
    Let $e_{u,k}$ denote the item allocated to agent $u \in U$ from $B_k$, and set $D_u := \{e_{u,k} : k \in [K]\}$.

    Clearly, bundles $D_u$, $u \in U$ are disjoint.
    To see that $v_u(D_u) \geq 1$, we observe that for every $k \in [K]$, each item in $B_{k + 1}$ has value at most $v_u(e_{u,k})$ of the item $e_{u,k}$ allocated from $B_k$.
    So,
    \[
        v_u(B_{k + 1})\leq L\cdot v_u(e_{u,k}),\qquad\qquad \forall k \in [K].
    \]
    Summing over all $k \in [K]$, we get
    \[
        v_u(D_u) \geq \frac{1}{L}\sum_{k = 1}^Kv_u(B_{k + 1}) = \frac{T_u - v_u(B_1)}{L} \geq \frac{T_u - L}{L} \geq 1.
    \]

    Next, for any agent $i \in N$, consider the value removed by $D = \cup_{u \in U}D_u$.
    In each full block $k \in [K]$, set $D\cap B_k$ consists of the last $|U|$ items.
    Then, since every item in $B_k - D$ is worth at least the average value of the items in $D\cap B_k$, we have
    \[
        v_i(D\cap B_k) \leq \frac{|U| }{L}\cdot v_i(B_k).
    \]
    Summing over all $k \in [K]$ we get 
    \[
        v_i(D) \leq \frac{|U|}{L}\sum_{k = 1}^Kv_i(B_k) \leq \frac{|U|}{L}\cdot T_i.
    \]
    Finally, for every $i \in N$ with $\lambda_i \geq 1/2$, \cref{commonhead} gives
    \[
        v_i(D) \leq  \frac{|U|}{L}\cdot T_i \leq \frac{|U| q(16\Lambda + 1)}{L} \leq \frac{1}{2q^2}.
    \]
\end{proof}

Assuming that number of agents $n$ is sufficiently large, we apply \cref{poor} to the set $U$ of head-poor agents, and obtain an allocation $\{D_u\}_{u \in U}$.
At this point, head-poor agents $U$ are done: every $u \in U$ keeps their allocated bundle $D_u$ until the very end of the algorithm as part of the final allocation.
So, we will remove agents $U$ from $N$ and items $D = \bigcup_{u \in U}D_u$ from $M$, and proceed to construct an allocation giving every remaining agent value at least $1 - 2/q$.

Observe that, for every agent $i \in N \setminus U$ wtih MMS bundles $B_{i,j}$, $j \in [n]$, if $\lambda_i \geq 1/2$ then
\[
    v_i(B_{i,j}\setminus D) \geq v_{i}(B_{i,j}) - v_i(D) \geq 1 - \frac{1}{2q^2}.
\]
For every agent $i\in N \setminus U$, replace their MMS bundles $B_{i,j}$, $j \in[n]$, with $B_{i,j}\setminus D$, relabeling $B_{i,j}\setminus D \to B_{i,j}$.
In addition, we will also relabel remaining agents $N- U\to N$, and remaining items $M\setminus D\to M$, while preserving order, as well as the common prices $w(e)$ for $e \in M$.

From now on $N$ denotes the $d:= |N| = n - |U| \geq n - 2\Lambda$ remaining agents (lower bound is due to \cref{poorsize}), $M$ denotes the $m = |M| \leq \Gamma n$ remaining items.
Each of the remaining $d$ agents still has a personal partition $\{B_{i,1}, \ldots, B_{i,n}\}$ of $M$ into $n$ parts.
Note that some of these parts potentially may contain very little value or even be empty.

Note that we \textbf{do not} renormalize the instance, our goal is still to give each of the remaining agents $i \in N$ a bundle of $v_i$-value at least $1 - 2/q$.
By \cref{poor}, the common head $\{1, \ldots, h\}$ still lies completely in $M$, and by definition of $U$, now for \textit{every} $i \in N$ it holds
\[
    v_i([h]) \geq \left(1 - \frac{1}{q}\right)n.
\]
By \cref{poorsize} $G\cap U = \varnothing$, so $G\subseteq N$ is unchanged, as well as its head-value guarantees.

Consider some remaining agent $i \in N$ with his ``MMS'' partition $\{B_{i,1}, \ldots, B_{i,n}\}$ of $M$.
By construction, for every $j \in [n]$ we have $v_i(B_{i,j})\leq 1$, and by \cref{poor} if $\lambda_i \geq 1/2$ then 
\[
    1 \geq v_i(B_{i,j}) \geq 1 - \frac{1}{2q^2},\qquad\qquad \forall j \in [n],
\]
so every ``MMS'' bundle of these $i$ keeps almost all of its original value.
In addition, since head $[h]$ stays unchanged, by \cref{commonhead} for every $i \in N$ with $\lambda_i \geq 1/2$ we have a sharper bound
\[
    v_i([h]) \geq n - q(16\Lambda + 1).
\]

Since every good agent $i \in G$ for sure has at least $n$ nonempty bundles, the number of remaining items satisfies $m \geq n$.
As already mentioned, item deletions also preserve the original upper bound of \cref{deltammsproof}, so $m \leq \Gamma n$.

Next, the common prices $w$ keep their order $w(1) \geq \ldots \geq w(m)$ w.r.t items $M$.
For every subset $B\subseteq M$ of remaining items, none of the items of $B$ were in the removed set $D$.
Hence, the value of every $B\subseteq M$ stayed exactly the same, as well as the price.
So, for every $B\subseteq M$,
\[ 
    v_i(B) \geq 1 - \frac{2}{q}\quad\implies \quad w(B)\geq \lambda_i.
    \]
Since items were removed, $w(M)\leq n$, i.e equality no longer needs to hold.
Finally, we have
\[
    \sum_{i \in N}(1 - \lambda_i)\leq \Lambda \leq 2\rho(\Gamma n, (\Gamma  + 1)n),
\]
by using \cref{commonhead} and substituting an upper bound $\Gamma n$ on the original number of items.

Combining all these derivations above, we obtain the proof of the following corollary.
\begin{corollary}\label{notpoor}
    Suppose that $n \geq \max\{q^2(16\Lambda + 1), 8q^4\Lambda(16\Lambda + 1) + 2q\}$, where $\Lambda$ retains its original meaning from \cref{commonhead}.
    After preallocating head-poor agents $U$ with items $D$, the remaining instance with agents $N$ and items $M = [m]$ has the following properties.
    \begin{itemize}
        \item For $d := |N|$ it holds $n\geq d \geq n - 2\Lambda$, and for $m:= |M|$ it holds $n \leq m \leq \Gamma n$.
        \item The common head $h$ satisfies $qn \geq h \geq n - q(8\Lambda + 1)$, and for every $i \in N$ it holds 
        \[
            v_i([h]) \geq \left(1 - \frac{1}{q}\right)n.
        \]
        \item For every agent $i \in N$, there is a partition $\{B_{i,1}, \ldots, B_{i,n}\}$ of $M$, where $v_i(B_{i,j}) \in [0, 1]$ for all $j \in [n]$, and some bundles $B_{i,j}$ potentially may be empty.
        If in addition $\lambda_i \geq 1/2$, then
        \[
            \forall j \in [n]:\; v_i(B_{i,j})\geq 1-\frac{1}{2q^2},\qquad \text{and}\qquad v_i([h]) \geq n - q(16\Lambda + 1).
        \]
        \item The set of good agents $G = \{i\in N: n(1 - \lambda_i)\leq 4\Lambda\}$ is unchanged.
        In particular, $|G|\geq 3n/4$, $G\subseteq \{i \in N : \lambda_i \geq 1/2\}$, and for every $i \in G$ it holds
        \[
            \forall e \leq h :\; v_i(e) \geq \frac{1}{q}.
        \]
        \item Common prices satisfy $w(1) \geq \ldots \geq w(m) \geq 0$, as well as $w(e) \geq 1/q$ for all $e \leq h$.
        In addition, $w(M) \leq n$, and for every $i \in N$ and $B\subseteq M$,
        \[ 
    v_i(B) \geq 1 - \frac{2}{q}\quad\implies \quad w(B)\geq \lambda_i.
    \]
    \item For every $i \in N$ we have $\lambda_i \in [0, 1]$, and $\sum_{i \in N}(1 - \lambda_i)\leq \Lambda \leq 2\rho(\Gamma n, (\Gamma  + 1)n)$.
    \end{itemize}
\end{corollary}

Our goal now is to prove the following theorem.
\begin{theorem}\label{qmms}
    Let $\Gamma  \geq 1$ and $q \geq 9$ be integers.
    There exists $n(\Gamma , q)$ such that for every  $n \geq n(\Gamma , q)$ the following holds.
    Consider a fair allocation instance with $m$ items $m$ and $d$ agents $N$ satisfying \cref{notpoor}, where for every $i \in N$ valuation $v_i$ is additive.
    Then there exists an allocation $(B_1, \ldots, B_d)$ of $M$ to $N$ such that for every agent $i \in N$ we have $v_i(B_i) \geq 1 - 2/q$.
    Furthermore, it suffices to take $n(\Gamma , q)$ as
    \[
        n(\Gamma , q) = 2^{4q^2\log^2(2q)}\cdot (\log(2\Gamma ))^{2q}.
    \]
\end{theorem}
It is easy to see that, combining \cref{qmms} with \cref{notpoor} and \cref{poor}, we obtain an allocation for all original agents that gives each agent a bundle of value at least $1 - 2/q$, thus proving \cref{deltammsproof}.
\section{Allocating common head items}

\label{sec:partition}

We will now focus on finding an allocation of remaining $m$ items $M$ to remaining $d = |N|$ agents $N$ such that each agent $i \in N$ receives a bundle of value at least $1-2/q$.
Our approach will be similar to the one used in \cref{sec:dmms}.
That is, we would like to find a collection of profiles $\Pi_N = \{\pi_i\}_{i \in N}$ for \textit{all} agents, such that for every $i \in N$ we have $v(\pi_i) \geq 1 - 2/q$, and for every prefix $0 \leq \ell \leq m$ it holds
\[
    \Pp(\Pi_N, \ell) \leq \ell.
\]
Then, we can apply \cref{loadequiv} and give all $i \in N$ disjoint bundles of value at least $1 - 2/q$.

As before, we will first find a collection $\Pi_N$ of head profiles $\pi_i \in \Pi_N$, and then complete it to full profiles with a collection of tail profiles $\psi_i \in \Psi_N$, so that for every $i \in N$ we have $v(\pi_i) + v(\psi_i) \geq 1- 2/q$, and agent $i$ gets $\{\pi_i, \psi_i\}$.
The main difficulty is that, under conditions of \cref{notpoor}, even though all agents $i \in N$ share a common head $h$ that they all value highly, not all items $e \leq h$ are worth high enough for every agent.
Because of this, the distinction between head and tail profiles becomes a little bit obscure.

In the proof of \cref{dmms}, it was of no difference for us how many different head profiles there are for agents $i \in N$, and how large these profiles can be.
Each head demand could take essentially arbitrary value in $(0, 1]$, and there could be up to $O(n^2)$ different demands in a head profile.
However, for the proof of \cref{qmms}, it will be crucial for us that 1) each head profile has at most constant size, and 2) all head profiles used in the initial rounding have demands from some finite set, whose size depends only on $q$.
Later we may replace them by smaller demands in a controlled way.

We achieve this by rounding down the high-valued items to powers of $(1 - 1/q)$.
\begin{definition}
    For an integer $q > 0$, a set $\Qq$ of all possible \textbf{rounded high demands} is
    \[
        \Qq = \left\{\frac{1}{q}\right\}\cup \left\{\left(1 - \frac{1}{q}\right)^k : \; k \in \Z_{\geq 0}, \left(1 - \frac{1}{q}\right)^k > \frac{1}{q}\right\}.
    \]
    That is, a \textbf{rounded high demand} is a positive value $\alpha \in (0, 1]$ such that either $\alpha = 1/q$ or $\alpha = (1 - 1/q)^k$ for integer power $k$ satisfying $(1 - 1/q)^k > 1/q$.
    A \textbf{rounded high profile} is a finite multiset $\pi = \{\alpha_1, \ldots, \alpha_{|\pi|}\}$ of rounded high demands.
    
    Similarly, a \textbf{low demand} is a positive value $\sigma \in (0, 1]$ such that $\sigma < 1/q$.
\end{definition}
We will sometimes omit the word ``rounded'' and just refer to such profiles as high profiles.
Denote $Q:=|\Qq|$.
It is easy to see that
\[
    Q := |\Qq| = 1 + \left\lceil \frac{\ln q}{-\ln(1 - 1/q)}\right\rceil \leq 2 + q\ln q.
\]
Note that for every agent $i \in N$, a bundle $B\subseteq M$ such that $v_i(B) \leq 1$ can contain at most $q$ items of value at least $1/q$.
So, a high profile induced by such $B$ has size at most $q$.

\begin{definition}
    For $t \in [0, 1]$, define the \textbf{rounding function} $g$ as
    \[
        g(t) := \begin{cases}
            t,& t < 1/q;\\
            \max\{\alpha \in \Qq : t \geq \alpha\}, &t \geq 1/q.
        \end{cases}
    \]
    Observe that for every $t \in [0, 1]$ it holds
    \[
        t \geq g(t) \geq \left(1 - \frac{1}{q}\right)t.
    \]
    For every agent $i \in N$, his \textbf{rounded valuation} is $\wv_i$ such that
    \[
        \wv_i(e) := g(v_i(e)).
    \]
    It is easy to see that for every $\alpha \in \Qq$ we have $\wv_i(e) \geq \alpha \iff v_i(e) \geq \alpha$, therefore
    \[
        |\{e \in M : \wv_i(e) \geq \alpha\}| = |\{e \in M: v_i(e) \geq \alpha\}| = m_i(\alpha).
    \]
    So, breakpoints w.r.t rounded $\wv_i$ for $\alpha \in \Qq$ are exactly the same as original breakpoints w.r.t $v_i$.
\end{definition}
\begin{observation}\label{roundedhead}
    For every agent $i \in N$ it holds
    \[
        \wv_i([h]) \geq \left(1 - \frac{1}{q}\right)v_i([h]) \geq \left(1 - \frac{1}{q}\right)^2n.
    \]
    For every good agent $i \in G$, $\wv_i(e) \in \Qq$ for every $e \leq h$, thus $\wv_i(e)\geq 1/q$ for all $e \leq h$.
\end{observation}

\subsection{Head partition and buffer set}

In \cref{sec:dmms} and the proof of \cref{dmms}, in order to complete a collection $\Pi_N$ of head profiles simultaneously satisfiable with $\{1, \ldots, h\}$ to full with tail profiles $\Psi_N$, we needed to ensure the total load of $\Pi_N$ had enough slackness at every prefix.

For now, let us not worry about slackness for the tail profiles $\Psi_N$ at all, and try to at least get some collection $\Pi_N = \{\pi_i\}_{i \in N}$ of just head profiles that satisfies the prefix load constraints.
That is, we want that for every $0\leq \ell \leq m$,
\[
    \sum_{i \in N}\Pp_i(\pi_i, \ell) \leq \ell.
\]
In addition, we restrict ourselves only to head profiles that are \textit{high}, i.e for every $i \in N$, we want for every demand $\alpha \in \pi_i$ that $\alpha \in \Qq$.

Following the approach of \cref{sec:dmms} and the proof of \cref{dmms}, for every agent $i \in N$ we could define MMS head profiles $\pi_{i,j}$, $j \in [n]$, while rounding the item values to $\Qq$:
\[
    \pi_{i,j} = \{\wv_i(e) : e \in B_{i,j} \cap [h]\}.
\]
Of course, not every item $e \leq h$ has $\wv_i(e) \geq 1/q$ for every agent $i \in N$, but let us for now assume that all demands of $\pi_{i,j}$ are in $\Qq$, i.e every $\pi_{i,j}$ is a \textit{high profile}.
Then, taking  common profile set 
\[\Pi := \{\varnothing, \bot\} \cup \bigcup_{i \in N}\{\pi_{i,j} : j \in [n]\},\] we could consider fractional profiles $(x, \Pi)$, where for $\pi \in \Pi\setminus\bot$, for every agent $i \in N$ we set
\[
        x_i(\pi) = \frac{|\{j \in [n]: \pi_{i,j} = \pi\}|}{n},
    \]
and then apply the rounding of \cref{etarounding} to $(x, \Pi)$, to obtain some integral collection.

Unfortunately, the rounding is not good enough just by itself.
Even though it leaves only a small fraction of agents without a profile, as discussed in \cref{sec:dual} it does not provide strong enough guarantees to give an acceptable bundle to every unsassigned agent.
Unlike \cref{sec:dmms}, we want to assign some head profile $\pi_i$ to \textit{every} agent $i \in N$.

\subsubsection{$E/S$-partition of the common head}

\label{sec:genES}

Our approach will be precisely the extension of \cref{sec:integer}.
Let $(x, \Pi)$ be some collection of fractional profiles defined on $\Pi$.
Suppose that, after rounding $(x, \Pi)$ to some integral collection of profiles $(y, \Pi)$, we obtained a large subset of agents $I\subseteq N$ with assigned integral profiles $\Pi_I = \{\pi_i\}_{i \in I}$, where each $\pi_i$ is a rounded head profile $\pi_i = \pi_{i,j}$ for some $j \in [n]$.

We would like to choose a large subset $A\subseteq I$ of agents that will keep their assigned profiles, and give every $u \in N\setminus A$ some new profile $\pi_u$ (not necessarily of the form $\pi_{u,j}$ for some $j \in [n]$, and even not necessarily a high profile), so that for every $0 \leq \ell \leq m$,
\[
    \sum_{i \in A}\Pp_i(\pi_i, \ell) + \sum_{u \in N\setminus A}\Pp_u(\pi_u, \ell) \leq \ell.
\]

Similarly to \cref{sec:integer}, we will partition the head items $[h]$ into $[h] = E\sqcup S$, where set $E$ contains most of the head items, and set $S$ is very small.
We define decompositions into $E$- and $S$-sub-profiles  exactly the same way.
\begin{definition}
    Consider an arbitrary head partition $[h] = E \sqcup S$.
    An $E/S$-decomposition of a demand profile $\pi$ is a pair $\{\pi^E, \pi^S\}$ of sub-profiles such that \[
        \pi = \pi^E \uplus \pi^S,
    \] where $\uplus$ is union with multiplicities.
    Demands in $\pi^E$ are designated to be satisfied exclusively by items of $E$, and demands in $\pi^S$ are designated to be satisfied exclusively by items of $S$.
\end{definition}
\begin{remark}
    Note that, since we work with demand profiles and not actual bundles, in reality each demand of sub-profiles $\pi^E, \pi^S$ may be also satisfied by many items outside of $E, S$ respectively, and even outside the head $[h]$.
    However, for the purpose of the approach, our intension is to try and restrict $\pi^E, \pi^S$ only to items from their corresponding sets $E, S$, and find the collection of appropriate sub-profile pairs that indeed can be satisfied exclusively by respectively $E, S$ only.
\end{remark}

The sufficient conditions  for profiles $\{\pi_i\}_{i \in A}$, $\{\pi_u\}_{u \in N\setminus A}$ of agents $A, N\setminus A$ respectively to satisfy all prefix-load constraints are that each of these profiles has some $E/S$-decomposition, such that for every $0 \leq \ell \leq m$ it holds
\begin{gather}
    \sum_{i \in A}\Pp_i(\pi_i^E, \ell) + \sum_{u \in N\setminus A}\Pp_u(\pi_u^E, \ell) \leq |E\cap [\ell]|;\tag{E}\label{genE}\\
    \sum_{i \in A}\Pp_i(\pi_i^S, \ell) + \sum_{u \in N\setminus A}\Pp_u(\pi_u^S, \ell) \leq |S\cap [\ell]|.\tag{S}\label{genS}
\end{gather}
Indeed, $\Pp_i(\pi, \ell) = \Pp_i(\pi^E, \ell) + \Pp_i(\pi^S, \ell)$, and the same holds for $S$, so together (\ref{genE}) and (\ref{genS}) imply
\[
     \sum_{i \in A}\Pp_i(\pi_i, \ell) + \sum_{u \in N\setminus A}\Pp_u(\pi_u, \ell)\leq |E\cap [\ell]| + |S\cap [\ell] = \min\{h, \ell\} \leq \ell.
\]

Similarly to \cref{sec:integer}, we can bound
\[
    \sum_{u \in N \setminus A}\Pp_u(\pi_u^E, \ell) \leq \sum_{u \in N \setminus A}|\pi_u|,
\]
and since $A\subseteq I$, bounding the total load of $A$ by the total load of $I$ gives
\[
    \sum_{i \in A}\Pp_i(\pi_i^E, \ell) +  \sum_{u \in N \setminus A}\Pp_u(\pi_u^E, \ell) \leq \sum_{i \in I}\Pp_i(\pi_i^E, \ell) +  \sum_{u \in N \setminus A}|\pi_u|.
\]
Thus, for (\ref{genE}) to hold, it suffices that for some integer $\nu \geq 1$, the old profiles $\pi_i$ for $i \in I$ and the new profiles  $\pi_u$ for $u \in N\setminus A$ satisfy
\begin{align}
    &&&\sum_{i \in I}\Pp_i(\pi_i^E, \ell) \leq |E\cap [\ell]| - \frac{\nu}{q},&\forall 0\leq \ell\leq m\tag{E-$\nu$}\label{genEbad}\\
    &&&\sum_{u \in N \setminus A}|\pi_u|\leq \frac{\nu}{q} \tag{A-$\nu$}\label{genUbad}.
\end{align}
Once again, we cannot expect uniform slackness (\ref{genEbad}) to hold at every prefix $0 \leq \ell \leq m$, and we will mitigate this issue by construct a small \textit{buffer set} $F\subseteq E$.

\subsubsection{Buffer set construction}

As in \cref{sec:integer}, our set $S\subseteq [h]$ will be of the form
\[
    S = \{a_1, a_1 + 1,\ldots, b_1\} \cup \ldots \cup \{a_t, a_t + 1, \ldots, b_t\},
\]
where each $[a_s, b_s]$ for $s \in [t]$ is a maximal consecutive interval of items from $S$.
We will have $t = O(q\log q)$,  and the length of each consecutive interval and the total size of $S$ are both at most $O_{\Gamma ,q}(\log^2 n)$.
In addition, it will hold that $a_1 = 1$ and $b_t = h$, i.e $S$ includes some prefix $\{1, \ldots, b_1\}$ and some suffix $\{a_t, \ldots, h\}$ of the common head $[h]$.

Take $E = [h]\setminus S$.
We will have a \textit{buffer set} $F\subseteq E$ defined by $S$.
After every interval $[a_s, b_s]\subset S$, $s \in [t]$, we take another interval $\{b_s + 1, \ldots, b_s + \nu\}$ of length $\nu = O_{\Gamma ,q}(\log^2 n)$, and
\[
    F = E\cap (\{b_1 + 1, \ldots, b_1 + \nu\} \cup \ldots \cup \{b_t + 1, \ldots, b_t + \nu\}),
\]
or equivalently, setting $c_s := \min\{b_s + \nu, a_{s + 1} - 1\}$ for $s < t$, and using $b_t = h$,
\[
    F = \{b_1 + 1, \ldots, c_1\} \cup \ldots \cup \{b_{t - 1} + 1, \ldots, c_{t -1}\}.
\]
The set $S$ and parameter $\nu \geq 1$ are picked in such a way that
\begin{itemize}
    \item for every $\ell \in E \setminus F$, the whole interval $(\ell - \nu, \ell] = \{\ell - \nu + 1, \ldots, \ell\}\subseteq E$;
    \item at least $|N|/q$ agents have \textit{no} rounded demand breakpoints in $(\ell - \nu, \ell]$.
\end{itemize}
For the agents with no rounded demand breakpoints in $(\ell - \nu, \ell]$, the prefix $E$-load stays constant over the whole interval $(\ell - \nu, \ell]$ of length $\nu$.
At the same time, the available $E$-capacity at $\ell$ increases by $\nu$.
Averaging over all agents $N$ gives slackness at least $\nu/q$ at this prefix $\ell$.

We will now describe the construction of $S$.
Recall the set of rounded high demands
\[
        \Qq = \left\{\frac{1}{q}\right\}\cup \left\{\left(1 - \frac{1}{q}\right)^k : \; k \in \Z_{\geq 0}, \left(1 - \frac{1}{q}\right)^k > \frac{1}{q}\right\},\qquad Q := |\Qq|.
    \]
The following lemma is an analogue of \cref{buffer} from \cref{sec:integer}.
\begin{lemma}[Buffer set]\label{genbuffer}
    Let $\nu \geq 1$ be an integer and $\eps \in (0, 1)$.
    There exist sets $S\subseteq [h]$ and $F\subseteq E := [h]\setminus S$ of items, and set $W\subseteq N$ of agents with the following properties.
    \begin{itemize}
        \item $1, h \in S$, and $S$ is a union of at most $t_0=\lceil Q\cdot \frac{q}{q - 1}\rceil + 1$ maximal consecutive intervals:
        \item sets $S$ and $F$ will be of the form
        \[
    S = \bigcup_{s = 1}^t\{a_s, \ldots, b_s\},\quad\quad F = \bigcup_{s = 1}^{t - 1}\{b_s + 1, \ldots, c_s\},\quad \text{where}\quad c_s := \min\{b_s + \nu, a_{s + 1} - 1\};
\]
        \item the sizes of the sets satisfy
        \[
            |S|\leq t_0\left(1 + \frac{Q\nu}{\eps}\right),\qquad |F|\leq t_0\nu,\qquad |W|\leq \eps d;
        \]
        \item set $W$ contains every agent $i \in N$ with a rounded high-demand breakpoint in $F$, i.e
        \[
            W = \{i \in N: \; \exists \alpha \in \Qq\ \text{  s.t  }\ m_i(\alpha) \in F\};
        \]
        \item for every $\ell \in E\setminus F$, $(\ell - \nu, \ell]\subseteq E$, and the total number of high breakpoints in $(\ell - \nu, \ell]$ is
        \[
            \big|\big\{(i, \alpha) \in N\times \Qq :\; \ell - \nu < m_i(\alpha) \leq \ell\big\}\big| \leq \left(1 - \frac{1}{q}\right)d.
        \]
    \end{itemize}
    Here $d = |N|$, and all breakpoints $m_i(\alpha)$ are with respect to set $M$ of all items.
\end{lemma}
\begin{proof}
    For every agent $i \in N$, we list all possible positive high-demand breakpoints of $i$ that lie in $[h]$, i.e all $0 < m_i(\alpha) \leq h$ for $\alpha \in \Qq$, and combine them into a multiset $\Mm(\Qq)$:
\[
    \Mm(\Qq) := \biguplus_{i \in N}\biguplus_{\substack{\alpha \in Q\\0 <m_i(\alpha)\leq h}}\{(i, m_i(\alpha))\}.
\]
    Note that $|\Mm(\Qq)| \leq dQ$.
    We sort $\Mm(\Qq)$ in non-decreasing order w.r.t $m_i(\alpha)$, and mark every $\kappa$-th breakpoint, where
    \[
        \kappa := \left\lfloor \left(1 - \frac{1}{q}\right)d\right\rfloor + 1.
    \]
    Note that any sub-interval of $\Mm(\Qq)$ with no marked breakpoints has length at most $\kappa - 1$.
    Let set $S_0$ contain $\{1, h\}$ and every unique position in $M$ corresponding to a marked breakpoint, i.e
\[
    S_0 = \{1, h\}\cup \big\{s \in \{1, \ldots, h\} : \exists i \in N \text{ s.t }(i, s) = (i_{j\kappa},    m_{j\kappa}) \in \Mm(\Qq) \text{ for some }j\geq 1\big\}.
\]
Since $\kappa > (1 - 1/q)d$, the number of marked breakpoints is less than $qQ/(q -1)$, therefore
\[
    |S_0| \leq \left\lceil Q\cdot \frac{q}{q - 1}\right\rceil + 1 =:t_0.
\]
Any subinterval of $M$ disjoint from $S_0$ contains at most $\kappa - 1$ breakpoints across all agents.
Indeed, (possibly repeating) breakpoints that fall into a subinterval of $M$  form a consecutive interval in $\Mm(\Qq)$, and such interval without any marked breakpoint has length less than $\kappa$.

We construct a sequence of \textbf{protected regions}.
For integer radius $r \geq 1$, protected region $S(r)$ will contain all items of $[h]$ within distance $r$ from $S_0$ (to the right):
\[
        S(r) := [h]\cap \bigcup_{s \in S_0}\{s, s + 1, \ldots, s + r -1\}.
\]
We grow the radius in increments of $\nu$.
For $k = 0, 1, \ldots, K$, where $K := \lfloor Q/\eps\rfloor + 1$, put
\[
    r_k := 1 + k\nu.
\]
By construction, sets $S(r_{k + 1}) \setminus S(r_{k})$ are pairwise disjoint for all $k \geq 0$.
Observe that there must exist some $0 \leq k^* \leq  K - 1$ such that at most $\eps d$ breakpoints of $\Mm(\Qq)$ fall into $S(r_{k^* + 1})\setminus S(r_{k^*})$.
Indeed, otherwise the total number of breakpoints falling into any $S(r_{k + 1})\setminus S(r_k)$ is more than
\[
    K\cdot \eps d = \left(\left\lfloor \frac{Q}{\eps}\right\rfloor + 1\right) \cdot \eps d > dQ.
\]
Take this $k^*$ and define sets
\[
    S:= S(r_{k^*}),\qquad F := S(r_{k^* + 1})\setminus S(r_{k^*}),\qquad E := [h]\setminus S.
\]
In addition, let $W$ be the agents who have a breakpoint in $F$, then $|W|\leq \eps d$.

Clearly, every $s \in S_0$ contributes to $S$ an interval of length at most $1 + (K - 1)\nu$, and growing this interval once more contributes at most $\nu$ items to $F$.
Hence,
\[
    |S|\leq |S_0|(1 + (K - 1)\nu) \leq t_0\left(1 + \frac{Q\nu}{\eps}\right),\qquad\qquad |F|\leq |S_0|\nu \leq t_0\nu.
\]
It follows that $S$ is a union of $t\leq t_0$ maximal consecutive intervals, i.e
\[
    S = \bigcup_{s = 1}^t\{a_s, \ldots, b_s\},\quad \text{where}\quad a_1= 1 \text{ and }b_t = h,
\]
and the buffer set $F$ is precisely
\[
    F = E\cap \bigcup_{s = 1}^{t}\{b_s + 1, \ldots, b_s + \nu\} = \bigcup_{s = 1}^{t - 1}\{b_s + 1, \ldots, c_s\}\quad\text{where}\quad c_s := \min\{b_s + \nu, a_{s + 1} - 1\}.
\]

Finally, for $\ell \in E\setminus F$, let $b_s \leq \ell$ be the endpoint of an interval of $S$ immediately preceding $\ell$, such endpoint exists as $1 \in S$.
Since $\ell \notin F$, we must have $\ell > b_s + \nu$.
Hence,  $(\ell - \nu, \ell] \subseteq (b_s, \ell]\subseteq E$.
This interval does not intersect $S_0\subseteq S$, so it must contain at most $\kappa - 1$ breakpoints of $\Mm(\Qq)$.
The lemma follows.
\end{proof}

We will apply \cref{genbuffer} with some integer $\nu \geq 1$ and $\eps \in (0, 1)$, and obtain a partition $[h] = E\sqcup S$, as well as buffer set $F\subseteq E$ and a small set $W\subseteq N$ of agents whose high-demand breakpoints fall into $F$.
These agents $W$ will require extra attention when defining the fractional head profiles.
Specifically, we want to ``forbid'' the selected profiles of $i \in W$ from using any items of $F$, in order to guarantee enough slackness at every prefix.

\subsubsection{Protected prefixes}

We will now define the set of \textit{protected prefixes}, similarly to \cref{sec:integer}.
Recall that
\[
    S = \bigcup_{s = 1}^t[a_s, b_s],\quad \text{where}\quad a_1= 1,\; b_t = h \quad\text{and}\quad t \leq \left\lceil Q\cdot \frac{q}{q - 1}\right\rceil + 1.
\]
In addition, 
\[
    F = \bigcup_{s = 1}^{t - 1}(b_s, c_s]\quad\text{where}\quad c_s := \min\{b_s + \nu, a_{s + 1} - 1\}.
\]

Consider an interval $[a_s, b_s]\subset S$ for $s < t$.
Throughout this interval, the $E$-capacity $|E\cap [\ell]|$ stays constant, but the $E$-load of old/new profiles may increase.
However, we can control the load at the whole interval $[a_s, b_s]$ by controlling the load at a later \textit{protected prefix}.
Concretely, we can include the following buffer $(b_s, c_s]\subset F$, and preserve the $E$-load on the entire interval $[a_s, c_s]$ using a stronger prefix-load condition at the endpoint of $F$:
\[
    \sum_{i \in A}\Pp_i(\pi_i^E, c_s) + \sum_{u \in N\setminus A}\Pp_u(\pi_u^E, c_s) \leq |E\cap [b_s]|.
\]
Indeed, for every $\ell \in [a_s, c_s]$, the prefix-load at $\ell$ is at most the prefix-load at $c_s$.
Meanwhile
\begin{itemize}
    \item if $\ell \leq b_s$, we have $[\ell, b_s]\subset S$ and $E\cap [\ell] = E\cap [b_s]$,
    \item if $\ell \geq b_s$, then $[b_s]\subseteq [\ell]$, and $|E\cap [\ell]| \geq |E\cap [b_s]|$,
\end{itemize}
so the single constraint above enforces $E$-load feasibility at every prefix $\ell \in [a_s, c_s]$.

When $s = t$, i.e $[a_t, b_t] = [a_t, h]\subset S$, we need a slightly different endpoint to preserve the $E$-load on $[a_t, h]$.
Since this interval ends at $h$, every $e \in E$ satisfies $e \leq a_t$.
Hence, we can put
\[
    \sum_{i \in A}\Pp_i(\pi_i^E, m) + \sum_{u \in N\setminus A}\Pp_u(\pi_u^E, m) \leq |E|,
\]
i.e with endpoint $c_t = m$, to ensure that the total number of breakpoints of $E$-parts of the profiles does not exceed $|E|$.
This constraint controls the load at every $\ell \in [a_t, b_t]$, as well as every prefix $\ell > h$.
Note that using $m$ rather than $h$ is necessary, as breakpoints of a demand in a head profile may exceed $h$.

\begin{definition}
    For the given sets $S, F \subseteq [h]$ of the form above, obtained from \cref{genbuffer} applied to some integer $\nu \geq 1$ and $\eps \in (0, 1)$, define the set $\Ll_S$ of $t$ \textbf{protected prefixes} as
    \[
        \Ll_S := \{(b_1, c_1), \ldots, (b_{t - 1}, c_{t - 1}), (b_t, c_t)\},\qquad \text{where}\qquad (b_t, c_t) = (h, m).
    \]
    Note that $|\Ll_S| \leq \lceil Q\cdot \frac{q}{q - 1}\rceil + 1$.
\end{definition}
For agents $A\subseteq I$, retaining their old profiles, and agents $N\setminus A$ receiving new profiles, the \textit{protected conditions} for every protected prefix $(b_s, c_s) \in \Ll_S$ are
\begin{equation}
    \sum_{i \in A}\Pp_i(\pi_i^E, c_s) + \sum_{u \in N\setminus A}\Pp_u(\pi_u^E, c_s) \leq |E\cap [b_s]|.\tag{$\mathrm{E}_{\Ll_S}$}\label{genES'}
\end{equation}

The following corollary is an extension of \cref{protpref} from \cref{sec:integer}.
\begin{corollary}\label{genprotpref}
    Consider a partition $[h] = E\sqcup S$ obtained from \cref{genbuffer} applied to some integer $\nu \geq 1$ and $\eps \in (0, 1)$.
    For every agent $i \in N$, let $\{\pi_{i,j}^E : j \in [n]\}$ be a collection of high sub-profiles with demands in $\Qq$ (allowing empty profiles), such that for every $0 \leq \ell \leq m$,
    \[
        \sum_{j = 1}^n\Pp_i(\pi_{i,j}^E, \ell) \leq |E\cap [\ell]|.
    \]
    Suppose that agents $I\subseteq N$ are assigned profiles $\{\pi_i\}_{i \in I}$ satisfying for every $0 \leq \ell \leq m$:
    \[
        \sum_{i \in I}\Pp_i(\pi_i^E, \ell) \leq \frac{1}{n}\sum_{i \in N}\sum_{j = 1}^n\Pp_i(\pi_{i,j}^E, \ell).
    \]
    Let $A\subseteq I$, and to every agent $u \in N\setminus A$ give a new profile $\pi_u$.
    If $\{\pi_i\}_{i \in A}$ and $\{\pi_u\}_{u \in N\setminus A}$ satisfy
    \begin{align}
    &&&\sum_{i \in A}\Pp_i(\pi_i^E, c_s) + \sum_{u \in N\setminus A}\Pp_u(\pi_u^E, c_s) \leq |E\cap [b_s]|,&\forall (b_s, c_s) \in \Ll_S;\tag{$\mathrm{E}_{\Ll_S}$}\label{genES}\\
    &&&\sum_{u \in N \setminus A}|\pi_u|\leq \frac{\nu}{q} \tag{A-$\nu$}\label{genAnu};
\end{align}
then for every prefix $0 \leq \ell \leq m$ it holds
\begin{equation}
    \sum_{i \in A}\Pp_i(\pi_i^E, \ell) +  \sum_{u \in N \setminus A}\Pp_u(\pi_u^E, \ell) \leq |E\cap [\ell]|.\tag{E}\label{genEprot}
\end{equation}
In particular, if $|\pi_u|\leq q$ for every $u \in N\setminus A$, it suffices that (\ref{genES}) holds with $|N\setminus A|\leq \nu/q^2$.
\end{corollary}
\begin{proof}
    First, take $\ell \in E\setminus F$, and consider the interval $(\ell - \nu, \ell]$.
    By \cref{genbuffer}, $(\ell - \nu, \ell]\subseteq E$, and at least $d/q = |N|/q$ agents have no high-demand breakpoint in $(\ell - \nu, \ell]$.
    Denote these agents by $N_\ell$, by assumption for every $i \in N_\ell$ all demands of profiles $\pi_{i,j}$ belong to $\Qq$, so
    \[
        \sum_{j = 1}^n\Pp_i(\pi_{i,j}^E, \ell) = \sum_{j = 1}^n\Pp_i(\pi_{i,j}^E, \ell - \nu) \leq |E\cap [\ell - \nu]| = |E\cap[\ell]| - \nu.
    \]
    For every other $i\in N\setminus N_\ell$, the total $\ell$-prefix load of $E$-subprofiles $\pi_{i,j}^E$ is at most $|E\cap [\ell]|$.
    Hence,
    \begin{multline*}
        \frac{1}{n}\sum_{i \in N}\sum_{j = 1}^n\Pp_i(\pi_{i,j}^E, \ell) = \frac{1}{n}\sum_{i \in N_\ell}\sum_{j = 1}^n\Pp_i(\pi_{i,j}^E, \ell) + \frac{1}{n}\sum_{i \in N\setminus N_\ell}\sum_{j = 1}^n\Pp_i(\pi_{i,j}^E, \ell)\\
        \leq \frac{|N_\ell|}{n}\cdot (|E\cap[\ell]| - \nu) + \frac{|N\setminus N_\ell|}{n}\cdot |E\cap [\ell]| = \frac{|N|}{n}\cdot |E\cap[\ell]| - \frac{|N_\ell|}{n}\cdot \nu \\
        \leq \frac{d}{n}\left(|E\cap [\ell]| - \frac{\nu}{q}\right) \leq |E\cap [\ell]| - \frac{\nu}{q}.
    \end{multline*}
    The last inequality uses $d\leq n$ and $|E\cap [\ell]| \geq \nu$.
    It follows from assumptions that, for any $A\subseteq I$, under condition (\ref{genAnu}) it holds for this prefix $\ell$:
    \begin{multline*}
        \sum_{i \in A}\Pp_i(\pi_i^E, \ell) +  \sum_{u \in N \setminus A}\Pp_u(\pi_u^E, \ell) \leq \sum_{i \in I}\Pp_i(\pi_i^E, \ell) + \sum_{u \in N \setminus A}|\pi_u|\\
        \leq \frac{1}{n}\sum_{i \in N}\sum_{j = 1}^n\Pp_i(\pi_{i,j}^E, \ell) + \frac{\nu}{q} \leq |E\cap [\ell]| - \frac{\nu}{q} + \frac{\nu}{q} = |E\cap [\ell]|.
    \end{multline*}
    It remains to consider prefixes $\ell \in S\cup F$.
    Denote $\Pp^E(\ell) := \sum_{i \in A}\Pp_i(\pi_i^E, \ell) +  \sum_{u \in N \setminus A}\Pp_u(\pi_u^E, \ell)$ for short.
    Take the interval representation of $S$ and $F$.
    If $\ell \in [a_s, c_s]$ for  $s < t$, (\ref{genES}) ensures
    \[
        \Pp^E(\ell) \leq \Pp^E(c_s) \leq |E\cap [b_s]|\leq |E\cap [\ell]|,
    \]
    as $[\ell, b_s]\subset S$ for $\ell \leq b_s$, and $[b_s]\subseteq [\ell]$ for $\ell \geq b_s$.
    If $a_t \leq \ell \leq m$, the last protected prefix gives
    \[
        \Pp^E(\ell) \leq  \Pp^E(m) \leq |E| = |E\cap [\ell]|.
    \]
    Finally, since $1 \in S$ and $[b_1] \subseteq S$, we have $E\cap [b_1] = \varnothing$ and the first protected prefix gives $\Pp^E(c_1) \leq |E\cap [b_1]| = 0$.
    So, for $\ell = 0$ we have $\Pp^E(0)\leq \Pp^E(c_1) = 0$.
    The corollary follows.
\end{proof}

For every agent $i \in N$, we  will obtain a suitable collection $\{\pi_{i,j} : j \in [n]\}$ of high profiles from their MMS partitions.
By \cref{genprotpref}, to guarantee (\ref{genEprot}), we need to find
\begin{itemize}
    \item a set of agents $I\subseteq N$ with profiles $\{\pi_i\}_{i \in I}$ such that their total $E$-load is at most the average total $E$-load of profiles $\{\pi_{i,j} : j \in [n]\}$ for $i \in N$.
    We will find such set $I$ by rounding a specific collection of \textit{fractional head profiles} using \cref{rounding}, as in \cref{sec:integer}.
    \item a set $A\subseteq I$ of agents who keep old profiles, and new profiles $\pi_u$ for every $u \in N\setminus A$, satisfying (\ref{genES}) and (\ref{genAnu}).
    This will be done later.
\end{itemize}

\subsection{Fractional head profiles}

In this subsection, for every agent $i \in N$ we provide a collection $\{\pi_{i,j} : j \in [n]\}$ of \textit{rounded head profiles} obtained from their MMS partitions $\{B_{i,1}, \ldots, B_{i,n}\}$, similarly to \cref{sec:integer}.
We will take $B_{i,j}$ for $j \in [n]$, consider only its head items $B_{i,j}\cap [h]$, and induce head demands through rounded valuation $\wv_i$, creating $\pi_{i,j}$.
However, we will be interested only in head profiles $\pi_{i,j}$ that 1) are acceptable, i.e have value at least $1 - 2/q$, and 2) contain only rounded high demands, i.e for every $\alpha \in \pi_{i,j}$ it must hold $\alpha \geq 1/q$.
This limits the amount of available head profiles for every agent.
Despite the fact that all $i \in N$ value head $[h]$ highly, different MMS bundles $B_{i,j}$ have different overlaps with $[h]$, and only for good agents $i \in G$ we have a guarantee that every head item is worth at least $1/q$.
This makes our construction slightly less straightforward.

\begin{definition}
    For every agent $i \in N$, consider the partition $\{B_{i,1},\ldots, B_{i,n}\}$ of $M$.
    For every bundle $j \in [n]$, define the \textbf{rounded head profile} $\pi_{i,j}$ as follows.
    \begin{itemize}
        \item if $v_i(B_{i,j}\cap [h]) \geq 1 - 1/q$ and $v_i(e) \geq 1/q$ for all $B_{i,j}\cap [h]$, then
        \[
            \pi_{i,j} = \{\wv_i(e) : \; e \in B_{i,j} \cap [h]\}.
        \]
        \item otherwise, $\pi_{i,j} = \varnothing$, we will call such profile \textbf{invalid}.
    \end{itemize}
    Let $\theta_i := |\{j \in [n]:\; \pi_{i,j} = \varnothing\}|$ be the number of invalid head profiles of agent $i$.
\end{definition}
Setting a profile $\pi_{i,j} = \varnothing$ makes $\pi_{i,j}$ not acceptable for agent $i$, so it will never be assigned to $i$.
However, we may still select $i$ into set $I$ in the future LP rounding, and agent $i$ still may use items from $B_{i,j}$ in his future potential allocation. 

\begin{observation}\label{roundprofile}
    For every $i \in N$, if $\pi_{i,j}\neq \varnothing$ then $|\pi_{i,j}|\leq q$, every $\alpha \in \pi_{i,j}$ belongs to $\Qq$, and
    \[
        1 \geq v(\pi_{i,j}) \geq 1 - \frac{2}{q} .
    \]
\end{observation}
\begin{proof}
    By construction, every valid profile $\pi_{i,j}$ satisfies
    \[
        v(\pi_{i,j}) = \sum_{e \in B_{i,j}\cap [h]}g(v_i(e)) \geq \left(1 - \frac{1}{q}\right)v_i(B_{i,j}\cap [h])\geq \left(1 - \frac{1}{q}\right)^2 = 1 - \frac{2}{q} + \frac{1}{q^2},
    \]
    as well as $v(\pi_{i,j}) \leq v_i(B_{i,j}\cap [h]) \leq v_i(B_{i,j})\leq 1$.
    Since every $e \in B_{i,j}\cap [h]$ is worth at least $1/q$ for $i$, we have $|\pi_{i,j}|\leq q$ as well as $\alpha \in \Qq$ for all $\alpha \in \pi_{i,j}$.
\end{proof}

We will now bound the number of invalid head profiles $\theta_i$ for every agent $i \in N$, and show that on average, the number of invalid profiles is relatively small.
This will require two types of estimates, corresponding to the two conditions on rounded head profile $\pi_{i,j}$ to be valid.
First, the head-portion $B_{i,j}\cap [h]$ of bundle $B_{i,j}$ must have high value.
We will use the bounds on the total head value from \cref{notpoor} to show that failing this condition cannot happen too often.
Second, every head-item of $B_{i,j}$ must have value at least $1/q$.
We will use common prices $w$ to show that, if some item $e \in B_{i,j}$ has $v_i(e) < 1/q$, then removing it from $B_{i,j}$ still leaves an acceptable bundle, therefore it must have high common price.
And, in turn, the total number of bundles with high common price cannot be too large.

\begin{lemma}\label{invalid}
    For every agent $i \in N$ the number $\theta_i$ of invalid rounded head profiles satisfies
    \[
        \theta_i \leq q^2(16\Lambda + 1) + qn(1 - \lambda_i).
    \]
    In addition, for every $i \in G$ it holds $\theta_i \leq q^2(16\Lambda + 1)$.
    Consequently, taking $\theta := q^2(16\Lambda + 1) + q\Lambda$, 
    \[
        \frac{1}{n}\sum_{i \in N}\theta_i \leq \theta.
    \]
\end{lemma}
\begin{proof}
    First consider agent $i \in N$ with $\lambda_i \geq 1/2$.
    Since bundles $\{B_{i,1}, \ldots, B_{i,n}\}$ partition $M$,
    \[
        \sum_{j = 1}^n(1 - v_i(B_{i,j}\cap [h])) = n - v_i([h]) \leq q(16\Lambda + 1),
    \]
    where we used \cref{notpoor}.
    Since $v_i(B_{i,j})\leq 1$ for every $j \in [n]$, all summands are non-negative.
    Each bundle $B_{i,j}$ such that $v_i(B_{i,j}\cap [h]) < 1 - 1/q$ contributes at least $1/q$ to the sum above, so the number of $j \in [n]$ with $v_i(B_{i,j}\cap [h]) < 1 - 1/q$ can be at most $q^2(16\Lambda + 1)$.

    Next, for the same agent $i$, suppose that $B_{i,j}$, $j \in [n]$, contains item $e\leq h$ with $v_i(e) < 1/q$.
    By \cref{notpoor}, 
    \[
        v_i(B_{i,j} - \{e\}) > 1 - \frac{1}{2q^2} - \frac{1}{q} \geq 1 - \frac{2}{q},
    \]
    so $B_{i,j} - \{e\}$ is acceptable for $i$.
    Therefore, \cref{notpoor} guarantees us
    \[
        w(B_{i,j} - \{e\}) \geq \lambda_i \quad \implies \quad w(B_{i,j}) - \lambda_i \geq w(e) \geq \frac{1}{q}.
    \]
    Since every $B_{i,j}$, $j \in [n]$, is acceptable for $i$, using $w(M)\leq n$ we get
    \[
        0 \leq \sum_{j = 1}^n(w(B_{i,j}) - \lambda_i) = w(M) - n\lambda_i \leq n(1 - \lambda_i).
    \]
    Any bundle $B_{i,j}$ containing item $e$ with $v_i(e) < 1/q$ contributes at least $1/q$ to the sum above.
    Therefore, the total number of such bundles is at most $qn(1-\lambda_i)$.
    Combining the two conditions,
    \[
        \theta_i  \leq q^2(16\Lambda + 1) + qn(1 - \lambda_i).
    \]
    Note that for $i \in G$ we always have $v_i(e) \geq 1/q$ for all $e\leq h$, so there is no summand $ qn(1 - \lambda_i)$.

    If $\lambda_i < 1/2$, we simply use $\theta_i \leq n\leq qn(1-\lambda_i)$, so the bound above holds for all agents $i \in N$.
    Finally, the total number of agents in $N$ is $d \leq n$, so by $\sum_{i \in N}(1 - \lambda_i)\leq \Lambda$ from \cref{notpoor},
    \[
        \frac{1}{n}\sum_{i \in N}\theta_i \leq \frac{d}{n}\cdot q^2(16\Lambda + 1) + q\sum_{i \in N}(1 -\lambda_i) \leq q^2(16\Lambda + 1) + q\Lambda.
    \]
\end{proof}

We will apply \cref{genbuffer} with some integer $\nu \geq 1$ and $\eps \in (0, 1)$, and obtain head partition $[h] = E\sqcup S$ and sets $F\subseteq E$ and $W\subseteq N$.
Each defined $\pi_{i,j}$ now has a natural $E/S$-decomposition.
\begin{definition}
    Consider the partition $[h] = E \sqcup S$ from \cref{genbuffer}.
    For an agent $i \in N$ with partition $\{B_{i,1}, \ldots, B_{i,n}\}$ of $M$, for every valid rounded head profile $\pi_{i,j} \neq \varnothing$ define sub-profiles
    \[
        \pi_{i,j}^E := \{\wv_i(e) :\; e \in B_{i,j} \cap E\}, \qquad\qquad
        \pi_{i,j}^S := \{\wv_i(e) :\; e \in B_{i,j}\cap S\}.
    \]
    For every invalid $\pi_{i,j} = \varnothing$, set $(\pi_{i,j}^E, \pi_{i,j}^S) = (\varnothing, \varnothing)$.
    Clearly, for all $\pi_{i,j}$ it holds $\pi_{i,j} = \pi_{i,j}^E \uplus \pi_{i,j}^S$.
\end{definition}
\begin{remark}
    The partition $[h] = E\sqcup S$ imposes additional restriction on the selected profiles.
    From now on, for a profile $\pi$ to be equal to $\pi_{i,j}$, it must also coincide on both $E$- and $S$-parts:
    \[
        \pi = \pi_{i,j} \quad \iff \quad (\pi^E, \pi^S) = (\pi_{i,j}^E, \pi_{i,j}^S).\]
    So, if $\pi$ and $\pi_{i,j}$ are the same as multisets, but have a different $E/S$-decomposition, we treat them as two different profiles.
    With this stronger equality notion, we define the common collection
    \[\Pi := \{\varnothing, \bot\} \cup \bigcup_{i \in N}\{\pi_{i,j} : j \in [n]\}.\]
\end{remark}
\begin{observation}\label{genmmspart}
    For every $i \in N$ and every prefix $0 \leq \ell \leq m$ it holds
    \[
        \sum_{j = 1}^n\Pp_i(\pi_{i,j}^E, \ell) \leq |E\cap [\ell]|,\qquad\qquad \sum_{j = 1}^n\Pp_i(\pi_{i,j}^S, \ell) \leq |S\cap [\ell]|.
    \]
\end{observation}
\begin{proof}
    Similar to \cref{etafracload}, every demand in $\pi_{i,j}$ comes from a unique item of $B_{i,j}$, and MMS partition $\{B_{i,1},\ldots, B_{i,n}\}$ of $M$ cannot use more than $\ell$ items in every $\ell$-prefix.
    Hence, there are at most $|E\cap [\ell]|$ $E$-demands and at most $|S\cap[\ell]|$ $S$-demands across all profiles $\{\pi_{i,j} : j \in [n]\}$ in every $\ell$-prefix.
    The observation follows by the fact that $\wv_i$ and $v_i$ have the same breakpoints.
\end{proof}
Now, for every agent $i \in N$ we will define a fractional head profile $(\wx_i, \Pi)$ by giving some weight $\wx_i(\pi)\geq 0$ to every profile $\pi \in \Pi$.
Similar to \cref{sec:integer}, each MMS-profile $\pi_{i,j}$ will be picked uniformly, i.e with probability $1/n$.
However, our construction will need several modifications.
\begin{itemize}
    \item If profile $\pi_{i,j}$ of agent $i \in N$ is \textit{invalid}, i.e $\pi_{i,j} = \varnothing$, then its uniform weight of $1/n$ will go to $\wx_i(\bot)$.
    This will force the rounding algorithm to select only profiles of $i$ that are 1) acceptable, and 2) contain only high demands, as needed in \cref{genprotpref}.
    As shown in \cref{invalid}, on average the number of invalid profiles removed from selection is small.
    
    \item In order to obtain stronger $E$-load bounds at protected prefixes (\ref{genES}), we will omit some additional profiles from the fractional selection.
    Concretely, if agent $i \in W$ and his MMS bundle $B_{i,j}$ contains items from the buffer $F$, we will not use the corresponding profile $\pi_{i,j}$, and also move its uniform weight $1/n$ to $\wx_i(\bot)$.
    This will allow us to satisfy the protected prefix conditions (\ref{genES}) fractionally for $(\wx, \Pi)$.
    By \cref{genbuffer}, both sets $W$ and $F$ are small, so omitting the corresponding profiles still leaves a lot of total weight $\Ii(\wx, \Pi)$.
\end{itemize}

\begin{definition}
    For every agent $i \in N$, define the following fractional head profile $(\wx_i, \Pi)$.
    \begin{itemize}
        \item If $i \in N \setminus W$, 
        set $\wx_i(\varnothing) = 0$, $\wx_i(\bot) = \theta_i/n$, and for every $\pi \in \Pi\setminus\{\varnothing, \bot\}$ let
        \[
            \wx_i(\pi) = \frac{|\{j \in [n]:\; \pi_{i,j} = \pi\}|}{n}.
        \]
        \item if $i \in W$, set $\wx_i(\varnothing) = 0$, for every $\pi \in \Pi\setminus \{\varnothing, \bot\}$, let
        \[
            \wx_i(\pi) = \frac{|\{j \in [n]:\; \pi_{i,j} = \pi \text{ and }B_{i,j}\cap F = \varnothing\}|}{n},
        \]
        as well as
        \[
            \wx_i(\bot) = \frac{\theta_i}{n} + \frac{|\{j \in [n] :\; \pi_{i,j}\neq \varnothing \text{ and } B_{i,j} \cap F\neq \varnothing\}|}{n}.
        \]
    \end{itemize}
    We will refer to $(\wx_i, \Pi)$ as a \textbf{fractional rounded head profile} of agent $i$.
\end{definition}
Thus, in the fractional collection $(\wx, \Pi)$ every invalid profile $\pi_{i,j}$, as well as every valid $\pi_{i,j}$ that corresponds to a bundle $B_{i,j}$ of some agent $i \in W$ that contains items from the buffer $F$, contributes its weight $1/n$ to $\bot$ instead.

\begin{definition}
    Consider the partition $[h] = E \sqcup S$ from \cref{genbuffer}.
    For the fractional profiles $(\wx, \Pi)$ and a prefix $0 \leq \ell \leq m$, define the total fractional $E$-load and $S$-load as respectively
    \begin{gather*}
        \Hh^E(\wx, \Pi, \ell) := \sum_{i \in N}\sum_{\pi \neq \bot}\wx_i(\pi)\cdot \Pp_i(\pi^E, \ell);\\
        \Hh^S(\wx, \Pi, \ell) := \sum_{i \in N}\sum_{\pi \neq \bot}\wx_i(\pi)\cdot \Pp_i(\pi^S, \ell).
    \end{gather*}
\end{definition}

We will next show that the fractional profiles $(\wx, \Pi)$ satisfy the requirements of \cref{genprotpref}, as well as condition (\ref{genS}), fractionally.
\begin{lemma}\label{genfracload}
    The number $\wX$ of strictly fractional coordinates in $\wx$ satisfies $\wX \leq dn$. 
    Moreover,
    \[
        \Ii(\wx, \Pi) \geq  d - \theta - \eps |F|,
    \]
    where $\theta = q^2(16\Lambda + 1) + q\Lambda$ from \cref{invalid}.
    In addition, for every prefix $0 \leq \ell \leq m$ it holds
    \begin{gather*}
        \Hh^E(\wx, \Pi, \ell) \leq \frac{1}{n}\sum_{i \in N}\sum_{j = 1}^n\Pp_i(\pi_{i,j}^E, \ell) \leq \frac{d}{n}|E\cap [\ell]|;\\
        \Hh^S(\wx, \Pi, \ell) \leq \frac{1}{n}\sum_{i \in N}\sum_{j = 1}^n\Pp_i(\pi_{i,j}^S, \ell) \leq \frac{d}{n}|S\cap [\ell]|;
    \end{gather*}
    and for every protected prefix $(b_s, c_s)  \in \Ll_S$ it holds
    \[
        \Hh^E(\wx, \Pi, c_s) \leq \frac{d}{n}|E\cap [b_s]|.
    \]
\end{lemma}
\begin{proof}
    For every agent $i \in N$, there are at most $n$ positive weights $\wx_i(\pi)$ for $\pi = \pi_{i,j}$.
    If $\wx_i(\bot) > 0$, at least one profile $\pi_{i,j}$ contributes its individual weight to $\bot$ (note that some $\wx_i(\pi_{i,j})$ may still be positive, as multiple $B_{i,j}$ can induce the same profile).
    There are at most $n - 1$ other $\pi_{i,j}$ contributing to non-$\bot$ coordinates, with a total of at most $n$ positive $\wx_i(\pi)$.
    Thus, $\wX \leq  dn$.
    
    Next, by construction, we have
    \begin{multline*}
         \Ii(\wx, \Pi) = \sum_{i \in N}\sum_{\pi\neq \bot}\wx_i(\pi) = |N| - \sum_{i \in N}\wx_i(\bot) \\
         = d - \frac{1}{n}\sum_{i \in N}\theta_i  - \frac{1}{n}\sum_{i \in W}|\{j \in [n] : \pi_{i,j}\neq \varnothing \text{ and }B_{i,j} \cap F \neq \varnothing\}|.
    \end{multline*}
    For every agent $i \in W$, there are at most $|F|$ MMS bundles that contain any item from $F$.
    Hence, using the bound on $|W|$ from \cref{genbuffer},
    \[
        \frac{1}{n}\sum_{i \in W}|\{j \in [n] : \pi_{i,j}\neq \varnothing \text{ and }B_{i,j} \cap F \neq \varnothing\}| \leq \frac{|W|}{n}\cdot |F| \leq \frac{\eps d}{n}\cdot |F| \leq \eps |F|. 
    \]
    Combining this bound with $\frac{1}{n}\sum_{i \in N}\theta_i \leq \theta$ from \cref{invalid}, we get
    \[
        \Ii(\wx, \Pi) \geq d - \theta - \eps|F|.
    \]

    Moving to the prefix load, each $\pi_{i,j}$ contributes at most $1/n$ to fractional load, so
    \[
         \Hh^E(\wx, \Pi, \ell) = \sum_{i \in N}\sum_{\pi \neq \bot}\wx_i(\pi)\cdot \Pp_i(\pi^E, \ell)\leq \frac{1}{n}\sum_{i \in N}\sum_{j = 1}^n\Pp_i(\pi_{i,j}^E, \ell) \leq \frac{d}{n}|E\cap [\ell]|,
    \]
    where the last inequality follows from \cref{genmmspart}.
    The bounds for $S$ are analogous.

    Finally, fix a protected prefix $(b_s, c_s)  \in \Ll_s$ for $s < t$.
    For agent $i \in N$, consider his MMS profile $\pi_{i,j} \neq \varnothing$, and let $\alpha \in \pi_{i,j}^E$ be its demand counted at prefix $c_s$, i.e $m_i(\alpha) \leq c_s$.
    If $e \in B_{i,j}\cap E$ is the originating item of $\alpha$, then $\alpha = \wv_i(e) \leq v_i(e)$ and $e \leq m_i(\alpha) \leq c_s$.
    Hence,
    \begin{itemize}
        \item if $i \notin W$, breakpoint $m_i(\alpha)$ cannot belong to $(b_s, c_s]\subset F$, so $m_i(\alpha)\leq b_s$ and thus $e \leq b_s$;
        \item if $i \in W$ and $B_{i,j}\cap F = \varnothing$, then $e\notin F$, and as $e \leq c_s$ and $(b_s, c_s]\subset F$, necessarily  $e \leq b_s$.
    \end{itemize}
    Thus, whenever whenever $i \notin W$ or $i \in W$ and $B_{i,j}\cap F = \varnothing$, we must have $e\leq b_s$.
    Therefore, by construction of $(\wx_i, \Pi)$, 
    every $E$-demand counted at prefix $c_s$, for a profile $\pi_{i,j}$ that does not contribute to $\bot$, must originate from an item in $E\cap [b_s]$.
    And, all these items must be distinct, since $\pi_{i,j}$ are induced by different MMS bundles $B_{i,j}$, $j \in [n]$.
    It follows that
    \[
        \sum_{\pi \neq \bot}\wx_i(\pi)\cdot \Pp_i(\pi^E, c_s) = \frac{1}{n}\sum_{j= 1}^n\Pp_i(\pi_{i,j}^E, c_s)\One\{i \notin W \text{ or }B_{i,j}\cap F = \varnothing\} \leq \frac{1}{n}\cdot |E\cap [b_s]|,
    \]
    and, summing over all $i \in N$,
    \[
        \Hh^E(\wx, \Pi, c_s) \leq \frac{d}{n}|E\cap [b_s]|.
    \]
    For the last protected prefix $(b_t, c_t) = (h, m)$, the earlier bound already gives
    \[
        \Hh^E(\wx, \Pi, c_t) = \Hh^E(\wx, \Pi, m) \leq \frac{d}{n}|E| = \frac{d}{n}|E\cap[h]| = \frac{d}{n}|E\cap[b_t]|.
    \]
    This finishes the proof.
\end{proof}

As promised earlier, we will now round the fractional profiles $(\wx, \Pi)$ to integral $(y, \Pi)$, that will define a large subset $I\subseteq N$ of selected agents with assigned profiles $\{\pi_i\}_{i \in I}$, satisfying the requirements of \cref{genprotpref}.
We will do it using \cref{rounding}, which we will restate here.

\begin{theorem}[Joint rounding]\label{genrounding}
    For a partition $[h] = E\sqcup S$, let $\Pi$ be a shared set of head profiles with well-defined $E/S$-decomposition, such that for every $\pi \in \Pi\setminus \bot$ it holds $|\pi|\leq q$ for some integer $q \geq 1$.
    For every $i \in N$, let $(x_i, \Pi)$ be a fractional profile, denote $x = \{x_i\}_{i \in N}$. 

    There exists a collection $y = \{y_i\}_{i \in N}$ of integer head profiles $(y_i, \Pi)$, such that for every $i, \pi$ we have $y_i(\pi) \in \{0, 1\}$ and $y_i(\pi) = 1$ for $\pi \neq \bot$ only if $x_i(\pi) > 0$, and
    \[\begin{aligned}
        & && \Hh^E(y, \Pi, \ell) \leq \Hh^E(x, \Pi, \ell),  & \forall 0 \leq \ell \leq m,\\
        & && \Hh^S(y, \Pi, \ell) \leq \Hh^S(x, \Pi, \ell),  & \forall 0 \leq \ell \leq m,\\
        &  && \Ii(y, \Pi)\geq \Ii(x, \Pi) - \rho_{ES}(q, X), 
    \end{aligned}\]
    where, for $X= |\{(i, \pi):0 < x_i(\pi) <1 \}|$,
    \[
        \rho_{ES}(q, X) =  8q\log \max\{1, X\}.
    \]
\end{theorem}
Its proof is analogous to \cref{etarounding} and is presented in \cref{appendix:rounding} (\cref{roundingproof}).
We note that for the purposes of this work, guarantees of \cref{etarounding} of $O_{q}(\log^2 n)$ rounding loss are sufficient, and the $O(q\log n)$-improvement of \cref{genrounding} will not be consequential.
We still present the alternative rounding of \cref{genrounding}, for potential use in future works.

\begin{corollary}\label{genESrounding}
    Let $[h] =E\sqcup S$, $F\subseteq E$ and $W\subseteq N$ be obtained from \cref{genbuffer} applied with some integer $\nu \geq 1$ and $\eps \in (0, 1)$, and let $(\wx, \Pi)$ be the fractional profiles defined earlier.

    There exist a set $I\subseteq N$ and profiles $\{\pi_{i}\}_{i \in I}$, where every $\pi_i$ has an $E/S$ decomposition $\pi_i = \pi_i^E \uplus \pi_i^S$, and satisfies $|\pi_i|\leq q$, $\alpha \in \Qq$ for all $\alpha \in \pi_i$, and $1 \geq v(\pi_i) \geq 1 - 2/q$.
    In addition,
    \[
        |I|\geq d - \theta - \eps|F| - \rho_{ES}(q, dn),\qquad\qquad \theta = q^2(16\Lambda + 1) + q\Lambda,
    \]
    and for every prefix $0 \leq \ell \leq m$ it holds
    \[
        \sum_{i \in I}\Pp_i(\pi_i^E, \ell) \leq \frac{1}{n}\sum_{i \in N}\sum_{j = 1}^n\Pp_i(\pi_{i,j}^E, \ell),\qquad \qquad \sum_{i \in I}\Pp_i(\pi_i^S, \ell) \leq |S\cap [\ell]|,
    \]
    as well as for every protected prefix $(b_s, c_s) \in \Ll_S$,
    \[
        \sum_{i \in I}\Pp_i(\pi_i^E, c_s) \leq |E\cap [b_s]|.
    \]
\end{corollary}
\begin{proof}
    By construction, profile $\pi \in \Pi\setminus \{\varnothing, \bot\}$ has a well-defined $E/S$ decomposition, and by \cref{roundprofile}  $|\pi|\leq q$.
    We apply \cref{genrounding} to $(\wx, \Pi)$ and obtain integral profiles $(y, \Pi)$.
    Let $I\subseteq N$ contain all $i \in N$ with $y_i(\bot)=  0$, and take $\pi_i \in \Pi\setminus \bot$ satisfying $y_i(\pi_i) = 1$ for this $i$.
    Since $y_i(\pi_i) > 0$ only if $\wx_i(\pi_i) > 0$, \cref{roundprofile} guarantees $1 \geq v(\pi_i) \geq 1 - 2/q$ for  $i \in I$.

    Next, using the bounds of \cref{genfracload} on $\wX\leq dn$ and $\Ii(\wx, \Pi)$, \cref{genrounding} gives us
    \[
        |I| = \Ii(y, \Pi) \geq \Ii(\wx, \Pi) - \rho_{ES}(q, \wX) \geq d - \theta - \eps|F| - \rho_{ES}(q, dn).
    \]

    Next, for  every prefix $0 \leq \ell \leq m$, we have by \cref{genrounding}:
    \[
        \sum_{i \in I}\Pp_i(\pi_i^E, \ell) = \Hh^E(y, \Pi, \ell) \leq \Hh^E(\wx, \Pi, \ell); \qquad
        \sum_{i \in I}\Pp_i(\pi_i^S, \ell) = \Hh^S(y, \Pi, \ell) \leq \Hh^S(\wx, \Pi, \ell).
    \]
    Combining these inequalities with \cref{genfracload}, and using $d\leq n$, proves
    \[
        \sum_{i \in I}\Pp_i(\pi_i^E, \ell) \leq \frac{1}{n}\sum_{i \in N}\sum_{j = 1}^n\Pp_i(\pi_{i,j}^E, \ell),\qquad \qquad \sum_{i \in I}\Pp_i(\pi_i^S, \ell) \leq |S\cap [\ell]|,
    \]
    and for every protected prefix $(b_s, c_s)  \in \Ll_S$,
    \[
        \sum_{i \in I}\Pp_i(\pi_i^E, c_s) \leq  \Hh^E(\wx, \Pi, c_s) \leq |E\cap [b_s]|.
    \]
\end{proof}
\section{Single supplier for the general case}

\label{sec:supplier}

Recall our original goals, described in \cref{sec:genES}.
Given a large subset of agents $I\subseteq N$ with assigned integral profiles $\Pi_I = \{\pi_i\}_{i \in I}$ defined on common head $\{1, \ldots, h\}$, we want to find a subset $A\subseteq I$ of agents that will keep their assigned profiles, and give a new profile $\pi_u$ to every agent $u \in N \setminus A$, so that for every $0 \leq \ell \leq m$ it holds
\[
    \sum_{i \in A}\Pp_i(\pi_i, \ell) + \sum_{u \in N\setminus A}\Pp_u(\pi_u, \ell) \leq \ell.
\]
As discussed earlier, to get this condition for every prefix it suffices that, for an arbitrary head partition $[h] = E\sqcup S$, for every $0 \leq \ell \leq m$ we have
\begin{align}
    &&&\sum_{i \in A}\Pp_i(\pi_i^E, \ell) + \sum_{u \in N\setminus A}\Pp_u(\pi_u^E, \ell) \leq |E\cap [\ell]|,&\forall 0 \leq \ell \leq m. \tag{E}\label{genE2}\\
    &&&\sum_{i \in A}\Pp_i(\pi_i^S, \ell) + \sum_{u \in N\setminus A}\Pp_u(\pi_u^S, \ell) \leq |S\cap [\ell]|,&\forall 0 \leq \ell \leq m. \tag{S}\label{genS2bad}
\end{align}

Suppose that we applied \cref{genbuffer} with some integer $\nu \geq 1$ and $\eps \in (0, 1)$, and obtained a head partition $[h] = E\sqcup S$, as well as protected prefixes $\Ll_S$.
Assume that the originally selected integral profiles $\Pi_I = \{\pi_i\}_{i \in I}$ are the ones from \cref{genESrounding}.
As shown in \cref{sec:genES}, \cref{genbuffer} and \cref{genprotpref}, to satisfy the conditions (\ref{genE2}) and (\ref{genS2bad}) above for every prefix $0 \leq \ell \leq m$, it suffices to find $A\subseteq I$ and profiles $\pi_u$ for $u \in N\setminus A$ such that
\begin{align}
    &&&\sum_{i \in A}\Pp_i(\pi_i^E, c_s) + \sum_{u \in N\setminus A}\Pp_u(\pi_u^E, c_s) \leq |E\cap [b_s]|,&\forall (b_s, c_s) \in \Ll_S;\tag{$\mathrm{E}_{\Ll_S}$}\label{genES2}\\
    &&&\sum_{u \in N \setminus A}|\pi_u|\leq \frac{\nu}{q} \tag{A-$\nu$}\label{genAnu2};\\
    &&&\sum_{i \in A}\Pp_i(\pi_i^S, \ell) + \sum_{u \in N\setminus A}\Pp_u(\pi_u^S, \ell) \leq |S\cap [\ell]|,&\forall 0 \leq \ell \leq m. \tag{S}\label{genS2}
\end{align}
By \cref{genprotpref}, conditions (\ref{genES2}) and (\ref{genAnu2}) applied to $I\subseteq N$ from \cref{genESrounding} give (\ref{genE2}).

By \cref{genESrounding}, the number of already assigned agents is $|I| \geq d - \theta - \eps|F| - \rho_{ES}(q, dn)$, which, applying the previously acquired bounds on $d$, $\theta$ and assuming that $|F| = O_{\Gamma ,q}(\log^2 n)$, is at least $n - O_{\Gamma ,q}(\log^2 n)$.
We would like to retain most of these profiles $\{\pi_i\}_{i \in I}$, and replace only a small number of them, say $O_{\Gamma ,q}(\log^2 n)$, and assign new profiles to the remaining $O_{\Gamma ,q}(\log^2 n)$ ones, to satisfy (\ref{genES2}), (\ref{genAnu2}) and (\ref{genS}).
This way, each agent $i \in N$ will receive a profile, so that the whole collection of profiles is simultaneously satisfiable by distinct items of the head $[h]$.

We will inevitably encounter several issues that we briefly mentioned in \cref{sec:partition}, but set aside earlier.
First, we would like to assign to every agent $i \in N$ a profile that is acceptable, i.e of value at least $1 - 2/q$.
The already assigned profiles $\pi_i$ for $i \in I$ satisfy $v(\pi_i) \geq 1 - 2/q$, and we need to guarantee the same for every new profile $\pi_u$, satisfying just (\ref{genES2}), (\ref{genAnu2}) and (\ref{genS}) is not enough.
What complicates this more is that, even though for every agent $i \in N$ we have $\wv_i([h]) \geq (1 - 1/q)^2n \geq (1 - 2/q)n$ (by \cref{roundedhead}), it does not mean that all items $e \leq h$ agent $i$ values highly, i.e $\wv_i(e) \geq 1/q$.
So, it may not even be possible to get a feasible collection of acceptable profiles satisfiable by just the head $\{1, \ldots, h\}$, and we cannot even treat the common tail $M\setminus [h]$ as a set of ``small-value'' items.

This obstacle is not that critical, as long as we manage to select $A\subseteq I$ and give every $u \in N\setminus A$ profile $\pi_u$, such that 1) the total deficit (relative to the rounding loss) of all new profiles $\pi_u$ is sufficiently small, and 2) all profiles can be simultaneously satisfied by a subset of items, such that every agent who ends up with a non-acceptable profile still values the unused items high enough to cover his deficit.
Then, we could use an analogue of tail completion \cref{etatailcomplete} and complete every assigned profile to total value of $1 - 2/q$.

This leads us to the second major issue: coordination.
Choosing a new profile for every agent independently provides little control over the combined load, as well as which items are required to simultaneously satisfy these profiles.
For example, if we create a new profile for $i \in N$ using agent $i$'s personal MMS partition, different agents may require the same items in order to satisfy their profiles, which may violate (\ref{genES2}) and (\ref{genS}).
Furthermore, to be able to use some analogue of \cref{etatailcomplete}, we need to ensure that all the assigned profiles use only items up to some common threshold, otherwise we cannot properly define a ``tail'' to complete the assignment --- existing $M\setminus [h]$ may not work.

\subsection{Suggested profiles}

As in \cref{sec:intsupplier}, we will resolve both the high value / common threshold issue, as well as the coordination issue, using the following approach.
We will pick a special agent $u_* \in N$, which we will refer to as \textbf{supplier}.
This agent will use his personal MMS partition $\{B_{u_*,1}, \ldots, B_{u_*, n}\}$ (as in \cref{notpoor}) in order to create a \textbf{suggested profile} $\hpi_{u_*,i}$ for every agent $i \in N$.
The supplier's MMS partition will provide a point of reference, which will allow us to simultaneously control the load associated with different bundles.
The suggested profiles $\hpi_{u_*,i}$ may reorganize the items in the partition $\{B_{u_*,1}, \ldots, B_{u_*, n}\}$ and adapt to every individual agent $i \in N$, while preserving the prefix feasibility constraints.

Recall that every good agent $i \in G$ satisfies $v_i(e) \geq 1/q$ for every item $e \leq h$, and set $W\subseteq N$ contains all agents $i$ that have a high-demand breakpoint in the buffer $F$.
\begin{definition}
    We refer to agents $G\setminus W$ as \textbf{supplier candidates}.
    For each $u \in G\setminus W$, let
    \[
        h_u := m_u(1/q) = |\{e \in M: v_u(e) \geq 1/q\}|
    \]
    be the number of items agent $u$ values at least $1/q$, we call it agent $u$'s \textbf{head threshold}.
    Note that for every $u \in G\setminus W$ it holds $h_u \geq h$.
    Define the \textbf{residual head} of agent $u$ as
    \[
        H_u^+ := \{h + 1, \ldots, h_u\}.
    \]
\end{definition}
Each supplier candidate $u \in G\setminus W$ provides a partition
\[
    M = E\sqcup S\sqcup H_u^+ \sqcup \{h_u + 1, \ldots, m\}.
\]
The reason we introduce $h_u$ is that not every item outside of the common head $[h]$ needs to be small (i.e, value less than $1/q$) for the supplier.
The additional set $H_u^+$ contains precisely those extra items that remain high-value for agent $u$, and $v_u(e) < 1/q$ for all $e > h_u$.
Now, with a given supplier $u \in G\setminus W$, our goal is to obtain a feasible collection of profiles that are simultaneously satisfiable by distinct items of $[h_u] = E\sqcup S\sqcup H_u^+$.
Then, we will be able to extend the tail completion \cref{etatailcomplete} to the supplier tail $\{h_u + 1, \ldots, m\}$.

\begin{observation}\label{uhead}
    For every supplier candidate $u \in G\setminus W$, the residual head $H_u^+$ is small, i.e
    \[
        h_u - h \leq q^2(16\Lambda + 1).
    \]
\end{observation}
\begin{proof}
    By definition of $H_u^+$, we have $v_u(H_u^+) \geq |H_u^+|/q$.
    Then, by \cref{notpoor},
    \[
        \frac{h_u - h}{q} \leq v_u(H_u^+)\leq v_u(M\setminus [h]) \leq n - v_u([h]) \leq q(16\Lambda + 1). 
    \]
    The observation follows.
\end{proof}

\begin{definition}
    Fix a supplier $u_* \in G\setminus W$.
    For each agent $i \in N$, his \textbf{suggested profile} is
    \[
        \hpi_{u_*,i} := \hpi_{u_*,i}^E \uplus \hpi_{u_*,i}^S \uplus \hpi_{u_*,i}^+,
    \]
    where each part is intended to be satisfied using exclusively distinct items of $E, S, H_{u_*}^+$ respectively.
\end{definition}

We will now partially explain how supplier $u_*$ constructs suggested profiles for agents $i \in N$.
\begin{definition}
    For a supplier $u_* \in G\setminus W$ with MMS partition $\{B_{u_*,1}, \ldots, B_{u_*, n}\}$, we call a bundle $B_{u_*,j}$, $j \in [n]$, \textbf{exceptional}, if
    \[
        B_{u_*,j}\cap S \neq \varnothing \qquad\text{or}\qquad v_{u_*}(B_{u_*,j}\cap [h]) < 1 - 1/q.
    \]
\end{definition}
\begin{observation}\label{genexcep}
    For a supplier $u_* \in G\setminus W$, the total number of exceptional bundles is at most
    \[
        |S| + q^2(16\Lambda + 1).
    \]
\end{observation}
\begin{proof}
    All bundles $B_{u_*,1}, \ldots, B_{u_*, n}$ are disjoint, so at most $|S|$ of them intersect $S$.
    Moreover, every bundle satisfying $v_{u_*}(B_{u_*,j}\cap [h]) < 1 - 1/q$ induces an invalid rounded head profile $\pi_{u_*,j}$.
    Since $u_* \in G$, by \cref{invalid} the total number of such bundles is $\theta_{u_*} \leq q^2(16\Lambda + 1)$, hence the total number of exceptional bundles is at most
    \[
        |S| + q^2(16\Lambda + 1).
    \]
\end{proof}
When constructing suggested profiles for agents $i \in N$, actions of the supplier $u_*$ differ based on which bundles of $\{B_{u_*,1}, \ldots, B_{u_*, n}\}$ are exceptional and which are not.
Concretely:
\begin{itemize}
    \item For every exceptional bundle $B_{u_*,j}$, we will choose a distinct recipient $i \in N$, and create suggested profile parts $\hpi_{u_*,i}^E$, $\hpi_{u_*,i}^S$ and $\hpi_{u_*,i}^+$ based off only values of items of $B_{u_*,j}$.
    The resulting suggested profile $\hpi_{u_*,i}$ may not be acceptable for agent $i$.
    If $i$ ends up using a non-acceptable $\hpi_{u_*,i}$, he will receive an appropriate tail profile later in the tail completion.
    \item All non-exceptional bundles $B_{u_*,j}$ satisfy $B_{u_*,j}\cap [h] \subseteq E$.
    For the remaining agents, we will reorganize the $E$-parts $B_{u_*,j}\cap E$ of non-exceptional bundles, obtaining a collection of suggested profiles $\hpi_{u_*,i}$ such that $\hpi_{u_*,i}^S = \varnothing = \hpi_{u_*,i}^+$, i.e $\hpi_{u_*,i} = \hpi_{u_*,i}^E$, and for every recipient $i$ its suggested profile $\hpi_{u_*,i}$ is already acceptable (thus needs no tail completion).
\end{itemize}

To give some idea of how exactly supplier $u_*$ creates suggested profiles for agents $i \in N$ and why such profiles will be feasible, we will present the construction of $S$- and $H_{u_*}^+$-parts of suggested profiles, assuming that we already know which exceptional bundle goes to which agent.
\begin{definition}\label{gensuggestpart}
    Fix a supplier $u_* \in G\setminus W$, and designate exceptional bundles of supplier's MMS partition $\{B_{u_*, 1}, \ldots, B_{u_*,n}\}$ to distinct recipients $i \in N$, call these agents \textbf{exceptional}.

    For every exceptional agent $i$ with designated bundle $B_{u_*,j}$, define
    \[
        \hpi_{u_*,i}^S:= \{\wv_{u_*}(e):\; e \in B_{u_*,j}\cap S\},\qquad\qquad \hpi_{u_*,i}^+ := \{\wv_{u_*}(e):\; e \in B_{u_*,j}\cap H_{u_*}^+\}.
    \]
    For every non-exceptional agent $i$, set $\hpi_{u_*,i}^S = \varnothing = \hpi_{u_*,i}^+$.
\end{definition}
Note that the values in suggested $S/H^+$-parts  $\hpi_{u_*,i}^S$ and $\hpi_{u_*,i}^+$ are according to $u_*$ and not to $i$.
A natural question is, whether assigning such a profile to agents $i \in N$ is still feasible, i.e they will be able to simultaneously satisfy all profiles using distinct items.
We show that, for every suggested profile $\hpi_{u_*,i}$, as long as the recipient $i$ is ``compatible'' with its designated exceptional bundle $B_{u_*,j}$, the $S$-parts and $H_{u_*}^+$-parts of the suggested profiles together are feasible in $S$ and $H_{u_*}^+$ respectively for any subset of agents $N$.
\begin{lemma}\label{gencompatible}
    For the suggested parts $\hpi_{u_*,i}^S$ and $\hpi_{u_*,i}^+$, $i \in N$, defined above, for every $0 \leq \ell \leq m$:
    \[
        \sum_{i \in N}\Pp_{u_*}(\hpi_{u_*,i}^S, \ell) \leq |S\cap [\ell]|,\qquad\qquad \sum_{i \in N}\Pp_{u_*}(\hpi_{u_*,i}^+, \ell) \leq |H_{u_*}^+\cap [\ell]|.
    \]
    Furthermore, if additionally every agent $i \in N$ satisfies
    \begin{itemize}
        \item \textbf{$S$-compatibility}: for every $\alpha \in \hpi_{u_*,i}^S$ it holds $v_i(\max(S\cap[m_{u_*}(\alpha)])) \geq \alpha$;
        \item \textbf{$H^+$-compatibility}: for every $\alpha \in \hpi_{u_*,i}^+$ it holds $v_i(m_{u_*}(\alpha)) \geq \alpha$;
    \end{itemize}
    then for every prefix $0 \leq \ell \leq m$ we have
    \[
        \sum_{i \in N}\Pp_i(\hpi_{u_*,i}^S, \ell) \leq |S \cap [\ell]|,\qquad\qquad \sum_{i \in N}\Pp_i(\hpi_{u_*,i}^+, \ell) \leq |H_{u_*}^+ \cap [\ell]|.
    \]
\end{lemma}
\begin{proof}
    Since $\hpi_{u_*,i}^S$ are constructed from supplier's MMS partition, every demand $\alpha$ counted by supplier's prefix load $\Pp_{u_*}(\hpi_{u_*,i}^S, \ell)$ has an originating item $e \in S$ satisfying $e \leq m_{u_*}(\alpha) \leq \ell$.
    And, all these items are distinct across all suggested $S$-profiles.
    This proves the the first inequality on $S$-parts, and the proof for $H_{u_*}^+$-parts is completely analogous.

    Next, consider a demand $\alpha = \wv_{u_*}(e)$ for some item $e \in B_{u_*,j} \cap H_{u_*}^+$.
    Since $v_{u_*}(e) \geq \alpha \geq 1/q$, 
    \[
        h < e \leq m_{u_*}(\alpha) \leq h_{u_*},
    \]
    i.e the breakpoint lies in $H_{u_*}^+$.
    If for $i \in N$ we have $v_i(m_{u_*}(\alpha)) \geq \alpha$, then $m_i(\alpha) \geq m_{u_*}(\alpha)$.
    Consequently, for every $i \in N$ and every prefix $0 \leq \ell \leq m$, 
    \[
        \Pp_i(\hpi_{u_*,i}^+, \ell) \leq \Pp_{u_*}(\hpi_{u_*,i}^+, \ell),
    \]
    which directly implies
    \[
        \sum_{i \in N}\Pp_i(\hpi_{u_*,i}^+, \ell) \leq \sum_{i \in N}\Pp_{u_*}(\hpi_{u_*,i}^+, \ell) \leq |H_{u_*}^+\cap [\ell]|.
    \]

    Finally, consider a demand $\alpha = \wv_{u_*}(e)$ for some item $e \in B_{u_*,j}\cap S$, such that $\alpha \in \hpi_{u_*,i}^s$ for agent $i \in N$.
    Since $v_{u_*}(e) \geq \alpha$, we have $e \in S\cap [m_{u_*}(\alpha)]$, so it is non-empty.
    Then, by assumption $v_i(\max(S\cap[m_{u_*}(\alpha)])) \geq \alpha$ and therefore
    \[
        e \leq \max(S\cap[m_{u_*}(\alpha)]) \leq m_i(\alpha).
    \]
    Consequently, if $\alpha$ is counted in $\Pp_i(\hpi_{u_*,i}^S, \ell)$, its originating item $e$ satisfies $e \leq m_i(\alpha) \leq \ell$.
    Thus, every demand counted in the total $S$-prefix load at $\ell$ has an originating item in $S\cap [\ell]$, and all these items are distinct across all suggested $S$-profiles.
    It follows that for every $0 \leq \ell \leq m$,
    \[
        \sum_{i \in N}\Pp_i(\hpi_{u_*,i}^S, \ell) \leq |S \cap [\ell]|.
    \]
\end{proof}
We will pick a supplier $u_*$ and designated recipients of his exceptional bundles so that the compatibility conditions of \cref{gencompatible} do hold.
The construction of $E$-parts of suggested profiles is significantly more complicated, and will be covered later.

\subsubsection{Bounded replacement}

Suppose for now that we chose some supplier $u_* \in G\setminus W$ and constructed some collection $\{\hpi_{u_*,i}\}_{i \in N}$ of suggested profiles (with $S$- and $H_{u_*}^+$-parts according to \cref{gencompatible}), where agent $i \in N$ is suggested $\hpi_{u_*,i}$.
Now, as discussed earlier in the section, our task is to choose a set $A\subseteq I$ of agents that retain their old profiles $\pi_i$, and give a suggested profile $\hpi_{u_*,i}$ to every other agent $i \in N\setminus A$, so that conditions (\ref{genES2}), (\ref{genAnu2}) and (\ref{genS2}) hold.
More precisely, we want that
\begin{align}
    &&&\sum_{i \in A}\Pp_i(\pi_i^E, c_s) + \sum_{i \in N\setminus A}\Pp_i(\hpi_{u_*,i}^E, c_s) \leq |E\cap [b_s]|,&\forall (b_s, c_s) \in \Ll_S;\tag{$\mathrm{E}_{\Ll_S}$}\label{genES3}\\
    &&&\sum_{i \in N \setminus A}|\hpi_{u_*,i}|\leq \frac{\nu}{q} \tag{A-$\nu$}\label{genAnu3};\\
    &&&\sum_{i \in A}\Pp_i(\pi_i^S, \ell) + \sum_{i \in N\setminus A}\Pp_i(\hpi_{u_*,i}^S, \ell) \leq |S\cap [\ell]|,&\forall 0 \leq \ell \leq m. \tag{S}\label{genS3}
\end{align}

Note that the residual-head parts of profiles $\hpi_{u_*,i}$ impose no additional restriction on the choice of set $A$, as old profiles $\pi_i$ for $i \in I$ use only head $[h]$, and by \cref{gencompatible} for any $A\subseteq I$,
\[
    \sum_{i \in N\setminus A}\Pp_i(\hpi_{u_*,i}^+, \ell) \leq \sum_{i \in N}\Pp_i(\hpi_{u_*,i}^+, \ell)\leq |H_{u_*}^+\cap [\ell]|, \qquad\forall 0 \leq \ell \leq m.
\]

The constraints of (\ref{genS3}) are more problematic, as we need to simultaneously control load at $m + 1$ different prefixes. 
Fortunately, we can obtain a similar reduction as we did earlier with conditions (\ref{genE2}) and (\ref{genES2}), and preserve (\ref{genS3}) at only a constant number of different prefixes, corresponding to the high-demand breakpoints of the supplier $u_*$.
\begin{definition}
    Let $I\subseteq N$ and $\{\pi_i\}_{i \in I}$ be obtained from \cref{genESrounding}.
    An \textbf{$S$-certificate} for profiles $\{\pi_i^S\}_{i \in I}$ is a family of pairwise-disjoint sets $S_i\subseteq S$ for $i \in I$, such that $|S_i| = |\pi_i^S|$ and $S_i$ satisfies profile $\pi_i^S$ for agent $i$.
    By \cref{genESrounding} and \cref{loadequiv}, such certificate exists.
\end{definition}
The sets $S_i$ for $i \in I$ record which items of set $S$ satisfy the $S$-part of the old profile $\pi_i^S$.
If the agent $i \in I$ keeps his old profile, i.e $i \in A$, we ``reserve'' its certificate $S_i$, otherwise the items of $S_i$ become available for supplier $u_*$'s suggested profiles.

Note that for every $i \in I$ it holds $\Pp_i(\pi_i^S, \ell) \leq |S_i\cap [\ell]|$ for every prefix $\ell$.
We will use this bound in order to reduce condition (\ref{genS3}) on both profiles of agents $A$ and suggested supplier profiles of $N\setminus A$, to a condition on supplier profiles $\hpi_{u_*,i}$ only.
This way, instead of checking (\ref{genS3}) at every relevant breakpoint of profiles $\pi_{i}^S$ for $i \in A$ and $\hpi_{u_*,i}^S$ for $i \in N\setminus A$, the number of which can depend on the number of agents/items, it will suffice to check a supplier-only version of (\ref{genS3}) at every possible supplier breakpoint, which is a constant amount since $|\Qq| = Q = O(q\log q)$.
This will add an additional set of \textit{protected prefix} constraints w.r.t items $S$.

The following lemma is an analogue of \cref{sprot} from \cref{sec:intsupplier}.
\begin{lemma}\label{gensprot}
    Fix a supplier $u_*\in G\setminus W$ and construct suggested profiles $\hpi_{u_*,i}^S$, $i \in N$, and suppose they satisfy the $S$-compatibility condition of \cref{gencompatible}.
    If choice $A\subseteq I$ satisfies
    \begin{equation}
        \sum_{i \in A}|S_i\cap [m_{u_*}(\alpha)]| + \sum_{i\in N\setminus A}\Pp_{u_*}(\hpi_{u_*,i}^S, m_{u_*}(\alpha)) \leq |S\cap [m_{u_*}(\alpha)]|\tag{S$*$}\label{genS*}
    \end{equation}
    for every possible supplier breakpoint $m_{u_*}(\alpha)$ for $\alpha \in \Qq$, then for every prefix $0 \leq \ell \leq m$,
    \begin{equation}
        \sum_{i \in A}\Pp_i(\pi_i^S, \ell) + \sum_{i \in N\setminus A}\Pp_i(\hpi_{u_*,i}^S, \ell) \leq |S\cap [\ell]|.\tag{S}\label{genSstar}
    \end{equation}
\end{lemma}

\begin{proof}
    Reserve bundle $S_i$ for every $i \in A$, these agents will retail their old $S$-profiles.
    We will show that the suggested $S$-profiles $\hpi_{u_*,i}^S$ of all other agents $i\in N\setminus A$ can be simultaneously satisfied using distinct remaining items.

    Consider the condition (\ref{genS*}) for an arbitrary prefix $\ell$.
    Observe that we can rearrange it as 
    \begin{equation}
        \sum_{i\in N\setminus A}\Pp_{u_*}(\hpi_{u_*,i}^S, \ell) \leq |S\cap [\ell]|- \sum_{i \in A}|S_i\cap [\ell]|.\tag{$*$}\label{gen*}
    \end{equation}
    The right-hand side of (\ref{gen*}) counts the items of $S$ in prefix $[\ell]$ that remain available after reserving the certificates for $A$.
    Indeed, since all $S_i$ are pairwise-disjoint subsets of $S$, we have
    \[
        |S\cap [\ell]|- \sum_{i \in A}|S_i\cap [\ell]| = \left|\left(S\setminus\bigcup_{i \in A}S_i\right)\cap [\ell]\right|.
    \]
    In particular, this quantity is non-negative and non-decreasing in $\ell$, and when we increase $\ell$ by $1$, the remaining capacity in (\ref{gen*}) increases by $1$ only if the item $\ell + 1$ is an unreserved item of $S$.

    On the other hand,  the left-hand side of (\ref{gen*}) increases only at supplier's breakpoints $m_{u_*}(\alpha)$, $\alpha \in \Qq$.
    Therefore, in order to satisfy (\ref{gen*}) for every prefix $0 \leq \ell \leq m$, it suffices to presetve the inequality only at prefixes $\ell = m_{u_*}(\alpha)$ for $\alpha \in \Qq$.
    Indeed, if load at $\ell$ is positive, and $m_{u_*}(\alpha) \leq \ell$ is the last $S$-breakpoint of $\alpha \in \hpi_{u_*,i}^S$ preceding $\ell$, then load at  $m_{u_*}(\alpha)$ and $\ell$ is the same, while the remaining capacity at $\ell$ is at least the one at $m_{u_*}(\alpha)$.
    Hence, enforcing (\ref{gen*}) at $m_{u_*}(\alpha)$, which is exactly (\ref{genS*}), implies it at $\ell$ as well.
    If the load is $0$, then inequality holds regardless.

    By assumption, (\ref{genS*}) holds for every supplier breakpoint $m_{u_*}(\alpha)$, therefore (\ref{gen*}) holds at every prefix $0 \leq \ell \leq m$.
    Then, by \cref{loadequiv} we can simultaneously satisfy all suggested profiles $\{\hpi_{u_*,i}^S\}_{i \in N\setminus A}$ w.r.t valuation $v_{u_*}$ using distinct items of 
    \[
        S\setminus\bigcup_{i \in A}S_i.
    \]
    
    We can view this allocation as a matching between demands $\alpha \in \hpi_{u_*,i}^S$ and corresponding items $e$ satisfying $\alpha$, i.e with $v_{u_*}(e)\geq \alpha$.
    We claim that for every agent $i \in N\setminus A$ and every demand $\alpha \in \hpi_{u_*,i}^S$ matched to item $e$, it also holds $v_i(e) \geq \alpha$.
    Indeed, since $v_{u_*}(e)\geq \alpha$, we have $e \in S\cap[m_{u_*}(\alpha)]$, hence $e\leq \max\{S\cap[m_{u_*}(\alpha)]\}$ and under $S$-compatibility of \cref{gencompatible},
    \[
        v_i(e)\geq v_i(\max\{S\cap [m_{u_*}(\alpha)]\}) \geq \alpha.
    \]
    It follows that, together with certificates $S_i$ for $i \in A$, we obtain a matching between items of $S$ and demands of all profiles $\{\pi_i^S\}_{i \in A}$ and $\{\hpi_{u_*,i}^S\}_{i\in N\setminus A}$, such that for every item $e$ matched to a demand $\alpha$ in the profile of agent $i$ it holds $v_i(e)\geq \alpha$.
    But then, every demand counted in the total load  at prefix $\ell$ of agents $A$ and $N\setminus A$ together must be matched to a distinct item of $S\cap [\ell]$.
    As a result, for every prefix $0\leq \ell \leq m$ it must hold
    \[
        \sum_{i \in A}\Pp_i(\pi_i^S, \ell) + \sum_{i \in N\setminus A}\Pp_i(\hpi_{u_*,i}^S, \ell) \leq |S\cap [\ell]|.
    \]
\end{proof}

It follows from \cref{gencompatible} and \cref{gensprot} that, for the chosen subset $A\subseteq I$ of agent who keep their assigned profiles $\{\pi_{i}\}_{i \in A}$, and for the remaining agents $i \in N\setminus A$ who receive a supplier suggestion $\hpi_{u_*,i}$, to satisfy for every prefix $0 \leq \ell \leq m$
\[
    \sum_{i \in A}\Pp_i(\pi_i, \ell) + \sum_{i \in N\setminus A}\Pp_i(\hpi_{u_*,i}, \ell) \leq \ell,
\]
it suffices that the following conditions hold on $A\subseteq I$ and suggested collection $\{\hpi_{u_*,i}\}_{i \in N}$.
\begin{align}
    &&&\sum_{i \in A}\Pp_i(\pi_i^E, c_s) + \sum_{i \in N\setminus A}\Pp_i(\hpi_{u_*,i}^E, c_s) \leq |E\cap [b_s]|,&\forall (b_s, c_s) \in \Ll_S;\tag{$\mathrm{E}_{\Ll_S}$}\label{genES4}\\
    &&&\sum_{i \in N \setminus A}|\hpi_{u_*,i}|\leq \frac{\nu}{q} \tag{A-$\nu$}\label{genAnu4};\\
    &&& \sum_{i \in A}|S_i\cap [m_{u_*}(\alpha)]| + \sum_{i\in N\setminus A}\Pp_{u_*}(\hpi_{u_*,i}^S, m_{u_*}(\alpha)) \leq |S\cap [m_{u_*}(\alpha)]|, & \forall \alpha \in \Qq.\tag{S$*$}\label{genS4}
\end{align}

Note that now the total number of constraints that we need to take care of is $|\Ll_S| + |\Qq| + 1$, which is only a \textbf{constant} depending on $q$, as $|\Ll_S| \leq \lceil Q\cdot \frac{q}{q - 1}\rceil + 1$.

Let us first focus on trying to find a set $A\subseteq I$ that satisfies just (\ref{genES4}) and (\ref{genS4}).
Suppose that the suggested supplier collection $\{\hpi_{u_*,i}\}_{i \in N}$ satisfies (\ref{genES4}) and (\ref{genS4}) with  $A = \varnothing$, that is,
\begin{align}
    &&&\sum_{i \in N}\Pp_i(\hpi_{u_*,i}^E, c_s) \leq |E\cap [b_s]|,&\forall (b_s, c_s) \in \Ll_S;\tag{$\mathrm{E}_{\Ll_S}$-$u$}\label{genEsu}\\
    &&& \sum_{i\in N}\Pp_{u_*}(\hpi_{u_*,i}^S, m_{u_*}(\alpha)) \leq |S\cap [m_{u_*}(\alpha)]|, & \forall \alpha \in \Qq.\tag{S$*$-$u$}\label{genSu}
\end{align}
This way, if we replace every old profile $\{\pi_i\}_{i \in I}$ with suggested profile, we obtain a feasible collection with respect to protected prefix constraints (\ref{genES4}) and (\ref{genS4}).
The problem with such replacement is that the total number of added demands $\sum_{i \in N}|\hpi_{u_*,i}|$ may be unbounded, while we seek a replacement of size at most $\nu/q$, which will be roughly $O_q(\log^2 n)$.
Our aim is to show that only a bounded number of old profiles need to be replaced with suggested profiles, in order to get a feasible $A\subseteq I$ satisfying (\ref{genES4}) and (\ref{genS4}).
This is done similarly to \cref{sec:intsupplier}, using Dickson's lemma \cite{Dickson13}.

The following theorem is a complete analogue of \cref{supplierchange}.
\begin{theorem}[Bounded replacement]\label{gensupplierchange}
    For every pair of integers $t, q \geq 1$, there exists a global constant $C_{t,q} \geq 1$, depending only on $t$ and $q$, such that the following holds.

    Let $I\subseteq N$ and $\{\pi_i\}_{i \in I}$ be obtained from \cref{genESrounding}.
    For every $i \in I$, fix an $S$-certificate $\{S_i\}_{i \in I}$, and let the number of protected $E$-prefixes be $|\Ll_S| = t$.
    Fix a supplier $u_* \in G\setminus W$, and for every agent $i \in N$ a corresponding suggested profile $\hpi_{u_*,i} = \hpi_{u_*,i}^E \uplus \hpi_{u_*,i}^S\uplus \hpi_{u_*,i}^+$.
    
    Suppose that for every $i \in N$, $|\hpi_{u_*,i}|\leq q$, every $S$-demand $\alpha \in \hpi_{u_*,i}^S$ belongs to $\Qq$, as well as 
    \begin{align}
    &&&\sum_{i \in N}\Pp_i(\hpi_{u_*,i}^E, c_s) \leq |E\cap [b_s]|,&\forall (b_s, c_s) \in \Ll_S;\tag{$\mathrm{E}_{\Ll_S}$-$u$}\label{genESu2}\\
    &&& \sum_{i\in N}\Pp_{u_*}(\hpi_{u_*,i}^S, m_{u_*}(\alpha)) \leq |S\cap [m_{u_*}(\alpha)]|, & \forall \alpha \in \Qq.\tag{S$*$-$u$}\label{genSu2}
    \end{align}

    Then there exists $A\subseteq I$ such that $|N\setminus A|\leq C_{t,q}(d - |I|)$ and
    \begin{align}
    &&&\sum_{i \in A}\Pp_i(\pi_i^E, c_s) + \sum_{i \in N\setminus A}\Pp_i(\hpi_{u_*,i}^E, c_s) \leq |E\cap [b_s]|,&\forall (b_s, c_s) \in \Ll_S;\tag{$\mathrm{E}_{\Ll_S}$}\label{genES5}\\
    &&& \sum_{i \in A}|S_i\cap [m_{u_*}(\alpha)]| + \sum_{i\in N\setminus A}\Pp_{u_*}(\hpi_{u_*,i}^S, m_{u_*}(\alpha)) \leq |S\cap [m_{u_*}(\alpha)]|, & \forall \alpha \in \Qq.\tag{S$*$}\label{genS5}
\end{align}
    In particular,
    \[
        \sum_{i \in N\setminus A}|\hpi_{u_*,i}|\leq qC_{t,q}(d - |I|),
    \]
    so if integer $\nu$ satisfies $\nu \geq q^2C_{t,q}(d - |I|)$, condition (\ref{genAnu4}) holds for the same choice of $A$.
\end{theorem}
We prove \cref{gensupplierchange} in \cref{appendix:supplier} (\cref{supplierchangeproof}).
We apply its proof with $Q := |\Qq|$ and $C_{t,q} := C_{t,Q,q}$.
\cref{genESrounding} guarantees initial protected $E$-prefix inequalities, while the fixed $S$-certificates of \cref{gensprot} give initial inequalities on $S$.
We use only the $E/S$-parts of suggested profiles, the combined size remains at most $q$.

\cref{gensupplierchange} allows to complete the initial collection $\{\pi_i\}_{i \in I}$ of profiles for agents $I\subseteq N$ to a full collection of profiles for all agents $N$, while replacing only a constant number of old profiles per every new agent added.
We will now show how we can combine this theorem with previously established reductions in \cref{sec:partition} in order obtain feasibility at every prefix $0 \leq \ell \leq m$.

Recall that \cref{genESrounding} leaves at most 
\[
    d - |I| \leq \theta + \eps |F| + \rho_{ES}(q, dn)
\]
agents unassigned.
Therefore, \cref{gensupplierchange} introduces at most
\[
    \sum_{i \in N\setminus A}|\hpi_{u_*,i}|\leq qC_{t,q}(\theta + \eps |F| + \rho_{ES}(q, dn))
\]
additional demands.
We want this quantity to be at most $\nu/q$, so that \cref{genprotpref} applies.
Since $|F|\leq (\lceil Q\cdot \frac{q}{q - 1}\rceil + 1)\cdot \nu$, increasing $\nu$ also increases our bound on the number of unassigned agents caused by the buffer $F$.
So, in order to control this contribution, we will first choose sufficiently small $\eps \in (0, 1)$, and then choose integer $\nu \geq 1$ to accommodate the remaining loss.

\begin{definition}
    For integer $q \geq 9$, define the global constant $C_q$ as
    \[
        C_q := \max\left\{C_{t, q} : 1 \leq t \leq \left\lceil Q\cdot \frac{q}{q - 1}\right\rceil + 1\right\},
    \]
    where $C_{t, q}$ are obtained from \cref{gensupplierchange}.
\end{definition}
\begin{corollary}\label{genepsnuconst}
    In the setting of \cref{notpoor}, choose $\eps \in (0, 1)$ and integer $\nu \geq 1$ satisfying
    \[
        \eps \leq \frac{1}{2q^2C_q\left(\left\lceil Q\cdot \frac{q}{q - 1}\right\rceil + 1\right)}, \qquad\qquad \nu \geq 2q^2C_q(\theta + \rho_{ES}(q, dn)).
    \]
    Apply \cref{genbuffer} with these values of $\eps, \nu$ to obtain a partition $[h] = E\sqcup S$, sets $F \subseteq E$, $W\subseteq N$ and protected prefixes $\Ll_S$, and then use \cref{genESrounding} to get the set $I\subseteq N$ of agents with assigned profiles $\{\pi_i\}_{i \in I}$, with the following properties:
    \begin{itemize}
        \item for every $i \in I$, $\pi_i$ has a decomposition $\pi_i = \pi_i^E \uplus \pi_i^S$, with $|\pi_i|\leq q$, $v(\pi_i) \geq 1 - 2/q$ and 
        \[
            d - |I|\leq \theta + \eps |F| + \rho_{ES}(q, dn);
        \]
        \item their $E$-prefix loads satisfy
        \begin{align}
        &&&\sum_{i \in I}\Pp_i(\pi_i^E, \ell) \leq |E\cap [\ell]| - \frac{\nu}{q},&\forall \ell \in E\setminus F;\tag{$\mathrm{E}$-$\nu$}\label{genEnu6}\\
        &&& \sum_{i \in I}\Pp_i(\pi_i^E, c_s) \leq |E\cap [b_s]|, & \forall (b_s, c_s)\in \Ll_S.\tag{$\mathrm{E}_{\Ll_S}$}\label{genES6}
        \end{align}
        \item there are disjoint sets $S_i\subseteq S$, $i \in I$, such that $S_i$ satisfies $\pi_i^S$ for agent $i$ and $|S_i| = |\pi_i^S|$.        
    \end{itemize}    
    Suppose there is a supplier $u_*\in G\setminus W$ and profiles $\hpi_{u_*,i} = \hpi_{u_*,i}^E\uplus \hpi_{u_*,i}^S \uplus \hpi_{u_*,i}^+$, $i \in N$, such that
    \begin{itemize}
        \item for every $i \in N$, $|\hpi_{u_*,i}|\leq q$, and $\hpi_{u_*,i}^S$, $\hpi_{u_*,i}^+$ are as in \cref{gensuggestpart};
        \item the $E$-parts of $\{\hpi_{u_*,i}\}_{i \in N}$ satisfy protected prefix inequalities:
        \begin{align}
        &&&\sum_{i \in N}\Pp_i(\hpi_{u_*,i}^E, c_s) \leq |E\cap [b_s]|,&\forall (b_s, c_s) \in \Ll_S;\tag{$\mathrm{E}_{\Ll_S}$-$u$}\label{genESu6}
        \end{align}
        \item every $i \in N$ is \textbf{$S$-compatible}: for every $\alpha \in \hpi_{u_*,i}^S$  it holds $v_i(\max(S\cap[m_{u_*}(\alpha)])) \geq \alpha$;
        \item every $i \in N$ is \textbf{$H^+$-compatible}: for every $\alpha \in \hpi_{u_*,i}^+$  it holds $v_i(m_{u_*}(\alpha)) \geq \alpha$.
    \end{itemize}
    Then there exists set $A\subseteq I$ such that $|N\setminus A| \leq \nu/q^2$ and for every prefix $0 \leq \ell \leq m$,
    \[
        \sum_{i \in A}\Pp_i(\pi_i, \ell) + \sum_{i \in N\setminus A}\Pp_i(\hpi_{u_*,i}, \ell) \leq \min\{\ell, h_{u_*}\}.
    \]
    As a corollary, these profiles $\{\pi_i\}_{i \in A}$ and $\{\hpi_{u_*,i}\}_{i \in N\setminus A}$ can be simultaneously satisfied by distinct items of $[h_{u_*}]$, with their $E$-, $S$- and residual-head parts satisfied within $E, S, H_{u_*}^+$ respectively.
\end{corollary}
\begin{proof}
    Since $|F|\leq (\lceil Q\cdot \frac{q}{q - 1}\rceil + 1)\cdot \nu$, by the choice of parameters $\eps, \nu$ we have
    \[
        d - |I|\leq \theta + \rho_{ES}(q, dn) + \eps\left(\left\lceil Q\cdot \frac{q}{q - 1}\right\rceil + 1\right)\nu \leq \frac{\nu}{2q^2C_q} + \frac{\nu}{2q^2C_q} = \frac{\nu}{q^2C_q}.
    \]
    By \cref{gensuggestpart}, all demands in $\hpi_{u_*,i}^S$, $i \in N$, belong to $\Qq$.
    In addition, \cref{gencompatible} implies
    \begin{align}
    &&& \sum_{i\in N}\Pp_{u_*}(\hpi_{u_*,i}^S, m_{u_*}(\alpha)) \leq |S\cap [m_{u_*}(\alpha)]|, & \forall \alpha \in \Qq.\tag{S$*$-$u$}\label{genSu5}
    \end{align}
    Hence, we can apply \cref{gensupplierchange}, which gives $A\subseteq I$ with $|N\setminus A| \leq C_q(d - |I|) \leq\nu/q^2$ and
    \begin{align}
    &&&\sum_{i \in A}\Pp_i(\pi_i^E, c_s) + \sum_{i \in N\setminus A}\Pp_i(\hpi_{u_*,i}^E, c_s) \leq |E\cap [b_s]|,&\forall (b_s, c_s) \in \Ll_S;\tag{$\mathrm{E}_{\Ll_S}$}\label{genES7}\\
    &&& \sum_{i \in A}|S_i\cap [m_{u_*}(\alpha)]| + \sum_{i\in N\setminus A}\Pp_{u_*}(\hpi_{u_*,i}^S, m_{u_*}(\alpha)) \leq |S\cap [m_{u_*}(\alpha)]|, & \forall \alpha \in \Qq.\tag{$\mathrm{S}*$}\label{genSstar7}
\end{align}
In particular,
\[
    \sum_{i \in N\setminus A}|\hpi_{u_*,i}|\leq q|N\setminus A| \leq \frac{\nu}{q}.
\]
Applying \cref{genprotpref} to $A$ and $\{\hpi_{u_*,i}\}_{i \in N\setminus A}$ with (\ref{genES7}), we get that for every $0 \leq \ell \leq m$,
\begin{equation}
    \sum_{i \in A}\Pp_i(\pi_i^E, \ell) +  \sum_{i \in N \setminus A}\Pp_i(\hpi_{u_*,i}^E, \ell) \leq |E\cap [\ell]|.\tag{E}\label{genE7}
\end{equation}
Next, (\ref{genSstar7}) together with $S$-compatibility allows us to apply \cref{gensprot}, giving for $0 \leq \ell \leq m$:
\begin{equation}
        \sum_{i \in A}\Pp_i(\pi_i^S, \ell) + \sum_{i \in N\setminus A}\Pp_i(\hpi_{u_*,i}^S, \ell) \leq |S\cap [\ell]|.\tag{S}\label{genS7}
\end{equation}
Finally, \cref{gencompatible} and $H^+$-compatibility imply for every $0 \leq \ell \leq m$:
\begin{equation}
    \sum_{i \in N\setminus A}\Pp_i(\hpi_{u_*,i}^+, \ell) \leq  \sum_{i \in N}\Pp_i(\hpi_{u_*,i}^+, \ell)\leq |H_{u_*}^+\cap [\ell]|.\tag{$+$}\label{genPlus}
\end{equation}
Combining (\ref{genE7}), (\ref{genS7}) and (\ref{genPlus}),  and using $E\sqcup S\sqcup H_{u_*}^+ = [h_{u_*}]$, gives for every $0 \leq \ell \leq m$:
\[
        \sum_{i \in A}\Pp_i(\pi_i, \ell) + \sum_{i \in N\setminus A}\Pp_i(\hpi_{u_*,i}, \ell) \leq |E\cap[\ell]| + |S\cap[\ell]| + |H_{u_*}^+\cap [\ell]| = \min\{\ell, h_{u_*}\}.
\]
\cref{loadequiv} realizes each sub-profile within its respective set, which completes the proof.
\end{proof}

\subsection{Selecting the supplier and recipients}

\cref{genepsnuconst} identifies the conditions that supplier $u_*$ and its suggested profiles $\{\hpi_{u_*,i}\}_{i \in N}$ must satisfy in order to complete initial assignment $\{\pi_i\}_{i \in I}$ for agents $I\subseteq N$ to a full feasible assignment of all agents $N$ (potentially replacing some old profiles with suggested ones).
We will describe how to choose a supplier and recipients of his suggested profiles, that are \textit{gencompatible}, i.e giving these agents corresponding profiles satisfies condition (\ref{genESu6}), as well as $S$- and $H^+$-compatibility conditions of \cref{gencompatible}.

Recall that for supplier $u_*$ with partition $\{B_{u_*,1}, \ldots, B_{u_*,n}\}$, a bundle $B_{u_*,j}$ is \textit{exceptional} if
\[
    B_{u_*,j}\cap S \neq \varnothing \qquad\text{or}\qquad v_{u_*}(B_{u_*,j}\cap [h]) < 1 - 1/q.
\]
Each of these exceptional bundles will be designated to a distinct \textit{exceptional recipient} $i \in N$, and corresponding suggested $S$- and $H_{u_*}^+$-parts of $\hpi_{u_*,i}$ are constructed simply as
\[
    \hpi_{u_*,i}^S = \{\wv_{u_*}(e) :\; e \in B_{u_*,j}\cap S\},\qquad\qquad \hpi_{u_*,i}^+ = \{\wv_{u_*}(e):\; e \in B_{u_*,j} \cap H_{u_*}^+\}.
\]
According to \cref{gencompatible}, for a supplier $u_*$ and recipient $i$ of $\hpi_{u_*,i}$, we require:
\begin{itemize}
    \item \textbf{$S$-compatibility}: for every $\alpha \in \hpi_{u_*,i}^S$ it holds $v_i(\max(S\cap[m_{u_*}(\alpha)])) \geq \alpha$;
    \item \textbf{$H^+$-compatibility}: for every $\alpha \in \hpi_{u_*,i}^+$ it holds $v_i(m_{u_*}(\alpha)) \geq \alpha$;
\end{itemize}
This suggests associating with every exceptional bundle $B_{u_*,j}$ a set of compatible agents, which we will refer to as \textit{eligible recipients} of $B_{u_*,j}$.
Then, when $u_*$ needs to designate an exceptional bundle  $B_{u_*,j}$ (and construct a profile), he can pick an arbitrary eligible recipient $i$ of $B_{u_*,j}$.

Note that eligible recipients of different bundles may coincide or overlap, as they essentially break all agents into a relatively small number of (possibly non-disjoint) classes.
Hence, choosing an eligible recipient separately for each exceptional bundle does not guarantee that these agents will be distinct.
A sufficient condition would be that every exceptional bundle has at least as many eligible recipients as the supplier has exceptional bundles.
We can then designate recipients one at a time, always finding an eligible agent who has not been used yet.

This property need not hold for an arbitrary supplier.
We therefore will choose $u_*$ from a sufficiently large set of supplier candidates, that at the same time will also serve as a pool of potential recipients of exceptional bundles.
The key observation here is that every supplier candidate is eligible to receive its own exceptional bundles.
Consequently, a candidate having an exceptional bundle with too few eligible recipients must himself belong to one of these small eligible-recipient sets.
Then, choosing a supplier $u_*$ outside all such small eligible-recipient classes will guarantee that each of its exceptional bundles will have a large eligible-recipient set.

We will show that the total number of different compatibility classes that exceptional bundles can induce is bounded.
Together with a bound of \cref{genexcep} on the number of exceptional bundles per candidate, this will allow us to bound the total number of supplier candidates belonging to the small eligible-recipient sets.
Then, if the pool of supplier candidates is sufficiently large, we will be able to choose a supplier outside of all these small sets.
Every exceptional bundle of this supplier will then have enough eligible recipients to allow distinct assignments.

Observe that this way, we can choose the supplier and the recipients of its exceptional bundles within any sufficiently large set of supplier candidates.
This motivates us to employ the following three-stage process of selecting the supplier and recipients and constructing suggested profiles.
\begin{enumerate}
    \item Choose a sufficiently large set $U$ of supplier candidates, such that for any candidate $u \in U$ we can always construct a collection $\{\hpi_{u,i}^E\}_{i \in N}$ of suggested $E$-profiles satisfying
    \begin{align}
        &&&\sum_{i \in N}\Pp_i(\hpi_{u,i}^E, c_s) \leq |E\cap [b_s]|,&\forall (b_s, c_s) \in \Ll_S.\tag{$\mathrm{E}_{\Ll_S}$-$u$}\label{genESu7}
\end{align}
\item Within this large set $U$ of supplier candidates, choose a supplier $u_* \in U$ whose exceptional bundles all have a large number of eligible recipients.
\item Having chosen the supplier $u_*$, designate every exceptional bundle of this supplier to a distinct eligible recipient, and construct $S$- and $H^+$-profiles according to \cref{gensuggestpart}.
Then, construct the corresponding suggested $E$-profiles $\{\hpi_{u_*,i}^E\}_{i \in N}$ satisfying (\ref{genESu7}).
\end{enumerate}

The construction of suggested $E$-profiles $\{\hpi_{u,i}^E\}_{i \in N}$ for a given supplier candidate $u$ will be explained in the next section.
The rest of this section will be dedicated to choosing the large set of supplier candidates $U$ and subsequent selection of the supplier $u_* \in U$.

\subsubsection{$E$-classes and supplier candidates}

In order to find a large set $U$ of supplier candidates, such that for any $u \in U$ there are suggested $E$-profiles  $\{\hpi_{u,i}^E\}_{i \in N}$ satisfying protected prefix inequalities (\ref{genESu7}),  we will partition all agents $N$ into classes, so that within each class, all agents are ``equivalent'' with respect to protected prefixes $(b_s, c_s) \in \Ll_S$.
We do it as follows.
Recall that set $S$ from \cref{genbuffer}, as well as internal protected prefix $c_s$, $s \in [t - 1]$, have the following form:
\[
    S = \bigcup_{s = 1}^t\{a_s, \ldots, b_s\},\qquad\text{and}\qquad c_s = \min\{b_s + \nu, a_{s + 1} - 1\},\; \forall s \in [t - 1].
\]
The last protected prefix is $(b_t, c_t) = (h, m)$ and will not affect our classification.
The set $S$ partitions items $E = [h]\setminus S$ into $t - 1$ intervals between consecutive components of $S$.
\begin{definition}
    For every $s \in [t - 1]$, define the $s$-th interval of $E$ as
    \[
    E_s := \{b_s + 1,\ldots, a_{s + 1} - 1\}.
    \]
    We associate each $e \in E$ with a \textbf{representative item} $c_s + 1$ corresponding to interval $E_s \ni e$.
    Then, we will classify agents $N$ based off their rounded values of these $t - 1$ representative items.
    
    For every agent $i \in N$ and every $s \in [t - 1]$, define the \textbf{$s$-th representative demand} as
    \[
        \alpha_{i,s} := \wv_{i}(c_s + 1).
    \]
    Note that $m_{i}(\alpha_{i,s}) \geq c_s + 1$ if $\alpha_{i,s} > 0$, so demand $\alpha_{i,s}$ contributes no load at protected prefix $c_s$.
\end{definition}
\begin{definition}
    For every agent $i \in N$, define his \textbf{$E$-signature} as 
    \[
        \fe_i := (\fe_{i}(1), \ldots, \fe_i(t - 1)),\quad \text{where}\quad \fe_i(s) := \begin{cases}
            \alpha_{i,s},& \alpha_{i,s} \geq 1/q;\\
            0, & \alpha_{i,s} < 1/q.
        \end{cases}
    \]
    Two agents belong to the same class if they have the same $E$-signature.
    For signature $\fe$, denote
    \[
        N_\fe := \{i \in N: \; \fe_i = \fe\}.
    \]
\end{definition}
The small representative demands $\alpha_{i,s} < 1/q$ will not be replaced by zeroes, we just use only high demands $\alpha_{i,s} \geq 1/q$ in order define the classes.

\begin{lemma}\label{gensignature}
    Let $\Phi = \{\fe_i : i \in N\}$ denote the set of all occurring $E$-signatures.
    Then
    \[
        |\Phi| \leq \phi := \binom{\lceil Q\cdot \frac{q}{q - 1}\rceil + Q}{Q}.
    \]
    Moreover, there exists a signature $\fe$, such that the set $U=  (G\setminus W)\cap N_\fe$ satisfies 
    \[
        |U|\geq \frac{n}{2\phi},
    \]
    for every $i \in N_\fe, s \in [t-1]$ we have $\alpha_{i,s} = \fe(s) \in \Qq$, and for every $u \in U, s \in [t -1]$ and $e \in E_s$,
    \[
        \wv_u(e) \leq \alpha_{u,s} = \fe(s).
    \]
    That is, rounded value of every $e \in E_s$ does not exceed that value of $c_s + 1$ for every agent $u \in U$.
\end{lemma}
\begin{proof}
    Since positions of representative items $c_s + 1$ increase with with $s$, for every agent $i \in N$ we have $\alpha_{i,1} \geq \alpha_{i,2} \geq \ldots \geq \alpha_{i, t - 1}$, and replacing small demands with $0$ does not affect the order.
    Hence, every signature $\fe_i$ is a non-increasing vector of length $t - 1$ with entries in $\Qq\cup \{0\}$.
    Every such vector is determined by the number of entries of each type, so by stars and bars the total number of possible signatures is (using the upper bound $t \leq \lceil Q\cdot \frac{q}{q - 1}\rceil + 1$ from \cref{genbuffer}):
    \[
        \binom{t - 1 + Q}{Q} \leq \binom{\lceil Q\cdot \frac{q}{q - 1}\rceil + Q}{Q} = \phi.
    \]

    Next, we find a signature class containing many supplier candidates.
    By \cref{notpoor}, there are at least $|G|\geq 3n/4$ good agents, and by \cref{genbuffer} we have $|W|\leq\eps d$.
    By the choice of parameters in \cref{genepsnuconst} we have $\eps \leq 1/4$, and since $d \leq n$,
    \[
        |G\setminus W| \geq |G| - |W| \geq \frac{3n}{4} - \eps d \geq \frac{n}{2}.
    \]
    Choose $\fe$ maximizing $|(G\setminus W)\cap N_\fe|$, and take $U := (G\setminus W)\cap N_\fe$.
    Since there are at most $\phi$ classes, it must hold that $|U|\geq n/(2\phi)$.

    We now show that agents $i \in N_\fe$ must fully agree on all representative demands, not just $\fe_i$-values.
    To see this, take $u \in U$.
    Bu construction, for every $s \in [t - 1]$ it holds $c_s + 1 \leq a_{s + 1} \leq h$.
    Since $u \in G$, by \cref{notpoor} we have $v_u(c_s + 1) \geq 1/q$, thus $\fe(s) = \alpha_{u,s} = \wv_u(c_s + 1) \in \Qq$.
    At the same time, for any other $i \in N_\fe$ we also must have $\alpha_{i,s} = \fe_i(s) = \fe(s) \in \Qq$.

    Finally, fix $u \in U, s \in [t - 1]$ and item $e \in E_s$, put $\alpha := \wv_u(e)$, note that $\alpha \in \Qq$.
    If $e \geq c_s + 1$, then we immediately have $\wv_u(e) \leq \wv_u(c_s + 1) = \alpha_{u,s}$.
    Otherwise, $b_s < e \leq c_s$ and $m_u(\alpha) \geq e > b_s$.
    Since $u \notin W$, none of $u$'s high-demand breakpoints belongs to $F$, in particular, $m_u(\alpha)\notin (b_s, c_s]$.
    Hence, $m_u(\alpha) \geq c_s + 1$ and therefore $\alpha_{u,s} = \wv_u(c_s + 1) \geq \alpha = \wv_u(e)$.
\end{proof}

We will show in the next section  that for every agent $u \in U$, where $U$ is chosen in \cref{gensignature}, it is possible to construct a collection $\{\hpi_{u,i}^E\}_{i \in N}$ of suggested $E$-profiles satisfying protected prefix inequalities (\ref{genESu7}).

\subsubsection{Compatibility classes and eligible recipients}

Fix signature $\fe$ and set $U = (G\setminus W)\cap N_\fe$, obtained in \cref{gensignature}.
All agents $u \in U$ agree on representative demands corresponding to the intervals $E_s$, $s \in [t - 1]$.
However, their values on $S$ as well as their high-demand breakpoints may differ.
We would like to find a supplier $u_* \in U$ such that each of $u_*$'s exceptional bundles has a lot of eligible recipients, i.e agents compatible with this bundle's $S$- and $H^+$-parts.
Then we will be able to designate all exceptional bundles of $u$ to distinct recipients, while preserving compatibility conditions of \cref{gencompatible}.

First, we will associate each exceptional bundle of every supplier candidate with a short description of the compatibility conditions the recipient of this bundle must satisfy.
\begin{definition}
    Consider an exceptional bundle $B_{u,j}$ of a supplier candidate $u \in U$.
    For every item $e \in B_{u,j}\cap (S\cup H_{u}^+)$, define the \textbf{reference item} of $e$ as
    \[
        r_u(e) := \begin{cases}
            \max(S\cap [m_u(\alpha)]),& e \in S;\\
            m_u(\alpha),& e \in H_u^+;
        \end{cases}\qquad\text{where}\qquad \alpha = \wv_u(e).
    \]
    So, for $e \in S, H_u^+$, the reference item $r_u(e)$ is the last item of respectively $S, H_u^+$ that can satisfy the demand induced by $e$ for agent $u$.
\end{definition}
\begin{observation}\label{genreference}
    For any item $e  \in B_{u,j}\cap (S\cup H_{u}^+)$, if $i \in N$ satisfies $v_i(r_u(e)) \geq \alpha$, then every item in the same set (i.e, $S$ or $H_u^+$) as $e$ that satisfies demand $\alpha$ for $u$ also satisfies it for $i$.

    Furthermore, there are at most $L := |S| + \lceil q^2(16\Lambda + 1)\rceil$ distinct reference items, and every reference item for every supplier candidate $u \in U$ belongs to the set
    \[
        S\cup \{h + 1, \ldots, \min(m, h + \lceil q^2(16\Lambda + 1)\rceil)\}.
    \]
\end{observation}
\begin{proof}
    In both $S$ and $H_u^+$, reference item $r_u(e)$ is the last item capable of satisfying $\alpha = \wv_u(e)$ for $u$.
    Then, if item $e'$ belongs to the same set ($S$ or $H_u^+$) as $e$ and $v_u(e') \geq \alpha$, then $e'\leq r_u(e)$, implying that $v_i(e') \geq v_i(r_u(e)) \geq \alpha$.

    Now, for $e \in S$ we also have $r_u(e) \in S$, and for $e\in H_u^+$ we have $\wv_u(e) \geq 1/q$, thus
    \[
        h < e \leq r_u(e) = m_u(\alpha) \leq m_u(1/q) = h_u,
    \]
    implying $r_u(e) \in H_u^+$.
    By \cref{uhead}, $h_u \leq h + q^2(16\Lambda + 1)$, so every reference item belongs to 
    \[
        S\cup \{h + 1, \ldots, \min(m, h + \lceil q^2(16\Lambda + 1)\rceil)\}
    \]
    regardless of $u$, and the total number of reference items is at most $L =  |S| + \lceil q^2(16\Lambda + 1)\rceil$.
\end{proof}
Note that by \cref{genexcep}, the total number of exceptional bundles of a single $u \in U$ is also at most $L$.
This observation will be useful later.

\begin{definition}
    For $u \in U$ and exceptional $B_{u,j}$, define its \textbf{compatibility record} as a multiset
    \[
        R_{u,j} := \big\{\big(r_u(e), \wv_u(e)\big):\; e \in B_{u,j}\cap (S\cup H_u^+) \big\}.
    \]   
    For $u, u'\in U$, their exceptional bundles $B_{u,j}$ and $B_{u', j'}$ are in the same \textbf{compatibility class} $R$ if $R_{u,j} = R = R_{u', j'}$.
    Let $\Rr$ denote the collection of all distinct compatibility records:
    \[
        \Rr = \{R_{u,j} :\; u \in U, j \in [n], B_{u,j}\text{ exceptional}\}.
    \]
\end{definition}
\begin{definition}
    For every compatibility record $R \in \Rr$, define its \textbf{eligible-recipient set} as 
    \[
        U(R) := \{i \in U : \; v_i(r) \geq \alpha \text{ for every }(r, \alpha) \in R\}.
    \]
    This way, agent $i \in U(R_{u,j})$ is an eligible recipient of the exceptional bundle $B_{u,j}$ of supplier $u$.
    Note that if $R_{u,j} = \varnothing$, then $U(R_{u,j}) = U$, i.e every agent is eligible for such bundle.
\end{definition}
\begin{observation}\label{geneligible}
    Fix some $u \in U$ and his exceptional bundle $B_{u,j}$.
    Then
    \begin{itemize}
        \item $u \in U(R_{u,j})$, i.e supplier is eligible for his own MMS bundles; and
        \item agent $i \in U$ is eligible for $B_{u,j}$ if and only if $i$ is $S/H^+$-compatible w.r.t respectively $\hpi_{u,i}^S/\hpi_{u,i}^+$.
    \end{itemize}
\end{observation}
\begin{proof}
    By construction, for every pair $(r, \alpha) \in R_{u,j}$ the reference item satisfies $r \leq m_u(\alpha)$.
    Hence, $v_u(r) \geq \alpha$, so $u \in U(R_{u,j})$ by definition.

    Next, by definition $i \in U(R_{u,j})$ requires that \begin{itemize}
        \item $v_i(\max(S\cap[m_u(\alpha)]))\geq \alpha$ for every demand $\alpha$ induced by $B_{u,j}\cap S$, and
        \item $v_i(m_u(\alpha)) \geq \alpha$ for every demand $\alpha$ induced by $B_{u,j}\cap H_u^+$.
    \end{itemize}
    These are exactly the $S/H^+$-compatibility conditions from \cref{gencompatible} with respect to corresponding profile parts $\hpi_{u,i}^S$ and $\hpi_{u,i}^+$ suggested by $u$ for agent $i$:
    \[
    \hpi_{u,i}^S = \{\wv_{u}(e) :\; e \in B_{u,j}\cap S\},\qquad\qquad \hpi_{u,i}^+ = \{\wv_{u}(e):\; e \in B_{u,j} \cap H_{u}^+\}.
    \]
\end{proof}
Note that that compatibility classes $\Rr$ classify exceptional bundles, whereas eligible-recipient sets $U(R)$ for $R \in \Rr$ are subsets of agents. 
Unlike $E$-classes introduced earlier, sets $U(R)$ need not be disjoint, and different compatibility records may even have the same eligible-recipient set.
This will not be an issue, as we will choose supplier $u_* \in U$ in such a way that every exceptional bundle of $u_*$ has a lot of eligible recipients.

Before explaining how to choose such supplier, we bound the number of compatibility classes.
\begin{lemma}\label{gencompnum}
    Every $R \in \Rr$ has size $|R| \leq q$, and contains at most $q - 1$ distinct reference items.
    Furthermore,
    \[
        |\Rr| \leq Q^q\sum_{r = 0}^{q - 1}\binom{L}{r}\binom{q}{r} \leq (2Q)^qL^{q - 1}.
    \]
\end{lemma}
\begin{proof}
    Fix some exceptional bundle $B_{u,j}$, each pair in $R_{u,j}$ corresponds to a distinct item in $B_{u,j} \cap (S\cup H_u^+)$.
    Since each of these items has value at least $1/q$, and $v_u(B_{u,j})\leq 1$, there are at most $q$ such items and $|R_{u,j}|\leq q$.
    
    Next, if $|R_{u,j}|\leq q - 1$, there are at most $q - 1$ reference items associated with $B_{u,j}$.
    Suppose that $|R_{u,j}| = q$, then every single positive-valued item in $B_{u,j}$ must have value exactly $1/q$.
    Consequently, all reference items of $B_{u,j}\cap S$ are the same, and all reference items of $B_{u,j}\cap H_u^+$ are the same, so there are at most two distinct reference items, which is at most $q - 1$.

    Now, consider a compatibility record $R$ with exactly $r$ distinct reference items, $0 \leq r \leq q - 1$.
    Since the total number of possible reference items is at most $L$, there are at most $\binom{L}{r}$ possible ways to choose reference items from $\Rr$ into record $R$.
    Once these reference items are fixed, there are at most $\binom{q}{r}$ possible ways to choose positive counts of reference items so they sum up to at most $q$.
    Finally, there are at most $Q^q$ options of assigning a value from $\Qq$ to each individual reference item occurrence.
    Summing over $0 \leq r\leq q - 1$, we get
    \[
        |\Rr| \leq Q^q\sum_{r = 0}^{q - 1}\binom{L}{r}\binom{q}{r}.
    \]
    Using standard binomial bounds, we have 
    \[
        \sum_{r = 0}^{q - 1}\binom{L}{r}\binom{q}{r} \leq L^{q - 1}\sum_{r = 0}^{q - 1}\binom{q}{r}\leq 2^qL^{q - 1},
    \]
    which finishes the proof.
\end{proof}
\cref{gencompnum} together with \cref{geneligible} imply that, if the set of supplier candidates $U$ is sufficiently large, we can always choose a supplier that will be able to designate all of his exceptional bundles to distinct agents.
Indeed, if $u \in U$ has an exceptional bundle with less than $L$ eligible recipients, then by \cref{geneligible} $u$ himself belongs to that small eligible-recipient set.
At the same time, the union of all possible eligible-recipient sets of size less than $L$ contains at most $(L - 1)|\Rr|$ agents.
Consequently, any $u_*$ outside of this union has at least $L$ eligible recipients for each of his exceptional bundles.
By \cref{genexcep}, $u_*$ has at most $L$ exceptional bundles, hence we can choose their recipients one a time from the corresponding eligible-recipient set, and there always will be enough recipients left.

We formalize this idea in the following theorem, establishing the choice of the supplier $u_* \in U$ and designation of its exceptional bundles to distinct recipients.
For future tail completion, it will be useful for us to assume that $U$ has some priority order, to guarantee that every designated recipient has priority at least as high as a supplier.
For now, the reader may assume that the order is arbitrary.
\begin{lemma}[Supplier choice]\label{genchoice}
    Fix any total priority order on $U$.
    If 
    \[
        |U|> (L - 1)|\Rr|,
    \]
    then there exists a supplier $u_* \in U$ and an assignment of its exceptional bundles $B_{u_*,j}$ to distinct recipients $i(j) \in U$ such that $i(j) \in U(R_{u_*,j})$ and every $i(j)$ has priority at least as high as $u_*$.
    In particular, the size condition above holds whenever
    \[
        n \geq 2\phi(2QL)^q.
    \]
\end{lemma}
\begin{proof}
    For every compatibility record $R \in \Rr$, let $U_L(R)\subseteq U(R)$ be the $L - 1$ highest-priority agents agents of $U(R)$, or all $U(R)$ if $|U(R)|< L - 1$.
    By \cref{gencompnum} and assumption on $|U|$,
    \[
        \left|\bigcup_{R \in \Rr}U_L(R)\right| \leq (L - 1)|\Rr|  < |U|.
    \]
    Choose an arbitrary agent $u_* \in U\setminus \bigcup_{R \in \Rr}U_L(R)$.
    Take an arbitrary exceptional bundle $B_{u_*,j}$, by \cref{geneligible}, $u_* \in U(R_{u_*,j})$.
    However, $u_*\notin U_L(R_{u_*,j})$, therefore $U_L(R_{u_*,j})$ must contain exactly $L - 1$ agents, as otherwise $U_L(R_{u_*,j})) = U(R_{u_*,j})$, leading to a contradiction.
    Moreover, every agent $u \in U_L(R_{u_*,j})$ has higher priority than $u_*$.
    So, the set $U_L(R_{u_*,j})\cup \{u_*\}$ contains exactly $L$ eligible recipients of priority at least as high as $u_*$.

    Now, by \cref{genexcep} $u_*$ has at most $L$ exceptional bundles.
    Process his exceptional bundles in order.
    For every exceptional $B_{u_*,j}$, consider $U_L(R_{u_*,j})\cup \{u_*\}$ and designate $B_{u_*,j}$ to an arbitrary agent of this set who has not yet been designated earlier.
    Since $|U_L(R_{u_*,j})\cup \{u_*\}| = L$, such agent will always exist, so every exceptional bundle of $u_*$ will eventually be assigned to a distinct recipient.

    Finally, by \cref{gensignature} we have $|U|\geq n/2\phi$, while by \cref{gencompnum} it holds
    \[
        (L - 1)|\Rr| \leq (L - 1)\cdot (2Q)^qL^{q - 1} < (2QL)^q.
    \]
    So $n \geq 2\phi (2QL)^q$ in combination with \cref{gencompnum} immediately implies $|U|> (L - 1)|\Rr|$.
\end{proof}

It follows that for large $n$ enough, we can choose a supplier $u_* \in U$ and construct $S$- and $H^+$-parts of suggested profile collection $\{\hpi_{u_*,i}\}_{i \in N}$, satisfying $S/H^+$-compatibility of \cref{genepsnuconst}.
\begin{corollary}\label{gensuplargen}
    Fix integers $\Gamma\geq 1, q\geq 9$, and choose parameters $\eps, \nu$ in \cref{genepsnuconst} as:
    \[
        \eps := \frac{1}{2q^2C_q\left(\left\lceil Q\cdot \frac{q}{q - 1}\right\rceil + 1\right)}, \qquad\qquad \nu := \lceil 2q^2C_q(\theta + \rho_{ES}(q, dn))\rceil.
    \]
    For sufficiently large $n$, there exists a supplier $u_* \in U$ and a designation of its exceptional bundles to distinct recipients in $U$, such that the suggested profile parts
    \[
        \hpi_{u_*,i}^S:= \{\wv_{u_*}(e):\; e \in B_{u_*,j}\cap S\},\qquad\qquad \hpi_{u_*,i}^+ := \{\wv_{u_*}(e):\; e \in B_{u_*,j}\cap H_{u_*}^+\}
    \]
    for every exceptional agent $i$ with designated bundle $B_{u_*,j}$, and $\hpi_{u_*,i}^S=\varnothing = \hpi_{u_*,i}^+$ for every non-exceptional agent $i$, satisfy 
    \begin{itemize}
        \item \textbf{$S$-compatibility}: for every $\alpha \in \hpi_{u_*,i}^S$ it holds $v_i(\max(S\cap[m_{u_*}(\alpha)])) \geq \alpha$;
        \item \textbf{$H^+$-compatibility}: for every $\alpha \in \hpi_{u_*,i}^+$ it holds $v_i(m_{u_*}(\alpha)) \geq \alpha$.
    \end{itemize}
    In addition, for any total priority order on $U$, supplier $u_*$ can be chosen so that every designated recipient $i$ of his exceptional bundles has priority at least as high as $u_*$.
\end{corollary}
\begin{proof}
    Since $C_q$ and $Q$ depend only on $q$, parameter $\eps > 0$ is a constant depending only on $q$.
    By \cref{notpoor}, $\Lambda = O_\Gamma(\log^2 n)$ and therefore
    \[
        \theta = q^2(16\Lambda + 1) + q\Lambda = O_{\Gamma,q}(\log^2 n).
    \]
    Furthermore, by $d\leq n$ we have $\rho_{ES}(q, dn) = O_q(\log n)$, hence our choice gives $\nu = O_{\Gamma,q}(\log^2 n)$.
    Then, by \cref{genbuffer}, we have
    \[
    |S| \leq \left(\left\lceil Q\cdot \frac{q}{q - 1}\right\rceil + 1\right)\left(1 + \frac{Q\nu}{\eps}\right) =  O_{\Gamma,q}(\log^2 n),
    \]
        implying
    \[
        L = |S| + \lceil q^2(16\Lambda + 1)\rceil = O_{\Gamma,q}(\log^2 n).
    \]
    By \cref{gensignature} and \cref{gencompnum}, it holds
    \[
        |U|\geq \frac{n}{2\phi},\qquad\text{and}\qquad (L -1 )|\Rr|\leq (2QL)^q = O_{\Gamma,q}(\log^{2q}n) = o(n),
    \]
    and since $\phi$ depends only on $q$, for all sufficiently large $n$ we have $|U| > (L - 1)|\Rr|$.
    It follows by \cref{genchoice} that there exists a supplier $u_* \in U$ and distinct eligible recipients $i(j) \in U$ of all exceptional bundles $B_{u_*,j}$ with priority not lower than of $u_*$.
    As shown in \cref{geneligible}, eligibility of each designated recipient implies precisely the $S/H^+$-compatibility conditions.
\end{proof}

\section{Suggested $E$-profiles for protected prefixes}
\label{sec:suggested}

In this section, we show that, for sufficiently large $n$, for any candidate $u \in U$ we can always construct a collection $\{\hpi_{u,i}^E\}_{i \in N}$ of suggested $E$-profiles satisfying
\begin{align}
        &&&\sum_{i \in N}\Pp_i(\hpi_{u,i}^E, c_s) \leq |E\cap [b_s]|,&\forall (b_s, c_s) \in \Ll_S.\tag{$\mathrm{E}_{\Ll_S}$-$u$}\label{genESu8}
\end{align}
This will give us the final missing condition of \cref{genepsnuconst}, and together with \cref{gensuplargen} establish that for the resulting supplier choice $u_* \in U$ and suggested collection $\{\hpi_{u_*,i}\}_{i \in N}$, there exists a set $A\subseteq I$ such that for every prefix $0 \leq \ell \leq m$ it holds
\[
        \sum_{i \in A}\Pp_i(\pi_i, \ell) + \sum_{i \in N\setminus A}\Pp_i(\hpi_{u_*,i}, \ell) \leq \min\{\ell, h_{u_*}\}.
\]

We will do it similarly to \cref{sec:intype} and \cref{sec:intsupplier}.
We will use repartition blocks of bounded size in order to construct, for every agent $i \in N$, a \textit{source bundle} $\hB_i$, which will define the $E$-part of the suggested profile $\hpi_{u_*,i}^E$.
For a fixed supplier $u_*$, for every agent $i \in N$ we will create a block repartition of the MMS partition $\{B_{u_*,1}, \ldots, B_{u_*,n}\}$ of the supplier, from which we will then extract disjoint source bundles using LP rounding for almost all agents.
As for the remaining agents, sufficiently large $n$ will guarantee that enough acceptable supplier bundles will remain for every agent, thus we can complete the allocation greedily.

The two main differences are that the MMS bundles $B_{i,j}$ of agents $i \in N$ may have arbitrary size, while the block-repartition relies on the fact that bundles are bounded.
Second, some of the rounded head profiles $\pi_{i,j}$ are invalid, which prevents us from using certain repartition blocks (in order not to give an agent a source bundle that is unacceptable).
We will address the first issue by constructing an auxiliary partition of items into bundles of bounded size, and the second by a similar block discarding method as in \cref{sec:intsupplier}.

\subsection{Bounded head partition and repartition blocks}

Throughout the construction, fix some supplier $u_* \in U$, together with a designation of all of his exceptional bundles to distinct recipients in $U$, as in \cref{genchoice}.
We will denote
\[
    \fe_* := \fe_{u_*}\qquad \text{and}\qquad V := N_{\fe_*},
\]
so $U\subseteq V$, and by \cref{gensignature}, $|V|\geq |U|\geq n/(2\phi)$.

\begin{definition}
    For agent $i \in N$, define the \textbf{representative valuation} by
    \[
        \hv_i(e) := \alpha_{i,s} = \wv_i(c_s +1),\qquad \qquad \forall e \in E_s,\; s \in [t - 1],
    \]
    and extending it additively.
    Set $\hv_i(e) = 0$ for $e \in S$ and any \textbf{dummy items} introduced below.

    For every $i \in N$ and $e \in E_s\setminus F$ we have $e \geq c_s + 1$, and therefore $\hv_i(e) \geq \wv_i(e)$.
    Additionally, by \cref{gensignature} every $e \in E$ satisfies $\hv_{u_*}(e) \geq \wv_{u_*}(e)$, including items of $F$.
\end{definition}

These representative valuations will determine demands of $E$-parts of suggested profiles.
They do not replace the original valuations, in particular, the assigned source bundle $\hB_i$ need not satisfy the profile $\hpi_{u_*,i}^E$ it defines.

We next construct an auxiliary partition of $[h]$ into bundles of size at most $q$.
\begin{observation}\label{aux}
    Let $\Omega$ be a set of $qn - h$ \textbf{dummy items}, disjoint from $M$, and put $\hM := [h]\sqcup \Omega$.
    For every agent $i \in N$, there exists a partition $\{\hB_{i,1}, \ldots, \hB_{i,n}\}$ of $\hM$ such that for every $j \in [n]$, $|\hB_{i,j}| = q$ and every item $e \in B_{i,j}\cap [h]$ with $v_i(e) \geq 1/q$ belongs to $\hB_{i,j}$.

    Consequently, if $\pi_{i,j}$ is valid, then $B_{i,j}\cap [h] \subseteq \hB_{i,j} \cap [h]$, and for every good agent $u \in G$ we additionally have $\hB_{u,j}\cap [h] = B_{u,j}\cap [h]$ for all $j \in [n]$.
\end{observation}
\begin{proof}
    Initially, start with $\hB_{i,j} := \{e \in B_{i,j}\cap [h]:\; v_i(e) \geq 1/q\}$, for every $j \in [n]$.
    Since $v_i(B_{i,j}) \leq 1$, clearly $|\hB_{i,j}|\leq q$.

    Suppose that $k$ common head items have been placed into auxiliary bundles this way.
    There are $qn - k$ available item slots and $h - k$ unallocated head items.
    Since $h \leq qn$, we can distribute all remaining head items among the bundles arbitrarily, and fill them in with dummy items $\Omega$, so that every bundle has size exactly $q$.

    Since every valid MMS profile $\pi_{i,j}$ of an agent contains only items worth at least $1/q$, the corresponding padded bundle $\hB_{i,j}$ preserves all the originating items of $B_{i,j}$.
    As a result, since for good agent $u \in G$ every item in $[h]$ has value at least $1/q$, every auxiliary bundle $\hB_{u,j}$ will receive all of the items $B_{u,j}\cap [h]$ first, before the remaining slots are filled.
\end{proof}

The dummy items are convenient, since they allow us to work with exactly $n$ nonempty bundles of size at most $q$ for every agent.
They serve only as a simplification tool, and will introduce no demands.

\begin{definition}
    We partition $\hM$ into the following $t + 1$ types $M_1, \ldots, M_t, M_{t + 1}$:
    \[
        \forall s \in [t- 1]:\; M_s := E_s\setminus F,\qquad M_t := S\cup F,\qquad M_{t + 1} = \Omega.
    \]
    Some of these types may be empty.
    For $\hB\subseteq \hM$, define its \textbf{pattern}
    \[
        p(\hB) := (|\hB \cap M_1|, \ldots, |\hB \cap M_{t + 1}|),\qquad\qquad |p| := \sum_{s =1}^{t + 1}p_s,
    \]
    and let $P := \{p \in \Z_{\geq 0}^{t + 1}: 1 \leq |p|\leq q\}$ contain all possible patterns.
    For $i \in N$ and $p \in P$, define
    \[
        \lambda_i(p) := |\{j \in [n]: \; p(\hB_{i,j}) = p\}|.
    \]
    In particular, we have $\sum_{p \in P}\lambda_i(p) = n$.
\end{definition}

Similarly to \cref{sec:intsupplier}, these types need not correspond to the original valuations.
Nevertheless, we can apply the bounded repartition theorem \cref{blockpartition}, as it relies only on the number of items of each type, the number of bundles and their size upper bound.
\begin{theorem}[\cref{blockpartition} restated]\label{genblock}
    For every pair of integers $t, q \geq 1$, there exists a global constant $D_{t + 1,q} \geq 1$, depending only on $t$ and $q$, such that the following holds.

    For a fixed supplier $u_*$ and every agent $i\in N$, there exists a sequence of $K_i \leq n$ blocks
    \[
        \beta_{i,1} = (x_{i,1}, y_{i,1}), \ldots, \beta_{i, K_i} = (x_{i, K_i}, y_{i, K_i}),
    \]
    such that for every $k \in [K_i]$ we have $|\beta_{i,k}|\leq D_{t + 1,q}$, and for every pattern $p \in P$ blocks satisfy
    \[
        \sum_{k = 1}^{K_i}x_{i,k}(p) = \lambda_{u_*}(p),\qquad \text{and}\qquad \sum_{k = 1}^{K_i}y_{i,k}(p) = \lambda_i(p).
    \]
    We can group auxiliary bundles of $u_*$ into disjoint blocks of size at most $D_{t + 1,q}$, so that every block can be repartitioned into the same number of bundles with patterns from $i$'s auxiliary partition.
\end{theorem}

We will choose a single representative agent from every non-empty $E$-class, different from the supplier class $V$.
We will choose representatives with only a few invalid profiles, in order to later bound the number of discarded blocks using \cref{invalid}.

\begin{definition}
    For every nonempty class $N_\fe$ with $\fe \in \Phi\setminus \fe_*$, choose \textbf{representative agent}
    \[
        i_\fe \in \argmin_{i \in N_\fe}\theta_i,\qquad\qquad n_\fe := |N_\fe|.
    \]
\end{definition}

For every class $\fe \in \Phi\setminus \fe_*$, we apply \cref{genblock} to the auxiliary partitions $\{\hB_{u_*,j}\}_{j = 1}^n$ of supplier $u_*$ and $\{\hB_{i,j}\}_{j = 1}^n$ of the representative agent $i_\fe$.
For every obtained block $\beta$, we will associate the output bundles with the corresponding auxiliary bundles $\hB_{i_\fe,j}$ that define the output pattern.
Note that each $j \in [n]$ will be used only once, since for every $p \in P$ the total number of produced bundles of pattern $p$ is exactly $\lambda_{i_\fe}(p)$.

\begin{definition}
    For every $\fe \in \Phi\setminus \fe_*$, let $\Bb(\fe)$ contain all blocks $\beta = \beta_{i_\fe,k}$ such that block $\beta$ \textbf{does not} produce a bundle $\hB_{i_\fe,j}$ such that profile $\pi_{i_\fe,j}$ is invalid.
    Denote the set
    \[
        \Bb = \bigsqcup_{\fe \in \Phi\setminus\fe_*}\Bb(\fe).
    \]
    Each block $\beta \in \Bb$ is associated with a corresponding signature class, as well as its corresponding index within that class, i.e we treat $\beta = \beta_{i_\fe,k} = (x, y)$ and $\beta'= \beta_{i_{\fe'}, k'} = (x', y')$ as different blocks when $\fe \neq \fe'$ or $k \neq k'$, even if $x = x'$ and $y = y'$.
\end{definition}

\begin{observation}\label{discardedblocks}
    For every $\fe \in \Phi\setminus \fe_*$ and a block $\beta = (x_{i_\fe,k}, y_{i_\fe,k})$ and pattern $p \in P$, let $\lambda_\beta(p) := x_{i_\fe,k}(p)$.
    Then, for every nonempty class $N_\fe$, $\fe\neq \fe_*$, and every pattern $p \in P$,
    \[
        \sum_{\beta \in \Bb(\fe)}\lambda_\beta(p) \leq \lambda_{u_*}(p).
    \]
    In addition, it holds
    \[
        n - D_{t + 1,q}\theta_{i_\fe} \leq \sum_{\beta \in \Bb(\fe)}|\beta| \leq n,\quad \text{and}\quad \frac{1}{n}\sum_{\fe\in \Phi\setminus \fe_*}n_\fe\cdot \theta_{i_\fe} \leq \theta.
    \]
\end{observation}
\begin{proof}
    Initially, before discarding ``invalid'' blocks, \cref{genblock} produces a collection that uses exactly $\lambda_{u_*}(p)$ copies of pattern $p \in P$, and the total size of all blocks together is exactly $n$.

    Since every output auxiliary bundle belongs to exactly one block, at most $\theta_{i_\fe}$ blocks will be discarded from $\Bb(\fe)$.
    Each of these blocks has size at most $D_{t + 1,q}$, hence discarding these blocks removes at most $D_{t+1,q}\theta_{i_\fe}$ output bundles.
    This gives us the corresponding bounds.

    Now, by the choice of $i_\fe$ it holds
    \[
        \theta_{i_\fe} \leq \frac{1}{n_\fe}\sum_{i \in N_\fe}\theta_i.
    \]
    Summing up over all non-supplier classes and applying \cref{invalid} gives us the desired bound.
\end{proof}

As in \cref{sec:intype}, each remaining block defines a repartition rule, and repeated copies of the same rule can simultaneously realize the corresponding output patterns via distinct supplier items.
The resulting output bundles may differ from the supplier's actual MMS bundles.

\subsection{Coordinating repartitions via LP}

Our next step is to coordinate the retained repartition rules across different $E$-classes.
As in \cref{sec:intype}, we preserve a fraction of every supplier bundle pattern, in order to later allocate the agents that are not covered by the rounded LP solution.
Define
\[
    \eta := \frac{n - d + |V|}{n},
\]
The $n - d$ term accounts for the fact that while there are $n$ bundles in each MMS partition, there are only $d$ remaining agents.
It holds
\[
    1 - \eta = \frac{d - |V|}{n}\qquad \text{and}\qquad \eta \geq \frac{|V|}{n} \geq \frac{1}{2\phi}.
\]

For every block $\beta \in \Bb$ we have a variable $z_\beta \geq 0$ that counts the number of realized copies of repartition rule $\beta$.
We consider the exact same LP as in \cref{sec:intype}.
\[
\begin{aligned}
    &\maximize&& \sum_{\beta \in \Bb}|\beta|\cdot z_\beta\\
    &\subto && \sum_{\beta \in \Bb}\lambda_\beta(p)\cdot z_\beta  \leq (1 - \eta)\lambda_{u_*}(p),&\forall p \in P;\\
    &&& \sum_{\beta \in \Bb(\fe)} |\beta|\cdot z_\beta \leq n_\fe,&\forall \fe \in \Phi \setminus \fe_*;\\
    &&& z_\beta \geq 0,&\forall \beta \in \Bb.
\end{aligned}\label{genlp}\tag{LP}
\]
Note that the optimum of (\ref{genlp}) no longer needs to be $d - |V|$, as we discarded certain blocks earlier.
Nevertheless, essentially the same proof as \cref{introunding} guarantees existence of an integral solution $\hz$ that allocates almost all non-supplier-class agents.
Specifically, the following lemma follows directly from the proof of \cref{introunding}.
\begin{lemma}\label{genintrounding}
    There exists an integral feasible solution $\hz$ to (\ref{genlp}) such that
    \[
    0 \leq d- |V| - \sum_{\beta \in \Bb}|\beta|\cdot \hz_\beta \leq D_{t+1,q}(\theta + |P| + |\Phi| - 1).
    \]
\end{lemma}
\begin{proof}
    For $\fe \in \Phi\setminus\fe_*$ and $\beta\in \Bb(\fe)$, set $z_\beta^0 = n_\fe/n$.
    By \cref{discardedblocks}, $z^0$ is a feasible solution of (\ref{genlp}) and 
    \[
        \sum_{\beta \in \Bb}|\beta|\cdot z_\beta^0 \geq d- |V| - \frac{D_{t+1,q}}{n}\sum_{\fe \in \Phi\setminus \fe_*}n_\fe\theta_{i_\fe} \geq d- |V| - D_{t+1,q}\theta.
    \]
    As in \cref{introunding}, the  optimal basic feasible solution to (\ref{genlp}) has at most $|P| + |\Phi| - 1$ positive coordinates.
    Rounding this optimal solution down to integral values loses at most $D_{t + 1,q}(|P| + |\Phi| - 1)$ in the objective.
    The lemma follows.
\end{proof}
Our next step is to realize these rules into actual source bundles for corresponding agents.
However, before describing the realization, we need to specify which supplier bundles must remain intact.
Concretely, we would like to preserve \textit{both} bundles intersecting $F$, for protected-prefix conditions, and also every exceptional bundle, in order to provide them to exceptional agents.
\begin{definition}
    For the supplier $u_*$ with MMS partition $\{B_{u_*,1},\ldots, B_{u_*,n}\}$, define
    \[
        \Jj := \{j \in [n]:\; B_{u_*,j}\text{ exceptional, or }B_{u_*,j}\cap F \neq \varnothing\}.
    \]
    Call $B_{u_*,j}$ and its corresponding auxiliary bundle $\hB_{u_*,j}$ \textbf{clean}, if $j \notin \Jj$.
    By \cref{genexcep},
    \[
        |\Jj|\leq L + |F|.
    \]
\end{definition}

By definition, every clean supplier bundle is disjoint from $M_t = S\cup F$.
\begin{corollary}\label{source}
    There exists a set $J\subseteq N\setminus V$ of disjoint source bundles $B_i\subseteq E\setminus F$, $i \in J$, and a set $Y\subseteq [n]$ of intact supplier bundles, such that
    \begin{itemize}
        \item for every $i \in J$ it holds $|B_i|\leq q$ and $\hv_i(B_i) \geq 1 - 2/q$;
        \item we have $|J| = d - |V| - \hZ$ where
        \[
            0 \leq \hZ \leq D_{t + 1,q}(\theta + |P| + |\Phi| -1 + L + |F|);
        \]
        \item the intact supplier bundles satisfy $|Y| = n - |J| = n - d + |V| + \hZ$, as well as $\Jj\subseteq Y$ and
        \[
            \bigcup_{i \in J}B_i = \bigcup_{j \in [n]\setminus Y}(B_{u_*,j}\cap [h]).
        \]
    \end{itemize}
    In addition, for every $p \in P$ the number of intact auxiliary supplier bundles of pattern $p$ is
    \[
        |\{j \in Y:\; p(\hB_{u_*,j}) = p\}| \geq \eta\cdot \lambda_{u_*}(p).
    \]
\end{corollary}
\begin{proof}
    Take the integral solution $\hz$ from \cref{genintrounding}.
    As in \cref{realization}, we can choose disjoint collections of auxiliary supplier bundles corresponding to every selected block copy.

    Similarly to \cref{partcon}, we discard every selected copy that contains supplier bundles from $\Jj$, this removes at most $|\Jj|$ selected block copies.
    Since each copy has size at most $D_{t+1,q}$, the resulting number of agents outside the supplier class $V$ that are covered by the remaining blocks is precisely $d - |V| - \hZ$ where
    \[
            0 \leq \hZ \leq D_{t + 1,q}(\theta + |P| + |\Phi| -1 + L + |F|).
    \]
    This will allow us to provide a source bundle for the $|J| = d - |V| - \hZ$ non-supplier class agents.
    Take a remaining block copy $\beta \in \Bb(\fe)$ for $\fe \in \Phi\setminus \fe_*$, the corresponding supplier bundles of $\beta$ do not intersect $M_t$.
    Hence, every bundle of type $p$ produced by realizing block $\beta$ must satisfy $p_t = 0$, for every $p \in P$.

    Use the type-preserving construction of \cref{bundleblock} to repartition the selected auxiliary supplier bundles of block $\beta$ into the corresponding output auxiliary bundles, and then remove all dummy items from each of them.
    Then, we obtain actual source bundles $B_i\subseteq E\setminus F$, each of size at most $q$, which we can assign simultaneously to distinct agents of $N_\fe$.
    We will now show that representative value of every source bundle is at least $1 - 2/q$.

    Consider the representative $i_\fe \in N_\fe$ for $\fe \in \Phi\setminus\fe_*$, and let $B_{i_\fe,j}$ be its MMS bundle that corresponds to a \textit{valid} profile $\pi_{i_\fe,j}$.
    Let $\hB_{i_\fe,j}$ be the corresponding auxiliary bundle from \cref{aux}, whose output pattern is prescribed by the selected block $\beta$.
    Then, by \cref{aux}, $B_{i_\fe,j}\cap [h]\subseteq \hB_{i_\fe,j}$.
    In addition, if $p = p(\hB_{i_\fe,j})$, then $p_t = 0$, so all items of $B_{i_\fe,j}\cap [h]$ lie outside $S\cup F$.
    For each such $e \in E_s\setminus F$,
    \[
        \alpha_{i_\fe,s} \geq \wv_{i_\fe}(e) \geq 1/q,
    \]
    therefore $\alpha_{i_\fe,s} = \fe(s)$ and its representative demand is shared by every agent of class $N_\fe$, i.e $\alpha_{i,s} \geq \fe(s)$ for all $i \in N_\fe$.
    It follows that for the source bundle $B_i$, obtained by realizing this pattern $p = p(\hB_{i_\fe,j})$ and removing dummy items,
    \[
        \hv_i(B_i) = \sum_{s = 1}^{t - 1}p_s\cdot \alpha_{i,s} \geq \sum_{s = 1}^{t - 1}p_s\cdot \fe(s) \geq \wv_{i_\fe}(B_{i_\fe,j}\cap [h]) \geq \left(1 - \frac{1}{q}\right)^2 \geq 1 - \frac{2}{q}.
    \]
    We use the fact that the profile $\pi_{i_\fe,j}$ is valid, as well as estimate for rounded high demands.

    Each non-discarded selected block repartitions the entire collection of corresponding supplier auxiliary bundles, into a collection of the exact same size.
    After removing dummy items, each produced bundle contains only items from the common head $[h]$ from the original supplier bundles, corresponding to the block.
    Therefore, 
    \[
            \bigcup_{i \in J}B_i = \bigcup_{j \in [n]\setminus Y}(B_{u_*,j}\cap [h]),
        \]
    and the number of intact supplier bundles is precisely $|Y| = n - d + |V| + \hZ$.
    By construction, none of the realized blocks used any bundles from $\Jj$.

    Finally, before discarding block copies, at least $\eta$-fraction of every supplier pattern $p$ remained intact.
    Discarding blocks only increases the number of unused bundles, completing the proof.
\end{proof}

\subsection{Completing the source-bundle assignment}

After applying \cref{source}, there are only $\hZ$ agents outside $V$ that have not been assigned a source bundle.
As in \cref{sec:intype}, we will give them acceptable bundles greedily from the pool of intact supplier MMS bundles.

We will first show that every agent $i \in N$ finds a lot of auxiliary bundles of the supplier's MMS partition acceptable w.r.t representative valuation $\hv_i$.
\begin{lemma}\label{repacc}
    For every agent $i \in N$, let $g_i$ be the number of auxiliary supplier bundles that are acceptable for $i$ w.r.t his representative valuation $\hv_i$:
    \[
        g_i := \big|\big\{j \in [n]:\; \hv_i(\hB_{u_*,j}\setminus M_t) \geq 1 - 2/q\big\}\big|.
    \]
    Then, it must hold that
    \[
        g_i \geq \frac{n/q^2 - |M_t|}{q - (1 - 2/q)}.
    \]
    In particular, if $n \geq 2q^2|M_t|$, then $g_i \geq n/(2q^3)$.
\end{lemma}
\begin{proof}
    Since for every $i \in N$ and every $e \in E_s\setminus F$ we have $\hv_i(e) \geq \wv_i(e)$, and dummy items have representative value $0$, \cref{aux} and \cref{roundedhead} gives us
    \[
        \sum_{j = 1}^n\hv_i(\hB_{u_*,j}\setminus M_t) = \hv_i(E\setminus F) \geq \wv_i(E\setminus F) \geq \wv_i([h]) - |M_t| \geq \left(1 - \frac{1}{q}\right)^2n - |M_t|.
    \]
    On the other hand, every auxiliary bundle has $q$ items, each of representative value at most $1$, so every summand is at most $q$.
    Since any summand not counted by $g_i$ is less than $1 - 2/q$,
    \[
       \left(1 - \frac{1}{q}\right)^2n - |M_t|\leq  \sum_{j = 1}^n\hv_i(\hB_{u_*,j}\setminus M_t) \leq g_iq + (n - g_i)\left(1 - \frac{2}{q}\right).
    \]
    The lemma follows.
\end{proof}

Note that for every $i \in N$, the acceptance condition in \cref{repacc} depends only on the pattern of the auxiliary supplier bundle:
\[
    \hv_i(\hB_{u_*,j}\setminus M_t) = \sum_{s = 1}^{t - 1}p_s(\hB_{u_*,j})\cdot \alpha_{i,s}.
\]
Hence, \cref{source} keeps at least $\eta g_i$ acceptable supplier bundles for every agent $i \in N$.

Note that some of these bundles may still belong to $\Jj$.
Hence, by removing at most $L + |F|$ bundles, we obtain a collection of \textit{clean} supplier MMS bundles for every agent.
This allows us to fully complete the construction of suggested $E$-profiles for large enough $n$.
\begin{theorem}\label{fullsuggested}
    Suppose that 
    \[
        n \geq 4\phi q^3D_{t+1,q}(\theta + |P| + |\Phi| - 1 + 2(L + |F|)).
    \]
    Then source bundles from \cref{source} extend to a collection of pairwise-disjoint source bundles $\{B_i\}_{i \in N}$, with $B_i\subseteq[h]$ for all $i \in N$, such that
    \begin{itemize}
        \item every exceptional recipient $i \in N$ with designated bundle $B_{u_*,j}$ has $B_i = B_{u_*,j}\cap [h]$;
        \item every non-exceptional recipient $i \in N$ has $B_i \subseteq E$, $|B_i|\leq q$ and $\hv_i(B_i) \geq 1 - 2/q$.
    \end{itemize}
    Furthermore, if we define the $E$-parts of suggested profiles $\hpi_{u_*,i}^E$ for $i \in N$ as 
    \begin{itemize}
        \item $\hpi_{u_*,i}^E = \{\wv_{u_*}(e):\; e \in B_{u_*,j}\cap E\}$ if $i$ is designated exceptional bundle $B_{u_*,j}$;
        \item $\hpi_{u_*,i}^E = \{\hv_i(e): \; e \in B_i, \hv_i(e) > 0\}$ if $i$ is non-exceptional,
    \end{itemize}
    and define $\hpi_{u_*,i}^S$ and $\hpi_{u_*,i}^+$ according to \cref{gensuggestpart}, then every suggested profile satisfies $|\hpi_{u_*,i}|\leq q$, every non-exceptional agent $i$ has $v(\hpi_{u_*,i}) \geq 1 - 2/q$, and the collection $\{\hpi_{u_*,i}^E\}_{i \in N}$ satisfies
    \begin{align}
        &&&\sum_{i \in N}\Pp_i(\hpi_{u_*,i}^E, c_s) \leq |E\cap [b_s]|,&\forall (b_s, c_s) \in \Ll_S.\tag{$\mathrm{E}_{\Ll_S}$-$u$}\label{genESu10}
\end{align}
\end{theorem}
\begin{proof}
    First, we will complete the remaining agents outside $V$.
    Since $|M_t| = |S| + |F|\leq L + |F|$, the bound on $n$ satisfies the requirements of \cref{repacc}, hence every agent accepts at least $n/(2q^3)$ auxiliary supplier bundles.
    \cref{source} guarantees that at least $\eta$-fraction of each pattern is intact.
    Thus, after discarding the bundles from $\Jj$, every remaining agent outside $V$ has at least
    \[
        \eta \cdot \frac{n}{2q^3} - (L + |F|) \geq \frac{n}{4\phi q^3} - (L + |F|)
    \]
    acceptable clean supplier bundles, which by the bound on $n$ is at least
    \[
        D_{t  + 1,q}(\theta + |P| + |\Phi| - 1 + L + |F|) \geq \hZ.
    \]
    Hence we can complete the $\hZ$ remaining agents outside $V$ greedily, assigning each a distinct acceptable clean supplier bundle $B_{u_*,j}$ and taking $B_i := B_{u_*,j}\cap [h]$.
    Since $B_{u_*,j}$ is clean, its head part lies completely in $E\setminus F$, and by \cref{repacc} its representative value is at least $1 - 2/q$.

    Next, we consider agents in $V$.
    After the previous step, exactly $n - d + |V|$ supplier bundles remain intact, including those corresponding to $\Jj$.
    For every exceptional agent $i \in V$ with designated bundle $B_{u_*,j}$, set $B_i := B_{u_*,j}\cap[h]$ and assigning it to $i$.
    Then, assign a distinct remaining non-exceptional bundles $B_{u_*,j}$ to each remaining agent $i \in V$, again setting  $B_i := B_{u_*,j}\cap[h]$.
    Note that we keep $n - d$ supplier bundles unused in this construction.

    First, let $i \in V$ be non-exceptional and receive and get assigned $B_{u_*,j}$.
    We take $B_i := B_{u_*,j}\cap [h]$, it must hold that $B_i\subseteq E$ and $v_{u_*}(B_i) \geq 1 - 1/q$.
    Since agents in $V$ have the same representative demands as $u_*$, by \cref{gensignature}
    \[
        \hv_i(B_i) = \hv_{u_*}(B_i) \geq \wv_{u_*}(B_i) \geq \left(1 - \frac{1}{q}\right)v_{u_*}(B_i) \geq \left(1 - \frac{1}{q}\right)^2 \geq 1 - \frac{2}{q}.
    \]
    All source bundles are pairwise disjoint, since we used disjoint collections of supplier bundles in our block repartitions, each belonging to $[h]$.

    Now, consider the suggested profiles $\{\hpi_{u_*,i}\}_{i \in N}$ as defined in the statement of the theorem.
    Every non-exceptional $i$ has at most one positive demand per every item of the source bundle, so $|\hpi_{u_*,i}|\leq |B_i|\leq q$ and $v(\hpi_{u_*,i}) = \hv_i(B_i) \geq 1 - 2/q$.
    As for the exceptional $i$ with designated bundle $B_{u_*,j}$, all its suggested parts together form
    \[
        \hpi_{u_*,i} = \{\wv_{u_*}(e) :\; e\in B_{u_*,j} \cap [h_{u_*}]\}.
    \]
    Every originating item in this profile has value at least $1/q$, while $v_{u_*}(B_{u_*,j})\leq 1$.
    Hence, the full suggested profile also satisfies $|\hpi_{u_*,i}|\leq q$.

    Finally, we will prove that these suggested profiles satisfy protected prefix inequalities.
    By construction, every suggested $E$-demand has a distinct originating item $e \in E$.
    Suppose that $e \in E_s$.
    If $i$ is not exceptional, then its demand is
    \[
        \alpha := \hv_i(e) = \alpha_{i,s} > 0 \quad \implies \quad m_i(\alpha) \geq c_s + 1.
    \]
    If $i$ is exceptional, then the demand of $e$ is $\alpha := \wv_{u_*}(e)$.
    Since $i \in V$, \cref{gensignature} gives $\alpha \leq \alpha_{u_*,s} = \alpha_{i,s}$, and once again $m_i(\alpha) \geq c_s + 1$.

    Now, for $r < t$, fix a protected pair $(b_r, c_r)$.
    A demand counted at prefix $c_r$ must originate from an item in some $E_s$ with $s < r$.
    Since all originating items are distinct,
    \[
        \sum_{i \in N}\Pp_i(\hpi_{u_*,i}^E, c_r) \leq \sum_{s < r}|E_s| = |E\cap [b_r]|.
    \]
    For the final pair $(b_t, c_t) = (h, m)$, there are at most $|E|$ suggested $E$-demands, each associated with distinct item of $E$.
    Thus,
    \[
        \sum_{i \in N}\Pp_i(\hpi_{u_*,i}^E, m) \leq |E| = |E\cap [h]|.
    \]
    This proves all protected prefix conditions.
\end{proof}

Note that the suggested $E$-parts $\hpi_{u_*,i}^E$ in \cref{fullsuggested} may contain demands that are worth less than $1/q$ for agent $i$.
This is still allowed, as \cref{gensupplierchange} requires demands from $\Qq$ only for the suggested $S$-parts.

Now, for fixed $\Gamma, q$, all the quantities $t, |P|, |\Phi|, D_{t + 1, q}$ are bounded constants of $\Gamma, q$, while all quantities $\theta, L, |F|$ are $O_{\Gamma, q}(\log^2 n)$.
Hence, the requirement of \cref{fullsuggested} holds for sufficiently large $n$.
We will show the exact dependence on $\Gamma, q$ in \cref{appendix:constant}.

We conclude this section with the following corollary.
\begin{corollary}[Feasibility on supplier's head]\label{finalhead}
    Choose $\eps, \nu$ in \cref{gensuplargen}.
    For sufficiently large $n$, and for any total priority order on $U$, there exist
    \begin{itemize}
        \item a supplier $u_* \in U$ and a designation of his exceptional bundles to distinct recipients in $U$ with priority at least that of $u_*$;
        \item a collection of suggested head profiles $\{\hpi_{u_*,i}\}_{i \in N}$ of size at most $q$; and 
        \item a set $A\subseteq I$ with $|N\setminus A|\leq \nu/q^2$,
    \end{itemize} 
    such that for every prefix $0 \leq \ell \leq m$ it holds
    \[
        \sum_{i \in A}\Pp_i(\pi_i, \ell) + \sum_{i \in N\setminus A}\Pp_i(\hpi_{u_*,i},\ell) \leq \min\{\ell, h_{u_*}\}.
    \]
    Consequently, these profiles can be simultaneously satisfied using distinct items of $[h_{u_*}]$, with their $E$-, $S$-, and residual-head parts satisfied within $E$, $S$, and $H_{u_*}^+$ respectively.
    
    In addition, every  $i \in A$ has $v(\pi_i) \geq 1 - 2/q$, every non-exceptional $i \in N\setminus A$ has $v(\hpi_{u_*,i}) \geq 1 - 2/q$; and every exceptional agent $i \in N\setminus A$ with designated supplier bundle $B_{u_*,j}$ satisfies
        \[
            v(\hpi_{u_*,i}) \geq \left(1 - \frac{1}{q}\right)\left(1 - \frac{1}{2q^2} - T_{u_*,j}\right),\quad \text{where}\quad T_{u_*,j} := v_{u_*}(B_{u_*,j}\setminus [h_{u_*}]).
        \]
        In particular, for these agents $i \in N\setminus A$ the corresponding \textbf{deficit} of $\hpi_{u_*,i}$ satisfies
        \[
             \left(1 - \frac{2}{q} - v(\hpi_{u_*,i})\right)_+ \leq \left(T_{u_*,j} - \frac{1}{q}\right)_+.
        \]
\end{corollary}
\begin{proof}
    We apply \cref{gensuplargen}, subject to the given total priority order.
    We obtain a supplier $u_*$ and designated recipients of $u_*$'s exceptional bundles, satisfying $S$- and $H^+$-compatibility conditions.
    Then, for sufficiently large $n$, \cref{fullsuggested} provides the suggested $E$-profiles $\{\hpi_{u_*,i}^E\}_{i \in N}$, while preserving the parts $\hpi_{u_*,i}^S$ and $\hpi_{u_*,i}^+$ of every exceptional agent $i$.
    These profiles have size at most $q$ and satisfy the protected prefix conditions
    \begin{align}
        &&&\sum_{i \in N}\Pp_i(\hpi_{u_*,i}^E, c_s) \leq |E\cap [b_s]|,&\forall (b_s, c_s) \in \Ll_S.\tag{$\mathrm{E}_{\Ll_S}$-$u$}\label{genESu11}
\end{align}
    Hence, we can apply \cref{genepsnuconst} to the collection $\{\hpi_{u_*,i}\}_{i \in N}$.
    We obtain a set $A\subseteq I$, such that $|N\setminus A|\leq \nu/q^2$ and for every prefix $0 \leq \ell \leq m$ it holds
    \[
        \sum_{i \in A}\Pp_i(\pi_i, \ell) + \sum_{i \in N\setminus A}\Pp_i(\hpi_{u_*,i},\ell) \leq \min\{\ell, h_{u_*}\}.
    \]
    For every agent $i \in A$ that keeps his old profile $\pi_i$, \cref{genESrounding} guarantees $v(\pi_i) \geq 1 - 2/q$.
    For every agent $i \in N\setminus A$ that is not exceptional, \cref{fullsuggested} assures that $v(\hpi_{u_*,i}) \geq 1 - 2/q$.

    Consider some exceptional agent $i$ that is designated a bundle $B_{u_*,j}$ of the supplier.
    By \cref{notpoor}, since supplier was chosen among the good agents $G$, it must hold that
    \[
        v_{u_*}(B_{u_*,j}) \geq 1 - \frac{1}{2q^2}.
    \]
    As a result,
    \[
        v(\hpi_{u_*,i}) = \wv_{u_*}(B_{u_*,j}\cap [h_{u_*}]) \geq \left(1 - \frac{1}{q}\right)v_{u_*}(B_{u_*,j}\cap [h_{u_*}]) \geq \left(1 - \frac{1}{q}\right)\left(1 - \frac{1}{2q^2} - T_{u_*,j}\right)
    \]
    where $T_{u_*,j} := v_{u_*}(B_{u_*,j}\setminus [h_{u_*}])$.
    Now, observe that if $T_{u_*,j} \leq 1/q$, then the inequality above guarantees
    \[
        v(\hpi_{u_*,i}) \geq \left(1 - \frac{1}{q}\right)\left(1 - \frac{1}{2q^2} - \frac{1}{q}\right) > 1 - \frac{2}{q}.
    \]
    If on the other hand $T_{u_*,j} \geq 1/q$, we get
    \[
        v(\hpi_{u_*,i}) + T_{u_*,j} - \frac{1}{q} \geq 1 - \frac{2}{q} + \frac{1}{q^2} - \frac{1 - 1/q}{2q^2} > 1 - \frac{2}{q}.
    \]
    These inequalities imply the upper bound on the missing value, which  finishes the proof.
\end{proof}
\section{Tail allocation and completion of the proof}
\label{sec:tail}

\cref{finalhead} establishes that for sufficiently large $n$, we can choose a supplier $u_* \in U$, set $A\subseteq I$ and suggested profiles $\{\hpi_{u_*,i}\}_{i \in N}$, so that for every prefix $0 \leq \ell \leq m$ it holds
\[
        \sum_{i \in A}\Pp_i(\pi_i, \ell) + \sum_{i \in N\setminus A}\Pp_i(\hpi_{u_*,i},\ell) \leq \min\{\ell, h_{u_*}\}.
\]
\begin{definition}
    For every agent $i \in N$, it will be convenient to define the assigned head profile
    \[
        \pi_i^* := \begin{cases}
            \pi_i,& i \in A;\\
            \hpi_{u_*,i}, &i \in N\setminus A.
        \end{cases}
    \]
    Thus, for every prefix $0 \leq \ell \leq m$,
    \[
        \sum_{i \in N}\Pp_i(\pi_i^*, \ell) \leq \min\{\ell, h_{u_*}\}.
    \]
\end{definition}
Using \cref{loadequiv}, this proves that we can simultaneously satisfy all these profiles within the supplier head $[h_{u_*}]$.
Every agent $i \in A$ is guaranteed that $v(\pi_i^*) \geq 1 - 2/q$.
However, not every agent from $N\setminus A$ is guaranteed to receive an acceptable suggested head profile $\hpi_{u_*,i}$, in particular, \cref{finalhead} ensures this only for non-exceptional agents $i \in N\setminus A$.
For every exceptional agent, however, we obtain a bound on the deficit of their suggested profile with respect to the supplier's tail value.
\begin{definition}
    Let $U \ni u_*$ be the the set of supplier candidates from \cref{gensignature}, and let $J\subseteq (N\setminus A)\cap U$ be the set of all exceptional agents that did not receive an acceptable profile:
    \[
        J := \{i \in (N\setminus A)\cap U:\; v(\pi_i^*) < 1 - 2/q\}.
    \]
    Note that by construction of \cref{finalhead}, every such agent must be an exceptional agent.
    For each $i = i(j) \in J$, we associate a corresponding exceptional bundle $B_{u_*,j}$.
\end{definition}
\begin{definition}
    In this section, we measure the \textbf{deficit} of a head profile $\pi$ as
    \[
        \Delta(\pi) := (1 - 1/q - v(\pi))_+.
    \]
\end{definition}
\begin{observation}\label{defbound}
    For every $i = i(j) \in J$ it holds
    \[
        \Delta(\pi_i^*) \leq T_{u_*,j}\qquad \text{where}\qquad T_{u_*,j} = v_{u_*}(B_{u_*,j}\setminus[h_{u_*}]).
    \]
    It immediately follows that
    \[
        \sum_{i \in J}\Delta(\pi_i^*) \leq T_{u_*} =: v_{u_*}(M\setminus [h_{u_*}]).
    \]
\end{observation}
\begin{proof}
    By \cref{finalhead}, for every exceptional agent $i$ with designated bundle $B_{u_*,j}$,
    \[
        \left(1 - \frac{2}{q} - v(\hpi_{u_*,i})\right)_+ \leq \left(T_{u_*,j} - \frac{1}{q}\right)_+.
    \]
    For $i = i(j) \in J$ we have $1 - 2/q - v(\pi_i^*) > 0$, which by \cref{finalhead} forces $T_{u_*,j} > 1/q$ and
    \[
        \Delta(\pi_i^*)  - \frac{1}{q}= 1 - \frac{2}{q} - v(\pi_i^*) \leq T_{u_*,j} - \frac{1}{q}.
    \]
    The observation follows by $\sum_{i = i(j) \in J}T_{u_*,j} \leq T_{u_*}$.
\end{proof}
Note that only agents $i \in J$ miss value to $1 - 2/q$.
For each of these agents, we will design an appropriate tail profile $\psi_i$, as in \cref{sec:dmms}, and give an empty tail profile for every other agent, so that for every $i \in N$ we have $v(\pi_i^*) + v(\psi_i) \geq 1 - 2/q$.
We will assume that $T_{u_*} > 0$, as otherwise $J = \varnothing$.

\subsection{Supplier selection for tail completion}

For every agent $i \in N$, denote $T_i := v_i(M\setminus [h_i])$, where $h_i = |\{e \in M: v_i(e) \geq 1/q\}|$.
Recall that in \cref{sec:dmms}, where we also needed to complete the tail, agent $i$ with profile $\pi$ tried to compensate for his deficit $\Delta(\pi)$ by taking a $\tau_i := \tau_i(\pi)$-fraction of every tail item, where
\[
        \tau_i(\pi) := \begin{cases}
            \Delta(\pi)/T_i,& \pi\neq \bot\text{ and }T_i > 0;\\
            0,& \text{otherwise}.
        \end{cases}
\]
In our previous work \cite{FG26}, we used a similar tail completion to \cref{sec:dmms}.
Concretely, we constructed a tail profile by taking the last item in each tail segment of length roughly $1/\tau_i$.
This costs at most the value of one tail item, which is at most $1/q$.
\cref{sec:dmms} takes the first item of each segment, which is allowed in that setting due to stronger slackness guarantees, so in that case no additional loss of $1/q$ occurs.

We would like to repeat this construction on the common tail $\{h_{u_*} + 1, \ldots, m\}$.
However, an item after $h_{u_*}$ may still be a high-value item for some agent $i \in J$, hence have value greater than $1/q$.
So, using the old construction directly risks losing more than $1/q$ value total.

We would like to construct tail profiles $\psi_i$ with demands at most $1/q$.
Then, from the perspective of tail profiles, all items of value more than $1/q$ are essentially the same.
This gives us the following idea, used in \cite{FG26}.
In order to keep loss at $1/q$, we will cap the values of all head items to $1/q$, just for the sake of the tail completion step.
Concretely, for every agent $i \in N$ and item $e\in M$, define valuation
\[
    \ov_i(e) := \min\{v_i(e), 1/q\},
\]
and extend it additively.
We will construct tail demands based off the item values according to capped valuations $\ov_i$.
Note that all the definitions of head profiles, breakpoints and prefix loads still refer to original uncapped valuations $v_i$.

Observe that for every $i \in N$ it holds
\[
    \ov_i(M) = \sum_{e = 1}^m\min\{v_i(e), 1/q\} = \frac{h_i}{q} + T_i.
\]
This is the total capped value of items $M$ for agent $i$ that can in theory be used for his tail profile, however, in our case we can only use items $\{h_{u_*} + 1, \ldots, m\}$ of the common tail.
Depending on $h_i$ and $h_{u_*}$, common tail may contain some high value items of agent $i$, or exclude some tail items of agent $i$.
The value $h_i/q + T_i$ in a sense captures the available ``tail completion budget'', and allows us to compare between the two cases.

At the same time, for every agent $i \in N$ the first $h_{u_*}$ items have capped value at most $h_{u_*}/q$.
So, the the tail completion budget of $i$, i.e the capped value of the common tail, is at least
\[
    \sum_{e > h_{u_*}}\ov_i(e) = \ov_i(M) - \ov_i(\{1, \ldots, h_{u_*}\}) \geq \frac{h_i}{q} + T_i - \frac{h_{u_*}}{q}.
\]
By \cref{defbound}, the total deficit of agents $i \in J$ satisfies $\sum_{i \in J}\Delta(\pi_i^*) \leq T_{u_*}$.
Hence, in order to guarantee that every agent $i \in J$ has large enough tail completion budget to cover the whole deficit, it suffices to ensure that he values the common tail $\{h_{u_*} + 1, \ldots, m\}$ at least $T_{u_*}$.
In particular, this is guaranteed when
\[
    \frac{h_i}{q} + T_i \geq \frac{h_{u_*}}{q} + T_{u_*}.
\]

This inequality specifies our total priority order on $U$ for \cref{finalhead}, we give a higher priority to $i \in U$ with larger $\ov_i(M) = h_i/q + T_i$.
Both \cref{genchoice} and \cref{finalhead} apply with any total priority order.
This gives us a corresponding supplier $u_* \in U$, the designation of all his exceptional bundles, and the assigned head profiles $\{\pi_i^*\}_{i \in N}$.
All the preceeding guarantees hold for these choices.
Now, every exceptional agent $i \in J$ has priority at least as high as $u_*$, thus satisfies
\[
    \ov_i(\{h_{u_*} + 1, \ldots, m\}) \geq  \frac{h_i}{q} + T_i - \frac{h_{u_*}}{q} \geq T_{u_*}.
\]
It follows that, in order to cover the deficit $\Delta(\pi_i^*) > 0$ of agent $i \in J$, instead of picking $\tau_i$-fraction of every common tail item, he can pick $\tau_{i,*}$-fraction, where
\[
    \tau_{i,*} := \begin{cases}
        \Delta(\pi_i^*)/T_{u_*},& i \in J;\\
        0, & i \notin J.
    \end{cases}
\]
Indeed, the derivations above imply
\[
    \sum_{e > h_{u_*}}\tau_{i,*}\cdot \ov_i(e)\geq \tau_{i,*}\cdot T_{u_*} = \Delta(\pi_i^*).
\]
Furthermore, \cref{defbound} guarantees us that
\[
    \sum_{i \in J}\tau_{i,*} = \sum_{i \in J}\frac{\Delta(\pi_i^*)}{T_{u_*}} \leq 1,
\]
so all fractional choices of all agents $i \in J$ fit within the common tail.

To keep the paper self-contained, we repeat the proof of the tail-completion from \cite{FG26}.
\begin{lemma}\label{commontailcomplete}
    Let $\Pi_N := \{\pi_i^*\}_{i \in N}$.
    There exists a collection $\Psi_N := \{\psi_i\}_{i \in N}$ of tail profiles, where agent $i$ is assigned $\psi_i$, such that
    \[
        \forall 0 \leq\ell \leq m:\; \Pp(\Pi_N, \ell) + \Pp(\Psi_N, \ell) \leq \ell
    \]
    and $v(\pi_i^*) + v(\psi_i) \geq 1 - 2/q$ for every $i \in N$.
\end{lemma}
\begin{proof}
    By assumption, $T_{u_*} > 0$.
    Consider $i \in N$, if $\tau_{i,*} = 0$, also take $\psi_i = \varnothing$.
    Since for $i \in J$ we have $\Delta(\pi_i^*) > 1/q$ and hence $\tau_{i,*} > 0$, $\tau_{i,*} = 0$ implies $i \notin J$.
    So, it remains to construct tail profiles for $i \in J$ with $\tau_{i,*} > 0$.

    Break the common tail $\{h_{u_*} + 1, \ldots, m\}$ into segments of length roughly $1/\tau_{i,*}$, and in each segment add the \textit{capped} value $\ov_i$ of the last item to $\psi_i$: for every $t = 1, \ldots, \lfloor \tau_{i,*}\cdot (m - h_{u_*})\rfloor$, we add $\ov_i(e_{i,t})$ to $\psi_i$ (as long as it is positive), where
    \[
        e_{i,t} = h_{u_*} + \left\lceil \frac{t}{\tau_{i,*}}\right\rceil.
    \]
    We can represent this construction equivalently as follows.
    For $1 \leq k \leq m - h_{u_*}$, let $\sigma_{i,k} := \ov_i(h_{u_*} + k)$.
    We also set $\sigma_{i,m-h_{u_*}+1} = 0$.
    Then, for every $k \in [m - h_{u_*}]$, we add exactly
    \[
        \lfloor\tau_{i,*}\cdot k \rfloor - \lfloor \tau_{i,*}\cdot (k - 1)\rfloor
    \]
    copies of $\sigma_{i,k}$ to the profile $\psi_i$ (if it is positive).
    This way, the first $k$ common tail items contain at most $\lfloor \tau_{i,*}\cdot k\rfloor$ selected items from $\psi_i$.
    Therefore,
    \begin{multline*}
        v(\psi_i) = \sum_{k = 1}^{m - h_{u_*}}\sigma_{i,k}\cdot \big(\lfloor \tau_{i,*}\cdot k\rfloor - \lfloor\tau_{i,*}\cdot (k - 1)\rfloor\big) = \sum_{k = 1}^{m - h_{u_*}}\lfloor \tau_{i,*}\cdot k\rfloor\cdot \big( \sigma_{i,k}- \sigma_{i,k + 1}\big)\\
        \geq \sum_{k = 1}^{m - h_{u_*}}(\tau_{i,*}k - 1)\cdot \big( \sigma_{i,k}- \sigma_{i,k + 1}\big) = \tau_{i,*}\sum_{k = 1}^{m - h_{u_*}}\sigma_{i,k} - \sigma_{i,1}.
    \end{multline*}
    Since $\tau_{i,*} > 0$ implies $i \in J$, we have $h_i /q + T_i \geq h_{u_*}/q + T_{u_*}$ and
    \[
        \tau_{i,*}\sum_{k = 1}^{m - h_{u_*}}\sigma_{i,k} = \tau_{i,*}\cdot \ov_i(\{h_{u_*} + 1,\ldots, m\}) \geq \tau_{i,*}\cdot T_{u_*} = \Delta(\pi_i^*),
    \]
    therefore, using $\sigma_{i,1} = \ov_i(h_{u_*} + 1) \leq 1/q$, and the fact that sequence is non-increasing, 
    \[
        v(\psi_i) \geq \Delta(\pi_i^*) - \sigma_{i,1} \geq \Delta(\pi_i^*) - 1/q.
    \]
    Since $\Delta(\pi_i^*) = (1 - 1/q - v(\pi_i^*))_+$, we get $v(\pi_i^*) + v(\psi_i) \geq 1 - 2/q$.

    It remains to bound the additional prefix load generated by $\Psi_N = \{\psi_i\}_{i \in N}$.
    For every $0 \leq \ell \leq h_{u_*}$ we have $\Pp_i(\psi_i, \ell) = 0$, for every agent $i \in N$.
    When $\ell > h_{u_*}$, the number of selected items from $\psi_i$ within the first $k = \ell - h_{u_*}$ common tail items is at most $\lfloor\tau_{i,*}\cdot k\rfloor \leq \tau_{i,*}\cdot k$.
    In addition, since $\sigma_{i,k} = \ov_i(h_{u_*} + k) \leq v_i(h_{u_*} + k)$, every selected tail demand that is nonzero satisfies $m_i(\sigma_{i,k}) \geq h_{u_*} + k$.
    Hence, only demands originating from items at most $\ell$ can contribute to the prefix load at $\ell$, and the additional load generated by tail profiles $\Psi_N$ at prefix $\ell$ is
    \[
        \Pp(\Psi_N, \ell)  = \sum_{i \in N}\Pp_i(\psi_i, \ell) \leq \sum_{i \in N}\tau_{i,*}\cdot (\ell - h_{u_*})_+ =(\ell - h_{u_*})_+\sum_{i \in J}\frac{\Delta(\pi_i^*)}{T_{u_*}}\leq (\ell - h_{u_*})_+.
    \]
    By \cref{finalhead}, all demands in $\Pi_N$ can be simultaneously satisfied by distinct items in $\{1, \ldots, h_{u_*}\}$, it holds 
    \[
        \Pp(\Pi_N, \ell) \leq \min\{\ell, h_{u_*}\}.
    \]
    for every $0 \leq \ell \leq m$. Consequently,
    \[
        \Pp(\Pi_N, \ell) + \Pp(\Psi_N, \ell) \leq \min\{\ell, h_{u_*}\} + (\ell - h_{u_*})_+ = \ell
    \]
    for every $0 \leq \ell \leq m$, as desired.
\end{proof}

\subsection{Finishing the proof}

With \cref{commontailcomplete} equipped, we are finally ready to prove \cref{qmms}.
\begin{proof}[Proof of \cref{qmms}]
    Fix integers $\Gamma \geq 1$ and $q \geq 9$, and take $n$ sufficiently large for all preceding constructions.
    For each supplier candidate $u \in U$, we give $u$ higher priority when the total capped value
    \[
        \ov_u(M) = \frac{h_u}{q} + T_u
    \]
    is larger, breaking ties arbitrarily.
    Apply \cref{finalhead} with this priority order, obtaining a supplier $u_*$, the designation of his exceptional bundles to corresponding recipients, and the assigned head profiles $\Pi_N = \{\pi_i^*\}_{i \in N}$.

    If $J = \varnothing$, every assigned head profile already has value at least $1 - 2/q$, so put tail profiles $\psi_i = \varnothing$ for all $i \in N$.
    Otherwise, by \cref{defbound} $T_{u_*} > 0$, and we can apply \cref{commontailcomplete}.
    In either case, we obtain a collection $\Psi_N = \{\psi_i\}_{i \in N}$ of tail profiles such that for every $0 \leq \ell \leq m$,
    \[
        \Pp(\Pi_N, \ell) + \Pp(\Psi_N, \ell) \leq \ell,
    \]
    and for every $i \in N$ it holds $v(\pi_i^*) + v(\psi_i) \geq 1 - 2/q$.
    Hence, applying \cref{loadequiv} to the collection of joint profiles $\{\pi_i^*\uplus \psi_i\}_{i \in N}$, we can simultaneously satisfy all these profiles using distinct items of $M$.
    Distribute all unused items arbitrarily.
    This gives us an allocation $\{B_i\}_{i \in N}$ of $M$ such that $v_i(B_i) \geq 1 - 2/q$ for every $i \in N$.

    All the statements hold for sufficiently large $n$ depending only on $\Gamma, q$.
    In \cref{appendix:constant} (\cref{finalconstant}) we show that specific choice
    \[
        n(\Gamma, q) = 2^{4q^2\log^2(2q)}(\log^{2q}(2\Gamma))
    \]
    suffices.
    This finishes the proof of \cref{qmms}.
    To recover the original \cref{deltammsproof}, combine this allocation with the allocation of head-poor agents from \cref{poor}.
    Both allocations use disjoint sets of items, therefore the combined allocation is valid.
\end{proof}

\bibliographystyle{alpha}
\bibliography{main}

\appendix
\crefalias{section}{appendix}

\section{Proofs of fractional profile rounding theorems}
\label{appendix:rounding}

\begin{theorem}[\cref{etarounding} restated]\label{etaroundingproof}
    Let $\Pi$ be a shared set of head profiles, such that $|\pi|\leq h$ for some integer $h \geq 1$.
    For every $i \in N$, let $(x_i, \Pi)$ be a fractional profile, denote $x = \{x_i\}_{i \in N}$.

    There exists a collection $y = \{y_i\}_{i \in N}$ of integer head profiles $(y_i, \Pi)$, such that for every $i, \pi$ we have $y_i(\pi) \in \{0, 1\}$ and $y_i(\pi) = 1$ for $\pi \neq \bot$ only if $x_i(\pi) > 0$, and
    \[\begin{aligned}
        & && \Hh(y, \Pi, \ell) \leq \Hh(x, \Pi, \ell),  & \forall 0 \leq \ell \leq m,\\
        & && \tau(y, \Pi) \leq\tau(x, \Pi),\\
        & && \Ii(y, \Pi) \leq \Ii(x, \Pi),\\
        &  && \Ii(y, \Pi)\geq \Ii(x, \Pi) - \rho(h, X), 
    \end{aligned}\]
    where $X = |\{(i, \pi) :\; 0 < x_i(\pi) < 1\}|$ and
    \[
        \rho(h, X) := 8 + 4\log \max\{1, X\}\cdot (1 + \ln h + \ln \max\{1, X\}).
    \]
\end{theorem}
\begin{proof}
If $X = 0$, $x$ is already integral, so we can take $y = x$.
From now on, assume $X\geq 2$.
We will round the starting fractional profiles $x$ (which we will refer to as a ``solution'') to an integral $y$.
The rounding algorithm will operate in phases, each phase $k$ will start with the current solution $x^k$ and obtain $x^{k + 1}$.
Every entry in $x^{k + 1}$ corresponding to some profile $\pi \neq \bot$ that reaches $0$ or $1$ will be fixed for subsequent phases.
Throughout the phase, we maintain for every $i \in N$ that $x_i^k(\pi) \geq 0$ for all $\pi \in \Pi$, and $\sum_{\pi \in\Pi}x_i^k(\pi) = 1$.
Let $\Xx_k := \{(i, \pi) : 0 < x_i^k(\pi) < 1\}$ be the set of strictly fractional entries in $x^k$ at the beginning of phase $k$, and let $X_k := |\Xx_k|$.
Initially, we start with $x^1 := x$ and $X_1 := X$.

Suppose that $X_k > 16$.
Before rounding some of the entries of $x^k$ to $0$ or $1$, we will create additional slackness in every prefix $0 \leq \ell \leq m$, that can absorb potential increase in the prefix load caused by the rounding.
Denote 
\[
    p_k := \frac{4(1 + \ln (hX_k))}{X_k}.
\]
We create an \textbf{intermediate solution} $\hx^k$ as follows.
For every $(i, \pi) \in \Xx_k$, where $\pi \neq \bot$, we transfer $\min\{x_i^k(\pi), p_k\}$ weight from $x_i^k(\pi)$ into $x_i^k(\bot)$.
That is,
\[
    \forall (i, \pi) \in \Xx_k, \pi \neq \bot:\;\hx^k_i(\pi) = x^k_i(\pi) - \min\{x_i^k(\pi), p_k\},\quad \hx_i^k(\bot) = x_i^k(\bot) + \sum_{\substack{(i, \pi) \in \Xx_k\\\pi\neq \bot}}\min\{x_i^k(\pi), p_k\}.
\]

Let $\hXx_k = \{(i, \pi) :0 < \hx^k_i(\pi) < 1\}$ be the strictly fractional entries in $\hx^k$.
If $\hXx_k = \varnothing$, then all entries of $\hx^k$ are integral.
Then we set $x^{k + 1} = \hx^k$ and move to the next phase.

If $\hXx_k \neq \varnothing$, every pair $(i, \pi) \in \hXx_k$ with $\pi\neq \bot$ had $x_i^k(\pi) > p_k$, and lost exactly $p_k$ when moving to $\hx^k$.
For $(i, \pi) \in \hXx_k$ with $\pi \neq \bot$, we list all the breakpoints of $i$ for all demands $\alpha \in \pi$, and combine them into a multiset $\Mm_k$:
\[
    \Mm_k := \biguplus_{\substack{(i, \pi) \in \hXx_k\\\pi \neq \bot}}\{m_i(\alpha) : \alpha \in \pi\}.
\]
\begin{lemma}\label{etatransfer}
    It holds $\Ii(\hx^k, \Pi) \geq \Ii(x^k, \Pi) - p_kX_k$ and $\tau(\hx^k, \Pi) \leq \tau(x^k, \Pi)$.
    Furthermore, 
    \[
        \forall 0 \leq \ell \leq m:\qquad \Hh(\hx^k, \Pi, \ell) + p_k\Pp_k(\ell) \leq \Hh(x^k, \Pi, \ell),
    \]
    where for every $-1 \leq \ell \leq m$,
    \[
        \Pp_k(\ell) = \sum_{(i, \pi) \in \hXx_k}\Pp_i(\pi, \ell)
    \]
    is the total number of breakpoints in $\Mm_k$ that are at most $\ell$, with a convention $\Pp_k(-1) = 0$.
\end{lemma}
\begin{proof}
    Since each non-$\bot$ entry in $x^k$ loses at most $p_k$, the bound $\Ii(\hx^k, \Pi) \geq \Ii(x^k, \Pi) - p_kX_k$ is immediate.
    Similarly, moving weight to $\bot$ can only decrease prefix loads, so $\tau(\hx^k, \Pi) \leq \tau(x^k, \Pi)$.

    Finally, by construction of $\hx^k$, for every $0 \leq \ell \leq m$ we have
\[
    \Hh(x^k, \Pi, \ell) - \Hh(\hx^k, \Pi, \ell) = \sum_{i \in N}\sum_{\pi\neq \bot}\big(x_i^k(\pi) - \hx_i^k(\pi)\big)\cdot \Pp_i(\pi, \ell) \geq \sum_{(i, \pi) \in \hXx_k}p_k\cdot \Pp_i(\pi, \ell),
\]
which is exactly $p_k\cdot \Pp_k(\ell)$ by definition.
\end{proof}

Next, we will create a small set of \textbf{head checkpoints}, i.e a subset of prefixes $\ell$ for which we would like to preserve the inequality $\Hh(x^{k + 1}, \Pi, \ell) \leq \Hh(x^k, \Pi, \ell)$ no matter what happens in the rounding.
We do it as follows.
First, sort all the breakpoints in $\Mm_k$ in non-decreasing order, and let $m_{i_j}$ denote the $j$-th breakpoint in $\Mm_k$.
For every $1 \leq j \leq |\Mm_k|$, we associate a weight $1/j$ with a breakpoint $m_{i_j}$.
We start with a virtual checkpoint $t = -1$.
Then, we find a smallest prefix $\ell \in \{0, \ldots, m\}$ such that $\ell > t$ and the total weight of breakpoints between $t + 1$ and $\ell$ in $\Mm_k$ is greater than $p_k$.
That is, $\ell > t$ is the smallest satisfying
\[
    \sum_{j = \Pp_k(t) + 1}^{\Pp_k(\ell)}\frac{1}{j} > p_k.
\]
We put a checkpoint at $\ell$, increase $t$ to $\ell$, and repeat the process until no such $\ell > t$ exists.
We denote the set of obtained checkpoints as $\Ll_k$ (we do not include $-1$).

\begin{lemma}\label{etacheckpoints}
    The total number of checkpoints at phase $k$ is $|\Ll_k|\leq X_k/4$.
    In addition, for any solution $z$ satisfying $z_i(\pi) = \hx_i^k(\pi)$ for all $(i, \pi)$ such that $\hx_i^k(\pi) \in \{0, 1\}$, the following holds.

    Suppose that for every $\ell_k \in \Ll_k$ we have $\Hh(z, \Pi, \ell_k) \leq \Hh(\hx^k, \Pi, \ell_k)$.
    Then, for every $0 \leq \ell \leq m$,
    \[
        \Hh(z, \Pi, \ell) \leq \Hh(\hx^k, \Pi, \ell) + p_k\Pp_k(\ell).
    \]
\end{lemma}
\begin{proof}
If $\Mm_k = \varnothing$, take $\Ll_k = \varnothing$, then all $(i, \pi) \in \hXx_k$ have no breakpoints at any prefix $\ell$, in which case lemma holds trivially.
From now on, assume $\Mm_k \neq \varnothing$, so $\Pp_k(m) \geq 1$.
We will start with the number of checkpoints.
By construction of $\Ll_k$, we have
\[
    |\Ll_k|\cdot p_k \leq \sum_{j = 1}^{\Pp_k(m)}\frac{1}{j} \leq 1 + \ln\Pp_k(m)\quad\implies\quad |\Ll_k|\leq \frac{1 + \ln\Pp_k(m)}{p_k}.
\]
By assumption, every profile $\pi \in \Pi - \bot$ has $|\pi|\leq h$.
In addition, every coordinate $(i, \pi) \in \hXx_k$ with $\pi \neq \bot$ was strictly fractional in solution $x^k$ as well.
Therefore
\[
    \Pp_k(m) = |\Mm_k| = \sum_{\substack{(i, \pi) \in \hXx_k\\\pi\neq\bot}}|\pi| \leq \sum_{\substack{(i, \pi) \in \Xx_k\\\pi\neq\bot}}|\pi| \leq hX_k,
\]
and substituting the definition of $p_k = (1/X_k)\cdot 4(1 + \ln (hX_k))$, we get
\[
    |\Ll_k| \leq \frac{1 + \ln\Pp_k(m)}{p_k} \leq \frac{X_k(1 + \ln(hX_k))}{4(1 + \ln (hX_k))} = \frac{X_k}{4}.
\] 

Next, for $0 \leq \ell \leq m$, let $\ell_k \in \Ll_k$ be the largest checkpoint preceding $\ell$, i.e satisfying $\ell_k \leq \ell$ (set $\ell_k = -1$ if none exists).
Since $z$ coincides with $\hx^k$ on all coordinates outside of $\hXx_k$, it holds
\begin{multline*}
    \Hh(z, \Pi, \ell) - \Hh(\hx^k, \Pi, \ell) = \Hh(z, \Pi, \ell_k) - \Hh(\hx^k, \Pi, \ell_k) \\
    + \sum_{(i, \pi) \in \hXx_k}\big(z_i(\pi) - \hx_i^k(\pi)\big)\cdot \big(\Pp_i(\pi, \ell) - \Pp_i(\pi, \ell_k)\big).
\end{multline*}
By assumption, $\Hh(z, \Pi, \ell_k) \leq \Hh(\hx^k, \Pi, \ell_k)$.
As both $z_i(\pi)$ and $\hx^k_i(\pi)$ lie in $[0, 1]$, we have
\begin{multline*}
    \Hh(z, \Pi, \ell) - \Hh(\hx^k, \Pi, \ell)\leq \sum_{(i, \pi) \in \hXx_k}|z_i(\pi) - \hx_i^k(\pi)|\cdot \big(\Pp_i(\pi, \ell) - \Pp_i(\pi, \ell_k)\big)\\
    \leq \sum_{(i, \pi) \in \hXx_k} \big(\Pp_i(\pi, \ell) - \Pp_i(\pi, \ell_k)\big) = \Pp_k(\ell) - \Pp_k(\ell_k).
\end{multline*}
If $\ell_k = \ell$ then $\Pp_k(\ell) - \Pp_k(\ell_k) = 0$, and we are done.
Otherwise, by construction of $\Ll_k$ we have
\[
    \sum_{j = \Pp_k(\ell_k) + 1}^{\Pp_k(\ell)}
    \frac{1}{j} \leq p_k.
\]
When $\Pp_k(\ell) > 0$, each summand above is at least $1/\Pp_k(\ell)$, hence 
\[
    \frac{\Pp_k(\ell) - \Pp_k(\ell_k)}{\Pp_k(\ell)} \leq \sum_{j = \Pp_k(\ell_k) + 1}^{\Pp_k(\ell)}
    \frac{1}{j} \leq p_k \quad\implies\quad \Pp_k(\ell) - \Pp_k(\ell_k) \leq p_k\Pp_k(\ell).
\]
We conclude that $\Hh(z, \Pi, \ell) - \Hh(\hx^k, \Pi, \ell) \leq p_k\Pp_k(\ell)$, which finishes the proof.
\end{proof}

It follows from \cref{etatransfer} and \cref{etacheckpoints} that, to preserve the prefix load constraints at every prefix $0 \leq \ell \leq m$ at phase $k$, it suffices to preserve only the checkpoint prefixes $\Ll_k$.

We will now construct an auxiliary polytope (P) with a relatively small number of constraints, and partially round the solution $\hx^k$ to one of its extreme points, i.e some $z^*$, that will satisfy \cref{etacheckpoints}.
The polytope (P) is defined by the following constraints.
\begin{equation}
        \begin{aligned}
        & && \sum_{\pi \in \Pi}z_i(\pi) =1,&\forall i \in N,\\
        & && z_i(\pi) = \hx^{k}_i(\pi),&\forall (i, \pi): \hx_i^k(\pi) \in \{0, 1\},\\
        & && \tau(z, \Pi) \leq \tau(\hx^k, \Pi),\\
        & && \Hh(z, \Pi, \ell_k) \leq \Hh(\hx^k, \Pi, \ell_k),&\forall \ell_k \in \Ll_k,\\
        & && 0 \leq z_i(\pi) \leq 1,&\forall i \in N, \pi \in \Pi.
    \end{aligned}\tag{P}\label{P}
    \end{equation}
Clearly, $z = \hx^k$ is feasible for (P).
We choose an extreme point $z^*$ of (P) maximizing 
\[
    \Ii(z^*, \Pi) = \sum_{i \in N}\sum_{\pi\neq \bot}z_i^*(\pi).
\]
Such an extreme point must exist, because (P) is nonempty and compact, i.e bounded and closed.
Then, by maximality $\Ii(z^*, \Pi) \geq \Ii(\hx^k, \Pi)$.
\begin{lemma}\label{etaextreme}
    Let $z^*$ be an extreme point of the polytope $\mathrm{(P)}$.
    Then, the number of strictly fractional entries in $z^*$ is at most $X_k/2 + 2$.
\end{lemma}
\begin{proof}
    Let $Z^* := |\{(i, \pi) : 0 < z_i^*(\pi) < 1\}|$ be the total number of strictly fractional coordinates in $z^*$.
    Note that if for some $i \in N$ we have $0 < z_i^*(\pi') < 1$ for some $\pi' \in \Pi$, there must be another $\pi''\in \Pi$ with $0 < z_i^*(\pi'') < 1$, as $\sum_{\pi \in \Pi}z_i^*(\pi)= 1$.
    Hence, there can be at most $Z^*/2$ different agents $i \in N$ with $0 < z_i^*(\pi) < 1$ for some $\pi \in \Pi$, and for the remaining agents $z_i^*(\pi) \in \{0, 1\}$ for every coordinate $(i, \pi)$.

    Suppose that $Z^* > X_k/2 + 2$.
    Restrict the equation-matrix of the polytope (P) to the entries $(i, \pi)$ with $0 < z_i^*(\pi) < 1$, it is a system of equations with $Z^*$ variables.
    It has at most $Z^*/2$ constraints of the form $\sum_{\pi \in \Pi}z_i^*(\pi) = 1$ for some $i \in N$.
    We have an additional constraint on $\tau(z^*, \Pi)$, and the number of constraints on $\Hh(z^*, \Pi, \ell_k)$ for $\ell_k \in \Ll_k$ is $|\Ll_k|\leq X_k/4$ by \cref{etacheckpoints}.
    So, if $Z^* > X_k/2 + 2$, the total number of constraints in this restriction is at most
    \[
        \frac{Z^*}{2} + 1 + |\Ll_k|\leq \frac{Z^*}{2}  + \frac{4 + X_k}{4} < \frac{Z^*}{2} + \frac{2Z^*}{4} = Z^*.
    \]
    Therefore, the restricted matrix has rank less than $Z^*$, implying that there is a non-zero solution supported on strictly fractional entries of $z^*$ that preserves all constraints.
    All these entries are strictly between $0$ and $1$, hence small perturbations of both signs still keep $z^*$ inside the polytope.
    This contradicts the extremality of $z^*$.
    The lemma follows.
\end{proof}
Take this extreme point $z^*$ of (P) and set $x^{k + 1} = z^*$.
By \cref{etatransfer} and \cref{etacheckpoints},
\[
    \forall 0 \leq \ell \leq m:\; \Hh(x^{k + 1}, \Pi, \ell)\leq \Hh(x^k, \Pi, \ell),\qquad\text{and}\qquad \tau(x^{k+ 1}, \Pi) \leq \tau(x^k, \Pi),
\]
as well as
\[
    \Ii(x^k, \Pi) -\Ii(x^{k + 1}, \Pi)\leq p_kX_k= 4(1 + \ln(hX_k)).
\]

By \cref{etaextreme}, $X_{k + 1} \leq X_k/2 + 2$, hence by induction on $k$ it holds for every $k \geq 1$ that
\[
    X_k \leq 4 + (X - 4)\cdot 2^{-(k -1)}.
\]
Then, if at phase $K + 1$ we have $X_{K + 1} \leq 16$, it must hold $K \leq \log X$.
At this phase, solution $x^{K + 1}$ has at most $16$ strictly fractional coordinates, hence there are at most $8$ agents with a strictly fractional coordinate.
We create a new solution $y$ as follows: for every $i \in N$ with $0 < x_i^{K + 1}(\pi) < 1$ for some $\pi \in \Pi$, set $y_i(\bot) = 1$, i.e throw these agents away.
For every other agent $i \in N$ set $y_i(\pi) = x_i^{K + 1}(\pi)$ for all $\pi \in \Pi$.
Clearly, $y$ is integral,  every head/tail load can only decrease, and
\[
    \Ii(y, \Pi) \geq \Ii(x^{K + 1}, \Pi) - 8.
\]
Summing up the loss in $\Ii(x^k, \Pi)$ over all phases $k \in [K]$, we get (using $X_k \leq X$)
\begin{multline*}
    \Ii(x, \Pi) - \Ii(y, \Pi) \leq 8 + \sum_{k = 1}^{K}\big(\Ii(x^k, \Pi) - \Ii(x^{k + 1}, \Pi)\big) \leq 8 + 4\sum_{k = 1}^K(1 + \ln(hX_k))\\
    \leq 8 + 4\log X\cdot (1 + \ln(hX)) = \rho(h, X).
\end{multline*}
To ensure that $\Ii(y, \Pi) \leq \Ii(x, \Pi)$, in case $\Ii(y, \Pi) > \Ii(x, \Pi)$ we retain any $\lfloor \Ii(x, \Pi)\rfloor$ agents assigned by $y$, and give $\bot$ to the rest.
Removing agents preserves prefix load constraints, and we still have at least $\Ii(x, \Pi) - 1$ selected agents.
Finally, by construction the agents $i \in N$ with $y_i(\pi) = 1$ for some $\pi \neq\bot$ must have had $x_i(\pi) > 0$.
This finishes the proof.
\end{proof}

\begin{theorem}[\cref{rounding}, \cref{genrounding} restated]\label{roundingproof}
    For a partition $[h] = E\sqcup S$ (where we assume $h = m$ unless specified), let $\Pi$ be a shared set of head profiles with well-defined $E/S$-decomposition, such that for every $\pi \in \Pi\setminus \bot$ it holds $|\pi|\leq q$ for some integer $q \geq 1$.
    For every $i \in N$, let $(x_i, \Pi)$ be a fractional profile, denote $x = \{x_i\}_{i \in N}$. 

    There exists a collection $y = \{y_i\}_{i \in N}$ of integer head profiles $(y_i, \Pi)$, such that for every $i, \pi$ we have $y_i(\pi) \in \{0, 1\}$ and $y_i(\pi) = 1$ for $\pi \neq \bot$ only if $x_i(\pi) > 0$, and
    \[\begin{aligned}
        & && \Hh^E(y, \Pi, \ell) \leq \Hh^E(x, \Pi, \ell),  & \forall 0 \leq \ell \leq m,\\
        & && \Hh^S(y, \Pi, \ell) \leq \Hh^S(x, \Pi, \ell),  & \forall 0 \leq \ell \leq m,\\
        &  && \Ii(y, \Pi)\geq \Ii(x, \Pi) - \rho_{ES}(q, X), 
    \end{aligned}\]
    where, for $X= |\{(i, \pi):0 < x_i(\pi) <1 \}|$,
    \[
        \rho_{ES}(q, X) =  8q\log \max\{1, X\}.
    \]
\end{theorem}
\begin{proof}
    The proof of this theorem is a straightforward extension of \cref{etaroundingproof}, and follows a similar structure.
    We start with a solution $x^1 = x$ and round it to integral $y$ over multiple phases, fixing non-$\bot$ entries of current solution $x^k$ that reach $0$ or $1$, and maintaining for every $i \in N$ that $x_i^k(\pi) \geq 0$ for all $\pi \in \Pi$ and $\sum_{\pi \in \Pi}x_i^k(\pi) = 1$.
    The key two differences are that 1) we preserve two head-load families, and 2) we use the bound of $|\pi|\leq q$ to obtain a constant loss of at most $8q$ per phase, which allows us to obtain total loss $O(q\log X)$.
    For the most part, we will keep this proof self-contained, with only minor references to the proof of \cref{etaroundingproof}.

    Once again, let $\Xx_k := \{(i, \pi) : 0 < x_i^k(\pi) < 1\}$ be strictly fractional entries of $x^k$, and $X_k := |\Xx_k|$, with $x^1 = x$ and $X_1 = X$.
    If $X = 0$, take $y = x$, otherwise $X \geq 2$.

    Suppose that $X_k > 0$.
    As in \cref{etaroundingproof}, we construct an intermediate solution $\hx^k$.
    For part $R \in \{E, S\}$ and profile $\pi \in \Pi$ with $\pi^R\neq \varnothing$, let
    \[
        b_i^R(\pi) := \min_{\alpha \in \pi^R}m_i(\alpha)
    \]
    be the earliest breakpoint of agent $i$ in sub-profile $\pi^R$.
    We perform the following operation, first for $R = E$ and then for $R = S$.
    We sort all pairs $(i, \pi) \in \Xx_k$ with $\pi \neq \bot$ and $\pi^R\neq \varnothing$ with respect to $b_i^R(\pi)$ in non-decreasing order.
    Then, starting from the first pair, we keep transferring weight $x_i^k(\pi)$ to $x_i^k(\bot)$ until either a total weight of $4q$ has been transferred, or all non-$\bot$ weights in this sorted list are set to $0$.
    Every transfer, except for possibly the last one, completely exhausts the corresponding coordinate $x_i^k(\pi)$.
    After doing so for $R = E$, apply the exact same process for $R = S$ to the solution obtained after $R = E$.
    We denote the resulting solution by $\hx^k$, and $\hXx_k := \{(i, \pi) : 0 < \hx_i^k(\pi) < 1\}$ and $\hX_k := |\hXx_k|$.

    As in the proof of \cref{etaroundingproof}, for every $R\in \{E, S\}$ and every $(i, \pi) \in \hXx_k$ with $\pi \neq \bot$, we list all the breakpoints of $i$ for all demands $\alpha \in \pi^R$, and combine them into a multiset $\Mm_k^R$:
    \[
        \Mm_k^R := \biguplus_{\substack{(i, \pi) \in \hXx_k\\\pi \neq \bot}}\{m_i(\alpha) : \alpha \in \pi^R\}.
    \]
    As before, for $-1 \leq \ell \leq m$ we denote by
    \[
        \Pp_k^R(\ell) = \sum_{(i, \pi)\in \hXx_k}\Pp_i(\pi^R, \ell)
    \]
    the total number of breakpoints in $\Mm_k^R$ that are at most $\ell$, with $\Pp_k^R(-1) = 0$.
\begin{lemma}\label{ktransfer}
        It holds $\Ii(\hx^k, \Pi) \geq \Ii(x^k, \Pi) - 8q$ and $\Hh^R(\hx^k, \Pi, \ell) \leq \Hh^R(x^k, \Pi, \ell)$ for every prefix $0 \leq \ell \leq m$ and every part $R \in \{E, S\}$.
        In addition, if $\Pp_k^R(\ell) > 0$, then
        \[
            \Hh^R(\hx^k, \Pi, \ell) + 4q \leq \Hh^R(x^k, \Pi, \ell).
        \]
    \end{lemma}
    \begin{proof}
        Every claim except for the last bound on head prefix loads is immediate.
        Fix $R\in \{E, S\}$.

        Let $\ell$ be a prefix such that $\Pp_k^R(\ell) > 0$.
        Then, there is $(i_*, \pi_*) \in \hXx_k$ such that $\Pp_{i_*}(\pi_*^R, \ell) > 0$, and the earliest breakpoint of $\pi^R_*$ satisfies $b_{i_*}^R(\pi_*) \leq \ell$.
        As $\hx^k_{i_*}(\pi_*) > 0$, we must have $x^k_{i_*}(\pi_*) > 0$, and every pair $(i, \pi) \in \Xx_k$ from which positive weight was transferred into $\bot$ w.r.t $R$-part must have occurred before $(i_*, \pi_*)$ (or ended at $(i_*, \pi_*))$.
        Then, for every such $(i, \pi)$ it holds $b_{i}^R(\pi) \leq b_{i_*}^R(\pi_*) \leq \ell$, and hence $\Pp_i(\pi^R, \ell) \geq 1$.

        Since $\hx_{i_*}^k(\pi_*) > 0$, the $R$-transfer stopped before zeroing-out all weights, and exactly $4q$ total weight was transferred to $\bot$.
        As we have just shown, each unit of this weight decreased the $R$-load by at least $1$.
        Therefore, the $R$-transfer decreased that load by at least $4q$, and the other transfer could only decrease it further.
        Consequently, 
        \[
            \Hh^R(x^k, \Pi, \ell) - \Hh^R(\hx^k, \Pi, \ell) = \sum_{i \in N}\sum_{\pi\neq \bot}\big(x_{i}^k(\pi) - \hx^k_i(\pi)\big)\cdot \Pp_i(\pi^R, \ell) \geq 4q.
        \]
    \end{proof}
    
    As in \cref{etaroundingproof}, for each $R \in \{E, S\}$, we construct \textbf{head checkpoints}.
    Instead of using harmonic checkpoints, bounded profile size allows us to take uniformly separated breakpoints.
    
    For $R \in \{E, S\}$, sort $\Mm_k^R$ in non-decreasing order, and then choose every $4q$-th breakpoint in $\Mm_k^R$ as a head checkpoint.
    Let $\Ll_k^R$ be the set of chosen checkpoints.

\begin{lemma}\label{headcheckpoints}
    The total number of head checkpoints at phase $k$ satisfies 
    \[
      |\Ll_k^E| + |\Ll_k^S| \leq \frac{X_k}{4}.  
    \]
    In addition, for any solution $z$ satisfying $z_i(\pi) = \hx_i^k(\pi)$ for all $(i, \pi)$ with $\hx_i^k(\pi) \in \{0, 1\}$, the following holds.
    Suppose that for each $R \in \{E, S\}$ and every $\ell_k^R \in \Ll_k^R$ we have $\Hh^R(z, \Pi, \ell_k^R) \leq \Hh^R(\hx^k, \Pi, \ell_k^R)$.
    Then, for each $R \in \{E, S\}$, for every prefix $0 \leq \ell \leq m$,
    \[
        \Hh^R(z, \Pi, \ell) \leq \Hh^R(\hx^k, \Pi, \ell) + 4q\One\{\Pp_k^R(\ell) > 0\}.
    \]
\end{lemma}
\begin{proof}
    This is an analogue of earlier \cref{etacheckpoints}.
    By construction, every fractional non-$\bot$ coordinate in $\hx^k$ was fractional in $x^k$, and newly created $\bot$-coordinates contribute no breakpoints.
    Hence, using the assumption $|\pi|\leq q$ for $\pi \in \Pi$, we have
    \[
        |\Mm_k^E| + |\Mm_k^S| = \sum_{\substack{(i, \pi) \in \hXx_k\\\pi\neq\bot}}(|\pi^E| + |\pi^S|) \leq qX_k \quad\implies\quad |\Ll_k^E| + |\Ll_k^S| \leq \frac{|\Mm_k^E| + |\Mm_k^S|}{4q} \leq \frac{X_k}{4}.
    \]

    Next, fix $R \in \{E, S\}$ and let $0 \leq \ell \leq m$.
    If $\Pp_k^R(\ell) = 0$, every fractional entry of $\hx^k$ contributes $0$ to $\Hh^R(\hx^k, \Pi, \ell)$.
    Since $z$ coincides with $\hx^k$ on all coordinates outside of $\hXx_k$, we must have $\Hh^R(z, \Pi, \ell) = \Hh^R(\hx^k, \Pi, \ell)$.

    If $\Pp_k^R(\ell) > 0$, let $\ell_k^R \in \Ll_k^R$ be the largest checkpoint preceding $\ell$, i.e satisfying $\ell_k^R \leq \ell$ (set $\ell_k^R = -1$ if none exists).
    The exact same reasoning as in \cref{etacheckpoints} gives
    \[
        \Hh^R(z, \Pi, \ell) - \Hh^R(\hx^k, \Pi, \ell) \leq \Pp_k^R(\ell) - \Pp_k^R(\ell_k^R).
    \]
    If $\ell_k^R = \ell$, $\Pp_k^R(\ell) - \Pp_k^R(\ell_k^R) = 0$ and we are done.
    Otherwise, by construction of $\Ll_k^R$ there are at most $4q$ breakpoints of $\Mm_k^R$ in $(\ell_k^R, \ell]$.
    Thus, $\Pp_k^R(\ell) - \Pp_k^R(\ell_k^R) \leq 4q$, and the lemma follows.
\end{proof}
It follows from \cref{ktransfer} and \cref{headcheckpoints} that, in order to preserve all load constraints at every prefix $0 \leq \ell \leq m$ at phase $k$, it suffices for each $R\in \{E, S\}$ to preserve the constraints on $\Hh^R$ only at head checkpoints $\Ll_k^R$.

Similar to \cref{etarounding}, we construct a polytope (P):
    \begin{equation}
        \begin{aligned}
        & && \sum_{\pi \in \Pi}z_i(\pi) =1,&\forall i \in N,\\
        & && z_i(\pi) = \hx^{k}_i(\pi),&\forall (i, \pi): \hx_i^k(\pi) \in \{0, 1\},\\
        & && \Hh^E(z, \Pi, \ell_k^E) \leq \Hh^E(\hx^k, \Pi, \ell_k^E),&\forall \ell_k^E \in \Ll_k^E,\\
        & && \Hh^S(z, \Pi, \ell_k^S) \leq \Hh^S(\hx^k, \Pi, \ell_k^S),&\forall \ell_k^S \in \Ll_k^S,\\
        & && 0 \leq z_i(\pi) \leq 1,&\forall i \in N, \pi \in \Pi.
    \end{aligned}\tag{P}\label{P}
    \end{equation}
    Clearly, $z = \hx^k$ is feasible for (P).
    We choose an extreme point $z^*$ of (P) maximizing
    \[
        \Ii(z^*, \Pi) = \sum_{i \in N}\sum_{\pi \neq \bot}z_i^*(\pi) .
    \]
    It exists, since (P) is nonempty and compact.
    By maximality, $\Ii(z^*, \Pi) \geq \Ii(\hx^k, \Pi)$.

    \begin{lemma}\label{zextreme}
        Let $z^*$ be an extreme point of the polytope $(\mathrm{P})$. Then, the number of strictly fractional entries in $z^*$ is at most $X_k/2$.
    \end{lemma}
    \begin{proof}
        The proof is completely analogous to the proof of \cref{etaextreme}.
        If $Z^*$ is the total number of strictly fractional coordinates in $z^*$, then there can be at most $Z^*/2$ different agents $i \in N$ with fractional $z_i^*(\pi)$ for some $\pi \in \Pi$.

        Suppose that $Z^* > X_k/2$ and restrict the equation-matrix of (P) to the fractional entries of $z^*$.
        There are $Z^*$ variables, at most $Z^*/2$ constraints of the form $\sum_{\pi \in \Pi}z_i(\pi) = 1$ for some $i \in N$ and at most $|\Ll_k^E| + |\Ll_k^S|\leq X_k/4$  load constraints.
        Hence, the number of constraints in this restriction is
        \[
            \frac{Z^*}{2} + \frac{X_k}{4} < Z^*,
        \]
        where we used $Z^* >X_k/2$.
        As in \cref{etaextreme}, this contradicts extremality of $z^*$.
    \end{proof}
    We take this extreme point $z^*$ of (P) and set $x^{k + 1} = z^*$.
    It follows from the lemmas above that $X_{k + 1}\leq X_k/2$, $\Ii(x^{k + 1}, \Pi) \geq \Ii(x^k, \Pi) - 8q$, as well as for each $R \in \{E, S\}$, for every $0 \leq \ell \leq m$ that $ \Hh^R(x^{k + 1}, \Pi, \ell) \leq \Hh^R(x^k, \Pi, \ell)$.    

    We repeat the process until at some phase $K + 1$ we have $X_{K + 1} = 0$.
    For this $K$,
    \[\begin{aligned}
        &&& \Ii(x^{K + 1}, \Pi) \geq \Ii(x^1, \Pi) - 8qK,\\\
        & && \Hh^E(x^{K + 1}, \Pi, \ell) \leq \Hh^E(x^1, \Pi, \ell),  & \forall 0 \leq \ell \leq m,\\
        & && \Hh^S(x^{K + 1}, \Pi, \ell) \leq \Hh^S(x^1, \Pi, \ell),  & \forall 0 \leq \ell \leq m.
    \end{aligned}\]
    Since the last phase $K$ starts with 
    \[
        2 \leq X_K \leq X\cdot 2^{-(K-1)},
    \]
    we have $K \leq \log X$.
    Take $y := x^{K + 1}$, solution $y$ is integral, and $\Ii(y, \Pi) \geq \Ii(x, \Pi) - 8q\log X$.
    By construction, agents $i \in N$ with $y_i(\pi) = 1$ for some $\pi \neq \bot$ must have had $x_i(\pi) > 0$.
\end{proof}
\section{Proofs of supplier-related theorems}

\label{appendix:supplier}

\begin{theorem}[\cref{supplierchange} restated]\label{supplierchangeproof}
    For every pair of integers $t, Q, q \geq 1$, there exists a global constant $C_{t,Q,q} \geq 1$, depending only on $t, Q, q$, such that the following holds.

    Let $I\subseteq N$ and $\{\pi_i\}_{i \in I}$ be obtained from \cref{ESrounding}.
    For every $i \in I$, fix an $S$-certificate $\{S_i\}_{i \in I}$, and let the number of protected $E$-prefixes be $|\Ll_S| = t$.
    Fix a supplier $u_*$, and for every agent $i \in N$ a corresponding suggested profile $\hpi_{u_*,i} = \hpi_{u_*,i}^E \uplus \hpi_{u_*,i}^S$ for a partition $[h] = E\sqcup S$ (where we assume $h = m$ unless specified).
    
    Suppose for every $i \in N$, $|\hpi_{u_*,i}|\leq q$, every $S$-demand $\alpha \in \hpi_{u_*,i}^S$ belongs to $\Qq$, $|\Qq| = Q$, and 
    \begin{align}
    &&&\sum_{i \in N}\Pp_i(\hpi_{u_*,i}^E, c_s) \leq |E\cap [b_s]|,&\forall (b_s, c_s) \in \Ll_S;\tag{$\mathrm{E}_{\Ll_S}$-$u$}\label{ESuend}\\
    &&& \sum_{i\in N}\Pp_{u_*}(\hpi_{u_*,i}^S, m_{u_*}(\alpha)) \leq |S\cap [m_{u_*}(\alpha)]|, & \forall \alpha \in \Qq.\tag{S$*$-$u$}\label{Suend}
    \end{align}

    Then there exists $A\subseteq I$ such that $|N\setminus A|\leq C_{t,Q,q}(|N| - |I|)$ and
    \begin{align}
    &&&\sum_{i \in A}\Pp_i(\pi_i^E, c_s) + \sum_{i \in N\setminus A}\Pp_i(\hpi_{u_*,i}^E, c_s) \leq |E\cap [b_s]|,&\forall (b_s, c_s) \in \Ll_S;\tag{$\mathrm{E}_{\Ll_S}$}\label{ESend}\\
    &&& \sum_{i \in A}|S_i\cap [m_{u_*}(\alpha)]| + \sum_{i\in N\setminus A}\Pp_{u_*}(\hpi_{u_*,i}^S, m_{u_*}(\alpha)) \leq |S\cap [m_{u_*}(\alpha)]|, & \forall \alpha \in \Qq.\tag{S$*$}\label{Send}
\end{align}
    In particular,
    \[
        \sum_{i \in N\setminus A}|\hpi_{u_*,i}|\leq qC_{t,Q,q}(|N| - |I|),
    \]
\end{theorem}
\begin{proof}
    The key idea to proving this theorem is that, even though the number of agents $d = |N|$ and their breakpoints are quite large, the actual number of possible \textbf{load configurations} with respect to constraints (\ref{ESend}) and (\ref{Send}) is significantly smaller, i.e only a constant number.

    Write $\Ll_S = \{(b_s, c_s) :\; s \in [t]\}$ in increasing order, and enumerate the high demands as $\Qq = \{\alpha_1 > \ldots > \alpha_Q\}$, so $m_{u_*}(\alpha_1) \leq \ldots \leq m_{u_*}(\alpha_Q)$.
    For each $i \in I$, his \textbf{old load configuration} is a vector $p_i$ of length $t + Q$ of the following form:
    \[
        p_i := \big\{\Pp_i(\pi_i^E, c_1), \ldots, \Pp_i(\pi_i^E, c_t), |S_i\cap [m_{u_*}(\alpha_1)]|, \ldots, |S_i\cap [m_{u_*}(\alpha_Q)]|\big\}.
    \]
    Similarly, for every $i \in N$, define the \textbf{suggested load configuration} as 
    \[
        p_{u_*,i} := \big\{\Pp_i(\hpi_{u_*,i}^E, c_1), \ldots, \Pp_i(\hpi_{u_*,i}^E, c_t), \Pp_{u_*}(\hpi_{u_*,i}^S, m_{u_*}(\alpha_1)), \ldots, \Pp_{u_*}(\hpi_{u_*,i}^S, m_{u_*}(\alpha_Q))\big\}.
    \]
    The first $t$ coordinates record loads at the protected $E$-prefixes, and the other $Q$ coordinates record usage of items from $S$: for an old profile $\pi_i$, we count items occupied by certificate $S_i$, for a suggested profile $\hpi_{u_*,i}$ we count its $S$-demands at supplier's breakpoints.
    These are exactly the contributions appearing at (\ref{ESend}) and (\ref{Send}).

    We will refer to the first $t$ coordinates as the $E$-block, and the last $Q$ coordinates as the $S$-block.
    Note that the coordinates within each block are non-decreasing in every load configuration.
    Moreover, the sum of the two last coordinates of these blocks is at most $q$.
    Indeed,
    \[
        \Pp_i(\pi_i^E, c_t) + |S_i\cap [m_{u_*}(\alpha_Q)]| \leq |\pi_i^E| + |\pi_i^S| = |\pi_i|\leq q
    \]
    for an old configuration $p_i$ of agent $i \in I$, and
    \[
        \Pp_i(\hpi_{u_*,i}^E, c_t) + \Pp_{u_*}(\hpi_{u_*,i}^S, m_{u_*}(\alpha_Q)) \leq |\hpi_{u_*,i}^E| + |\hpi_{u_*,i}^S| \leq |\hpi_{u_*,i}|\leq q
    \]
    for a suggested configuration $p_{u_*,i}$ of agent $i \in N$.
    So, all configurations belong to the finite set
    \[
        P :=
        \left\{ p \in \Z_{\geq 0}^{t+Q}:\; \begin{array}{l}
        p_1 \leq \ldots\leq p_t;\\
        p_{t + 1} \leq \ldots \leq p_{t + Q};\\
        p_t + p_{t + Q} \leq q;
        \end{array}
        \right\}.
    \]
    The set $P$ depends only on $t, Q, q$.
    In particular, it does not depend on the number of agents $N$ nor on the locations of the protected prefixes.
    We will also define the \textbf{load capacity} vector
    \[
        \Cc := \big\{|E\cap [b_1],\ldots, |E\cap[b_t]|, |S\cap [m_{u_*}(\alpha_1)]|, \ldots, |S\cap [m_{u_*}(\alpha_Q)]|\big\}.
    \]
    
    For disjoint sets $A\subseteq I$ and $J\subseteq N$, we consider the \textbf{cumulative load} vector
    \[
        \Pp(A, J) := \sum_{i \in A}p_i + \sum_{j \in J}p_{u_*,j}.
    \]
    So, $A$ consists of agents $i$ that keep their old profile $\pi_i$, and $J$ consists of agents $j$ that will use their suggested profile $\hpi_{u_*,j}$.
    We will call  a pair $(A, J)$ feasible, if $\Pp(A, J) \leq \Cc$, where inequality is component-wise. 
    \cref{ESrounding} as well as disjointness of certificates $S_i$ for $i \in I$ guarantee us
    \[
        \Pp(I, \varnothing) \leq \Cc,
    \]
    while by assumption of the theorem suggested profiles $\{\hpi_{u_*,j}\}_{j \in N}$ satisfy
    \[
        \Pp(\varnothing, N)\leq \Cc.
    \]
    So, we have two feasible pairs, the initial $(I, \varnothing)$ and the suggested $(\varnothing, N)$.
    We will move from the initial pair $(I, \varnothing)$ towards $(\varnothing, N)$, getting $(A, J)$ such that $A\sqcup J = N$, while swapping only a constant number of profiles of agents $I$ for every initially unassigned agent $j \in N\setminus I$.

    We do this iteratively.
    At the beginning of iteration $k \geq 1$, we have disjoint sets $A^k\subseteq I$ and $J^k\subseteq N$, satisfying $\Pp(A^k, J^k) \leq \Cc$.
    Initially, $A^1 = I$ and $J^1 = \varnothing$.
    Note that while $A^k$ contains only the initially assigned agents from $I$, set $J^k$ can contain both the unassigned agents $j \in N\setminus I$ that received a new profile from the supplier, as well as assigned agents $i \in I$ that swapped their old profile $\pi_i$ to the supplier's suggestion $\hpi_{u_*,i}$.
    Thus, at every iteration $I\subseteq A^k\sqcup J^k$, and every agent outside of $A^k\sqcup J^k$ has not been previously assigned.

    Suppose that $A^k \sqcup J^k \neq N$.
    We choose some unassigned agent $j \in N\setminus (A^k \sqcup J^k)$.
    We would like to give agent $j$ profile $\hpi_{u_*,j}$ and include $j$ into the next feasible pair, possibly swapping some of the old profiles to the corresponding suggestions.
    Note that a feasible replacement always exists: swap the profile for every agent in $A^k$ and add $j$, then
    \[
        \Pp(\varnothing, J^k \sqcup A^k \sqcup \{j\}) \leq \Pp(\varnothing, N) \leq \Cc.
    \]
    However, such replacement may swap too many profiles.
    We need one of only a constant size.
    \begin{lemma}\label{dicksonchoice}
        For a global constant $C_{t,Q,q}$, there exists $A'\subseteq A^k$ such that $|A'| + 1 \leq C_{t,Q,q}$ and
        \[
            \Pp(A^k\setminus A', J^k \sqcup A'\sqcup \{j\}) \leq \max\big\{\Pp(A^k, J^k), \Pp(\varnothing, J^k\sqcup A^k \sqcup \{j\})\big\} \leq \Cc,
        \]
        where maximum is taken component-wise.
    \end{lemma}
    \begin{proof}
        Replacing agent $i$'s old profile by a suggested one changes the total load by $p_{u_*,i} - p_i$.
        We will index each replacement by a pair $(p, p_*) \in P \times P$.
        Define a vector $x \in \Z_{\geq 0}^{P\times P}$, where coordinate $x_{p,p*}$ counts how many agents with old configuration $p$ and suggested configuration $p_*$ we switch.
        For an arbitrary suggested configuration $\sigma \in P$ that we would like to add, define
        \[
            z(\sigma, x) := \sigma + \sum_{(p, p_*) \in P\times P}x_{p,p_*}\cdot (p_* - p), \qquad\text{and}\qquad z_+(\sigma, x) := (z(\sigma, x))_+,
        \]
        where $(\cdot)_+ = \max\{0, \cdot\}$ is taken component-wise.
        Here, $z(\sigma, x)$ is the total load change after adding one unassigned agent with suggested configuration $\sigma$ and performing the replacement according to $x$.
        Vector $z_+(\sigma, x)$ records only positive increases in the load.
        The reason we keep track of the increases in load is because just making fewer exchanges does not necessarily preserve the load constraints, and controlling $z_+(\sigma, x)$ ensures that the new replacements are feasible.

        Now, consider an abstract set 
        \[
            \Xx_\sigma(t, Q, q) := \big\{(x, z_+(\sigma, x)): \; x \in \Z_{\geq 0}^{P\times P}\big\}.
        \]
        It is a subset of $\Z_{\geq 0}^{|P|^2 + t + Q}$, and depends only on $t, Q, q, \sigma$.
        At iteration $k$, we have a pair $(A^k, J^k)$ and would like to add an unassigned $j$ with suggested configuration $p_{u_*,j}$.
        With $\sigma = p_{u_*,j}$, this corresponds to taking some element $(x, z_+(\sigma, x)) \in \Xx_\sigma(t, Q, q)$ and swapping some old profiles of agents $A^k$ to suggested ones according to $x$, while increasing the total load by $z_+(\sigma, x)$.

        The  earlier ``full exchange'', i.e swapping profiles of every agent in $A^k$ and obtaining a pair $(\varnothing, J^k\sqcup A^k\sqcup \{j\})$, corresponds to the following element $(x^*, z_+(\sigma, x^*)) \in \Xx_\sigma(t, Q, q)$ with $\sigma = p_{u_*,j}$:
        \[
            \forall(p, p_*) \in P\times P:\quad x^*_{p,p_*} = |\{i\in A^k:\; p_i = p\text{ and } p_{u_*,i} = p_*\}|, 
        \]
        with load increase
        \[
            z(\sigma, x^*) = p_{u_*,j} + \sum_{i \in A^k}(p_{u_*,i} - p_i) = \Pp(\varnothing, J^k\sqcup A^k \sqcup \{j\}) - \Pp(A^k, J^k).
        \]
        Hence, with $z^*_+ := (\Pp(\varnothing, J^k\sqcup A^k \sqcup \{j\}) - \Pp(A^k, J^k))_+$, we have $(x^*, z^*_+) \in \Xx_\sigma(t, Q, q)$.
        Unfortunately, the replacement $(x^*, z^*_+)$ may swap too many profiles, i.e the sum 
        \[
            1 + \sum_{p, p_*}x^*_{p,p_*} = 1 + \sum_{i \in A^k}1 = 1 + |A^k|  
        \] 
        can be unbounded, so we need to choose a different replacement.
        Concretely, we need to find an element $(x, z_+(\sigma, x)) \in \Xx_\sigma(t, Q, q)$ that 1) $x$ matches our particular instance, i.e sets $A^k$ and $J^k$, 2) the added load $z_+(\sigma, x)$ fits into the $\Cc$-bound, and 3) the total number of performed swaps $\sum_{p,p_*}x_{p,p_*}$ is only a constant.
        Dickson's lemma says that such element necessarily must exist.
        \begin{lemma}[\cite{Dickson13, Cox15}]\label{dickson}
            For $K \geq 1$, every subset of $\Z^K_{\geq0}$ has finitely many component-wise minimal elements, and every element of the subset dominates some minimal element.
        \end{lemma}
        For completeness, a short proof of \cref{dickson} is given below (\cref{dicksonproof}).

        Since $\Xx_\sigma(t, Q, q)\subseteq \Z_{\geq 0}^{|P|^2 + t + Q}$, by \cref{dickson} set $\Xx_\sigma(t, Q, q)$ has finitely many component-wise minimal elements, let them be $\Ee_\sigma$.
        Since $\Xx_\sigma(t, Q, q)$ depends only on $t, Q, q, \sigma$, and every possible configuration $\sigma \in P$ by definition depends only on $t, Q, q$, then the number of minimal elements $|\Ee_\sigma|$, as well as the coordinates of every minimal element $(\hx, \hz_+) \in \Ee_\sigma$  and the sum $\sum_{p,p_*}\hx_{p,p_*}$, are also bounded by some function of $t, Q, q$ (otherwise we would obtain an infinite sequence of minimal elements).
        We take
        \[
            C_{t, Q, q} := \max_{\sigma \in P}\max_{(\hx, \hz_+) \in \Ee_\sigma}\left(1 + \sum_{p,p_*}\hx_{p,p_*}\right),
        \]
        which is a constant depending only on $t, Q, q$ (the additional $1$ accounts for the newly added agent).
        Then, for every minimal element $(\hx, \hz_+) \in \Ee_\sigma$ we have $1 + \sum_{p,p_*}\hx_{p,p_*} \leq C_{t,Q,q}$.

        Now, take the element $(x^*, z^*_+) \in \Xx_\sigma(t, Q, q)$ we considered earlier, corresponding to the replacement $(A^k, J^k) \to (\varnothing, J^k \sqcup A^k \sqcup \{j\})$.
        By \cref{dickson}, there exists a minimal element $(\hx, \hz_+) \in \Ee_\sigma\subseteq \Xx_\sigma(t, Q, q)$ such that $(\hx, \hz_+) \leq (x^*, z^*_+)$ component-wise, and $1 + \sum_{p,p_*}\hx_{p,p_*} \leq C_{t,Q,q}$.
        Since $\hx \leq x^*$, for every replacement pair $(p, p_*) \in P\times P$ we have $\hx_{p,p_*}\leq x^*_{p,p_*}$.
        In other words, if $\hx$ swaps an old profile with configuration $p$ to a suggested profile with configuration $p_*$, it is allowed to swap only profiles of agents from $A^k$ (as zeros stay zeros), and it cannot swap more profiles with that exchange configuration pair than the number of corresponding agents in $A^k$.
        Then, there exists a subset $A'\subseteq A^k$ containing exactly $\hx_{p,p_*}$ agents of $A^k$ with exchange configuration $(p, p_*)$, for every $(p, p_*) \in P\times P$, and
        \[
            1 + |A'| = 1 + \sum_{p,p_*}\hx_{p,p_*} \leq C_{t,Q,q}.
        \]
        At the same time, the total added load $\hz_+$ of this replacement satisfies
        \[
            \hz_+ \leq z^*_+  =(\Pp(\varnothing, J^k\sqcup A^k \sqcup \{j\}) - \Pp(A^k, J^k))_+.
        \]
        It follows that, after swapping the old profiles of agents $A'$ to the suggested ones, and adding the unassigned agent $j$, the total load satisfies
        \begin{multline*}
            \Pp(A^k \setminus A', J^k\sqcup A'\sqcup \{j\}) = \Pp(A^k, J^k) + z(\sigma, \hx) \leq \Pp(A^k, J^k) + \hz_+\\
            \leq \Pp(A^k, J^k) + z^*_+  = \Pp(A^k, J^k) + (\Pp(\varnothing, J^k\sqcup A^k \sqcup \{j\}) - \Pp(A^k, J^k))_+\\
            = \max\big\{\Pp(A^k, J^k), \Pp(\varnothing, J^k\sqcup A^k \sqcup \{j\})\big\} \leq \Cc.
        \end{multline*}
    \end{proof}

    Returning to the proof of \cref{supplierchangeproof}, we can now explicitly describe every iteration of our replacement procedure.
    At iteration $k$, starting with a pair $(A^k, J^k)$ such that $A^k\sqcup J^k\neq N$, we choose an unassigned agent $j \in N\setminus (A^k\sqcup J^k)$ and apply \cref{dicksonchoice}.
    This gives us a subset $A'\subseteq A^k$ such that $|A'| + 1 \leq C_{t,Q,q}$ and $\Pp(A^k \setminus A', J^k\sqcup A'\sqcup \{j\})\leq \Cc$.

    Define the new sets $A^{k + 1}:= A^k\setminus A'$ and $J^{k + 1} := J^k\sqcup A'\sqcup \{j\}$, these sets are disjoint.
    Every agent in $i \in A'$ swaps to a corresponding suggested profile $\hpi_{u_*,i}$, and the previously unassigned agent $j$ gets a profile $\hpi_{u_*,j}$.
    Since $A^{k + 1}\sqcup J^{k + 1} = A^k \sqcup J^k \sqcup \{j\}$, the number of agents that are assigned a profile increases by one, while $\Pp(A^{k + 1}, J^{k + 1})\leq \Cc$.
    At the same time, the number of agents that received a suggested profile during this iteration (either $i \in I$ swapping old $\pi_i$ to a suggested $\hpi_{u_*,i}$, or agent $j$ receiving $\hpi_{u_*,j}$) is at most $C_{t,Q,q}$, so $|J^{k + 1}| - |J^k|\leq C_{t,Q,q}$.

    After exactly $|N| - |I|$ iterations, every agent $i \in N$ is assigned some profile.
    For the final disjoint sets $A\subseteq I$ and $J\subseteq N$ it holds $A\sqcup J = N$ and 
    \[
        |N\setminus A| = |J| = \sum_{k = 1}^{|N| - |I|}(|J^{k + 1}| - |J^k|) \leq C_{t,Q,q}(|N| - |I|),\qquad\text{as well as}\qquad \Pp(A, J)\leq \Cc.
    \]
    Expanding the inequality $\Pp(A, J)\leq \Cc$ gives precisely conditions (\ref{ESend}) and (\ref{Send}).
\end{proof}

\begin{lemma}[Short proof of Dickson's lemma, \cite{Dickson13, Cox15}]\label{dicksonproof}
        For $K \geq 1$, every subset of $\Z^K_{\geq0}$ has finitely many component-wise minimal elements, and every element of the subset dominates some minimal element.
\end{lemma}
\begin{proof}
    Let $S\subseteq \Z^K_{\geq 0}$.
    Element $y\in S$ is minimal if there is no $x \in S-\{y\}$ satisfying $x \leq y$.

    It is well-known that every infinite sequence of non-negative integers has an infinite non-decreasing subsequence.
    Then, every infinite sequence $x^1, x^2, \ldots \in \Z^K_{\geq 0}$ has an infinite subsequence that is non-decreasing in every coordinate: choose an infinite subsequence non-decreasing in the first coordinate, then its infinite sub-sequence non-decreasing in the second coordinate, and so on, up until the $K$-th coordinate.
    This gives an infinite sub-sequence $x^{i_1}\leq x^{i_2} \leq \ldots \in \Z^K_{\geq 0}$.

    Suppose that $S$ has infinitely many component-wise minimal elements.
    Choose an infinite sequence of distinct minimal elements.
    Then, we can obtain its infinite sub-sequence non-decreasing in every coordinate.
    This subsequence then contains two distinct minimal elements $x, y \in S$ with $x \leq y$, which is not possible.
    Hence, $S$ must have finitely many minimal elements.

    Next, fix $x \in S$ and consider $S_x := \{y \in S : y \leq x\}$.
    This set is nonempty ($x \in S_x$) and finite, as every coordinate of $y \in S_x$ is an integer between $0$ and corresponding coordinate of $x$, and there are $K$ coordinates.
    Choose $y \in S_x$ minimizing $\sum_{k = 1}^Ky_k$.
    If some $z \in S - \{y\}$ satisfied $z \leq y$, then $z \leq x$ and $z \in S_x$.
    Furthermore, every coordinate of $z$ would be at most the corresponding coordinate of $y$, with at least one strict inequality.
    Hence, $\sum_{k = 1}^Kz_k <\sum_{k = 1}^Ky_k$, leading to a contradiction.
    Thus, $y$ is minimal in $S$, and $y \leq x$.
    The lemma follows.
\end{proof}

\section{Constant estimate}
\label{appendix:constant}

\begin{theorem}\label{constant1}
    For integers $t,Q,q\geq 1$, the constant $C_{t,Q,q}$ in \cref{supplierchangeproof} can be chosen as
    \[
        C_{t,Q,q} = \binom{t + Q + q}{q}^2(t + Q + 1)q^{t + Q}.
    \]
\end{theorem}
\begin{proof}
    Recall the set $P$ of all possible load configurations:
    \[
        P :=
        \left\{ p \in \Z_{\geq 0}^{t+Q}:\; \begin{array}{l}
        p_1 \leq \ldots\leq p_t;\\
        p_{t + 1} \leq \ldots \leq p_{t + Q};\\
        p_t + p_{t + Q} \leq q;
        \end{array}
        \right\}.
    \]
    Taking the differences of consecutive coordinates of $p \in P$ separately in each block gives us a list of $t + Q$ non-negative integers that sum up to at most $q$.
    Hence, using stars-and-bars, 
    \[
        |P| = \binom{t + Q + q}{q}.
    \]
    
    Recall that for a fixed configuration $\sigma \in P$, the set $\Xx_\sigma(t, Q, q)\subseteq  \Z_{\geq 0}^{|P|^2 +t + Q}$ contains all pairs $(x, z_+(\sigma, x))$ such that $x \in \Z_{\geq 0}^{P\times P}$ and
    \[
        z_+(\sigma, x) := (z(\sigma, x))_+\qquad \text{where}\qquad z(\sigma, x) := \sigma + \sum_{(p, p_*) \in P\times P}x_{p,p_*}\cdot (p_* - p).
    \]
    To prove the theorem, it suffices to show that every minimal element $(\hx, \hz_+)$ of $\Xx_\sigma(t, Q, q)$ satisfies
    \[
        1 + \sum_{(p, p_*)\in P\times P}\hx_{p,p_*} \leq \binom{t + Q + q}{q}^2(t + Q + 1)q^{t + Q}.
    \]
    While Dickson's lemma gives us that the left-hand side sum is bounded by a constant, it does not provide us the explicit bound we need.
    To get an explicit constant, we will reformulate the problem of finding a minimal element in $\Xx_\sigma(t, Q, q)$ as a problem of finding the \textit{smallest} basic feasible solution of a certain linear program.
    Then, minimality of $(\hx, \hz_+)$ would imply that the corresponding solution must have bounded norm, with a bound computed explicitly from the linear program itself.
    We will use the following result on indecomposable kernel vectors.
    \begin{theorem}[\cite{Onn2010}, Lemma 3.20]\label{onn}
    Let $M$ be a rank-$r$ integer matrix with $k$ columns.
    Define 
    \[
        \Delta(M) := \max\{|\det M'|:\; M'\text{ is a square submatrix of }M\}.
    \]
    Suppose some $v \in \Z_{\geq 0}^k$, $v\neq 0$ satisfies $Mv = 0$ and is indecomposable, meaning that if for some $u \in \Z_{\geq 0}^k$ such that $0 \leq u \leq v$ it also holds $Mu = 0$, then  $u = 0$ or $u = v$.
    Then
    \[
        \|v\|_1 \leq (k - r)(r + 1)\Delta(M).
    \]
\end{theorem}
We will define a certain matrix $M$ with $k \leq |P|^2 + t + Q$ columns and rank $r = t + Q$, such that $\Delta(M) \leq q^{t + Q}$.
Then, we will show that every minimal element of $\Xx_\sigma(t, Q, q)$ corresponds to some indecomposable kernel vector of $M$, which will allow us to use the bound of \cref{onn}.

Fix the minimal element $(\hx, \hz_+) \in \Xx_\sigma(t, Q, q)$.
First, observe that for every $p \in P$ we have $\hx_{p,p} = 0$, otherwise we can remove every identity replacement $p \to p$ while keeping $\hz_+$ unchanged, which gives us a smaller element.
Recall that $\hz_+ = (\hz)_+$, where
\[
    \hz = z(\sigma, \hx) = \sigma + \sum_{p\neq p_*}\hx_{p,p_*}\cdot (p_* - p).
\]
For every pair $p \neq p_* \in P$, define vectors $a_{p,p_*} = p_* - p$, and $a_0 = \sigma$, of length $t + Q$.
The vectors $a_{p,p_*}$ record the load changes caused by swapping an old configuration $p$ to a suggested $p_*$, and $a_0$ records the extra load caused by adding agent with $\sigma$.
In addition, for every $s \in [t + Q]$, let $e_s$ be the $s$-th standard basis vector, and define
\[
    b_s := \begin{cases}
        -e_s,&\hz_s > 0;\\
        e_s,&\hz_s \leq 0;
    \end{cases}\qquad \text{then}\qquad \sum_{s= 1}^{t + Q}|\hz_s|\cdot b_s = -\hz.
\]
These additional columns allow us to express the replacement as a kernel vector, balancing the total load sum to $0$ coordinate by coordinate.

Define the following matrix $M_0$ with $k = |P|(|P| - 1) + t + Q + 1$ columns, and vector $v$ as
\[
    M_0 = [\{a_{p,p_*}\}_{p\neq p_*}, a_0, b_1, \ldots, b_{t + Q}],\qquad \qquad v = [\{\hx_{p,p_*}\}_{p\neq p_*}, 1, |\hz_1|,\ldots, |\hz_{t + Q}|]^T.
\]
By construction, $v$ is non-zero and $M_0v = 0$, and the rank of $M_0$ is $r = t + Q$.
\begin{lemma}\label{indecom}
    Vector $v$ is indecomposable w.r.t $M_0$ (as per \cref{onn}).
\end{lemma}
\begin{proof}
    Suppose the contrary, i.e that $v = v' + v'' $, where both $v', v'' \neq 0$ are non-negative integer vectors satisfying $M_0v' = 0 = M_0v''$.
    We will show that this must contradict the minimality of $(\hx, \hz_+)$, by presenting a pair $(\wx, \wz_+) \in \Xx_\sigma(t, Q, q)$ strictly smaller than $(\hx, \hz_+)$.

    Note that the $a_0$-coordinate equals to $1$ in exactly one of $v', v''$.
    Without loss of generality,
    \[
        v'= (x', 1, y'), \qquad\text{and}\qquad v'' = (x'', 0, y'').
    \]
    Extend $x', x''$ with $0$ to $(p, p)$, $p \in P$.
    Now, $x'+ x'' = \hx$ and $y'+ y'' = (|\hz_1|, \ldots, |\hz_{t + Q}|)$.
    Define
    \[
        z':= \sigma + \sum_{p\neq p_*}x'_{p,p_*}\cdot (p_* - p).
    \]
    Writing $z'_+ = (z')_+$, we have, $(x', z'_+) \in \Xx_\sigma(t, Q, q)$.
    Since $M_0v'= 0$, then $z' = -\sum_{s = 1}^{t + Q}y_s'\cdot b_s$, therefore for every $s \in [t + Q]$,
    \[
        z'_s = \begin{cases}
            y'_s,& \hz_s > 0;\\
            -y'_s,& \hz_s \leq 0.
        \end{cases}
    \]
    Then, $0 \leq y'_s \leq |\hz_s|$ and $z'_+ \leq \hz_+$, thus $(x', z'_+) \leq (\hx, \hz_+)$.

    We claim that $x'' \neq 0$.
    Indeed, otherwise $v''$ would be non-zero only at coordinates corresponding to $b_1, \ldots, b_{t + Q}$.
    Since they are linearly independent, together with $M_0v'' = 0$ this would imply $v'' = 0$, contradicting the initial assumption.

    As a result, $x'= \hx - x''$ is strictly smaller than $\hx$ in at least one coordinate.
    So, $(x', z'_+)$ is an element of $\Xx_\sigma(t, Q, q)$ strictly smaller than $(\hx, \hz_+)$, contradicting the minimality of $(\hx, \hz_+)$. 
\end{proof}

We will now transform $M_0$ into a matrix with bounded submatrix determinants.
Define an invertable matrix $L$ so that for any vector $u$,
\[
    Lu = (u_1, u_2 - u_1, \ldots, u_{t} - u_{t - 1}, u_{t + 1}, u_{t + 2} - u_{t + 1}, \ldots, u_{t + Q} - u_{t + Q - 1}),
\]
and set $M = LM_0$.
The rank of $M$ is still $t + Q$, and $\ker M = \ker M_0$.
So, vector $v$ remains indecomposable w.r.t matrix $M$ as well.
\begin{lemma}\label{subdet}
    For any square submatrix $M'$ of $M$ it holds $|\det M'| \leq q^{t + Q}$.
\end{lemma}
\begin{proof}
Let $e_0 := 0 \in \R^{t + Q}$.
We will call a vector of the form $e_i - e_{i'}$ for $0 \leq i,i' \leq t + Q$ an \textit{elementary exchange}.
We will show that every column of matrix $M$ is a sum of $q$ elementary exchanges (allowing zeros), which will allow us to represent any $\det M'$ for submatrix $M'$ of $M$ as a sum of at most $q^{t + Q}$ determinants of totally-unimodular matrices.

Note that for every $p \in P$, vector $Lp$ has non-negative integer coordinates, and
\[
    \sum_{s = 1}^{t + Q}(Lp)_s = p_t + p_{t + Q} \leq q,
\]
so we can choose (possibly repeating) indices $i_1, \ldots, i_q \in \{0, 1, \ldots, t + Q\}$ such that
\[
    Lp = \sum_{j = 1}^qe_{i_j}.
\]
By definition, for every $(p, p_*) \in P\times P$ we have $La_{p,p_*} = Lp_* - Lp$.
Hence, representing
\[
    Lp =\sum_{j = 1}^qe_{i_j} \qquad \text{and}\qquad  Lp_* = \sum_{j = 1}^qe_{i_j^*},
\]
for some $\{i_1, \ldots, i_q\}$ and $\{i_1^*, \ldots, i_q^*\}$ respectively, we get
\[
    La_{p,p_*} =  \sum_{j = 1}^q(e_{i_j^*} - e_{i_j}),
\]
and $La_0 = L\sigma = \sum_{j = 1}^q(e_{i_j^\sigma} - e_0)$ for some $\{i_1^\sigma, \ldots, i_q^\sigma\}$.
Finally, for every $s \in [t + Q]$, $Lb_s = \pm Le_s$, and by definition of $L$,
\[
    Le_s = \begin{cases}
        e_s - e_{s + 1},& s \notin \{t, t + Q\},\\
        e_s - e_0 ,& s \in \{t, t + Q\}.
    \end{cases}
\]
Hence, $\pm Le_s$ also can be expressed as $q$ elementary exchanges (by adding zero exchanges).

Next, consider a collection of non-zero elementary exchanges $e_{i} - e_{i'}$.
Construct a directed graph on $\{0, 1, \ldots, t + Q\}$, where for every elementary exchange $e_i - e_{i'}$ we draw a directed edge $(i'\to i)$.
After deleting the row corresponding to vertex $0$, the directed incidence matrix of this graph has precisely $e_i - e_{i'}$ as its columns.
Restricting the elementary exchanges to a common set of coordinates amounts to deleting more matrix rows.
Since directed incidence matrices are totally-unimodular (see \cite{Schrijver2003}, Theorem 13.9), every square matrix obtained this way has determinant in $\{-1, 0, 1\}$.
The same holds if some column is zero.

Now, let $M'$ be a square submatrix of $M$ of size $\ell \leq t + Q$.
By the elementary-exchange decomposition above, for $i \in [\ell]$ columns of $M'$ can be written as 
\[
    c_i := \sum_{j = 1}^qc_{i,j},
\]
where each $c_{i,j}$ is a restriction of an elementary exchange to the rows of $M'$.
Thus,
\[
    \det M' = \sum_{(j_1, \ldots, j_\ell) \in [q]^\ell}\det[c_{1,j_1},\ldots , c_{\ell,j_\ell}].
\]
As discussed earlier, every summand $\det[c_{1,j_1},\ldots , c_{\ell,j_\ell}]$ belongs to $\{-1, 0, 1\}$.
Since there are $q^\ell$ summands, triangle inequality gives us $|\det M'| \leq q^\ell \leq q^{t + Q}$, as desired.
\end{proof}
Thus, $\Delta(M) \leq q^{t + Q}$, and applying \cref{onn} to $M$ and $v$, we get for $(\hx, \hz)$ that
\begin{multline*}
    1 + \sum_{(p, p_*) \in P\times P}\hx_{p,p_*}\leq \|v\|_1 \leq (k - r)(r + 1)\Delta(M)\\ \leq |P|^2(t + Q + 1)q^{t + Q} 
    =  \binom{t + Q + q}{q}^2(t + Q + 1)q^{t + Q}.
\end{multline*}
This bound holds for every $\sigma \in P$ and every minimal element of $\Xx_\sigma(t, Q, q)$.
\end{proof}

\begin{corollary}\label{blockpartitionconst}
    For integers $t , q\geq 1$, the constant $D_{t,q}$ in \cref{blockpartition} can be chosen as
    \[
        D_{t,q} = \left(\binom{t + q}{q} - 1\right)(t + 2)q^{t}.
    \]
\end{corollary}
\begin{proof}
    Recall that the set $\Bb(t, q)$ consists of repartition blocks $\beta = (x, y) \in \Z_{\geq 0}^P \times \Z^P_{\geq 0}$, where $P = \{p \in \Z_{\geq 0}^t : 1 \leq |p|\leq q\}$, $|P| = \binom{t + q}{q} - 1$, with $\beta$ satisfying 
    \[
    \sum_{p\in P}x(p)
    =
    \sum_{p\in  P}y(p)
    \geq 1,
    \qquad
    \sum_{p\in  P}x(p)\cdot p
    =
    \sum_{p\in  P}y(p)\cdot p.
    \]
    To prove the statement, it suffices that every minimal element $\hbeta = (\hx, \hy) \in \Bb(t, q)$ satisfies
    \[
        |\hbeta| = \sum_{p \in P}\hx(p) \leq |P|\cdot (t + 2)\cdot q^t.
    \]

    Fix a component-wise minimal element $\hbeta = (\hx, \hy) \in \Bb(t, q)$.
    For every $p \in P$, define the column $a_p := (1, p) \in \Z^{t + 1}_{\geq 0}$.
    Its first coordinate corresponds to the number of bundles in the block, and the remaining coordinates correspond to the number of items of each type.
    Let
    \[
        M := [\{a_p\}_{p \in P}, \{-a_p\}_{p \in P}],\quad \text{and}\quad v := (\hx, \hy)^T.
    \]
    By construction, $M$ has $k = 2|P|$ columns, rank $r \leq t + 1$, and it holds
    \[
        Mv = 0 \quad \text{and}\quad \|v\|_1 = 2|\hbeta|.
    \]
    We will first show that $v$ is indecomposable w.r.t $M$, in the sense of \cref{onn}.
    Suppose that $u = (x', y') \in \Z_{\geq 0}^P \times \Z_{\geq 0}^P$ satisfies $0 \neq u \leq v$ and $Mu = 0$.
    Then, the first row implies
    \[
    \sum_{p\in P}x'(p)
    =
    \sum_{p\in P}y'(p).
    \]
    Note that this sum is at least $1$, as otherwise $u = 0$.
    The remaining rows of $Mu = 0$ imply 
    \[
    \sum_{p\in P}x'(p)\cdot p
    =
    \sum_{p\in P}y'(p)\cdot p.
    \]
    Thus, $(x', y') \in \Bb(t, q)$, while $(x', y') \leq (\hx, \hy)$, which contradicts the minimality of $(\hx, \hy)$.
    Thus, it can only be that $u = v$, i.e $v$ is indecomposable.

    We next show that $\Delta(M) \leq q^t$, similarly to \cref{subdet}, we will use the fact that determinant is multilinear in order to reduce it to a simple collection of columns.

    Let $e_1, \ldots, e_t$ be the standard basis vectors, and define auxiliary columns
    \[
        b_0 := (1, 0)\qquad \text{and}\qquad \forall s \in [t]:\; b_s = (1, q\cdot e_s).
    \]
    Since $|p|\leq q$, we can represent each column $a_p$ as 
    \[
        a_p = \left(1 - \frac{|p|}{q}\right)b_0 + \sum_{s = 1}^t\frac{p_s}{q}\cdot b_s.
    \]
    Note that all the coefficients before $b_0, b_1, \ldots, b_t$ are non-negative and sum up to $1$, so it is a convex combination.

    Now, consider an arbitrary square submatrix $M'$ of $M$ of size $\ell$, without loss of generality we can assume that its columns are restrictions of columns $a_p$.
    Then, substituting the expressions for $a_p$ above and expanding the determinant, we can express $\det M'$ as a convex combination of determinants whose columns are restrictions of $b_0, b_1, \ldots, b_t$ to the exact same set of rows.
    Thus, to bound $\det M'$, it suffices to bound the determinats w.r.t auxiliary columns.

    First, consider the case that the first row of $M$ is absent.
    Then, each restricted column is either $0$ or a $q$-multiple of the standard basis vector.
    It follows that the determinant is either $0$ or $\pm q\ell$, the absolute value of which is at most $q^t$.

    Next, assume that the first row of $M$ is present.
    We move it to the top and divide every other row by $q$.
    We obtain the set of $\ell$ columns of the form 
    \[
        (1, 0), \quad (1, e_1), \quad\ldots\quad (1, e_{\ell - 1}).
    \]
    If two of these columns coincide, the determinant is $0$.
    Otherwise, all $\ell$ types of columns are present.
    Hence, by subtracting $(1, 0)$ column from every other gives a matrix with determinant $1$.
    Thus, the absolute value of the original submatrix determinant was $q^{\ell - 1} \leq q^t$.

    It follows that $|\det M'|\leq q^t$ as a convex combination, which proves $\Delta(M) \leq q^t$.
    Applying \cref{onn} to $M$ and $v$ gives
    \[
        2|\hbeta| = \|v\|_1 \leq (k - r)(r + 1)\Delta(M) \leq 2|P|(t + 2)q^t.
    \]
    Dividing by $2$ we obtain the desired bound.
\end{proof}

\begin{remark}\label{constremark}
    In \cref{sec:suggested}, we have a constant $D_{t + 1,q}$ for $t \leq t_0 := \lceil Q\cdot \frac{q}{q - 1} \rceil + 1$.
    Let $s := t_0 + Q$, then our choice of constant $C_q$ from \cref{sec:supplier} (using \cref{constant1}) is
    \[
        C_q = \binom{s + q}{q}^2(s + 1)q^{s}.
    \]
    Since $Q \geq 2$ in our setting, $t + 1 \leq s - 1$ and $t + 3 \leq s +1$.
    As a result, \cref{blockpartitionconst} gives
    \[
        D_{t+ 1,q} \leq \left(\binom{t + 1 + q}{q} - 1\right)(t + 3)q^{t + 1} \leq \binom{s + q}{q}(s + 1)q^{s} \leq C_q.
    \]
\end{remark}

\begin{theorem}\label{finalconstant}
    For integers $\Gamma \geq 1$ and $q \geq 9$, the value $n(\Gamma, q)$ in \cref{qmms} can be taken
    \[
        n(\Gamma, q) := 2^{4q^2\log^2(2q)}(\log(2\Gamma))^{2q}.
    \]
\end{theorem}
\begin{proof}
    Recall the exact definitions
    \[
        Q = 1 + \left\lceil \frac{\ln q}{-\ln(1 - 1/q)}\right\rceil,\qquad\qquad t_0 = \left\lceil Q \cdot \frac{q}{q - 1}\right\rceil + 1.
    \]
    It will be convenient for us to use the following conventions:
    \[
    s:=t_0+Q,
    \qquad
    \ell:=\log(2q),
    \qquad
    H:=2^{q\ell^2}.
    \]
    Since \(q\ge9\), we have \(\ell\ge4\).
    Using \cref{constant1}, take 
    \[
C_q=\binom{s+q}{q}^{\!2}(s+1)q^s,
\]
 and select $D:= D_{t+1, q}$ according to \cref{blockpartitionconst}.
 By \cref{constremark}, $D \leq C_q$ for $t \leq t_0$.

    First, we claim that $qC_q \leq H^{3/2}$.
    Indeed, convexity implies
    \[
        -\ln(1 - 1/q) \geq \frac{1}{q - 1/2 }\quad \implies \quad Q \leq 2 + \left(q - \frac{1}{2}\right)\ln q,
    \]
    and by definition of $t_0$ and $s$,
    \[
s
\le
2q\ln q+6+\frac{\ln q+4}{2(q-1)}
\le2q\ln q+7.
\]
For $q \geq 9$, we have $\ln q \leq q/4$, hence $s \leq (7/5)q(\ell - 1) + 7$, implying
\[
    s + q \leq \frac{q^2}{2} + q + 7 \leq \frac{7}{10}q^2.
\]
Using standard binomial estimates, we have 
\[
    \binom{s + q}{q} \leq \left(\frac{e(s + q)}{q}\right)^q \leq (2q)^q.
\]
Since $s + 1 \leq q^2$, we get
\[
    \log(qC_q) = (s + 1)\log q + \log(s + 1) + 2\log\binom{s + q}{q} \leq \frac{7}{5}q(\ell-1)^2 + 2q\ell + 10(\ell - 1).
\]
    Diving by $q$ and using $q \geq 9$ yields
    \[
        \frac{\log(qC_q)}{q} \leq \frac{7}{5}\ell^2 + \frac{14}{45}\ell + \frac{13}{45} \leq \frac{3}{2}\ell^2,
    \]
    where we used $\ell \geq 4$. 
    This proves $qC_q \leq H^{3/2}$.

    With our choices of parameters, we have
    \[
        Q, t_0 \leq q^2, \quad D\leq \frac{H^{3/2}}{q}, \quad |P|\leq H^{1/4}, \quad |\Phi|\leq \phi \leq H^{1/2}, \quad (2Q)^q \leq H^{1/2}.
    \]
    Indeed, by definition of $P$ and $\phi$ from \cref{sec:suggested}, and using $s \leq 2q\ell$
    \[
|P|
\le\binom{s+q}{q}
\le2^{q\ell}
\le H^{1/4},\qquad\text{and}\qquad \phi = \binom{t_0 - 1 + Q}{Q} \leq 2^{s - 1} \leq H^{1/2}.
\]
Finally, 
\[
(2Q)^q\le(2q^2)^q
=2^{q(2\ell-1)}
\le H^{1/2}.
\]

Now, choose the parameters of \cref{gensuplargen}:
\[
\varepsilon=\frac1{2q^2C_qt_0},
\qquad
\nu=
\left\lceil
2q^2C_q\bigl(\theta+\rho_{ES}(q, dn)\bigr)
\right\rceil.
\]
We need to assure that
\begin{itemize}
    \item for head-poor agents allocation \cref{notpoor}, 
    \[
n\ge
\max\left\{
q^2(16\Lambda+1),\
8q^4\Lambda(16\Lambda+1)+2q
\right\};
\]
\item for  supplier selection \cref{genchoice},
\[
n\ge2\phi(2QL)^q;
\]
\item for suggested profile completion \cref{fullsuggested},
\[
n\ge
4\phi q^3D
\bigl(
\theta+|\mathcal P|+|\Phi|-1+2(L+|F|)
\bigr).
\]
\end{itemize}
Note that the condition $n\geq 2q^2|S\cup F|$ from \cref{sec:suggested} follows from the latter, as $|S\cup F|\leq L + |F|$.

Define 
\[
Y:=1+\log\bigl((\Gamma+1)(n+1)\bigr).
\]
From the bound on \(\Lambda\) in \cref{notpoor} and the definition of \(\rho\),
\[
\Lambda\le2\rho(\Gamma n,(\Gamma+1)n)\le32Y^2.
\]
In addition,
\[
\rho_{ES}(q,dn)\le16q\log n,
\qquad
\theta=q^2(16\Lambda+1)+q\Lambda,
\]
which implies
\[
    \theta+\rho_{ES}(q,dn) \leq \bigl(32(16q^2+q)+q^2+16q\bigr)Y^2 \leq 2^{10}q^2Y^2.
\]
By definition of $\nu$,
\[
\nu\le2^{12}q^4C_qY^2.
\]
Now, recall that in our construction,
\[
|S|\le t_0\left(1+\frac{Q\nu}{\varepsilon}\right),
\qquad
|F|\le t_0\nu,
\qquad
L=|S|+\left\lceil q^2(16\Lambda+1)\right\rceil.
\]
Substituting the estimates above gives
\[
    L+|F| \le t_0\left(1+\frac{Q\nu}{\varepsilon}\right)
+t_0\nu+q^2(16\Lambda+1)+1 
\le 2^{15}q^6Qt_0^2C_q^2Y^2 = 2^{15}q^4Qt_0^2(qC_q)^2Y^2.
\]
Next, since \(Q,t_0\le q^2\),  and $q\ell \geq 36$,
\[
2^{15}q^4Qt_0^2
\le2^{15}q^{10}
\le(2q)^{12}
\le H^{1/2} \qquad \implies\qquad L + |F| \leq H^{7/2} Y^2.
\]

We can now bound all requirements on $n$ directly.
First, for \cref{poor},
\[
\max\left\{
q^2(16\Lambda+1),\
8q^4\Lambda(16\Lambda+1)+2q
\right\}
\le2^{18}q^4Y^4
\le HY^4.
\]
Next, substituting the earlier estimates, and observing $\theta \leq HY^2$, gives
\[
\theta+| P|+|\Phi|-1+2(L+|F|)
\le5H^{7/2}Y^2
\le H^4Y^2.
\]
Therefore,  in \cref{fullsuggested}
\[
    4\phi q^3D
\bigl(
\theta+|\mathcal P|+|\Phi|-1+2(L+|F|)
\bigr) \leq 
4H^{1/2}q^3\frac{H^{3/2}}q\,H^4Y^2
=
4q^2H^6Y^2
\le H^7Y^2.
\]
Finally, for \cref{gensuplargen} we get
\[
    2\phi(2QL)^q \le 
2H^{1/2}H^{1/2}
\bigl(H^{7/2}Y^2\bigr)^q \le  H^{(7/2)q+2}Y^{2q}.
\]
Because $q \geq 9$ and $Y \geq 1$, it suffices for $n$ to satisfy only the last requirement.
Thus, we need
\[
n\ge H^{(7/2)q+2}
\left(1+\log((\Gamma+1)(n+1))\right)^{2q}.
\]

Now, define
\[
b:=\log(2\Gamma)\ge1,
\qquad
n_0:=2^{4q^2\ell^2}b^{2q}.
\]
We will show that every $n \geq n_0$ satisfies the inequality above.
At \(n=n_0\), using \(\Gamma+1\le2\Gamma\) and \(n_0+1\le2n_0\), substituting this value into $Y = Y(n)$ gives
\[
    Y(n_0)\le2+b+4q^2\ell^2+2q\log b\le b\bigl(4q^2\ell^2+2q+3\bigr)\le8q^2\ell^2b.
\]
For \(\ell\ge4\), we have \(\log\ell\le\ell/2\), and therefore
\[
\log(8q^2\ell^2)
=1+2\ell+2\log\ell
\le1+3\ell
\le\frac72\ell,
\]
which implies for $q \geq 9$ and $\ell \geq 4$ that
\[
    \log\left(
H^{(7/2)q+2}Y(n_0)^{2q}
\right)\le
\left(\frac72q^2+2q\right)\ell^2
+7q\ell+2q\log b \le
4q^2\ell^2+2q\log b= \log n_0,
\]
as desired.
Finally, the function 
\[
\frac{n}{Y(n)^{2q}}
\]
is increasing whenever \(Y(n)>2q/\ln2\).
For every $n \geq n_0$, we have 
\[
Y(n)\ge4q^2\ell^2>\frac{2q}{\ln2}.
\]
Hence, the bound holds for every $n \geq n_0$, which finishes the proof.
\end{proof}

\end{document}